\documentclass[12pt,a4paper,dvipsnames,BCOR=5mm,version=3.19a,toc=listof,parskip=never]{scrbook}
\setkomafont{disposition}{\normalcolor\bfseries}
\setkomafont{descriptionlabel}{\normalcolor\bfseries}
\usepackage[utf8]{inputenc}
\usepackage[T1]{fontenc}

\usepackage{lmodern}

\usepackage{amsmath,amssymb,url}
\usepackage{mathtools}
\usepackage[breakable, skins]{tcolorbox}
\usepackage{graphicx}
\usepackage{tikz}
\usepackage{tikz-feynman} 
\tikzfeynmanset{compat=1.0.0}
\usetikzlibrary{arrows,arrows.meta,matrix,fpu}
\usepackage{tabularx}
\usepackage{booktabs}
\usepackage{array}
\usepackage{rotating}
\usepackage{xspace}
\usepackage{units}
\usepackage{xcolor}
\definecolor{CBSkyBlue}{RGB}{86,180,233}
\definecolor{CBRedPurple}{RGB}{204,121,167}
\definecolor{CBOrange}{RGB}{230,159,0}
\definecolor{CBBlueishGreen}{RGB}{0,158,115}
\usepackage{slashed}
\usepackage{acronym}
\usepackage{enumitem}
\usepackage{microtype}
\usepackage{listings}
\usepackage{pdflscape}
\usepackage[singlespacing]{setspace}

\usepackage[caption=false,subrefformat=simple,labelformat=simple,listofformat=subsimple]{subfig}

\numberwithin{equation}{section}

\usepackage{multirow}
\usepackage[super]{nth}
\usepackage[titletoc]{appendix}

\usepackage[section]{placeins}     

\usepackage{scrhack}
\usepackage{setspace}

\usepackage{hyperref}

\hypersetup{%
	colorlinks,
	citecolor=black,
	filecolor=black,
	linkcolor=black,
	urlcolor=black
}
\usepackage{bookmark}
\usepackage{cleveref}
\usepackage[acronym,toc]{glossaries-extra}
\usepackage{glossary-mcols}

\usepackage[babel]{csquotes}
\usepackage[style=numeric,
maxnames=3,
minnames=3,
backend=biber,
citestyle=numeric-comp,
url=false,
maxbibnames=6,
sorting=none,
doi=false,
isbn=false
]{biblatex}

\newcommand*\circled[1]{\tikz[baseline=(char.base)]{
		\node[shape=circle,draw,inner sep=0.1pt] (char) {#1};}}
\DeclareFieldFormat{journaltitle}{\mkbibemph{#1},}
\DeclareFieldFormat[inbook,thesis]{title}{{#1}\addperiod}
\DefineBibliographyStrings{english}{%
	backrefpage  = {see p.}, 
	backrefpages = {see pp.} 
}
\DeclareFieldFormat[article]{volume}{\textbf{#1}\space}
\AtEveryBibitem{\clearfield{month}}

\renewbibmacro{in:}{}

\usepackage{pgf}
\usepackage{pgfkeys}
\addtocontents{toc}{\protect\thispagestyle{empty}}
\addtocontents{toc}{\protect\pagestyle{empty}}
\begin{document}
	\title{Automated NLO Electroweak Corrections to Processes at Hadron and Lepton Colliders}
	\subtitle{Dissertation zur Erlangung des Doktorgrades an der Fakult\"at f\"ur Mathematik, Informatik und Naturwissenschaften\\Fachbereich Physik\\der Universit\"at Hamburg}
	\date{}
	
	\author{vorgelegt von\\Pia Mareen Bredt}
	\publishers{Hamburg\\2022}
	\maketitle
	\chapter*{}
	\thispagestyle{empty}
	\vspace{9cm}
	\begin{minipage}[t]{0.57\textwidth}
		Gutachter/innen der Dissertation:\\
		\quad\\
		\quad\\
		Zusammensetzung der Pr\"ufungskommission: \\
		\quad\\
		\quad\\
		\quad\\
		\quad\\
		\quad\\
		Vorsitzender der Pr\"ufungskommission: \\
		\quad\\
		Datum der Disputation: \\
		\quad\\
		Vorsitzender des Fach-Promotionsausschusses\\ PHYSIK: \\
		\quad\\
		Leiter des Fachbereichs PHYSIK:\\
		\quad\\
		Dekan der Fakult\"at MIN:
	\end{minipage}
	\begin{minipage}[t]{0.43\textwidth}
		Dr.~J\"urgen~Reuter\\
		Prof.~Dr.~Gudrid Moortgat-Pick\\
		\quad\\
		Dr.~J\"urgen~Reuter\\
		Prof.~Dr.~Gudrid~Moortgat-Pick\\
		Prof.~Dr.~Sven-Olaf~Moch\\
		Dr.~Markus~Diehl\\
		Prof.~Dr.~Elisabetta~Gallo\\
		\quad\\
		Prof.~Dr.~Sven-Olaf~Moch\\
		\quad\\
		21.~11.~2022\\
		\quad\\
		\quad\\
		Prof. Dr. Wolfgang J. Parak\\
		\quad\\
		Prof. Dr. G\"unter H. W. Sigl\\
		\quad\\
		Prof. Dr.-Ing. Norbert Ritter
	\end{minipage}
	\chapter*{Abstract}
	\thispagestyle{empty}
	In order to search for new physics at collider experiments it is necessary for Monte-Carlo (MC) event generators to simulate Standard Model (SM) physics to the highest possible level of precision. The current precision frontier for predictions from the SM is at an unprecedented accuracy level for observables at the Large Hadron Collider (LHC): Perturbative corrections to at least next-to-next-to-leading order (NNLO) in the strong coupling constant $\alpha_s$ and next-to-leading order (NLO) in the electromagnetic coupling constant $\alpha$ must be included in the calculations. Furthermore, electroweak (EW) corrections play an essential role for predictions at lepton colliders -- the favoured candidates for future colliders -- in order to match the projected experimental accuracies.
	
	The aim of this thesis is the completion of the NLO automated framework of the MC program \texttt{WHIZARD}, accounting for NLO corrections in the full SM for cross sections and distributions of processes at hadron and lepton colliders.
	Specifically, it builds on the implemented FKS subtraction scheme for NLO QCD calculations, and extends it to automated NLO EW and QCD-EW mixed corrections. To that end, the implemented FKS scheme is generalised to systematically subtract QED and QCD infrared (IR) divergences in mixed coupling expansions. The NLO QCD-EW mixed corrections particularly are relevant for hadron-collider processes with dominant QCD interactions.
	The automated computation of NLO contributions in mixed coupling expansions with \texttt{WHIZARD} is validated by cross-checks with reference MC tools for a set of benchmark processes at the LHC, including e.~g. $t\bar{t}~(+H/W/Z)$ production.
	
	Cross-checks for $e^+e^-$ processes likewise show that \texttt{WHIZARD} can be used for predictions at lepton colliders including fixed $\mathcal{O}(\alpha)$ corrections universally. The NLO phase-space construction is performed taking massive emitters for initial-state radiation (ISR) into account.
	This framework is applied to the study of NLO EW cross sections and differential distributions for multi-boson processes at a future multi-TeV muon collider. Corresponding results are computed for the first time in this work.
	Further applications of this framework would be the combination of NLO EW corrections with beam structure features in \texttt{WHIZARD} such as polarisation.
	
	For some high-energy lepton-collider setups, tiny initial-state masses jeopardise the reliability of fixed-order expansions of observables in QED perturbation theory.
	In order to recover meaningful predictions for NLO EW observables, QED parton-distribution functions (PDFs) must be applied guaranteeing a universal treatment of collinear ISR effects. This however causes computational challenges for MC integration measures. Methods to cope with these difficulties are presented in this thesis.
	
	With the automation of NLO EW corrections for hadron and lepton collision processes, \texttt{WHIZARD} offers a powerful tool for EW precision studies at current and future colliders and provides the desired accuracy level of predictions for new physics searches.
	\chapter*{Zusammenfassung}
	F\"ur die Suche nach neuer Physik an Collider-Experimenten wird h\"ochste Pr\"azision f\"ur Simulations-Rechnungen durch Monte-Carlo (MC) Event-Generatoren ben\"otigt. Die derzeitige Pr\"azision f\"ur Vorhersagen aus dem Standard-Modell (SM) f\"ur Messgr\"o\ss en am LHC liegt auf einem hohen Niveau:
	St\"orungstheoretische Korrekturen mindestens zur n\"achst-n\"achst-f\"uhrenden Ordnung (NNLO) in der starken Kopplungskonstante $\alpha_s$ und in der n\"achst-f\"uhrenden Ordnung (NLO) in der elektromagnetischen $\alpha$ m\"ussen in die Rechnungen einbezogen werden.
	Des Weiteren spielen elektroschwache (EW) Korrekturen in theoretischen Vorhersagen f\"ur Lepton-Collider-Prozesse eine essentielle Rolle, um die erwartete experimentelle Genauigkeit nutzen zu k\"onnen.
	
	Das Ziel dieser Arbeit ist die Erweiterung des MC-Programms \texttt{WHIZARD} zu automatisierten Berechnungen von NLO Korrekturen im vollen SM bez\"uglich Wirkungsquerschnitten und differentiellen Verteilungen von Prozessen an Hadron- und Lepton-Collidern. Aufbauend auf dem implementierten FKS Subtraktionsschema f\"ur NLO QCD Rechnungen beinhaltet dies die Erweiterung hinsichtlich automatisierter NLO EW und gemischter NLO QCD-EW Korrekturen. Zu diesem Zwecke wurde das implementierte FKS Schema verallgemeinert, so dass QED und QCD Infrarot-Divergenzen in einer gemischten Kopplungsentwicklung systematisch regularisiert werden. Gemischte NLO QCD-EW Korrekturen k\"onnen insbesondere f\"ur Prozesse mit dominanten QCD Wechselwirkungen an Hadron-Collidern relevante Beitr\"age liefern. Die Validierung der automatisierten Berechnung von NLO Beitr\"agen in einer gemischten Kopplungsent-wicklung erfolgte durch Cross-Checks mit Referenzwerten bekannter MC-Generatoren f\"ur Benchmark-Prozesse am LHC wie z.~B. $t\bar{t}~(+H/W/Z)$ Produktion.	\thispagestyle{empty}
	
	In gleicher Weise zeigen Cross-Checks f\"ur $e^+e^-$-Prozesse, dass sich \texttt{WHIZARD} als universelles MC-Programm f\"ur Vorhersagen an Lepton-Collidern unter Ber\"ucksichtigung von festen $\mathcal{O}(\alpha)$ Korrekturen erweist. F\"ur diese Art von Rechnung werden die Anfangszust\"ande, welche Photonen abstrahlen, f\"ur die Phasenraum-Konstruktion zur n\"achst-h\"oheren Ordnung in $\alpha$ als massiv betrachtet. Unter Anwendung dieser Methode wurden im Rahmen dieser Arbeit das erste Mal st\"orungstheoretische NLO EW Korrekturen zu Wirkungsquerschnitten und differentiellen Verteilungen f\"ur Multi-Boson-Prozesse an einem zuk\"unftigen Multi-TeV-Muon-Collider berechnet. Ein weiteres Anwendungsfeld der automatisierten NLO Rechnungen in \texttt{WHIZARD} im Zusammenhang mit Lepton-Collider Studien ist die Kombination von NLO EW Korrekturen mit Beam-Struktur-Eigenschaften, wie z. B. Polarisation der Anfangszust\"ande.
	
	F\"ur manche Lepton-Collider mit hohen Collider-Energien f\"uhren verschwindend kleine Massen der leptonischen Anfangszust\"ande zu einer Abschw\"achung der Zuverl\"assigkeit der QED-St\"orungsentwicklung in fester Ordnung. Um aussagekr\"aftige Vorhersagen f\"ur NLO EW Observablen zu garantieren, m\"ussen kollineare Abstrahlungen von Anfangszust\"anden universell betrachtet werden. Dies wird durch die Anwendung von QED-Partondistributionen (QED-PDFs) f\"ur Lepton-Kollisionen erm\"oglicht. Die Ein-bettung~dieser PDFs in NLO-Rechnungen erfordert besondere MC-Integrationsmethoden, welche formal in dieser Arbeit pr\"asentiert werden.
	
	Insgesamt bietet die Automatisierung von NLO-EW-Rechnungen f\"ur beliebige Prozesse an Hadron- und Lepton-Collidern in \texttt{WHIZARD} die M\"oglichkeit zu elektro-schwachen Pr\"azisionsstudien an derzeitigen und zuk\"unftigen Collidern. Diese Rech-nungen erreichen ein Genauigkeitsniveau f\"ur theoretische Vorhersagen des SMs, das f\"ur die Suche nach neuer Physik unverzichtbar ist.
	\thispagestyle{empty}
		\begin{refsection}
		\nocite{Bredt:2022zpn}
		\nocite{Bredt:2022dmm}
		\nocite{Campbell:2022qmc}
		\nocite{Stienemeier:2021cse}
		\nocite{WhizardNLO}
		\defbibnote{myPrenote}{\noindent
			Parts of this thesis contributed to the following publications:
		}
		\printbibliography[
		title={List of Publications},
		prenote=myPrenote
		]
	\end{refsection}
	\thispagestyle{empty}
	\tableofcontents
	\thispagestyle{empty}
	\chapter{Introduction}
	Ten years after the discovery of the Higgs boson at the LHC, the SM of particle physics has been proven as the most accurate model which governs our understanding of elementary particles and interactions.
	However, as the SM describes nature only in an incomplete way, e.~g. leaving aside the question of fundamental interactions describing gravity or dark matter, and open theoretical issues such as the hierachy problem remain, the search for evidence of an ultimate theory continues. For its discovery, the investigation of the Higgs sector related to EW and top quark physics represents the most promising direction. For this sector, experimental data at the moment are not yet at a precision level to make theoretical advancement beyond the SM (BSM) possible.
	Driven by the goal of discovering new physics, high-energy collider experiments aim to detect new particles or interactions from two sides. The first involves identifying signals in final-state searches, while the second entails the exploration of large mass scales of BSM physics.
	
	At present, the capabilities of the LHC have been proven insufficient on both fronts: the former in the nearest future will be addressed by increasing the statistics through the high-luminosity upgrade at the LHC (HL-LHC); the latter is the primary goal of future collider experiments which extend the energy frontiers of particle interactions. A next-generation hadron collider pursuing this goal for example would be the FCC-hh \cite{FCC:2018byv,FCC:2018vvp}.
	Crucially, future lepton colliders with approximately fixed beam momenta of colliding elementary particles at TeV scales allow measurements of fundamental parameters of the EW sector such as Higgs boson couplings at the per mille level \cite{Campbell:2022qmc}.
	Among suggested future projects considered to be part of the European Strategy Update (ESU), first and foremost are $e^+e^-$ colliders with modern acceleration technologies which act as Higgs/Top/EW (HTE) factories. These include the proposed ILC \cite{Baer:2013cma,Behnke:2013lya}, CLIC \cite{CLIC:2016zwp,Aicheler:2012bya}, CEPC \cite{CEPCStudyGroup:2018ghi} and FCC-ee \cite{FCC:2018evy}. Furthermore, the European Committee for Future Accelerators (ECFA) has initiated studies addressing future $e^+e^-$ colliders as HTE factories. The need for EW precision calculations with respect to these studies provides an important aspect motivating this thesis.
	As a further option, for exploring the multi-TeV range, a muon collider has recently been proposed \cite{Delahaye:2019omf,Bartosik:2020xwr,Long:2020wfp,Aime:2022flm}. The large mass of the muon as beam particle compared to that of the electron is advantageous on both sides at such a lepton collider: High centre-of-mass energies -- beyond $10$ TeV -- could be targeted and effects diluting leptonic collisions such as bremsstrahlung are reduced. Moreover, there is a `no-lose theorem' for discovering new physics at a muon collider due to the discrepancy of SM predictions and measurements related to $(g-2)_{\mu}$~\cite{Capdevilla:2021rwo}.

	An important issue to address is that already at the LHC, but much more prominently at $e^+e^-$ HTE factories, measurements of fundamental parameters are limited by theoretical and not by experimental uncertainties.
	Therefore, from the theorists' side, ambitions lie in reaching the highest possible level of accuracy for predictions from the SM by means of higher order perturbative calculations. For hadron collision observables of LHC processes this demands in particular accounting for perturbative corrections of Quantum-Chromodynamics (QCD) up to next-to-next-to-leading order (NNLO) in the strong coupling constant $\alpha_s$. As the electromagnetic coupling numerically scales as $\alpha\sim \alpha_s^2$, perturbative corrections of the EW theory to the next-to-leading order (NLO) in $\alpha$ are of a similar numerical size as NNLO QCD corrections and therefore must be taken into account as well. Especially, in high-energy tails of differential distributions, EW corrections have a significant impact on the predictions. For lepton collisions EW corrections play an indispensable role for precision calculations, while QCD corrections influence the predictions only if there are coloured final-state particles.
	Since the principles of higher-order perturbative corrections are of universal nature, methods have been developed to include these corrections in predictions for arbitrary processes and observables.
	
	For several decades now, this has been the motivation of physicists for developing tools which provide computations of physical quantities for comparison with experimental data.
	For these tools, on the one hand, highest possible levels of precision for predictions from theoretical models -- matching at least those of experimental data -- are targeted.
	On the other hand, a flexible usage of these calculations is desired in order to study the broad range of physical processes at colliders to minimal amount of work.
	Monte-Carlo (MC) event generators represent indispensable tools of this kind as they are able to simulate physics in a universal manner with a predictive power at the level of exclusive data. Well-known multi-purpose MC event generators providing such a process-independent setup are \texttt{MG5\_aMC@NLO} \cite{Alwall:2011uj,Alwall:2014hca}, \texttt{SHERPA} \cite{Gleisberg:2008ta} and \texttt{WHIZARD} \cite{Kilian:2007gr,Moretti:2001zz}. Building on the latter framework, for which the automation of NLO QCD corrections has been achieved recently \cite{Stienemeier:2021cse,Rothe:2021sml,WhizardNLO}, the aim of this thesis is the extension to automated NLO EW corrections to arbitrary processes. This includes not only the automation for $pp$~collision processes, which so far has been technically realised by MC tools such as \texttt{MG5\_aMC@NLO}~\cite{Frederix:2018nkq}, \texttt{SHERPA}~\cite{Schonherr:2017qcj} and \texttt{MUNICH/MATRIX} \cite{Kallweit:2014xda,Buonocore:2021rxx,Bonciani:2021zzf}, but also for lepton collision processes.
	
	Both classes, hadron and lepton collider processes, entail intricacies in the technical achievement of theoretical predictions including NLO EW corrections. The two cases are essentially different to each other due to the corresponding beam character.
	
	For $pp$~processes, QCD interactions are dominant due to the (mostly) coloured partons in the initial-state. Consequently, for many processes additional non-zero leading-order (LO) contributions for coupling powers sub-leading in the strong coupling constant $\alpha_s$ emerge. This causes NLO contributions which contain overlapping QCD and EW corrections such that the subtraction of infrared (IR) divergent terms must account for both correction types simultaneously in a systematic way. In addition, photon-induced processes play a non-negligible role and significantly affect certain observables at NLO EW. Including EW corrections for these cases a proper treatment of photons entering the hard process regarding prescriptions of the EW renormalisation scheme choice is key.
	
	For lepton collisions, the initial-state can be described purely by perturbative means taking a finite lepton mass into account. Formally, this can lead to logarithmically enhanced terms in NLO EW observables due to factors $\alpha \log (Q^2/m^2)$ induced by initial-state radiation (ISR) off leptons with small masses $m$ compared to the energy scale of the process $Q$; the lower this initial-state mass, the less well-behaved the respective QED perturbative expansion gets. This implies the need for the factorisation and resummation of these logarithmic terms into parton-distribution functions (PDFs) analogous to their QCD counterparts, the PDFs of hadron collider processes.
	Leptonic PDFs for processes at a lepton collider contain a peaked structure with an integrable singularity at $x=1$, where $x$ is the fraction of the radiated beam energy. 
	Due to this, the phase-space construction for ISR in MC integration methods must contain an appropriate mapping of random numbers of the unit-square in order to cover the full area of the cross section differential in the radiated beam energy. In order to achieve NLO EW results with the application of lepton PDFs, a technical approach combining such a mapping with the implemented IR subtraction scheme in \texttt{WHIZARD} is formally presented in this thesis.
	Another way to achieve precise predictions for lepton collider observables at NLO EW is in a strictly fixed-order approximation. Initial-state emitters therefore are treated as massive on-shell particles in the NLO phase-space construction. For several beam setups of current and future lepton colliders this can be considered as a reliable approach. This thesis covers the validation of NLO EW results for $e^+e^-$ collisions achieved with this approach in \texttt{WHIZARD} and results on cross sections and differential distributions at NLO EW of multi-boson processes at a muon collider.
	
	This work is divided into two parts: in the first, theoretical aspects of automated NLO EW corrections in the MC framework \texttt{WHIZARD} are discussed. The second part contains numerical results for the proof of its validity for hadron and lepton collider processes as well as for benchmark processes at a future muon collider.
	
	Sec.~\ref{secPerturbativeCalculations} of part~I first gives an overview of prerequisites and quantum-field theoretical building blocks of perturbative calculations in the SM. The formalism of the Frixione-Kunszt-Signer (FKS) scheme generalised for the infrared subtraction in NLO QCD, EW and mixed corrections as well as the phase-space construction in this scheme for massless and massive initial-state emitters is presented in Sec.~\ref{secFKSscheme}. Sec.~\ref{secEWprecisioncalculations} contains the formal description of collinear ISR in lepton collisions incorporated in the form of lepton PDFs and methods for an efficient MC integration of LO and NLO cross sections applying these PDFs. In Sec.~\ref{secWhizardFrameworkNLO} the framework of the event generator \texttt{WHIZARD} and technical details on its automation of NLO calculations including the extension to NLO EW and NLO mixed corrections are presented.
	
	Part~II is further subdivided into two sections: Sec.~\ref{secHadronCollisions} contains numerical cross-checks on cross sections and differential distributions in NLO mixed-coupling expansions for benchmark processes at the LHC with reference results of the MC tools \texttt{MUNICH/MATRIX} and \texttt{MG5\_aMC@NLO}. In Sec.~\ref{secLeptonCollisions} the validation of calculations including NLO EW corrections for lepton collision processes is shown. Furthermore, results for cross sections and differential distributions at NLO EW in multi-boson production at a muon collider -- computed for the first time in this work -- are presented therein.
	\part{Theoretical Framework}
	\chapter{Perturbative calculations in the Standard Model}
	\label{secPerturbativeCalculations}
	First I will briefly introduce the SM of particle physics, the theory describing experimental observations with the smallest number of degrees of freedom, which lays the foundation of this work. Preliminaries on electroweak renormalisation and regularisation of infrared divergences will be discussed too, since these are essential methods of the higher order calculations which are performed for this thesis. A more detailed overview on the topics of this chapter is given for example in \cite{Bohm:2001yx,Denner:1991kt,Denner:2019vbn,Aoki:1982ed,Kallweit:2008dnd,Ellis:1996mzs}.
	\section{Standard Model of particle physics}
	\label{secStandardmodel}
	By means of the Lagrangian formalism, i. e. the invariance of the action under small variations according to Hamilton's principle
	\begin{equation}
		\delta \int d^4x \mathcal{L} = 0
	\end{equation}
	and the resulting Euler-Lagrange equations of motion, it is possible to formulate the theory of the SM of particle physics. The invariance of the Lagrangian of the SM on the one hand is with respect to Poincare transformations such that space-time symmetries are obeyed. On the other hand, the SM respects symmetries related to internal symmetry transformations which can be global or local, i.~e. the symmetry group has space-time dependent group elements and representations.
	The principle of minimal gauge-invariant coupling of matter to a gauge field defines the interaction in a charge-symmetric theory. For the simplest example, the electromagnetic interaction described by the theory of QED symmetric in the electric charge $Q$, we find the Lagrangian
	\begin{equation}
		\mathcal{L}_{\text{QED}}= -\frac{1}{4}F_{\mu\nu}(x)F^{\mu\nu}(x) + \overline{\psi}(x)(i\slashed{D}-m)\psi(x)
		\label{QEDLagrangian}
	\end{equation}
	which is invariant under local abelian gauge transformations of the group $U(1)_{\text{em}}$
	\begin{align}
	\centering
	    &\psi(x) \rightarrow \text{e}^{-iQ_fe\theta(x)}\psi(x), 	\quad\quad    \overline{\psi}(x) \rightarrow \overline{\psi}(x)\text{e}^{iQ_fe\theta(x)} \\
		&A_{\mu}(x) \rightarrow A_{\mu}(x) + \partial_{\mu}\theta(x).
	\end{align}
	$\psi$ ($\overline{\psi}$) denotes the charged fermion (anti-fermion) field. $A_{\mu}$ is the gauge field which enters the electromagnetic field strength tensor $F_{\mu\nu}$ and the covariant derivative $D_{\mu}$ as
	\begin{align}
		F_{\mu\nu} &=\partial_{\mu}A_{\nu}(x)-\partial_{\nu}A_{\mu}(x) \label{abelFST}\\
        D_{\mu} &=\partial_{\mu} +iQ_feA_{\mu}(x).	
	\end{align}
	The same principle can be applied to Yang-Mills (YM) gauge theories, i.~e. quantum field theories with non-abelian charge algebra, for which symmetry transformations are given by representations $D_g = \text{e}^{i\theta^aT^a}$ of a group G with elements $g\in G$ and generators $T^a$, $a=1,\ldots,\dim G$, the charges of the group.
	In the non-abelian generalisation the field strength tensor 
 has the anti-symmetric form
	\begin{align}
		&\mathcal{F}_{\mu\nu}(x)=\partial_{\mu}\mathcal{A}_{\nu}(x)-\partial_{\nu}\mathcal{A}_{\mu}(x)-i[\mathcal{A}_{\mu}(x),\mathcal{A}_{\nu}(x)], &\mathcal{A}_{\mu}(x)=A_{\mu}^a(x)T^a
	\end{align}
	with a bilinear term in the gauge field. It can be shown that the transformation behaviour of the gauge field $\mathcal{A}_{\mu}$ is inhomogenous, i.~e.
	\begin{align}
		i\mathcal{A}'_{\mu}(x)=D_g(x)(i\mathcal{A}_{\mu}(x)-\partial_{\mu})D_g^{-1}(x),
	\end{align}
	such that $\mathcal{A}_{\mu}$ has to be treated as a dynamical quantity. However, with $\mathcal{F}_{\mu\nu}$ transforming as
	\begin{align}
			\mathcal{F}_{\mu\nu}'(x)=D_g(x)\mathcal{F}_{\mu\nu}(x)D_g^{-1}(x)
			\label{Amugaugetrafo}
	\end{align}
	 we find gauge invariance for the gauge part of the Lagrangian written in the form
	\begin{equation}
		\mathcal{L}_{\text{gauge}} =\frac{1}{2g^2}\text{Tr}\left(\mathcal{F}_{\mu\nu}(x)\mathcal{F}^{\mu\nu}(x)\right)=-\frac{1}{4}F^{a}_{\mu\nu}(x)F^{a,\mu\nu}(x)
		\label{gauge-non-ab}
	\end{equation}
	which is analogous to the first term of Eq.~(\ref{QEDLagrangian})
	with $g$ a fixed physical constant. The field strength tensor, written in components, depends on $A^a_{\mu}$ as
	\begin{align}
	F^{a}_{\mu\nu}=\partial_{\mu}A^a_{\nu}(x)-\partial_{\nu}A^a_{\mu}(x) + g f^{abc}A^b_{\mu}(x)A^c_{\nu}(x)
	\label{fieldstrengthYM}
	\end{align}
	where $f^{abc}$ are structure constants of the Lie algebra.
	
	With these considerations we can formulate the Lagrange density of Quantum Chromodynamics (QCD) from which the strong interaction of quark fields to gluon gauge fields can be derived. This is based on the colour charge symmetry of the $SU(3)$ gauge group. The representation matrices of this symmetry group are the Gell-Mann matrices $\lambda^a/2$ which transform as adjoint representations of $SU(3)$ and allow for eight colour degrees of freedom for gluon fields. Quark fields are described by the fundamental representation which allow for three degrees of freedom. The gauge part of the QCD Lagrangian reads
	\begin{align}
		\mathcal{L}_{\text{QCD,gauge}}=-\frac{1}{4}\sum_a G^{a}_{\mu\nu}(x)G^{a,\mu\nu}(x) 
     \label{QCDlagrangian}
	\end{align}
    where the QCD field strength tensor according to the general YM case of Eq.~(\ref{fieldstrengthYM}) reads
	\begin{align}
		G^{a}_{\mu\nu}=\partial_{\mu}G^a_{\nu}(x)-\partial_{\nu}G^a_{\mu}(x) + g_s f^{abc}G^b_{\mu}(x)G^c_{\nu}(x)
	\end{align}
	with $a,b,c=1,\ldots,8$. The covariant derivative including QCD interactions can be formulated as
	\begin{align}
	&D_{\mu,kl}=\partial_{\mu}\delta_{kl}-ig_s(\lambda^a)_{kl}G_{\mu}^a/2, &k,l=1,2,3.
	\label{covariantderQCD}
	\end{align}
	Another nonabelian gauge theory of the SM is described by the unified electroweak interaction which is based on the Glashow-Salam-Weinberg model \cite{glashow:1961,weinberg:1967,salam:1968}. This theory is based on the non-simple group $SU(2)_W \times U(1)_{Y}$, the product of the weak isospin group $SU(2)_W$ and the group $U(1)_Y$. The four vector fields which transform according to the adjoint representations of this gauge group are the isotriplet $W^i_{\mu}$, $i=1,2,3$ with the $SU(2)$ generators $I^i_W$ and the isosinglet $B_{\mu}$ with the $U(1)_Y$ generator $Y_W$, i.~e. the weak hypercharge. The gauge part of the Lagrangian for the electroweak theory can thus be formulated as
	\begin{align}
		\mathcal{L}_{\text{EW,gauge}}=-\frac{1}{4}\sum_i W^{i}_{\mu\nu}W^{i,\mu\nu}-\frac{1}{4}B_{\mu\nu}B^{\mu\nu}
		\label{EWlagrangian}
	\end{align}
	whereby the field strength tensor $W^i_{\mu\nu}$ according to the general nonabelian one of Eq.~(\ref{fieldstrengthYM}) reads
	\begin{align}
		W^{i}_{\mu\nu}=\partial_{\mu}W^i_{\nu}-\partial_{\nu}W^i_{\mu} + g_2 \varepsilon^{ijm}W^j_{\mu}W^m_{\nu}.
	\end{align}
	where $\varepsilon^{ijm}$ are the structure constants of $SU(2)_W$.
	The form of the field-strength tensor $B_{\mu\nu}$ corresponds to the one of abelian theories as e.~g. the form of (\ref{abelFST}),
	\begin{align}
		B_{\mu\nu} =\partial_{\mu}B_{\nu}-\partial_{\nu}B_{\mu}.
	\end{align}
	Taking not only the QCD but also the EW interactions into account the covariant derivative of Eq.~(\ref{covariantderQCD}) can be rewritten as
	\begin{align}
		D_{\mu}=\partial_{\mu}-ig_2I^i_WW^i_{\mu}+ig_1\frac{Y_W}{2}B_{\mu}-ig_s\frac{\lambda^a}{2}G_{\mu}^a.
		\label{covariantderEW}
	\end{align}
	The electric charge $Q$, the weak isospin component $I^3_W$ and the hypercharge $Y_W$ are related to each other by the Gell-Mann Nishijima relation
	\begin{align}
		Q=I_W^3+\frac{Y_W}{2}.
	\end{align}
	By means of the projectors $\omega_{\pm}=(1\pm \gamma^5)/2$ with chirality operator $\gamma^5$ fermion fields can be grouped into left-handed leptons ($L^L$) and quarks ($Q^L$), i.~e. $SU(2)_W$ doublets with $I^i_W=\tau^i/2$ and Pauli matrices $\tau^i$,
	\begin{align}
		&L^L= \omega_-L=\left(
		\begin{array}{c}
		\nu^L\\
		l^L\\
		\end{array}
		\right)
		&Q^L= \omega_-Q=\left(
		\begin{array}{c}
		q_u^L\\
		q_d^L\\
		\end{array}
		\right)
	\end{align}
	and right-handed leptons ($l^R$) and quarks ($q^R$), i.~e. singlets with $I^i_W=0$,
	\begin{align}
		l^R=\omega_+l \quad\quad q_u^R=\omega_+q_u \quad\quad q_d^R=\omega_+q_d.
	\end{align}
	The quantum numbers $Q$, $I_W$, $I^3_W$ and $Y$ of all three generations of leptons and quarks are collected in Table \ref{leptonsquarks}.
	\begin{table}
		\centering
		{	\onehalfspacing
		\begin{tabularx}{0.88\linewidth}{c|c|c|c|c|c|c|c|c}
			\multicolumn{2}{c}{}&I&II&III&$Q$&$I_W$&$I_W^3$&$Y_W$\\
			\hline
			\text{leptons}&$L^L$&$\left(
			\begin{array}{c}
				\nu_e^L\\
				e^L\\
			\end{array}\right)$&$\left(
			\begin{array}{c}
			\nu_{\mu}^L\\
			{\mu}^L\\
			\end{array}\right)$&$\left(
			\begin{array}{c}
			\nu_{\tau}^L\\
			{\tau}^L\\
			\end{array}\right)$&$\begin{array}{c}
			0\\
			-1\\
			\end{array}$&$\begin{array}{c}
			\frac{1}{2}\\
			\frac{1}{2}\\
			\end{array}$&$\begin{array}{c}
			+\frac{1}{2}\\
			-\frac{1}{2}\\
			\end{array}$&$\begin{array}{c}
			-1\\
			-1\\
			\end{array}$\\
			&$l^R$&$e^R$&$\mu^R$&$\tau^R$&$-1$&$0$&$0$&$-2$\\
			\hline
			\text{quarks}&$Q^R$&$\left(
			\begin{array}{c}
			u^L\\
			d^L\\
			\end{array}\right)$&$\left(
			\begin{array}{c}
			c^L\\
			s^L\\
			\end{array}\right)$&$\left(
			\begin{array}{c}
			t^L\\
			b^L\\
			\end{array}\right)$&$\begin{array}{c}
			+\frac{2}{3}\\
			-\frac{1}{3}\\
			\end{array}$&$\begin{array}{c}
			\frac{1}{2}\\
			\frac{1}{2}\\
			\end{array}$&$\begin{array}{c}
			+\frac{1}{2}\\
			-\frac{1}{2}\\
			\end{array}$&$\begin{array}{c}
			+\frac{1}{3}\\
			+\frac{1}{3}\\
			\end{array}$\\
			&$q_u^R$&$u^R$&$c^R$&$t^R$&$+\frac{2}{3}$&$0$&$0$&$+\frac{4}{3}$\\
			&$q_d^R$&$d^R$&$s^R$&$b^R$&$-\frac{1}{3}$&$0$&$0$&$-\frac{2}{3}$\\
		\end{tabularx}}
	\caption{{Electric charge $Q$, weak isospin $I_W$, third compond of the weak isospin $I^3_W$ and hypercharge $Y_W$ of the three generations of fermions of the SM}}
	\label{leptonsquarks}
	\end{table}
    With the covariant derivative of Eq.~(\ref{covariantderEW}) the resulting fermionic part of the Lagrangian which describes all interactions of fermions with gauge bosons reads
    \begin{align}
        	\mathcal{L}_{\text{F,int}}=\sum_{g}(\bar{L}^L_gi\slashed{D}L_g^L+\bar{Q}_g^Li\slashed{D}Q_g^L+\bar{l}_g^Ri\slashed{D}l_g^R+\bar{q}_{u,g}^Ri\slashed{D}q_{u,g}^R+\bar{q}_{d,g}^Ri\slashed{D}q_{d,g}^R)
        	\label{fermionsintlagrangian}
    \end{align}
	with generations $g$. So far, no masses of the fermions and gauge bosons are assumed which would lead to violation of gauge invariance of the Lagrangian otherwise. However, since there is experimental evidence for their masses, the Higgs mechanism was proposed which induces the sponaneous symmetry breaking of the electroweak theory \cite{Englert:1964et,Higgs:1964pj,Higgs:1964ia,Guralnik:1964eu,Higgs:1966ev}. To this end, a complex scalar $SU(2)_W$ doublet field with $Y_W=1$, i.~e. the Higgs field, is introduced,
	\begin{align}
		\Phi(x)=\left(
		\begin{array}{c}
		\phi^+(x)\\
		\phi^0(x)\\
		\end{array}\right)
		\label{scalardoublet}
	\end{align}
	which couples to the gauge fields via Eq.~(\ref{covariantderEW}) and has a self coupling. This results in the Lagrangian
	\begin{align}
		&\mathcal{L}_{\text{Higgs}}=(D_{\mu}\Phi)^{\dagger}(D^{\mu}\Phi) - V(\Phi), &V(\Phi)=\frac{\lambda}{4}(\Phi^{\dagger}\Phi)^2-\mu^2(\Phi^{\dagger}\Phi)
		\label{Higgslagrangian}
	\end{align}
	whereby $\mu^2,\lambda>0$ such that there is a minimum for the potential $V(\Phi)$ for a non-vanishing Higgs field, i.~e. the non-zero vacuum expectation value (vev) $|\langle\Phi \rangle|=v/\sqrt{2}\neq0$ with the constant number $v>0$. By the condition that $U(1)_{\text{em}}$ symmetry has to remain unbroken the ground state $\Phi_0=(0, v/\sqrt{2})^{\top}$ can be found which leads to a parametrisation of the Higgs field,
	\begin{align}
		&\Phi(x)=\left(
		\begin{array}{c}
		\phi^+(x)\\
		\frac{1}{\sqrt{2}}[v+H(x)+i\chi(x)]\\
		\end{array}\right), &\phi^-(x)=(\phi^+(x))^{\dagger}
		\label{higgsfieldparam}
	\end{align}
	with the physical Higgs field $H$ and the unphysical would-be Goldstone fields $\phi^+$, $\phi^-$ and $\chi$. These Goldstone bosons can be eliminated by choosing the unitary gauge which renders $\phi^{\pm}=\chi=0$, but however is non-renormalisable due to power-counting reasons. By inserting Eq.~(\ref{higgsfieldparam}) into Eq.~(\ref{Higgslagrangian}) gauge boson mass terms arise in the Lagrangian due to the coupling of the vev to the gauge bosons. In consequence, $SU(2)_W\times U(1)_Y$ spontaneously breaks while the electromagnetic $U(1)_{\text{em}}$ symmetry is kept exact. To get fermion mass terms in the Lagrangian, Yukawa interactions of the scalar doublet with the fermions have to be added,
	\begin{align}
		\mathcal{L}_{\text{F,Yukawa}}=-\sum_{g,g'}\left(\bar{L}^L_gG^l_{gg'}l^R_{g'}\Phi+\bar{Q}^L_gG^u_{gg'}q^R_{u,g'}\tilde{\Phi}+\bar{Q}^L_gG^d_{gg'}q^R_{d,g'}\Phi+h.c. \right)
		\label{Yukawalagrangian}
	\end{align}
	with the charge conjugated field $\tilde{\Phi}=(\phi^{0*},-\phi^-)^{\top}$ and the Yukawa coupling matrices $G^l_{gg'}$, $G^u_{gg'}$ and $G^d_{gg'}$. Finally, the classical Lagrangian in the symmetric basis can be obtained by summing the parts of Eq.~(\ref{QCDlagrangian}), (\ref{EWlagrangian}), (\ref{fermionsintlagrangian}), (\ref{Higgslagrangian}) and (\ref{Yukawalagrangian}),
	\begin{align}
		\mathcal{L}_{\text{class}}=\mathcal{L}_{\text{QCD,gauge}}+\mathcal{L}_{\text{EW,gauge}}+\mathcal{L}_{\text{Higgs}}+\mathcal{L}_{\text{F,int}}+\mathcal{L}_{\text{F,Yukawa}}.
	\end{align}
    The physical gauge boson and fermion fields can be obtained by diagonalising the mass matrices occuring in $\mathcal{L}_{\text{class}}$ such that the mass (and charge) eigenstates read
	\begin{align}
		\begin{split}
		&W_{\mu}^{\pm}=\frac{1}{\sqrt{2}}\left(W^1_{\mu}\mp iW^2_{\mu}\right),\quad\quad \left(
		\begin{array}{c}
		Z_{\mu}\\
		A_{\mu}\\
		\end{array}\right)=
		\begin{pmatrix}
		&c_W &s_W\\
		&-s_W &c_W\\
		\end{pmatrix}
		\left(
		\begin{array}{c}
		W^3_{\mu}\\
		B_{\mu}\\
		\end{array}\right)\\
		&f'^L_g=\sum_{g'}U_{gg'}^{f,L}f^L_{g'},\quad\quad\quad\quad\quad f'^R_g=\sum_{g'}U_{gg'}^{f,R}f^R_{g'}
		\end{split}
	\end{align}
	for physical fermion fields $f'$ and using the shortcuts $c_W=\cos \theta_W$ and $s_W=\sin \theta_W$ with the weak mixing angle $\theta_W$. The following masses of the SM particles can thus be extracted
	\begin{align}
		\begin{split}
	    &M_W=\frac{1}{2}g_2 v,\quad\quad M_Z=\frac{1}{2}\sqrt{g_1^2+g_2^2}v,\quad\quad m_{\gamma}=0\\
		&M_H=\sqrt{2\mu^2}=\sqrt{\frac{\lambda}{2}}v,\quad\quad m_{f',g}=\frac{v}{\sqrt{2}}\sum_{h,l}U_{gh}^{f',L}G^{f'}_{hl}U_{lg}^{f',R\dagger}
		\end{split}
	\end{align}
	with the unitary matrices $U_{gh}^{f',L}$ and $U_{gh}^{f',R}$. Additionally, the following parameter relations can be found
	\begin{align}
		&c_W=\frac{M_W}{M_Z} &\sqrt{4\pi\alpha}=e=\frac{g_1g_2}{\sqrt{g_1^2+g2^2}}=g_2s_W=g_1 c_W
		\label{weinbergangle}
	\end{align}
	with the fine-structure constant $\alpha$. For quark-$W$-boson interactions flavour-changing currents are allowed which becomes manifest by the unitary quark-mixing matrix, the Cabibbo-Kobayashi-Maskawa (CKM) \cite{Cabibbo:1963yz,Kobayashi:1973fv} matrix,
	\begin{align}
		V=U^{u',L}U^{d',L\dagger}
	\end{align}
	which, due to the suppression of non-diagonal compared to diagonal entries, is approximated as diagonal for the calculations of this work. Also, neutrino masses are neglected due to their smallness such that the right-handed neutrinos can be excluded and $U^{\nu',L}=U^{l',L}$. However if they are included analogously to the CKM matrix a neutrino mixing matrix occurs in the lepton sector, the Pontecorvo-Maki-Nakagawa-Sakata (PMNS) \cite{Pontecorvo:1957qd,Maki:1962mu} matrix.\\
	In order to compute physical quantities for the SM built up so far, e.~g. S-matrix elements, it is necessary to quantise the theory. This is commonly done by means of the path integral formalism. For any perturbatively quantisable gauge theory however it is necessary to impose a gauge fixing condition. By this, the path integration is prevented from diverging when integrating over physically equivalent field configurations which are related to each other by gauge transformations, for example according to Eq.~(\ref{Amugaugetrafo}). 
	For higher order perturbative computations a gauge fixing is thus required which renders the theory multiplicatively renormalisable, i.~e. all ultraviolet (UV) singularities occuring due to a perturbative expansion of the theory can be absorbed into redefined parameters of the Lagrangian. This for example can be achieved in an elegant way by the so-called 't Hooft gauge \cite{tHooft:1971,tHooft2:1971} for which gauge boson propagators have the form $1/k^2$ if the gauge parameters $\xi_a$ are chosen finite and consequently lead to a multiplicatively renormalisable Lagrangian. For the quantisation of nonabelian theories as $SU(3)$ and $SU(2)_W\times U(1)_Y$ it is necessary to introduce gauge-fixing terms and corresponding Faddeev-Popov ghost terms in the Lagrangian. The former can be chosen as
	\begin{align}
		\mathcal{L}_{\text{fix}}=-\frac{1}{2}\left[(F^G)^2+(F^Z)^2+(F^{\gamma})^2+2F^+F^-\right]
		\label{fixlagrangian}
	\end{align}
	with the gauge-fixing functionals
	\begin{align}
		\begin{split}
		&F^G=\frac{1}{\sqrt{\xi^G}}\partial^{\mu}G_{\mu},\quad\quad F^{Z}=\frac{1}{\sqrt{\xi_1^Z}}\partial^{\mu}Z_{\mu}- iM_Z\sqrt{\xi_2^Z}\chi,\\
		&F^{\gamma}=\frac{1}{\sqrt{\xi^{\gamma}}}\partial^{\mu}A_{\mu},\quad\quad F^{\pm}=\frac{1}{\sqrt{\xi_1^W}}\partial^{\mu}W_{\mu}^{\pm}\mp iM_W\sqrt{\xi_2^W}\phi^{\pm}.
		\end{split}
	\end{align}
	with arbitrary gauge parameters $\xi^{\alpha}$. By fixing $\xi^{W,Z}_1=\xi^{W,Z}_2$ terms of $\mathcal{L}_{\text{class}}$ proportional to $V_{\mu}\partial^{\mu}\phi$ are cancelled by mixing terms appearing in (\ref{fixlagrangian}). Furthermore, for higher order calculations the most convenient choice for the gauge fixing parameters is the 't~Hooft Feynman gauge $\xi^G=\xi^{\gamma}=\xi^Z_{1,2}=\xi^W_{1,2}=1$. Additionally, due to the gauge-fixing condition the Faddeev-Popov determinant appears as a factor in the path integral which gives rise to an additional term in the Lagrangian, the Fadeev-Popov ghost term,
	\begin{align}
		\mathcal{L}_{\text{FP}}=\bar{u}^{\alpha}(x)\frac{\delta F^{\alpha}}{\delta \theta^{\beta}(x)}u^{\beta}(x)
	\end{align}
	with the variation of the functionals with an infinitesimal gauge transformation $\delta F^{\alpha}/\delta \theta^{\beta}$ and the massless anticommuting (anti-) ghost fields $u^{\alpha}(x)$ ($\bar{u}^{\alpha}(x)$).
	Finally, we arrive at the perturbatively quantised Lagrangian describing the theory of the SM,
	\begin{align}
		\mathcal{L}_{\text{SM}}=\mathcal{L}_{\text{class}}+\mathcal{L}_{\text{fix}}+\mathcal{L}_{\text{FP}}.
		\label{totallagrangian}
	\end{align}
	By adding $\mathcal{L}_{\text{fix}}$ and $\mathcal{L}_{\text{FP}}$ to the classical Lagrangian, $\mathcal{L}_{\text{SM}}$ is not anymore invariant under local gauge transformations but instead obeys the symmetry of Becchi-Rouet-Stora (BRS) transformations \cite{Becchi:1974xu,Becchi:1974md}, thereby guaranteeing unitarity of the S-matrix. With this, the Higgs mechanism can be understood in a generic way, i.~e. independently from the choice of the gauge. Due to the BRS symmetry an auxiliary massless scalar field $B^a$ is introduced into $\mathcal{L}_{\text{fix}}$ which leaves the Lagrangian invariant under transformations $\delta_{\text{BRS}}$. The massless Goldstone fields $\phi^{\pm}$ and $\chi$ in the BRS formalism form three quartets with these fields $B^a$, and ghost and anti-ghost fields $u^{\alpha}$, $\bar{u}^{\alpha}$, the so-called Nambu Goldstone (NG) modes \cite{Nambu:1961tp,Goldstone:1961eq}. These NG modes have zero-norm and effectively decouple from the physical states. For any gauge it can be observed from the Lagrangian that the three degrees of freedom which come along with the introduction of the would-be Goldstone bosons in the scalar doublet of Eq.~(\ref{scalardoublet}) are absorbed into the mass degrees of freedom of the gauge bosons. In other words, the NG modes are absorbed into the formation of three massive spin-1 bosons, i.~e. the longitudinal components of the gauge bosons.
	\newpage
	\section{Electroweak renormalisation}
	In this section I will outline aspects of the renormalisation of the electroweak theory which are prerequisite for the content of this thesis. Concretely, renormalisation schemes for on-shell and unstable particles, different electroweak-input schemes to compute the electromagnetic coupling $\alpha$ with renormalised parameters as well as prescriptions for EW higher order computations of processes with external photons are discussed.
	\subsection{On-shell and complex-mass scheme}
	\label{secOnshellCMS}
	In general in order to perform higher order perturbative calculations the renormalisability of the considered theory has to be guaranteed. Renormalisability means that in each order of the perturbative expansion of Greens functions and S-matrix elements the UV divergences can be absorbed into renormalised constants for parameters and fields of the theory. For the SM, in particular, for nonabelian gauge theories with spontaneous symmetry breaking this has been proven by 't~Hooft and Veltman \cite{tHooft:1971,tHooft2:1971,tHooft:1972tcz}. The EW bare parameters, i.~e. masses and couplings (denoted by a subscript `0'), can be written in terms of renormalised ones $M_W$, $M_Z$, $M_H$, $m_{f,i}$, $V_{ij}$ and $e$ as
	\begin{align}
		\begin{split}
		&M^2_{0,W}=M^2_{W}+\delta M^2_W, \quad M^2_{0,Z}=M^2_{Z}+\delta M^2_Z,\quad M^2_{0,H}=M^2_{H}+\delta M^2_H,\\
		&m_{0,f,i}=m_{f,i}+\delta m_{f,i}, \quad V_{0,ij}=V_{ij}+\delta V_{ij}, \quad e_0=(1+\delta Z_e)e=e+\delta e.
		\end{split}
		\label{baremasses}
	\end{align}
    Analogously to the parameters the fields have to be renormalised. The bare physical fields read
	\begin{align}
		\begin{split}
		&W^{\pm}_0=(1+\frac{1}{2}\delta Z_W)W^{\pm}, \quad \left(\begin{array}{c}
		Z_0\\
		A_0
		\end{array}\right)=\begin{pmatrix}
		&1+\frac{1}{2}\delta Z_{ZZ} &\frac{1}{2}\delta Z_{ZA}\\
		&\frac{1}{2}\delta Z_{AZ} &1+\frac{1}{2}\delta Z_{AA}
		\end{pmatrix}\left(\begin{array}{c}
		Z\\
		A
		\end{array}\right)\\
		&H_0=\left(1+\frac{1}{2}\delta Z_H\right)H, \quad f^{\lambda}_{0,i}=\sum_j\left(\delta_{ij}+\frac{1}{2}\delta Z_{ij}^{f,\lambda}\right)f_j^{\lambda}, \quad \lambda=\{L,R\}.
		\end{split}
	\end{align}
	In addition, as mentioned in the previous section due to the requirement of choosing a multiplicatively renormalisable gauge unphysical fields appear and thus this sector likewise must be renormalised. To this end, the bare would-be Goldstone fields read
	\begin{align}
		&\chi_0=\left(1+\frac{1}{2}\delta Z_{\chi}\right)\chi, &\phi^{\pm}_0=\left(1+\frac{1}{2}\delta Z_{\phi}\right)\phi^{\pm},
	\end{align}
	the bare Faddeev-Popov ghost fields
	\begin{align}
		&u^{\pm}_0=\left(1+\delta \tilde{Z}_{\pm}\right)u^{\pm}, &\left(\begin{array}{c}
		u^Z_0\\
		u^A_0
		\end{array}\right)=\begin{pmatrix}
		&1+\frac{1}{2}\delta \tilde{Z}_{ZZ} &\frac{1}{2}\delta \tilde{Z}_{ZA}\\
		&\frac{1}{2}\delta \tilde{Z}_{AZ} &1+\frac{1}{2}\delta \tilde{Z}_{AA}
		\end{pmatrix}\left(\begin{array}{c}
		u^Z\\
		u^A
		\end{array}\right)
	\end{align}
	and the bare gauge parameters
	\begin{align}
	 \xi^A_{0}=Z_{\xi^A}\xi^A,\quad \xi^W_{0,k}=Z_{\xi^W_{k}}\xi^W_{k}, \quad \xi^Z_{0,k}=Z_{\xi^Z_{k}}\xi^Z_{k},  \quad k=\{1,2\}.
	\end{align}
	With these forms of the bare quantities and the renormalisation constants written in the form $Z=1+\delta Z$ the bare Lagrangian of Eq.~(\ref{totallagrangian}) can be split up as
	\begin{align}
		\mathcal{L}(\Psi_0,p_0)=\mathcal{L}(\Psi,p)+\mathcal{L}_{\text{ct}}(\Psi,p,\delta Z)
	\end{align}
	In this way the Lagrangian depends on renormalised instead of unrenormalised parameters and fields and on the corresponding counterterms thereby retaining all symmetries, e.~g. its hermiticity, as a sum but not necessarily for each term $\mathcal{L}(\Psi,p)$ and $ \mathcal{L}_{\text{ct}}(\Psi,p,\delta Z) $ separately.\\
	By imposing renormalisation conditions, i.~e. choosing a certain renormalisation scheme, the renormalisation constants can be computed. The renormalised mass is defined via the location of the pole of the propagator which corresponds to the physical mass parameter.
	In the on-shell (OS) scheme proposed by Ross and Taylor \cite{RossTaylor:1979} the following renormalisation conditions are imposed for the renormalised two-point functions $\Gamma_R$ of OS fields for external EW gauge bosons
	\begin{align}
    	&{\left[\widetilde{\text{Re}}\Gamma_{R,\mu\nu}^{{V'}^{\dagger}V}(-k,k)\right]\varepsilon^{\nu}}\biggr\rvert_{k^2=M_V^2}=0, \quad V,V'=\{W,Z\}\label{OSconditions}\\ 
    	&\lim_{k^2\rightarrow M_V^2}\frac{1}{k^2-M_V^2}{\left[\widetilde{\text{Re}}\Gamma_{R,\mu\nu}^{{V'}^{\dagger}V}(-k,k)\right]\varepsilon^{\nu}(k)}=-\varepsilon_{\mu}(k)\label{OSconditionsfields}
	\end{align}
	$\widetilde{\text{Re}}$ is defined such that one precisely only takes the real part of the loop integral but does not act on complex couplings, e.~g. coming from a phase in the quark-mixing matrix.
	In order to just give a general overview on the renormalisation prescriptions, analogous conditions for the Higgs, the fermionic and the unphysical sector and their implications on the renormalisation constants will be left out in the following. Furthermore, the discussion on the renormalisation of the electric charge is postponed to the next section treating several different input schemes for the computation of the electromagnetic coupling.\\
    According to Eq.~(\ref{OSconditions}) the renormalised OS masses correspond to the physical ones. This means the mass renormalisation constants $\delta M_V^2$ equal the real part of the self-energy at the location of the poles,
	\begin{align}
		&\delta M_W^2=\widetilde{\text{Re}}\Sigma_{T}^W(k^2)\biggr\rvert_{k^2=M_W^2} &\delta M_Z^2=\widetilde{\text{Re}}\Sigma_{T}^{ZZ}(k^2)\biggr\rvert_{k^2=M_Z^2}.
	\end{align}
	The field renormalisation constants of the vector boson fields follow from inserting the corresponding two-point functions into Eq.~(\ref{OSconditionsfields}),
	\begin{align}
		\begin{split}
		&\delta Z_W=-\widetilde{\text{Re}}\frac{\partial\Sigma_{T}^W(k^2)}{\partial k^2}\biggr\rvert_{k^2=M_W^2},\quad \delta Z_{VV}=-\widetilde{\text{Re}}\frac{\partial\Sigma_{T}^{VV}(k^2)}{\partial k^2}\biggr\rvert_{k^2=M_V^2}, \quad V=\{A,Z\}\\
		&\delta Z_{AZ}=-2\widetilde{\text{Re}}\frac{\Sigma_{T}^{AZ}(M_Z^2)}{M_Z^2}, \quad\quad\quad \delta Z_{ZA}=2\frac{\Sigma_{T}^{AZ}(0)}{M_Z^2}.
		\end{split}
	\end{align}
	The mass and field renormalisation constants calculated in the OS scheme as shown above are required for higher order perturbative computations if the collider process in fact is associated with corresponding on-shell external particles.
	Treating particles of the SM as gauge bosons, the Higgs or fermions unstable in calculations for a process, i.~e. as intermediate states, the most realistic scenario is approximated. In this case, a different mass scheme than the OS scheme has to be used, for example the complex-mass scheme (CMS). In this scheme the masses of the unstable particles are considered as complex quantities which were first introduced in Ref.~\cite{Denner:1999gp} for LO calculations with $W$ and $Z$ bosons treated as resonances and in Ref.~\cite{Denner:2005fg,Denner:2006ic} for NLO calculations and general SM masses. The imaginary part of the mass of a particle $P$ coming into play already in LO calculations is the result of Dyson summing the imaginary parts of self-energy corrections at the position of the singularity of the propagator $[k^2-M_P^2]^{-1}$ for the particle treated as stable. This means this propagator changes to
	\begin{align}
		G_P(k^2)=i\left[k^2-M_P^2+\Sigma_R (k^2)\right]^{-1}
		\label{generalProp}
	\end{align}
	due to these corrections. By means of $\text{Re}~\Sigma_R (M^2_P) = 0$ according to the OS scheme for real masses $M_P$ and $\text{Im}~\Sigma_R (M_P^2)=M_P\Gamma_P$ due to the optical theorem in one-loop order the propagator for unstable particles near the pole can be approximated by
	\begin{align}
		G_P(M_P^2)\sim i \left[k^2-M_P^2+iM_P\Gamma_P\right]^{-1}
		\label{unstableProp}
	\end{align}
	with decay width $\Gamma_P>0$ such that a Breit-Wigner-like shape of squared amplitudes with these resonances emerge.
	The mass $M_P$ and decay width $\Gamma_P$ are scheme-dependent quantities. In particular the pole can be defined in the OS scheme by fixing the root of the inverse propagator of Eq.~(\ref{generalProp}) at $k^2=M_{P,\text{OS}}^2$
	\begin{align}
	 M_{P,\text{OS}}^2-M_{P,0}^2+\text{Re}~\Sigma(M^2_{P,\text{OS}})=0
	 \label{OSpoleloc}
	\end{align}
	or at the actual complex location of the pole $k^2=\mu^2_P$
	\begin{align}
      \mu^2_P-M_{P,0}^2+\Sigma(\mu^2_{P})=0.
      \label{comlexpoleloc}
	\end{align}
	The resulting quantities $M_P$ and $\Gamma_P$ starting from Eq.~(\ref{OSpoleloc}) can be translated into the pole definition according to Eq.~(\ref{comlexpoleloc}) for the gauge bosons $V=\{W,Z\}$ \cite{Bardin:1988xt,Beenakker:1996kn},
	\begin{align}
		&M_V=\frac{M_{V,\text{OS}}}{\sqrt{1+\Gamma^2_{V,\text{OS}}/M^2_{V,\text{OS}}}}, &		\Gamma_V=\frac{\Gamma_{V,\text{OS}}}{\sqrt{1+\Gamma^2_{V,\text{OS}}/M^2_{V,\text{OS}}}}.
	\end{align}
	For NLO calculations considering unstable particles, i.~e. resonances, the renormalisation of the masses, fields and couplings has to be performed in the CMS.
	In this scheme the renormalised masses are fixed at the position of the pole $\mu_P^2$ according to Eq.~(\ref{comlexpoleloc}) such that these and the corresponding mass counterterms entering Eq.~(\ref{baremasses}) become complex quantities. In the same way the field renormalisation constants can get a complex phase. By substituting $M^2\rightarrow\mu^2$, $\delta M^2 \rightarrow \delta\mu^2$ and $\delta Z \rightarrow \delta \mathcal{Z}$ in order to emphasise the difference of real to complex-valued masses and renormalisation constants the renormalised transverse self-energies of the gauge bosons in the CMS read
	\begin{align}
	\begin{split}
		&\Sigma_{R,T}^W(k^2)=\Sigma_{T}^W(k^2)-\delta \mu^2_W+(k^2-\mu^2_W)\delta \mathcal{Z}_W, \\ &\Sigma_{R,T}^{ZZ}(k^2)=\Sigma_{T}^{ZZ}(k^2)-\delta \mu^2_Z+(k^2-\mu^2_Z)\delta \mathcal{Z}_{ZZ},\\
		&\Sigma_{R,T}^{AZ}(k^2)=\Sigma_{T}^{AZ}(k^2)+k^2\frac{1}{2}\delta \mathcal{Z}_{AZ}+(k^2-\mu^2_Z)\frac{1}{2}\delta \mathcal{Z}_{ZA},\\
		&\Sigma_{R,T}^{AA}(k^2)=\Sigma_{T}^{AA}(k^2)+k^2\delta \mathcal{Z}_{AA}.
	\end{split}
	\end{align}
	With a generalisation of the OS conditions, i.~e.
	\begin{align}
		\begin{split}
		&\Sigma_{R,T}^W(\mu_W^2)=0,\quad \Sigma_{R,T}^{ZZ}(\mu_Z^2)=0, \quad \Sigma_{R,T}^{AZ}(\mu_Z^2)=0, \quad \Sigma_{R,T}^{AZ}(0)=0,\\
		&\Sigma_{R,T}^{\prime W}(\mu_W^2)=0, \quad \Sigma_{ R,T}^{\prime ZZ}(\mu_Z^2)=0, \quad
		\Sigma_{R,T}^{\prime AA}(0)=0,
		\end{split}
	\end{align}
	the renormalisation constants for the CMS can be found,
	\begin{align}
	\begin{split}
	&\delta \mu^2_W=\Sigma_{T}^W(\mu^2_W), \quad \delta \mu^2_Z=\Sigma_{T}^{ZZ}(\mu^2_Z),\\
	&\delta \mathcal{Z}_{ZA}=\frac{2}{\mu^2_Z}\Sigma_{T}^{AZ}(0), \quad \delta \mathcal{Z}_{AZ}=-\frac{2}{\mu^2_Z}\Sigma_{T}^{AZ}(\mu^2_Z),\\
	&\delta \mathcal{Z}_{W}=-\Sigma_{T}^{\prime W}(\mu_W^2),\quad \delta \mathcal{Z}_{ZZ}=-\Sigma_{T}^{\prime ZZ}(\mu_Z^2), \quad \delta \mathcal{Z}_{AA}=-\Sigma_{T}^{\prime AA}(0),
	\end{split}
	\label{renormconstCMS}
	\end{align}
	whereby partial derivatives with respect to the arguments are denoted with a prime.
	Since the weak mixing angle is defined by the weak gauge boson masses according to Eq.~(\ref{weinbergangle}) this quantity becomes complex in the CMS as well. The renormalisation of the weak mixing angle for that case proceeds via
	\begin{align}
	\frac{\delta s_W}{s_W}=-\frac{c_W^2}{2s_W^2}\left(\frac{\delta \mu_W^2}{\mu_W^2}-\frac{\delta \mu_Z^2}{\mu_Z^2}\right).
	\end{align}
	For the actual computation of the renormalisation constants in the CMS self-energies entering Eq.~(\ref{renormconstCMS}) have to be evaluated with complex arguments, which in general requires analytical continuation of the two-point functions in the momentum variable to the Riemann sheet \cite{Passarino:2010qk}. A method to achieve this, practical for automated NLO EW calculations, was presented in Ref.~\cite{Frederix:2018nkq} coming along with the corresponding \texttt{MadLoop} implementation. Another approach according to Refs.~\cite{Beenakker:1996kn,Denner:2005fg} is the expansion of the self-energies around real arguments assuming $\Gamma_P/M_P \ll 1$ for unstable particles of the SM and thereby retaining one-loop accuracy of the correction factor to the amplitudes.
	With the approximation $k^2-\mu_P^2 \approx iM_P\Gamma_P$ in the vicinity of the pole at $k^2\approx M_P^2$ the expanded self-energies read
	\begin{align}
		\Sigma (\mu^2_P)=\Sigma (M_P^2)- iM_P\Gamma_P\Sigma^{\prime}(M_P^2)-ia\frac{\Gamma_P}{M_P}.
	\end{align}
	The last term with constant $a$ originates from contributions of self-energy diagrams with a branch point at $k^2=\mu_P^2$ of charged fields with photon exchange and has to be omitted for neutral fields. Thus, the gauge boson self-energies for complex pole masses $\mu_P^2=M_P^2-iM_P\Gamma_P$ can be obtained,
	\begin{align}
	\begin{split}
		&\Sigma_{T}^W(\mu^2_W)=\Sigma_{T}^W(M^2_W)+(\mu_W^2-M_W^2)\Sigma^{\prime W}_T(M_W^2)+c_T^W+\mathcal{O}(\alpha^3)\\
		&\Sigma_{T}^{ZZ}(\mu^2_Z)=\Sigma_{T}^{ZZ}(M^2_Z)+(\mu_Z^2-M_Z^2)\Sigma^{\prime ZZ}_T(M_Z^2)+\mathcal{O}(\alpha^3)\\
		&\frac{1}{\mu_Z^2}\Sigma_{T}^{AZ}(\mu^2_Z)=\frac{1}{\mu_Z^2}\Sigma_{T}^{AZ}(0)+\frac{1}{M_Z^2}\Sigma_{T}^{AZ}(M_Z^2)-\frac{1}{M_Z^2}\Sigma_{T}^{AZ}(0)+\mathcal{O}(\alpha^2)
		\end{split}
		\label{selfenCMS}
	\end{align}
	with constant
	\begin{align}
		c_T^W=\frac{i\alpha}{\pi}M_W\Gamma_W.
	\end{align}
	Inserting Eq.~(\ref{selfenCMS}) into Eq.~(\ref{renormconstCMS}) leads to the renormalisations constants in the CMS. This approximation for example is used for the EW renormalisation in the one-loop providers \texttt{OpenLoops} \cite{Cascioli:2011va,Buccioni:2019sur} and \texttt{Recola} \cite{Actis:2012qn,Actis:2016mpe,Denner:2017wsf}.
	
	\subsection{Electroweak-input schemes}
	For fixing the numerical value of the electromagnetic coupling $\alpha$ at LO and NLO EW basically three different schemes exist, depending on the scale of the coupling and input quantities used for its definition, i.~e. the $\alpha (0)$, the $\alpha(M_Z^2)$ and the $G_{\mu}$ scheme.\\
	The electroweak-input scheme for $\alpha$ determined in the Thomson limit, i.~e. for scales $Q^2\rightarrow 0$, is the standard QED definition and corresponds to the value of the fine-structure constant $\alpha (0)\approx 1/137$. The $\alpha (0)$ scheme is thus primarily used if `on-shell external' photons enter the considered processes. Photons declared as such denote those which enter the LO hard process with virtuality $Q_{\gamma}^2=0$. Conversely, for later reference, `off-shell external' photons represent legs of the LO hard process with virtuality $Q_{\gamma}^2\neq0$.
	
	Together with the definition of $\alpha$, the scale of the coupling corresponding to the photon virtuality $Q_{\gamma}^2$ must be considered for the renormalisation of the electric charge. For on-shell photon couplings, i.~e. virtualities $Q_{\gamma}^2\rightarrow0$, $\delta Z_{e,\alpha (0)}$ can be derived from renormalising the vertex function of the $\gamma f\bar{f}$ vertex for on-shell external particles in the Thomson limit and using Gordon and Ward identities \cite{Denner:1991kt,Bohm:2001yx}, yielding
	\begin{align}
	\delta Z_{e,\alpha (0)}=\frac{\delta e}{e}=-\frac{1}{2}\left(\delta Z_{AA}+ \frac{s_W}{c_W}\delta Z_{ZA}\right)=\frac{1}{2}\frac{\partial\Sigma_{T}^{AA}(k^2)}{\partial k^2}\biggr\rvert_{k^2=0}-\frac{s_W}{c_W}\frac{\Sigma_{T}^{AZ}(0)}{M_Z^2}
	\label{chargerenoOS}
	\end{align}
   	The charge renormalisation constant is fixed in the OS scheme in this equation by using the OS conditions of Eq.~(\ref{OSconditions}) and (\ref{OSconditionsfields}) for the input quantities. However it gets an imaginary part if computed in the CMS, substituting $M^2_P\leftrightarrow\mu_P^2$. By the definition of the bare charge as a real quantity in Eq.~(\ref{baremasses}) the renormalised charge becomes complex whereby its imaginary part is fixed by the one of $\delta Z_e$. This imaginary part however is formally of two-loop order since the self-energies entering Eq.~(\ref{chargerenoOS}) at $k^2=0$ do not lead to imaginary parts of real internal masses. Dropping this part the charge renormalisation constant in one-loop accuracy in the CMS can be rewritten as
   	\begin{align}
   		\delta Z_{e,\alpha (0)}=\text{Re}\left[\frac{1}{2}\frac{\partial\Sigma_{T}^{AA}(k^2)}{\partial k^2}\biggr\rvert_{k^2=0}-\frac{s_W}{c_W}\frac{\Sigma_{T}^{AZ}(0)}{\mu_Z^2}\right]
   	\end{align}
   	whereby $s_W$ and $c_W$ are calculated with complex masses $\mu_Z^2$ and $\mu_W^2$.
   	
   	For $\alpha (Q^2)$ computed at high scales, as the EW scale $Q^2\approx M_Z^2$, light fermion mass effects appear. These come from large logarithms of ratios of the fermion masses and $Q^2\gtrsim M^2_Z$ related to the running of the coupling from $Q^2=0$ to that high scale. For light fermions $f\neq t$, excluding bosonic contributions and first of all for the case that external off-shell photons are absent at LO, the running of $\alpha$ can be described in terms of the vacuum polarisation $\Pi^{AA}_{f\neq t}$ by the quantity
   	\begin{align}
   		\Delta \alpha (M_Z^2)=\Pi^{AA}_{f\neq t}(0)-\Pi^{AA}_{f\neq t}(M_Z^2)= \frac{\alpha (0)}{3\pi} \sum_{f\neq t}N_{C,f}Q_f^2\left[\ln \frac{M_Z^2}{m_f^2}-\frac{5}{3}\right]
   		\label{delalphasimple}.
   	\end{align}
   	$N_{C,f}$ thereby denotes the colour degrees of freedom and $Q_f$ the charge of the corresponding fermions. Furthermore, mass-suppressed terms of $\mathcal{O}(m_f^2/M_Z^2)$ have been omitted.
   	The light fermion mass logarithms occuring in (\ref{delalphasimple}) can be resummed as \cite{Marciano:1979yg,Sirlin:1983ys}
   	\begin{align}
   		\alpha (M_Z^2)=\frac{\alpha (0)}{1-\Delta \alpha (M_Z^2)}.
   		\label{alphaMZ}
   	\end{align}
    By choosing the $\alpha(M_Z^2)$ scheme, practically, $\alpha(M_Z^2)$ is a user-defined input which has to be set such that the resummation of all light fermion mass logarithms coming from the EW corrections is contained in that number. From this input the quantity $ \Delta \alpha (M_Z^2)$ can be derived by means of Eq.~(\ref{alphaMZ}) and used for the computation of renormalisation constants.
   	The EW renormalisation of the coupling in this scheme proceeds as
   	\begin{align}
   	\frac{\delta \alpha(M_Z^2)}{\alpha(M_Z^2)}= \frac{\delta \alpha(0)}{\alpha(0)}-\Delta \alpha (M_Z^2)
   	\end{align}
   	and thus the corresponding renormalisation constant of the charge can be deduced,
   	\begin{align}
   	\delta Z_{e,\alpha (M_Z^2)}=\delta Z_{e,\alpha (0)}-\frac{\Delta \alpha (M_Z^2)}{2}.
   	\label{chargerenoMZ}
   	\end{align}
   	By this, it is obvious that the first term of $\delta Z_{e,\alpha (0)}$ in Eq.~(\ref{chargerenoOS}) depending on the vacuum polarisation $\Pi^{AA}_{f\neq t}(0)=\Sigma^{\prime AA}_T(0)$ which contains the fermion mass logarithms is cancelled by the corresponding term entering Eq.~(\ref{chargerenoMZ}) via $\Delta \alpha (M_Z^2)/2$.\\
   	In the case that external off-shell photons enter the hard scattering process the term $ \Pi^{AA}_{f\neq t}(M_Z^2) $ receives divergent contributions from massless fermion loops and $\Delta \alpha (M_Z^2)$ is rewritten in the dimensionally regularised form \cite{Kallweit:2017khh}
   	\begin{align}
   		\Delta^{(\text{reg})} \alpha (M_Z^2)=\frac{\alpha }{2\pi}\gamma_{\gamma}\left[\frac{C_{\epsilon}}{{\epsilon}}+\ln \frac{\mu^2_{D}}{M_Z^2}+\frac{5}{3}\right]- \frac{\alpha}{3\pi} \sum_{f\neq t}N_{C,f}Q_f^2\left[\ln \frac{m_f^2}{M_Z^2}+\frac{5}{3}\right]
   		\label{delalpha}
   	\end{align}
   	where $0<m_f<M_Z$. The first term in this equation depends on the regularisation scale $\mu_{D}$ and a $1/\epsilon$ pole with a scheme-dependent constant $C_{\epsilon}$. It comes along with a factor $\gamma_{\gamma}$ which is the anomalous dimension of the photon accounting for all massless fermions in loop contributions,
   	\begin{align}
   		\gamma_{\gamma}=-\frac{2}{3}\sum_{f\neq t}N_{C,f}Q_f^2.
   	\end{align}
   	The $1/\epsilon$ pole of this term originates from the inclusion of $\gamma\rightarrow f\bar{f}$ splittings into the evolution of the coupling inducing collinear singularities for massless fermions.
   	Hence, in the case if fermions generally are treated massive, $\gamma_{\gamma}$ becomes zero, the first term of Eq.~(\ref{delalpha}) drops and $\Delta \alpha (M_Z^2)$ becomes finite. 
   	If massless fermions however are taken into account, the singularity in the first term in Eq.~(\ref{delalpha}) which enters $\delta Z_{e,\alpha (M_Z^2)}$ by Eq.~(\ref{chargerenoMZ}) is cancelled in the renormalisation factor of the matrix elements only by corresponding counterterms coming from the photon PDF renormalisation or analogous virtual contributions from final state $\gamma\rightarrow f \bar{f}$ splittings. A more detailed discussion on this subject follows in Sec.~\ref{secExternalphotons}.
   	For the same reasons as explained above for the $\alpha (0)$ scheme, the renormalised charge can be either deduced in the OS scheme or in the CMS. In the latter case, the real part of the renormalised charge  has to be taken accordingly.
   	In summary, the $\alpha (M_Z^2)$ scheme due to its derivation is an optimal choice for computations at high scales and with enhanced QED effects.
   	
   	Another electroweak-input scheme at high scales is the $G_{\mu}$ scheme defining the coupling $\alpha=\alpha_{G_\mu}$ via the input parameters $G_{\mu}$, the Fermi constant, and the gauge boson masses.
   	The $G_{\mu}$ constant is extracted experimentally from the coupling in the muon decay in Fermi theory and thus can be related to squared matrix elements with $W$ boson exchange for low energies as
   	\begin{align}
   	\biggr\lvert\frac{8}{\sqrt{2}}G_{\mu}\biggr\rvert^2=\biggr\lvert\frac{g_2^2}{\mu^2_W}\biggr\rvert^2.
   	\end{align}
   	The $SU(2)$ coupling $g_2$, introduced in section \ref{secStandardmodel}, with the relation of Eq.~(\ref{weinbergangle}) leads to the definition of the coupling in the $G_{\mu}$ scheme at LO
   	\begin{align}
   		\alpha_{G_\mu}=\frac{\sqrt{2}}{\pi}G_{\mu}\big\lvert \mu^2_W s_W^2\big\rvert=\frac{\sqrt{2}}{\pi}G_{\mu}\biggr\lvert \mu^2_W \left(1-\frac{\mu^2_W}{\mu^2_Z}\right) \biggr\rvert.
   		\label{alphaGmu}
   	\end{align}
   	At NLO $\alpha_{G_\mu}$ depends on $\Delta r^{(1)}$ which represents the NLO EW corrections to the muon decay \cite{Sirlin:1980nh,Denner:1991kt,Hollik:1988ii,Marciano:1980pb} as
   	\begin{align}
   		\alpha_{G_\mu}=\alpha(0)\big\lvert 1+\Delta r^{(1)}\big\rvert
   	\end{align}
   	whereby
   	\begin{align}
   	\begin{split}
   	  \Delta r^{(1)}=&\Pi^{AA}(0)-\frac{c_W^2}{s_W^2}\left(\frac{\Sigma^{ZZ}_T(\mu^2_Z)}{\mu^2_Z}-\frac{\Sigma^{W}_T(\mu^2_W)}{\mu^2_W}\right)+\frac{\Sigma^{W}_T(0)-\Sigma^{W}_T(\mu^2_W)}{\mu^2_W} \\
   	&+2\frac{c_W}{s_W}\frac{\Sigma^{AZ}_T(0)}{\mu^2_W}+\frac{\alpha (0)}{4\pi s_W^2}\left(6+\frac{7-4s_W^2}{2s_W^2}\ln c_W^2\right).
   	\end{split}
   	\end{align}
   	The charge renormalisation constant for this scheme reads
   	\begin{align}
   		\delta Z_{e,\alpha_{G_{\mu}}}=\delta Z_{e,\alpha (0)}-\frac{1}{2}\Delta r^{(1)}
   		\label{chargerenoGm}
   	\end{align}
   	In the same way as explained in the context of Eq.~(\ref{chargerenoMZ}) the term proportional to $\Pi^{AA}_{f\neq t}(0)$ containing light fermion mass logarithms drops out for the renormalisation constant $\delta Z_{e,\alpha_{G_{\mu}}}$ in (\ref{chargerenoGm}).
   	Since $\alpha_{G_\mu}$, similar to $\alpha (M_Z^2)$, is defined at an EW scale, it incorporates the running of the coupling such that $\Delta r$ can be written in a decomposed form as \cite{Denner:2019vbn}
   	\begin{align}
   		\Delta r=\Delta \alpha (M^2_Z)+\ldots
   	\end{align}
   	Thus, all effects concerning the fermion mass singularities in the renormalisation of the coupling and their regularisation prescriptions as explained above for the $\alpha (M_Z^2)$ scheme apply in the same way for the $G_{\mu}$ scheme. In addition, as for the other schemes, the imaginary part of the charge renormalisation constant can be omitted with respect to NLO calculations.
   	
   	The ${G_{\mu}}$ scheme includes the renormalisation of the weak mixing angle $s_W$. It is therefore preferred over the $\alpha (M_Z)$ scheme at high scales if an appropriate description of process observables with respect to $SU(2)$ interactions is phenomenologically needed.
	\subsection{External photons}
	\label{secExternalphotons}
	EW corrections for processes with external photons at Born-level are intricate from the point of view that the photons' virtuality $Q^2_{\gamma}$ which defines its nature, i.~e. `on-shell' for $Q^2_{\gamma}=0$ or `off-shell' for $Q^2_{\gamma}\neq0$, have to be treated consistently throughout all schemes.
	
	For each external photon which couples to the hard process squared amplitudes at NLO EW receive renormalisation factors containing coupling and photon wave function counterterms according to
	\begin{align}
		\delta K_{\gamma}=\frac{\delta \alpha}{\alpha}+\delta Z_{AA}+\delta Z_{\gamma,\text{off}}=2\delta Z_e+\delta Z_{AA}+\delta Z_{\gamma,\text{off}}\quad,
	\end{align}
	whereby $\delta Z_{\gamma,\text{off}}$ represents an extra counterterm coming from virtual contributions associated with $\gamma\rightarrow f\bar{f}$ splittings. Due to the absence of these splittings at NLO for on-shell photons this term becomes zero, i.~e. $ \delta Z_{\gamma,\text{off}}=0 $, and the photon couples to the hard process with virtuality $Q_{\gamma}=0$ corresponding to the coupling $\alpha (0)$. Concerning only light-fermionic contributions the overall renormalisation factor vanishes for this photon type according to
	\begin{align}
		\delta K_{\gamma}^{\text{(on)}}\big\rvert_{\text{light}}=\left[2\delta Z_{e,\alpha (0)}+\delta Z_{AA}\right]_{\text{light}}=0
	\end{align}
	using the charge renormalisation constant defined in Eq.~(\ref{chargerenoOS}).
	In contrast to this exact cancellation of light fermion mass singularities of $\alpha (0)$ coupling and photon field counterterms for on-shell photons, amplitudes with off-shell photons require an additional renormalisation prescription. Since the coupling is computed using an EW input scheme at high energy scales for off-shell photon couplings, the running of the coupling eventually leaves photon wave function counterterm singularities uncancelled. This is indicated by the $1/\epsilon$ pole in $\Delta^{(\text{reg})} \alpha (M_Z^2)$ of Eq.~(\ref{delalpha}).
	These light fermion mass singularities are cancelled only by the collinear singularities contained in $\delta Z_{\gamma,\text{off}}$ which represent photon PDF counterterms or analogous terms in virtual contributions originating from $\gamma\rightarrow f\bar{f}$ splittings of final state photons. The renormalisation factor for off-shell external photons in the $\alpha (M_Z^2)$ scheme can thus be written as
	\begin{align}
		\delta K_{\gamma}^{\text{(off)}}\big\rvert_{\text{light}}=\left[2\delta Z_{e,\alpha (M_Z^2)}+\delta Z_{AA}\right]_{\text{light}}+\delta Z_{\gamma,\text{off}}=-\Delta^{(\text{reg})} \alpha (M_Z^2)+\delta Z_{\gamma,\text{off}}\quad.
	\end{align}
	Analogously, the renormalisation factor is rendered finite through $Z_{\gamma,\text{off}}$ if the $G_{\mu}$ scheme is chosen.
	In this way, depending on the type of the external photon at Born-level, the renormalisation factors get different finite remainders. It is thus required for one-loop providers as \texttt{OpenLoops} to keep track of the type of the external photon associated with the considered process at Born-level.
	Concretely, in the case of \texttt{OpenLoops} external photons are labeled as off-shell photons with particle-ID (PID) `$-2002$' and on-shell photons with PID `$2002$'.
	
	To include on-shell photons in a higher order prediction however is non-trivial from the phenomenological point of view. Collider processes at NLO EW typically are studied at high energy scales.
	However, on-shell photons are defined at low energy scales and thus the process has to be described with external photon fields and couplings at two different scales. This contradiction can be overcome only by the condition that the photon as a physical on-shell final-state object must emerge from a fragmentation process and never from the pure hard process \cite{Frederix:2016ost}. Incorporating this into an NLO EW computation would demand an application of photon fragmentation functions and photon jet definitions, rendering fully observable photons in an IR safe way.
	
	Another technical issue for the computation of collider observables which has to be adressed if external photons enter the hard process is that real and virtual amplitudes have to be computed at the same order in $\alpha$ defined by a specific input scheme for the IR subtraction scheme to be consistent.
	The complication comes by the fact that due to the renormalisation prescriptions explained above
	the coupling $\alpha$ of each external photon to the hard process
	at the same time has to correspond to the specific photon type. Due to this, for the evaluation of squared amplitudes \texttt{OpenLoops} automatically rescales the couplings in the following way.
	By default, the coupling of an on-shell photon will be changed to $\alpha(0)$ and that of an off-shell photon to $\alpha_{G_\mu}$ if not chosen already at a high scale, e.~g. in the case of the $\alpha(M_Z^2)$ scheme. If no photons are present in the Born process, in order to not spoil the IR cancellation, \texttt{OpenLoops} supplies to register `unresolved' photons with PID `$22$' describing a radiated photon at NLO EW. The photon-coupling $\alpha$ is left unchanged at the value which is computed with the EW input scheme chosen by the user.
	However, taking external photons with respect to the hard process into account the freedom to choose an arbitrary EW input scheme for the NLO calculation is restricted since the number of external photons present at Born-level is not conserved for the corresponding real process due to possible $\gamma \rightarrow f\bar{f}$ splittings. For example, for processes with off-shell external photons this forbids the choice $\alpha=\alpha(0)$ since otherwise the order in $\alpha(0)$ is not conserved in the real amplitude relative to the factorising Born process for which the off-shell photon couples with $\alpha$ at high energy scales. Consequently, the NLO components are not of the same order in $\alpha(0)$ in the subtraction scheme.
	For an overview the different prescriptions for each photon type are given in Table \ref{photontypes}.
	\begin{table}
		\centering
		{	\onehalfspacing
		\begin{tabularx}{0.9\linewidth}{l|c|c|c|c}
			photon type & \texttt{OpenLoops} PID & $\gamma \rightarrow f\bar{f}$ & EW input scheme & $\Delta \alpha$\\
			\hline
			`off-shell'&$-2002$&on& $\alpha (M^2_Z)$, ${G_{\mu}}$ & $\Delta^{(\text{reg})} \alpha (M_Z^2)$\\
			`on-shell'&$2002$&off&$\alpha (0)$&$\Delta \alpha (M_Z^2)$\\
			`unresolved'&$22$&off& $\alpha (0)$, $\alpha (M^2_Z)$, ${G_{\mu}}$&$\Delta \alpha (M_Z^2)$
		\end{tabularx}}
	\caption{Technical prescriptions \texttt{OpenLoops} photon PIDs, information on existing splittings $\gamma \to f \bar{f}$ at NLO EW, EW input schemes and renormalisation constants with respect to different photon type definitions, i.~e. `off-shell', `on-shell' and `unresolved' (cf. definitions given in the main text)}
	\label{photontypes}
	\end{table}
	\section{Regularisation of infrared divergences}
	\label{secRegularisationIR}
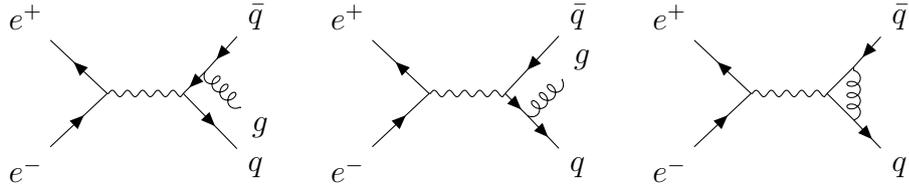
\begin{figure}
		\centering
			\begin{tikzpicture}
		\begin{feynman}
		\vertex(a);
		\vertex[above left=1cm of a] (i1){$e^+$};
		\vertex[below left=1cm of a] (i2){$e^-$};
		\vertex[right=1cm of a] (b);
		\vertex[above right=1cm of b] (f1){$\bar{q}$};
		\vertex[below right=1cm of b] (f2){${q}$};
		\vertex[above right=1em of b] (ii1);
		\vertex[below right=1.6em of ii1] (ii2){$g$};
		\diagram {
			(i2) -- [fermion,small] (a) -- [fermion,small] (i1),
			(a) -- [photon] (b),
			(f1) -- [fermion,small] (ii1) -- [fermion,small] (b) -- [fermion,small] (f2),
			(ii1) -- [gluon,small] (ii2),
		};
		\end{feynman}
		
		\end{tikzpicture}
		\quad
		\begin{tikzpicture}
		\begin{feynman}
		\vertex(a);
		\vertex[above left=1cm of a] (i1){$e^+$};
		\vertex[below left=1cm of a] (i2){$e^-$};
		\vertex[right=1cm of a] (b);
		\vertex[above right=1cm of b] (f1){$\bar{q}$};
		\vertex[below right=1cm of b] (f2){${q}$};
		\vertex[below right=1em of b] (ii1);
		\vertex[above right=1.6em of ii1] (ii2){$g$};
		\diagram {
			(i2) -- [fermion,small] (a) -- [fermion,small] (i1),
			(a) -- [photon] (b),
			(f1) -- [fermion,small]  (b)  -- [fermion,small] (ii1) -- [fermion,small] (f2),
			(ii1) -- [gluon,small] (ii2),
		};
		\end{feynman}
		\end{tikzpicture}
		\quad
			\begin{tikzpicture}
		\begin{feynman}
		\vertex(a);
		\vertex[above left=1cm of a] (i1){$e^+$};
		\vertex[below left=1cm of a] (i2){$e^-$};
		\vertex[right=1cm of a] (b);
		\vertex[above right=1cm of b] (f1){$\bar{q}$};
		\vertex[below right=1cm of b] (f2){${q}$};
		\vertex[below right=1.2em of b] (ii1);
		\vertex[above right=1.2em of b] (ii2);
		\diagram* {
			(i2) -- [fermion,small] (a) -- [fermion,small] (i1),
			(a) -- [boson] (b),
			(f1) -- [fermion,small] (ii2) -- [plain] (b) -- [plain] (ii1) -- [fermion,small] (f2),
			(ii1) -- [gluon,small] (ii2),
		};
		\end{feynman}
		\end{tikzpicture}
		\caption{Contributing diagrams of the NLO QCD corrections to $ e^+e^-\rightarrow q\bar{q}$}
		\label{figrealem}
	\end{figure}

	In order to increase the theoretical precision of an observable $O$ which represents a prediction for
	a physical quantity at weak coupling can be expanded in powers of said coupling as
	\begin{align}
		O=O^{(0)}+\alpha O^{(1)}+\alpha^2 O^{(2)}+\ldots \quad.
		\label{perturbationtheory}
	\end{align}
	Beyond the leading order infrared (IR) divergences occur due to real emission and virtual loop corrections which cancel each other order by order in perturbation theory. The real corrections originate from splittings into pairs of particles with at least one massless final state each whereas the virtual corrections come from loop diagrams with exchange of virtual particles between external state legs. For the process $e^+e^-\rightarrow q\bar{q}$ the corresponding NLO QCD real emission and virtual loop diagrams are shown in Figure \ref{figrealem}. Infrared divergences occur if the emission of real or virtual massless particles induces a collinear and/or soft splitting, i.~e. the propagator in the amplitudes takes the form
	\begin{align}
		\sim [k_i^0k_j^0(1-\cos\theta_{ij})]^{-1}
	\end{align}
	for a splitting particle pair $(i,j)$ and diverges if the polar angle between these particles $\theta_{ij}$ becomes zero and/or the energy $k^0_i$ or $k^0_j$ vanishes. Quantifying the higher order corrections according to finite coefficients of the expansion in Eq.~(\ref{perturbationtheory}) thus requires one to impose certain IR regularisation methods. In the last decades, a lot of progress has been achieved in developing such methods by making use of the Kinoshita-Lee-Nauenberg (KLN) theorem \cite{Kinoshita:1962ur, Lee:1964is}. It states that infrared divergences cancel in any unitary theory by summing over all states which are degenerate due to some resolution criteria. This implies that the infrared divergences of real and virtual corrections cancel in the sum.

	For the calculation of cross sections of particle processes the most common approach to go beyond the leading order in perturbation theory is to make use of dimensional regularisation \cite{tHooft:1972tcz}. This method, which analytically cancels IR divergences coming from loop and real emission corrections, requires an integration over dimensions of non-integer number, i.~e. $D=4-2\varepsilon$. Hence, applied in a purely analytical way, it is not suitable for observables exclusive in arbitrary kinematical quantities obtained by numerical integration means, such as Monte Carlo generators. For this purpose appropriate subtraction schemes based on the KLN theorem must be applied requiring the real and the virtual component to be integrated over different integer phase-space dimensions.
	There are subtraction schemes which rely on non-local counterterms as for example in the case of $q_T$-subtraction \cite{Catani:2007vq,Grazzini:2017mhc} or $N$-jettiness \cite{Gaunt:2015pea,Boughezal:2015dva} described by phase-space slicing methods. Due to the cancellation of the singularities in a global way these schemes are easily extendable in the theoretical precision, i.~e. to orders beyond NLO in perturbation theory. However since the cancellation happens at the numerical level the accuracy of the higher order prediction of an observable in these schemes is limited.
	Schemes with high numerical accuracy for NLO contributions  rely on the approximation of a cross section for a $2\rightarrow n$ scattering process by
	\begin{align}
		\sigma_{\text{NLO}}=\int d\Phi_n \mathcal{B}(\Phi_n) +\int d\Phi_{n+1}\left[\mathcal{R}(\Phi_{n+1})-S({\Phi}_{n+1})\right]+\int d\Phi_n\left[\mathcal{V}(\Phi_n)+I(\Phi_{n})\right]
		\label{KLNtheorem}
	\end{align}
	where $\mathcal{B}$ denotes the Born squared amplitude, $\mathcal{R}$ the real squared amplitude and $\mathcal{V}$ the virtual loop-tree interfering amplitudes, respectively.
	The subtraction terms $S(\Phi_{n+1})$ constructed to cancel the IR singularities of $\mathcal{R}$ at the non-integrated level take the integrated form $I(\Phi_{n})$ in terms of powers of $1/\varepsilon$ such that the poles appearing due to dimensional regularisation of the virtual contributions $\mathcal{V}(\Phi_n)$ are cancelled analytically.
	Formally, Eq.~(\ref{KLNtheorem}) describes the foundation of all local subtraction schemes which require local counterterms $S(\Phi_{n+1})$ for the cancellation of the real singularities. The most established schemes in this context are the Catani-Seymour (CS) subtraction \cite{Catani:1996vz,Catani:2002hc,Gleisberg:2007md} which uses universal dipole factors in the approximation of the counterterms and the Frixione-Kunszt-Signer (FKS) scheme \cite{Frixione:1995ms,Frixione:1997np,Frederix:2009yq} which partitions the phase-space into kinematic regions such that in each
	of these there is at most one collinear and one soft singularity. One significant benefit of the CS subtraction scheme lies in the generic construction of the real $n+1$ phase-space which leads to a large potential of efficiency improvements in the integration measures. This is different to the phase-space construction in the FKS scheme where the $n+1$ phase-space is restricted to a parametrisation in variables for which soft and collinear regions are well-defined.
	This is the main drawback of this scheme which is disadvantageous for example with respect to the regularisation of PDF singularities as in the case of electron PDFs with an appropriate phase-space mapping for the real emission process as outlined in Sec.~\ref{parametrisationPDFsing}. However with the emphasis on the subtraction the construction of the collinear and/or soft limits is simplified since this scheme relies on
	a $1 \rightarrow 2$ particle mapping for the description of the splitting pair
	instead of a $3\rightarrow2$ mapping outgoing from the real phase-space as in the CS scheme. In that way, in the FKS scheme the underlying Born configuration is basis of the construction such that the number of subtraction terms per real configuration is minimised. The cancellation of the FKS terms in the respective limits hence proceeds in an efficient way during phase-space integration which especially is of importance if the considered process is of high jet multiplicity. Also, if different coupling powers of $\alpha_s$ and $\alpha$ of the factorising Born processes for the case of NLO computations in mixed coupling expansions are taken into account this is a crucial point for the bookkeeping of subtraction terms on which more details follow in Sec.~\ref{mixedcouplingsSec}.
	
	All NLO calculations for the results of this thesis are performed using the FKS subtraction scheme on which the automation of NLO corrections in the MC event generator \texttt{WHIZARD} is based on.
	\chapter{Frixione-Kunszt-Signer subtraction for NLO SM corrections}
	\label{secFKSscheme}
	In this chapter the Frixione-Kunszt-Signer (FKS) subtraction scheme in a general form covering NLO QCD, EW and QCD-EW mixed corrections is presented. For clarification, speaking from `(QCD-EW) mixed corrections' in this thesis should not be mistaken with the same term used in literature in the context of $\mathcal{O}(\alpha_s\alpha)$ corrections, e.~g. by Refs.~\cite{Bonciani:2021zzf,Buonocore:2021rxx}. Here, this term is used for simplicity, denoting contributions at NLO level which for two different Born coupling orders can be seen either as NLO correction in $\alpha$ or in $\alpha_s$.
	
	The formulas and expressions used in this chapter are based on Refs.~\cite{Frixione:1995ms,Frixione:1997np,Frederix:2009yq,Alwall:2014hca,Frixione:2007vw,Alioli:2010xd} concerning pure QCD corrections with the extension to EW and mixed corrections given by Ref.~\cite{Barze:2012tt,Frederix:2018nkq}. Literature with respect to details of the \texttt{WHIZARD} implementation of the FKS scheme \cite{Weiss:2017qbj,ChokoufeNejad:2017rag,Rothe:2021sml,WhizardNLO} is considered in addition.
	
	The notation used throughout this chapter as well as the construction of the phase-space for initial-state radiation for both cases, massless and massive emitters, is given in Sec.~\ref{secFKSphasespace}. The FKS subtraction scheme in a mixed coupling expansion is described in detail in Sec.~\ref{mixedcouplingsSec}.
	\section{FKS phase-space and notation}
	\label{secFKSphasespace}
	For a $2\rightarrow n$ scattering process the Born phase-space is denoted by
	\begin{align}
		\bar{\Phi}_n=\{\bar{k}_{\oplus},\bar{k}_{\ominus},\bar{k}_{1+n_I},\ldots,\bar{k}_{N}\}
		\label{PSborn}
	\end{align}
	where $\bar{k}_{\oplus}$ and $\bar{k}_{\ominus}$ represent the initial state momenta of the Born process and $\bar{k}_l$, $l=1+n_I,\ldots,N$ the final state momenta with $n_I$ the total number of initial-state particles and $N=n+n_I$ the total number of external states, respectively. The phase-space of the real-emission process is parametrised by
	\begin{align}
		{\Phi}_{n+1}=\{\bar{\Phi}_n,{\Phi}_\text{rad}\}\quad,
		\label{PSreal}
	\end{align}
	and rewritten in external state momenta given by
	\begin{align}
		{\Phi}_{n+1}=\{{k}_{\oplus},{k}_{\ominus},{k}_{1+n_I},\ldots,k_{N+1}\}
	\end{align}
	with $k_{n+n_I+1}$ an additional final state momentum, i.~e. the momentum of the radiated particle. The momentum conservation rule applies to the phase-spaces $\bar{\Phi}_n$ and ${\Phi}_{n+1}$ as
	\begin{align}
		&\bar{k}_{\oplus}+\bar{k}_{\ominus}=\sum_{l=1+n_I}^{N}\bar{k}_l &{k}_{\oplus}+{k}_{\ominus}=\sum_{l=1+n_I}^{N}{k}_l
		\label{momentumconservation}\quad.
	\end{align}
	The parametrisation of the radiation phase-space $\Phi_{\text{rad}}$ in terms of the FKS variables follows according to
	\begin{align}
		\Phi_{\text{rad}}\rightarrow \{\xi,y,\phi\}
	\end{align}
	where $ \xi \in [0,\xi_{\text{max}}]$ denotes the emitted particles' energy defined as
	\begin{align}
		\xi =\frac{2k^0_{N+1}}{\sqrt{s}} \quad.
	\end{align}
	The maximal energy $\xi_{\text{max}}$ is defined through momentum conservation rules. While for final-state radiation (FSR) $\xi$ is trivially restricted by the energy of the emitter particle, the derivation of $\xi_{\text{max}}$ for ISR from massless or massive emitters is rather involved. Explicit definitions are given in Sec.~\ref{secMasslessISem} and \ref{secMassiveIS}.
	
	$y \in [-1,1]$ is the cosine of the polar angle between the splitting particles,
	\begin{align}
	y=\cos \theta_{ij} \quad,
	\end{align}
	where $i$ and $j$ denote the partons of the pair  $(i,j)$ with momenta $k_i$ and $k_j$ potentially inducing collinear and/or soft singularities in the squared matrix elements and for which the concrete definition is given in Sec.~{\ref{secSingularregions}}.
	$\phi \in [0,2\pi]$ denotes the azimuthal angle between $i$ and $j$ in the plane perpendicular to the beam axis.
	
	For the integration of the real squared amplitudes and subtraction terms over the real emission phase-space according to Eq.~(\ref{KLNtheorem}) it is important for the FKS scheme to factorise the phase-space element $d{\Phi}_{n+1}$ into
	\begin{align}
		d{\Phi}_{n+1}=d{\bar{\Phi}}_{n}d\Phi_{\text{rad}}=d{\bar{\Phi}}_{n} d\xi dy d\phi \mathcal{J}(\xi,y,\phi) \quad.
		\label{n+1PSparametr}
	\end{align}
	The Born-like phase-space measure $d{\bar{\Phi}}_{n}$ thereby can be written as
	\begin{align}
		d{\bar{\Phi}}_{n}=\prod^{N}_{l=1+n_I}\frac{d^3\bar{k}_l}{2\bar{k}^0_l (2\pi)^3}(2\pi)^4\delta^4\left(\bar{k}_{\oplus}+\bar{k}_{\ominus}-\sum_{l=1+n_I}^{N}\bar{k}_l\right)\quad.
	\end{align}
	The Jacobian $\mathcal{J}(\xi,y,\phi)$ in Eq.~(\ref{n+1PSparametr}) depends on the explicit definitions of the $n$ and $n+1$ phase-spaces which differ for ISR and FSR.
	
	Since the aim of this thesis is the automation of NLO EW corrections to processes in a manner which is independent of the collider beam setup, i.~e. hadron or lepton collisions, a special focus lies in the initial-state phase-space construction. For this reason the two cases, massless and massive ISR, are treated in more detail in the following. An overview on the construction of the FSR phase-space used for the FKS subtraction scheme is given in Ref.~\cite{Frixione:2007vw}.
	
	In order to label particles participating in the Born or real process based on the notation used in Ref.~\cite{Frederix:2009yq} generalised to QCD and QED interactions the following numbers for final-state particles are used
	\begin{align}
	\begin{split}
		n_{H}\text{ :} \quad \#&\text{ of coloured massive particles}\\
		n_{L}\text{ :} \quad \#&\text{ of coloured light particles}\\
		n_{e}\text{ :} \quad \#&\text{ of }n_{e(L)}\text{ light charged leptons and photons}\\
		&\text{ and }n_{e(H)}\text{ massive charged leptons}\\
		n_{W}\text{ :} \quad \#&\text{ of charged massive bosons}\\
		n_{\emptyset}\text{ :} \quad \#&\text{ of charge- and colour-neutral fermions and massive bosons}
	\end{split}
	\end{align}
	The ordering in which the particles occur in the process in detail is determined by the collider specific setup. Generally, considering processes associated with hadron-hadron or lepton-lepton collision setups the following definitions can be formulated,
	\begin{align}
	\begin{split}
		&1\le i \le n_I\quad  \text{: initial state}\\
		&n_I+1\le i \le n_I +n_e+n_L \quad \text{: charged leptons, massless quarks, gluons and photons}\\
		&n_I+n_e+n_L+1\le i \le n_I+n_e+n_L+n_H \quad \text{: heavy quarks}\\
		&n_I+n_e+n_L+n_H+1\le i \le n_I+n_e+n_L+n_H+n_W \quad \text{: $W^{\pm}$ bosons}\\
		&n_I+n_e+n_L+n_H+n_W+1\le i \le n_I+n_e+n_L+n_H+n_W+n_{\emptyset}\quad\text{: $\nu_l/\bar{\nu}_l/Z/H$}.
	\end{split}
	\label{particlepositions}
	\end{align}
	Furthermore, the process flavour structure of the Born $f_b$ as well as of the real $f_r$ is defined as
	\begin{align}
		&f_b=\{P_1,\ldots,P_{N}\} &f_r=\{P_1,\ldots,P_{N+1}\}
		\label{flavorstructure}
	\end{align}
	with the SM particle identity $P_i$ for the position $i$ declared in Eq.~(\ref{particlepositions}).
	We define the differential Born cross section as
	\begin{align}
		d\sigma^{n}_B(f_b)=\mathcal{B}(f_b)d\bar{\Phi}_n
	\end{align}
	using the quantity
	\begin{align}
		\mathcal{B}(f_b)=\mathcal{N}_{B}J_b\sum_{\text{colour}}\sum_{\text{spin}}\mathcal{A}(f_b)\mathcal{A}^{\dagger}(f_b)=J_b\mathcal{M}(f_b)
	\end{align}
	with $\mathcal{N}_B$ the normalisation factor averaging over initial state colours and spins and a function $J_b$ which takes phase-space cuts imposed on the Born final states into account.
	\subsection{Massless initial-state emitter}
	\label{secMasslessISem}
	For the description of the initial state in the massless limit collinear singularities are resummed and factorised in terms of PDFs. In this way, initial state partons entering the hard process receive modified momenta due to the fraction of energy which is radiated away from the beam particles. For theoretical predictions it is thus required to integrate over the new degrees of freedom coming from the two massless initial states. The phase-space construction for the massless initial state radiation presented in this section follows the formalism used in \cite{Frixione:2007vw}.
	
	Introducing the variables $x_{\oplus}$ and $x_{\ominus}$, the fractions of the corresponding beam energies $\sqrt{s}/2$, the initial state momenta are parametrised as
	\begin{align}
		&{k}_{\oplus}={x}_{\oplus}K_{\oplus} & {k}_{\ominus}={x}_{\ominus}K_{\ominus}
	\end{align}
	with the beam momenta $K_{\oplus}$ and $K_{\ominus}$. By retaining momentum conservation by Eq.~(\ref{momentumconservation}) the phase-space definitions of Eq.~(\ref{PSborn}) and (\ref{PSreal}) change to
	\begin{align}
		& \bar{\Phi}^{\prime}_n=\{\bar{x}_{\oplus},\bar{x}_{\ominus},\bar{k}_{1+n_I},\ldots,\bar{k}_{N}\} &{\Phi}^{\prime}_{n+1}=\{{x}_{\oplus},{x}_{\ominus},{k}_{1+n_I},\ldots,k_{N+1}\} \quad,
	\end{align}
	and the phase-space measures factorise as
	\begin{align}
		&d\bar{\Phi}^{\prime}_n=d\bar{x}_{\oplus}d\bar{x}_{\ominus}d{\bar{\Phi}}_{n} &d{\Phi}^{\prime}_{n+1}=d{x}_{\oplus}d{x}_{\ominus}d{\Phi}_{n+1}\quad.
		\label{factorPSmeasure}
	\end{align}
	For initial state radiation the momentum $k_{n+1}$ can be parametrised in terms of the FKS variables introduced above as
	\begin{align}
		k_{N+1}=k_{N+1}^0(1,\sqrt{1-y^2}\sin\phi,\sqrt{1-y^2}\cos\phi,y)
		\label{k+1FKSparametrisation}
	\end{align}
	with
	\begin{align}
		k_{N+1}^0=\frac{\sqrt{s}}{2}\xi
		\label{k+1energy}
	\end{align}
	which results in the integration variable
	\begin{align}
		\frac{d^3k_{N+1}}{2k^0_{N+1}(2\pi)^3}=\frac{s}{(4\pi)^3}\xi d\xi dyd\phi
		\label{dkn+1}
	\end{align}
	By the condition that the system of $n$ final-state particles before radiation of a parton, i.~e.
	\begin{align}
		\bar{k}_{n,f}=\sum_{l=1+n_I}^{N}\bar{k}_l=\bar{k}_{\oplus}+\bar{k}_{\ominus}=\bar{x}_{\oplus}K_{\oplus}+\bar{x}_{\ominus}K_{\ominus}\quad,
	\end{align}
	and after radiation, i.~e.
	\begin{align}
		{k}_{n,f}=\sum_{l=1+n_I}^{N}{k}_l={k}_{\oplus}+{k}_{\ominus}-k_{n+1}={x}_{\oplus}K_{\oplus}+{x}_{\ominus}K_{\ominus}-k_{n+1}\quad,
	\end{align}
	have the same invariant mass and rapidity, the relations of the rescaled momentum fraction ${x}_{\oplus}$ as a function of $\bar{x}_{\oplus}$ and ${x}_{\ominus}$ as a function of $\bar{x}_{\ominus}$, respectively, can be derived. They read
	\begin{align}
		&{x}_{\oplus}=\frac{\bar{x}_{\oplus}}{\sqrt{1-\xi}}\sqrt{\frac{2-\xi(1-y)}{2-\xi(1+y)}} & {x}_{\ominus}=\frac{\bar{x}_{\ominus}}{\sqrt{1-\xi}}\sqrt{\frac{2-\xi(1+y)}{2-\xi(1-y)}}
		\label{xplusxminus}
	\end{align}
	and are constrained by
	\begin{align}
		& {x}_{\oplus}\leq1, & {x}_{\ominus}\leq1\quad.
		\label{xle1conditions}
	\end{align}
	In addition, according to these conditions, the integration variables depend on each other by
	\begin{align}
		d{x}_{\oplus}d{x}_{\ominus}=\frac{d\bar{x}_{\oplus}d\bar{x}_{\ominus}}{1-\xi}\quad.
	\end{align}
	Using this together with the relations of Eq.~(\ref{factorPSmeasure}) and (\ref{dkn+1})
	the Jacobian $\mathcal{J}(\xi,y,\phi)$ can be extracted analogously to Eq.~(\ref{n+1PSparametr}) by factorising the Born phase-space measure $d\bar{\Phi}^{\prime}_{n}$ as
	 \begin{align}
	 	d{\Phi}^{\prime}_{n+1}=\frac{s}{(4\pi)^3}\xi d\xi dyd\phi d{x}_{\oplus}d{x}_{\ominus}d\bar{\Phi}_{n}=\frac{s}{(4\pi)^3}\frac{\xi}{1-\xi} d\xi dyd\phi d\bar{\Phi}^{\prime}_{n}\quad.
	 \end{align}
	The upper bound $\xi_{\text{max}}$ of the variable $\xi$ follows from the conditions of Eq.~(\ref{xle1conditions}), and thus with the relations of Eq.~(\ref{xplusxminus}) is defined as
	
	\begin{align}
	\begin{split}
		\xi_{\text{max}}=1-\max \left[
		\frac{2(1+y)\bar{x}_{\oplus}^2}{\sqrt{(1+\bar{x}_{\oplus}^2)^2(1-y)^2+16y\bar{x}_{\oplus}^2}+(1-y)(1-\bar{x}_{\oplus}^2)},\right. \\
		\left.\frac{2(1-y)\bar{x}_{\ominus}^2}{\sqrt{(1+\bar{x}_{\ominus}^2)^2(1+y)^2-16y\bar{x}_{\ominus}^2}+(1+y)(1-\bar{x}_{\ominus}^2)}\right].
		\label{ximax}
	\end{split}
	\end{align}
	\subsection{Massive initial-state emitter}
	\label{secMassiveIS}
	For lepton colliders the initial states can be considered as massive, thereby regularising all singularities from collinear ISR.
	For the construction of the phase-space for radiation off massive emitters a crucial criterion is that all external states before and after emission of a photon have to remain on-shell. The following discussion is therefore closely related to the construction of phase-space momenta of external states in a decay process based on the on-shell projection method presented in \cite{Dittmaier:2015bfe,Denner:2000bj}. This method originally is implemented in \texttt{WHIZARD} for its application to factorised processes with massive resonances~\cite{Bach:2017ggt,Weiss:2017qbj}. Treating the initial state as `one decaying particle', this method equivalently can be applied to the ISR phase-space construction for lepton collisions with massive emitters.
	
	First of all, the radiated photon momentum $k_{n+1}$ can be parametrised as in the case of massless emitters according to Eq.~(\ref{k+1FKSparametrisation}) using (\ref{k+1energy}). With the condition that the initial-state emitter momentum in the real process is equal to the momentum of the corresponding beam particle for the Born configuration, the sums of the initial-state momenta
	\begin{align}
		&k_{\text{in}}\equiv k_{\oplus} + k_{\ominus} &\bar{k}_{\text{in}}\equiv \bar{k}_{\oplus}+\bar{k}_{\ominus}\quad,
	\end{align}
	equal each other, i.~e.	$k_{\text{in}}=\bar{k}_{\text{in}}$. The momentum of the initial-state system after emission of a photon, i.~e. the recoiling system of the photon, hence can be written as
	\begin{align}
		k_{\text{rec},n}=\sum_{l=1+n_I}^{N}{k}_l=k_{\text{in}}-k_{{N+1}}=\bar{k}_{\text{in}}-k_{{N+1}}=\sum_{l=1+n_I}^{N}\bar{k}_l-k_{{N+1}}\quad,
	\end{align}
	giving an invariant mass $m_{\text{rec},n}^2=k_{\text{rec},n}^2$. By retaining the on-shell condition for all final-state particles before and after radiation which is represented by
	\begin{align}
		m_l^2=\bar{k}_l^2=k_l^2
		\label{onshellcondition}
	\end{align}
	the real momenta $k_l$, $l=\{1+n_I,\ldots,N\}$ are constructed by Lorentz boosts from the Born into the recoiling system. Practically, this proceeds as follows.
	
	The final-state Born momenta are decomposed into two systems, $\bar{k}_N$ with invariant mass $m_N^2$ and a recoiling system
	\begin{align}
	\bar{k}_{\text{rec},n-1}=\sum^{N-1}_{l=1+n_I}\bar{k}_l
	\label{bornrecsystem}
	\end{align}
	with invariant mass $m_{\text{rec},n-1}^2=\bar{k}_{\text{rec},n-1}^2$.
	Let $\Lambda^{-1}_{\text{rec},n}$ be the boost of the recoiling system $k_{\text{rec},n}$ into its rest frame, i.~e.
	\begin{align}
		\Lambda^{-1}_{\text{rec},n}k_{\text{rec},n}=\left(
		m_{\text{rec},n}, 0, 0, 0\right)\quad.
	\end{align}
	Defining the triangle function $\lambda(x,y,z)=x^2+y^2+z^2-2xy-2xz-2yz$, the magnitude of the three-momentum for both decay products is given by
	\begin{align}
		\lvert \vec{p}_n\rvert=\frac{\sqrt{\lambda\left(m_{\text{rec},n}^2,m_N^2,m_{\text{rec},n-1}^2\right)}}{2m_{\text{rec},n}}
	\end{align}
	such that the energies of the two resulting final-state systems by observing the on-shell condition of Eq.~(\ref{onshellcondition}) are given by
	\begin{align}
		&E_N^2=m_N^2+\lvert\vec{p}_n\rvert^2 & E_{\text{rec},n-1}^2=m_{\text{rec},n-1}^2+\lvert\vec{p}_n\rvert^2.
	\end{align}
	Subsequently, the four vectors
	\begin{align}
		&k^{\prime}_N=(E_N,\lvert \vec{p}_n\rvert\hat{n}_N) &k^{\prime}_{\text{rec},n-1}=(E_{\text{rec},n-1},\lvert \vec{p}_n\rvert\hat{n}_{\text{rec},n-1})
	\end{align}
	can be set up with spatial components rotated into the directions $\hat{n}_N$ and $\hat{n}_{\text{rec},n-1}$ with $\hat{n}_N=-\hat{n}_{\text{rec},n-1}$. These directions must equal those of the spatial components of the vectors
	\begin{align}
		&\bar{k}^{\prime}_N=\Lambda_n^{-1}\bar{k}_N &{\bar{k}}^{\prime}_{\text{rec},n-1}=\Lambda_n^{-1}\bar{k}_{\text{rec},n-1}
	\end{align}
	with $\Lambda_n^{-1}$ representing a boost into the rest frame of $(\bar{k}_N+\bar{k}_{\text{rec},n-1})$, i.~e. the Born rest frame.
	Finally, $k^{\prime}_N$ and $k^{\prime}_{\text{rec},n-1}$ are boosted by $\Lambda_{\text{rec},n}$ into the $n$-final-state recoiling system which gives the momenta
	\begin{align}
		&k_{N}=\Lambda_{\text{rec},n}k^{\prime}_n & k_{\text{rec},n-1}=\Lambda_{\text{rec},n} k^{\prime }_{\text{rec},n-1}
		\label{FSmomentarecsys}
	\end{align}
	By substituting $n\leftrightarrow n-1$ the steps from Eq.~(\ref{bornrecsystem}) to (\ref{FSmomentarecsys}) can be repeated down to $n=2$ for which all real final-state momenta in the recoiling system $\{k_{1+n_I},\ldots,k_{N}\}$ eventually are constructed.
	
	The Jacobian $\mathcal{J}$ results from the factorisation of the Born phase-space element according to Eq.~(\ref{n+1PSparametr}) \cite{Weiss:2017qbj,Whizard:2020},
	
		\begin{align}
		\begin{split}
			d\Phi_{n+1}=&\frac{\bar{k}^2_{\text{in}}}{k^2_{\text{rec},n}}\left(\frac{\lambda\left(k^2_{\text{rec},n},m_N^2,m_{\text{rec},n-1}^2\right)}{\lambda\left(\bar{k}^2_{\text{in}},m_N^2,m_{\text{rec},n-1}^2\right)}\right)^{1/2}\frac{s}{(4\pi)^3}\xi d\xi dyd\phi d\bar{\Phi}_{n}\quad.
		\end{split}
	\end{align}
	
	The integration bound to the FKS variable $\xi$ is determined in this approach recursively via \cite{Whizard:2020}
	\begin{align}
		\xi_{\text{max}}\equiv\xi^{(n)}_{\text{max}}=\min \left\{1-\frac{(m_N+m_{\text{rec},n-1})^2}{(\bar{k}_{N}+\bar{k}_{\text{rec},n-1})^2},\xi^{(n-1)}_{\text{max}}\right\}
	\end{align}
	with $\xi_{\text{max}}^{(2)}$ defined as
	\begin{align}
		\xi_{\text{max}}^{(2)}=1-\frac{(m_2+m_{1})^2}{(\bar{k}_{2}+\bar{k}_{1})^2}\quad.
	\end{align}
	\section{FKS subtraction scheme in mixed coupling expansions}
	\label{mixedcouplingsSec}
	In this section the FKS subtraction scheme for observables in mixed coupling expansions is described based on the considerations of Ref.~\cite{Frederix:2018nkq} and the general FKS formalism \cite{Frixione:1995ms,Frixione:1997np,Frederix:2009yq}. For $2\rightarrow n$ processes, observables such as total cross sections can be defined in fixed orders in the coupling constants $\alpha_s$ and $\alpha$ due to the contributing gauge invariant squared amplitudes being well-defined in the coupling powers, i.~e.
	\begin{align}
	\mathcal{A}_{(u_s,u_e)}\mathcal{A}^{\dagger}_{(v_s,v_e)}\propto g_s^{u_s+v_s}e^{u_e+v_e}\propto\alpha_s^{p_s}\alpha^{p_e}\quad.
	\end{align}
	Since no squared amplitudes with odd powers of $g_s$ or $e$ exist without exception $p_s=(u_s+v_s)/{2}$ and $p_e=(u_e+v_e)/{2}$ can be assumed.
	Thus the contributions to a fixed coupling order coming from squared and interfering matrix elements can be summed as
	\begin{align}
		\mathcal{M}(\alpha_s^{p_s}\alpha^{p_e})\propto \sum_{u_s,v_s}\sum_{u_e,v_e}\mathcal{A}_{(u_s,u_e)}\mathcal{A}^{\dagger}_{(v_s,v_e)}\delta_{p_s,(u_s+v_s)/2}\delta_{p_e,(u_e+v_e)/2}\quad.
	\end{align}
	This applies not only to the leading but also to any higher order computation.
	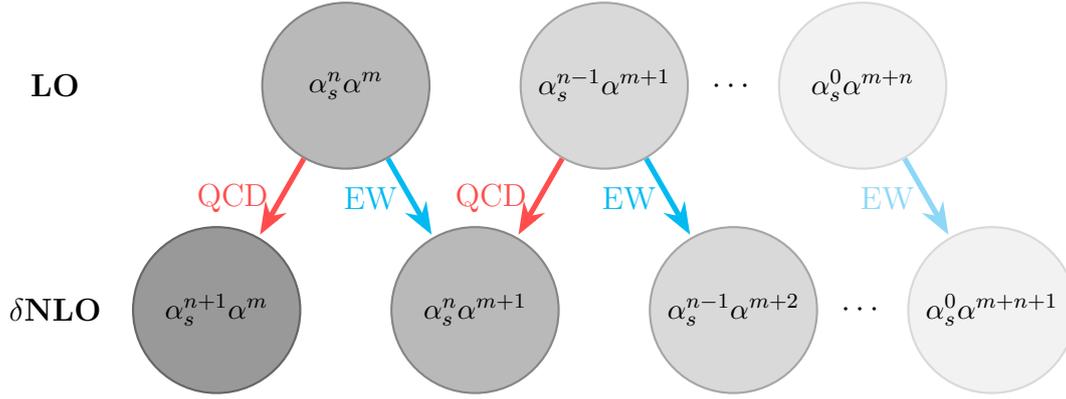
\begin{figure}
	\begin{tikzpicture}[scale=0.85,node distance=3cm]
		\node at (0,0) {\textbf{LO}};
		\node at (4.5,0) [circle,draw=black!47.5,fill=black!27.5,thick,minimum size=2.2cm](LO1){\small $\alpha_s^{n} {\alpha^m}$};
		\node at (8.5,0) [circle,draw=black!35,fill=black!15,thick,minimum size=2.2cm](LO2){\small $\alpha_s^{n-1} {\alpha^{m+1}}$};
		\node at (10.5,0){$\cdots$};
		\node at (12.5,0) [circle,draw=black!15,fill=black!5,thick,minimum size=2.2cm](LOlast){\small $\alpha_s^{0} {\alpha^{m+n}}$};
				\node at (0,-3.5) {$\delta$\textbf{NLO}};
		\node at (2.5,-3.5) [circle,draw=black!60,fill=black!40,thick,minimum size=2.2cm] (NLO1){\small $\alpha_s^{n+1} {\alpha^m}$};
		    		\draw (NLO1) [{Stealth[scale=1]}-,red!70,shorten <=1pt,thick, line width=2pt] to node [midway,left] {QCD} (LO1);
		\node at (6.5,-3.5) [circle,draw=black!47.5,fill=black!27.5,thick,minimum size=2.2cm](NLO2){\small $\alpha_s^{n} {\alpha^{m+1}}$};
				    \draw (NLO2) [{Stealth[scale=1]}-,red!70,shorten <=1pt,thick, line width=2pt] to node [midway,left] {QCD} (LO2);
				    \draw (NLO2) [{Stealth[scale=1]}-,cyan!80,shorten <=1pt,thick, line width=2pt] to node [midway,left] {EW} (LO1);
		\node at (10.5,-3.5) [circle,draw=black!35,fill=black!15,thick,minimum size=2.2cm](NLO3){\small $\alpha_s^{n-1} {\alpha^{m+2}}$};
				    \draw (NLO3) [{Stealth[scale=1]}-,cyan!80,shorten <=1pt,thick, line width=2pt] to node [midway,left] {EW}(LO2);
		\node at (12.5,-3.5) {$\cdots$};
		\node at (14.5,-3.5) [circle,draw=black!15,fill=black!5,thick,minimum size=2.2cm](NLOlast){\small $\alpha_s^{0} {\alpha^{m+n+1}}$};
					\draw (NLOlast) [{Stealth[scale=1]}-,cyan!40,shorten <=1pt,thick, line width=2pt] to node [midway,left] {EW}(LOlast);
	\end{tikzpicture}
		\caption{Connections of LO and NLO contributions to different coupling orders in a mixed coupling expansion with different grey shades of the blobs denoting the degree of highest QCD coupling power by the darkest blob and arrows indicating the correction type, QCD or EW.}
		\label{ziehharmonika}
	\end{figure}
    Already for NLO computations at fixed coupling orders gauge invariance potentially implies contributions of overlapping QCD-EW corrections. These can be considered either as QCD corrections to Born squared amplitudes of one order less in $\alpha_s$ or EW corrections to those of one order reduced in $\alpha$. Consequently, IR singularities are induced by both, QCD and QED splittings, which have to be regularised accordingly. Schematically, the connections of LO and NLO contributions explicit in the coupling powers with corresponding QCD or EW corrections are depicted in Fig.~\ref{ziehharmonika}. The contributions are ordered from left to right decreasing in the $\alpha_s$ and increasing in the $\alpha$ power, respectively. Naively, this ordering due to $\alpha_s \gg \alpha$ corresponds
    to contributions at LO and NLO decreasing in its magnitude with descending order in $\alpha_s$. In fact, for the most of the LHC processes this is true such that the typical labelling of the contributions is $\text{LO}_i$ and $\text{NLO}_j$ with $i\in\{1,\ldots,n+1\}$ and $j\in\{1,\ldots,n+2\}$ where $i=1$ and $j=1$ denotes the contribution at the highest order in $\alpha_s$. However there are exceptions spoiling this ordering, for example due to EW resonances leading to large amplitudes at certain collider energies and possibly occuring first at contributions $i>1$ or $j>1$. Numerical examples for this case are shown within the results of Sec.~\ref{secToppairLHC}. In order to treat NLO SM corrections due to this in a generic way, in this work the notation
    \begin{align}
    	&\text{LO}_{p_s,p_e}\propto \alpha_s^{p_s}\alpha^{p_e} & 0\leq p_s\leq n, \quad m\leq p_e \leq m+n, \quad p_s+p_e=n+m\label{loContrib}\\
    	&\delta\text{NLO}_{q_s,q_e}\propto \alpha_s^{q_s}\alpha^{q_e} & 0\leq q_s\leq n+1, \quad m\leq q_e \leq m+n+1, \quad q_s+q_e=n+m+1
    	\label{nloContrib}
    \end{align}
    is used with $n$ and $m$ referring to the coupling order $\mathcal{O}(\alpha_s^n\alpha^m)$ of the LO contribution for the considered process at highest order in $\alpha_s$. $\delta\text{NLO}_{q_s,q_e}$ denotes the NLO correction computed for a fixed coupling power combination $\{q_s,q_e\}$. According to this counting, speaking of NLO EW observables in this thesis means the sum of the contributions $\text{LO}_{n,m}$ and $\delta\text{NLO}_{n,m+1}$.
    
    Since for many LHC processes contributions associated with photon PDFs are non-negligible and come into play already for $\delta\text{NLO}_{n,m+1}$ it is recommended to define protons or jets including photons. Due to EW corrections coming along with $\gamma\rightarrow q\bar{q}$ splittings for $\delta\text{NLO}_{q_s,q_e}$ contributions with $q_s\le n-1$ and $q_e\ge m+2$ it is even mandatory to treat gluons and photons on a democratic basis in the proton and jet definition lest one spoils IR-safety for observables. The technical requirements for rendering observables IR-safe in view of MC integration, in particular with respect to a mixed coupling expansion of NLO observables, are treated in more detail in Sec.~\ref{secIRsafeobservables}.
	\subsection{NLO cross section formulation for $2\rightarrow n$ processes}
	\label{seccrosssecformulation}
	According to the general form of $\sigma_{\text{NLO}}$ introduced in Eq.~(\ref{KLNtheorem}), the NLO cross section in a mixed coupling expansion, written in elementary terms of the FKS subtraction scheme, is presented in this section.
	
	The physical differential cross section of any collider process with beams denoted by $B_{\oplus}$ and $B_{\ominus}$ can be written in a factorised form as
	\begin{align}
		d\tilde{\sigma}^{(B_{\oplus}B_{\ominus})}(K_{\oplus},K_{\ominus})=\sum_{ab}\int dx_{\oplus}dx_{\ominus}\Gamma^{(B_{\oplus})}_a(x_{\oplus})\Gamma^{(B_{\ominus})}_b(x_{\ominus})d{\sigma}_{ab}(x_{\oplus}K_{\oplus},x_{\ominus}K_{\ominus})
		\label{hadronicCrossSection}
	\end{align}
	with the PDFs $\Gamma^{(B_{\oplus})}_a$ and $\Gamma^{(B_{\ominus})}_b$ and the short-distance cross section $d{\sigma}_{ab}$. Note, that if the initial state is treated as massive the PDFs are just represented by $\delta$ functions and the cross sections $d\tilde{\sigma}^{(B_{\oplus}B_{\ominus})}$ and $d\hat{\sigma}_{ab}$ coincide. In order to simplify the formalism in the following the short-hand notation for the convolution product
	\begin{align}
		\Gamma \star d\sigma=\int dx \Gamma(x)d\sigma(xK)
	\end{align}
	is introduced. Furthermore, the indices for the initial states $a$ and $b$ are dropped if not explicitly needed.
	
	In general, for a precise prediction of observables, contributions from the whole tower of coupling power combinations indicated in Fig.~\ref{ziehharmonika} at LO and NLO, i.~e. Eqs.~(\ref{loContrib}) and (\ref{nloContrib}), must be summed. By specifying the correction type of an NLO perturbative expansion the hadronic cross section is well-defined. In this way, with
	\begin{align}
	\label{Borncomponent}
			d\left(\text{LO}_{p_s,p_e}\right)=d\tilde{\sigma}^{(B_{\oplus}B_{\ominus})}_{(p_s,p_e)}=\Gamma^{(B_{\oplus})}\star \Gamma^{(B_{\ominus})} \star d\sigma^{n}_{B,(p_s,p_e)}
	\end{align}
	we define a differential cross section at NLO in $\alpha$ as
	\begin{align}
	\label{nloEWcalc}
	d\left(\text{NLO}_{q_s,q_e}^{\text{EW}}\right)=d\left(\text{LO}_{q_s,q_e-1}\right) +d\left(\delta \text{NLO}_{q_s,q_e}\right)
	\end{align}
	and at NLO in $\alpha_s$ as
	\begin{align}
	\label{nloQCDcalc}
	d\left(\text{NLO}_{q_s,q_e}^{\text{QCD}}\right)=d\left(\text{LO}_{q_s-1,q_e}\right) +d\left(\delta \text{NLO}_{q_s,q_e}\right)\quad,
	\end{align}
	respectively.
	Written in terms of coupling powers the NLO correction to differential cross sections, appearing in both Eqs.~(\ref{nloEWcalc}) and (\ref{nloQCDcalc}), reads
	\begin{align}
		d\left(\delta \text{NLO}_{q_s,q_e}\right)=\Gamma^{(B_{\oplus})}\star \Gamma^{(B_{\ominus})} \star \left[d\sigma^{n+1}_{(q_s,q_e)}+d\sigma^{n}_{(q_s,q_e)}+d\bar{\sigma}^{n+1}_{(q_s,q_e)}\right] \quad.
		\label{sigmaNLOcorr}
	\end{align}
	The first term in the bracket denotes the subtracted real-emission contribution and can be written as
	\begin{align}
	\begin{split}
		d\sigma^{n+1}_{(q_s,q_e)}=d\bar{\Phi}_n\bigg[\sum_{E}d\Phi_{\text{rad}}^{(E)}(\xi,y)\sum_{\tilde{\alpha}}\delta_{EE_{\tilde{\alpha}}}\left( \mathcal{R}_{\tilde{\alpha},(q_s,q_e)}\left(\Phi_{n+1}^{(E)}\right)-S^{(E_{\tilde{\alpha}},{s}_{\tilde{\alpha}})}_{(q_s,q_e)}(\bar{\Phi}_n,\xi,y)\right)\bigg.\\
		\bigg.+d\Phi_{\text{rad}}\mathcal{R}_{\text{non-sing},(q_s,q_e)}\left(\Phi_{n+1}\right)\bigg]
	\end{split}
		\label{realsubtractedcontr}
	\end{align}
	with $\mathcal{R}_{\tilde{\alpha}}$ defined as the real emission squared amplitude weighted with respect to a kinematic factor corresponding to an FKS singular region $\tilde{\alpha}$ which will be treated more detailed in Sec.~(\ref{secSingularregions}). $S^{(E_{\tilde{\alpha}},{s}_{\tilde{\alpha}})}$ represents the subtraction term at the non-integrated level in the collinear and/or soft limit defined per emitter particle $E_{\tilde{\alpha}}\in \{1,\ldots,n_I+n\}$ which induces the splitting ${s}_{\tilde{\alpha}}$ into the associated FKS pair $(i,j)$.
	Correspondingly, $E_{\tilde{\alpha}}$ and $s_{\tilde{\alpha}}$ refer to a specific $\tilde{\alpha}$~singular region from a real flavour structure $f_r$ for which the particle content of the factorising Born process $f_b=f_{b,\tilde{\alpha}}$ as well as the correction type of the splitting, either QCD or QED, are uniquely defined. For this reason, $s_{\tilde{\alpha}}$ contains the information on the SM vertex of the splitting according to
	\begin{align}
	s_{\tilde{\alpha}}\equiv \{P_{E_{\tilde{\alpha}}},P_i,P_j\} \qquad\text{with}\quad P_i,P_j\in f_r\text{ and }P_{E_{\tilde{\alpha}}}\in f_{b,\tilde{\alpha}}\quad.
	\label{splittingdef}
	\end{align}
	The subtraction terms have a factorised form consisting of a kinematical factor corresponding to the collinear, soft or soft-collinear limit and a Born squared amplitude exclusive in the respective quantum numbers, spin and colour/charge. They are examined explicitly in Sec.~(\ref{secSubtractionterms}). The last term in Eq.~(\ref{realsubtractedcontr}) includes all contributions from interfering amplitudes of diagrams with $n+1$ final states which due to their coupling order do not exhibit singular regions and thus are treated separately. They will be discussed in particular in Sec.~(\ref{secnonsing}).
	
	Since according to the KLN theorem the collinear, soft and soft-collinear divergences of the virtual loop contributions $\mathcal{V}(\bar{\Phi}_{n})$ are cancelled analytically by the poles of
	\begin{align}
		I_{(q_s,q_e)}(f_b,\bar{\Phi}_{n})=\int d\Phi_{\text{rad}}(\xi,y)\sum_{E}S^{(E)}_{(q_s,q_e)}(f_b,\bar{\Phi}_n,\xi,y)\quad,
		\label{Isubtractionterms}
	\end{align}
	only the finite remainders $I^{(0)}$ and $\mathcal{V}^{(0)}$ are left to be integrated and summed over all Born flavour structures $f_b$ for the final result. The second term of Eq.~(\ref{sigmaNLOcorr}) thus contains all finite contributions to be integrated over the $n$ particle phase-space, i.~e.
	\begin{align}
	d\sigma^{n}_{(q_s,q_e)}(f_b,\bar{\Phi}_{n})=d\bar{\Phi}_n\left(I^{(0)}_{(q_s,q_e)}(f_b,\bar{\Phi}_{n})+\mathcal{V}^{(0)}_{(q_s,q_e)}(f_b,\bar{\Phi}_{n})\right)\quad,
	\label{finitecontrnphasespace}
	\end{align}
	where $\mathcal{V}^{(0)}$ denotes all finite contributions from Born-tree and virtual one-loop interfering amplitudes. $I^{(0)}$ can be further defined as
	\begin{align}
		I^{(0)}_{(q_s,q_e)}(f_b,\bar{\Phi}_{n})=d\sigma^{\text{FSR}}_{C,(q_s,q_e)}(f_b,\bar{\Phi}_{n})+d\sigma_{S,(q_s,q_e)}(f_b,\bar{\Phi}_{n})\quad.
	\end{align}
	Here, $d\sigma_{C}^{\text{FSR}}$ represents the remainder of the FSR collinear subtraction term in integrated form factorising as
	\begin{align}
	\begin{split}
		d\sigma_{C,(q_s,q_e)}^{\text{FSR}}(f_b,\bar{\Phi}_{n})=\mathcal{N}(\varepsilon)\sum_{E(f_b)}\left[\frac{\alpha_s}{2\pi}\mathcal{Q}^{\text{QCD}}_{E(f_b)}\mathcal{B}_{E(f_b),(q_s-1,q_e)}^{\text{QCD}}(f_b,\bar{\Phi}_{n})\right.\\
		\left.+\frac{\alpha}{2\pi}\mathcal{Q}^{\text{QED}}_{E(f_b)}\mathcal{B}_{E(f_b),(q_s,q_e-1)}^{\text{QED}}(f_b,\bar{\Phi}_{n})\right]
	\end{split}
		\label{collinearsubint}
	\end{align}
	with the kinematical factors $\mathcal{Q}^{\text{QCD}}$ and $\mathcal{Q}^{\text{QED}}$ in the limits of collinear QCD and QED splittings, the Born squared amplitudes $\mathcal{B}$ and a normalisation factor $\mathcal{N}$. Note, that the superscripts `QCD' and `QED' for $\mathcal{B}$ merely serve to affirm the difference in the coupling orders of these two factors which emerge from the splitting correction types. The Born squared matrix elements are non-zero only if the process flavour structure $f_b$ for the corresponding coupling order exists. An algorithm to detect allowed coupling orders to a given flavour structure $f_b$ or $f_r$ in a generic way will be presented in Sec.~\ref{CPCA}, concerning the technical details of the NLO SM automation in \texttt{WHIZARD}.
	The soft subtraction term $d\sigma_{S}$ in integrated form can be written as
	\begin{align}
	\begin{split}
		d\sigma_{S,(q_s,q_e)}(f_b,\bar{\Phi}_{n})=&\sum_{k,l}^n\mathcal{E}_{kl,0}^{(m_k,m_l)}\left(\frac{\alpha_s}{2\pi}\mathcal{B}^{\text{QCD}}_{kl,(q_s-1,q_e)}(f_b,\bar{\Phi}_{n})+\frac{\alpha}{2\pi}\mathcal{B}^{\text{QED}}_{kl,(q_s,q_e-1)}(f_b,\bar{\Phi}_{n})\right)\\
		&+d\sigma_{S,(q_s,q_e)}^{\text{ISR remn.}}
		\end{split}
		\label{softsubint}
	\end{align}
	where $\mathcal{E}_{kl,0}^{(m_k,m_l)}$ denotes the finite part of the eikonal factor, $\mathcal{B}^{\text{QCD}}_{kl}$ the colour-correlated Born squared matrix elements and $\mathcal{B}^{\text{QED}}_{kl}$ the charge-correlated, respectively.
	The explicit form of the contributing factors of Eq.~(\ref{collinearsubint}) and the first term of (\ref{softsubint}) will be further discussed in Sec.~(\ref{secIntegratedSubs}). The term $d\sigma_{S,(q_s,q_e)}^{\text{ISR remn.}}$ stems from finite remainders of PDF counterterms in the soft limit which will be elaborated in Sec.~(\ref{secDglapremnant}).

	The last term of Eq.~(\ref{sigmaNLOcorr}) is the degenerate $n+1$ phase-space contribution, also called DGLAP remnant, which arises as finite remainder of singularities coming from collinear initial-state radiation cancelled by those from the PDF evolution via the Dokshitzer-Gribov-Lipatov-Altarelli-Parisi (DGLAP) equations \cite{Dokshitzer:1977sg,Gribov:1972ri,Altarelli:1977zs}. In a factorised form these contributions can be written as
	\begin{align}
	\begin{split}
		d\bar{\sigma}^{n+1}_{ab,(q_s,q_e)}(\bar{\Phi}_{n})=\sum_d\left(\frac{\alpha_s}{2\pi}\left[\mathcal{K}^{\text{QCD}}_{da}\star\mathcal{B}^{\text{QCD}}_{db,(q_s-1,q_e)}(\bar{\Phi}_{n})+\mathcal{K}^{\text{QCD}}_{db}\star \mathcal{B}^{\text{QCD}}_{ad,(q_s-1,q_e)}(\bar{\Phi}_{n})\right]\right.\\
		\left. +\frac{\alpha}{2\pi}\left[\mathcal{K}^{\text{QED}}_{da}\star\mathcal{B}^{\text{QED}}_{db,(q_s,q_e-1)}(\bar{\Phi}_{n})+\mathcal{K}^{\text{QED}}_{db}\star \mathcal{B}^{\text{QED}}_{ad,(q_s,q_e-1)}(\bar{\Phi}_{n})\right]\right)
		\label{degeneraten+1contr}
	\end{split}
	\end{align}
	with $\mathcal{K}^{\text{QCD}}_{da}$ and $\mathcal{K}^{\text{QED}}_{da}$ containing the unregularised splitting functions and other kinematical factors as outlined in Sec.~(\ref{secDglapremnant}) and the Born squared matrix elements $\mathcal{B}_{db}$ summed up for a specific initial state $\{d,b\}$.
	\subsection{Singular regions in the $(n+1)$-body phase-space}
	\label{secSingularregions}
	The phase-space $\Phi_{n+1}$ as introduced in Sec.~(\ref{secFKSphasespace}) with $\Phi_{\text{rad}}$ parametrised in the FKS variables $\{\xi,y,\phi\}$ can be partitioned into disjoint regions $\tilde{\alpha}$, each containing at most one collinear and/or one soft singularity. The real-emission squared amplitude $\mathcal{R}$ in this way gets a kinematical weight factor $\mathcal{S}_{\tilde{\alpha}}$, i.~e. the partition function, such that
	\begin{align}
		\mathcal{R}=\sum_{\tilde{\alpha}}\mathcal{R}_{\tilde{\alpha}}=\sum_{\tilde{\alpha}}\mathcal{S}_{\tilde{\alpha}}\mathcal{R}\quad.
		\label{n+1phasespacepartition}
	\end{align}
	Each of these singular regions $\tilde{\alpha}$ depends on the corresponding splitting particle pair, the FKS pair
	\begin{align}
		&\mathcal{I}_{\tilde{\alpha}}=(i,j) \in \mathcal{P}_{\text{FKS}}(f_r)
	\end{align}
	where $\mathcal{P}_{\text{FKS}}(f_r)$ is the set of tuples $(i,j)$ for which the momenta $k_i$ and $k_j$ potentially induce a soft and/or collinear divergence. The particles indicated with integer numbers $i$ and $j$ to this end are restricted to those associated with $P_i$ and $P_j$ of $f_r$ defined in Eq.~(\ref{flavorstructure}). Physically, these pairs can arise from QCD or QED splittings, or both, which is allowed for certain $\delta\text{NLO}_{q_s,q_e}$ contributions with respect to $q\bar{q}$ pairs originating from gluons or photons. 
	The conditions for the splitting correction type in generalised NLO SM computations dependent on the coupling order $\{q_s,q_e\}$ for NLO contributions $\delta\text{NLO}_{q_s,q_e}$ can be listed as follows
	\begin{itemize}
		\item $q_s=n+1$ and $q_e=m$: \quad pure QCD splittings
		\item $q_s=0$ and $q_e=m+n+1$:\quad pure QED splittings
		\item all combinations $\{q_s,q_e\}$ with $1<q_s\le n$, $m+1<q_e\le m+n$ and $q_s+q_e=m+n+1$:\quad QCD and QED splittings
	\end{itemize}
	Excluded from the set of FKS pairs are thus all charge- and colour-neutral particles participating in the process by non-$U(1)$ interactions as the $Z$ and Higgs boson as well as neutrinos $\nu_l$ (anti-neutrinos $\bar{\nu}_l$). 
	The set $\mathcal{P}_{\text{FKS}}(f_r)$ thus can be written in a generalised form with respect to that of Ref.~\cite{Frederix:2009yq,Weiss:2017qbj} which is restricted to QCD interactions,
	\begin{align}
		\begin{split}
		\mathcal{P}_{\text{FKS}}(f_r)=\left\{(i,j)\text{ : }1\le i\le n_I+n_e+n_L+n_H+n_W, \right. \\
		\left. n_I+1\le j\le n_I+n_{e(L)}+n_L,\quad i\ne j,\right.\\
		\left.\mathcal{R}(f_r)\rightarrow\infty\quad\text{  if  }\quad k^0_j\rightarrow0\quad\text{  or  }\quad \vec{k}_i\parallel\vec{k}_j,\right.\\
		\left.(i,j)\leftrightarrow(j,i)\text{ non-redundancy conditions},\right.\\
		\left. \{q_s,q_e\}\text{ correction type conditions}\right\}
		\end{split}
		\label{PFKSset}
	\end{align}
	The $(i,j)\leftrightarrow(j,i)$ non-redundancy conditions are constructed by the following considerations.
	In order to not double-count contributions where by interchanging $(i,j)\leftrightarrow(j,i)$ for $i,j\ge n_I+1$ real squared matrix elements $\mathcal{R}_{\tilde{\alpha}}(f_r)$ are the same,  all pairs $(i,j)$ are hence removed from the set of FKS pairs which are redundant. This is done according to
	\begin{align}
		&P_j=g, P_i\ne g,\quad (j,i)\in \mathcal{P}_{\text{FKS}}\quad \Rightarrow\quad (i,j) \notin \mathcal{P}_{\text{FKS}}\quad \text{if}\quad n_I+1\leq i\\
		&P_j\neq g, P_i= g,\quad (j,i)\in \mathcal{P}_{\text{FKS}}\quad \Rightarrow\quad (i,j) \notin \mathcal{P}_{\text{FKS}}\quad \text{if}\quad n_I+1\leq i<j
	\end{align}
	if QCD splittings are allowed and
	\begin{align}
		&P_j=\gamma, P_i\ne \gamma,\quad (j,i)\in \mathcal{P}_{\text{FKS}}\quad \Rightarrow\quad (i,j) \notin \mathcal{P}_{\text{FKS}}\quad \text{if}\quad n_I+1\leq i\\
		&P_j\neq \gamma, P_i= \gamma,\quad (j,i)\in \mathcal{P}_{\text{FKS}}\quad \Rightarrow\quad (i,j) \notin \mathcal{P}_{\text{FKS}}\quad \text{if}\quad n_I+1\leq i<j
	\end{align}
	for the QED case, respectively. The pair $(g,g)$ is a special case for which $\mathcal{R}_{\tilde{\alpha}}$ is completely symmetric with respect to $(i,j)\leftrightarrow(j,i)$ if both gluons are final states. For this case a symmetry factor has to be applied. For final-state pairs $(P_i,P_j)=(f,\bar{f})$ due to $\gamma\rightarrow f\bar{f}$ splittings with $f$ denoting all charged fermions of the SM, where $(q,\bar{q})$ additionally can originate from $g\rightarrow q\bar{q}$ splittings, a similar symmetry factor as for $(g,g)$ has to be applied. Alternatively, and what is used in the following, an ordering of $i$ and $j$ according to the particles' identity, i.~e.
	\begin{align}
		P_j=f, P_i=\bar{f},\quad (j,i)\in \mathcal{P}_{\text{FKS}}\quad \Rightarrow\quad (i,j) \notin \mathcal{P}_{\text{FKS}}\quad \text{if}\quad n_I+1\leq i
	\end{align}
	must be imposed.
	
	All possible splittings allowed from SM vertices in terms of FKS pairs can be summarised in three groups with sets $G$ containing tuples $(P_i,P_j)$ with $(i,j)\in \mathcal{P}_{\text{FKS}}(f_r)$, each subdivided into two cases regarding QCD and QED splittings.\\
	\begin{itemize}
		\item[1.] 	The first group comprises the emission of gluons off quarks and photons off charged fermions or massive gauge bosons, i. e.
		\begin{align}
		&G_1^{\text{QCD}}=\{(q,g),\text{c.-c.}\} & G_1^{\text{QED}}=\{(f,\gamma),(W^{+},\gamma),\text{c.-c.}\}
		\label{GroupG1}
		\end{align}
		with `c.-c.' denoting the corresponding charge-conjugates of the previous tuples. In every tuple of these two sets at least one singularity exists which comes from either soft gluons or photons. For massless quarks for the QCD or massless fermions for the QED case there is an additional collinear singularity with respect to the FKS pairs $(q,g)$ or $(f,\gamma)$ and their corresponding charge-conjugates.
	\item[2.] The second group contains all splittings inducing both a soft and a collinear divergence of the real squared amplitudes. Since there are no (tree-level) self-interactions of photons in the SM $(\gamma,\gamma)$ pairs do not exist, and we are left with the sets
	\begin{align}
		&G_2^{\text{QCD}}=\{(g,g)\} &G_2^{\text{QED}}=\emptyset\quad.
	\end{align}
	\item[3.] All splittings inducing $f\bar{f}$ pairs belong to the third group for which collinear singularities in $\mathcal{R}_{\tilde{\alpha}}$ occur if $m_f=0$, but never a soft singularity. For this group the sets
	\begin{align}
		&G_3^{\text{QCD}}=\{(q,\bar{q})\} &G_3^{\text{QED}}=\{(f,\bar{f})\}
		\label{GroupG3}
	\end{align}
	can be defined. Note that if the third bullet point of the $\{q_s,q_e\}$ correction type conditions as stated above applies, the intersection $G_3^{\text{QCD}}\cap G_3^{\text{QED}}=\{(q,\bar{q})\}$ demands for subtraction of both, QCD and QED singularities.
		\end{itemize}
	
	The partition function $\mathcal{S}_{\tilde{\alpha}}$ introduced above is equivalent to $\mathcal{S}_{\mathcal{I}_{\tilde{\alpha}}}=\mathcal{S}_{(i,j)}$ with $\mathcal{I}_{\tilde{\alpha}}\in\mathcal{P}_{\text{FKS}}(f_r)$ since there is at most one FKS pair associated with one singular phase-space region $\tilde{\alpha}$.
	According to Eq.~(\ref{n+1phasespacepartition}) the condition
	\begin{align}
		\sum_{\tilde{\alpha}}\mathcal{S}_{\tilde{\alpha}}(f_r)=\sum_{(i,j)\in \mathcal{P}_{\text{FKS}}(f_r)}\mathcal{S}_{(i,j)}(f_r)=1
	\end{align}
	has to be imposed. With this constraint $\mathcal{S}_{(i,j)}$ can be explicitly constructed. To a certain extent, there is an arbitrariness in the concrete choice of the $\mathcal{S}_{\tilde{\alpha}}$ function describing the collinear and soft limits. However, according to Ref.~\cite{Frederix:2009yq}, independent of this choice, this function has to fulfil the following further conditions. First of all, it is required that $\mathcal{S}_{(i,j)}$ vanishes for all singular phase-space regions of FKS pairs distinct from $(i,j)$,
	\begin{align}
		&\lim_{\vec{k}_k\parallel\vec{k}_l}\mathcal{S}_{(i,j)}=0\qquad \forall (k,l)\ne (i,j) \quad\text{ with }\quad(k,l)\in \mathcal{P}_{\text{FKS}}\text{ and } m_k=m_l=0\\
		&\lim_{k^0_l\rightarrow 0}\mathcal{S}_{(i,j)}=0\qquad  \forall (k,l)\text{, }l\ne j\quad \text{ with }\quad P_l=g \text{ or }P_l=\gamma\text{ and }(k,l)\in \mathcal{P}_{\text{FKS}}.
	\end{align}
	In the collinear limit, the condition
	\begin{align}
		\lim_{\vec{k}_i\parallel\vec{k}_j}\mathcal{S}_{(i,j)}=h_{ij}(z_{ij}), \qquad z_{ij}=\frac{E_j}{E_j+E_i} \quad\text{ if }m_i=m_j=0
	\end{align}
	has to be fulfilled. $h_{ij}(z)$ is a symmetrisation factor defined in $0\le z\le 1$ with
	\begin{align}
		\lim_{z\rightarrow 0}h_{ij}(z)=1, \quad \lim_{z\rightarrow 1}h_{ij}(z)=0, \quad h_{ij}(z)+h_{ij}(1-z)=1,
	\end{align}
	which is thus a measure of the softness of the radiated particle $j$.
	In soft regions of the phase-space it is required that
	\begin{align}
		\lim_{k^0_j\rightarrow 0}\mathcal{S}_{(i,j)}=c_{ij} \qquad \text{ if } P_j=g\text{ or }P_j=\gamma \quad \text{ and } \sum_{i,(i,j)\in \mathcal{P}_{\text{FKS}}}c_{ij}=1
	\end{align}
	with $c_{ij}$ defined in $0<c_{ij}\le 1$. To meet these requirements the usual definition of $\mathcal{S}_{\tilde{\alpha}}$ used for example in \cite{Frixione:2007vw,Frederix:2009yq} is
	\begin{align}
		&\mathcal{S}_{(i,j)}=\frac{1}{\mathcal{D}}\frac{h_{ij}(z_{ij})}{d_{ij}}, \qquad (i,j)\in \mathcal{P}_{\text{FKS}} \label{SalphaDef}\\
		&\mathcal{D}\equiv \sum_{kl}d^{-1}_{kl}h_{kl}(z_{kl}),\quad (k,l)\in \mathcal{P}_{\text{FKS}}
	\end{align}
	where $d_{ij}$ must vanish if $k^0_i=0$, $k^0_j=0$ or $\vec{k}_i\parallel\vec{k}_j$, such that it can be expressed as
	\begin{align}
		d_{ij}=2k_i\cdot k_j \frac{k^0_ik^0_j}{(k^0_i+k^0_j)^2}\quad.
	\end{align}
	In the soft limit of $j$, i.~e. $k^0_j\rightarrow 0$, $\mathcal{D}d_{ij}$ which appears in the denominator of Eq.~(\ref{SalphaDef}) gets ill-defined. Therefore, for these regions the modified weights
	\begin{align}
		d_{ij}^{\text{soft}}=\frac{2k_i\cdot\hat{k}_j}{k^0_i} \qquad\text{ with }\quad\hat{k}_j=\frac{k_j}{k^0_j}
	\end{align}
	have to be used.
	The function $h_{ij}(z_{ij})$ can be chosen with respect to the asymmetrised set $\mathcal{P}_{\text{FKS}}$ of Eq.~(\ref{PFKSset}) as
	\begin{align}
		h_{ij}(z_{ij})=
		\begin{cases}
		z_{ij}=\frac{k^0_j}{k^0_j+k^0_i} & \text{ if }(P_i,P_j)=(g,g)\\
		1 & \text{ else}\quad.
		\end{cases}
	\end{align}
	\subsection{Subtraction of real-emission singularities}
	\label{secSubtractionterms}
	In this section the subtraction of singularities to real squared amplitudes in the infrared limits within the real subtracted contribution $d\sigma^{n+1}_{(q_s,q_e)}$ introduced in Eq.~(\ref{realsubtractedcontr}) is defined.
	In general, we note that the singularities in the FKS scheme appear in the limits $\xi\rightarrow0$ for soft radiation and $y\rightarrow1$ for collinear radiation off the final state. Both limits at the same time describes soft-collinear radiation. Due to this, real squared amplitudes  can be expressed in a regularised form, $\hat{\mathcal{R}}_{\tilde{\alpha}}$, by writing
	\begin{align}
		{\mathcal{R}}_{\tilde{\alpha}}=\frac{1}{\xi^2}\frac{1}{(1-y)}\left(\xi^2(1-y){\mathcal{R}}_{\tilde{\alpha}}\right)\equiv\frac{1}{\xi^2}\frac{1}{(1-y)}\hat{\mathcal{R}}_{\tilde{\alpha}}\quad.
		\label{regularisedRalpha}
	\end{align}
	If both initial state particles are potential emitters of a radiated particle $j$ these initial states have to be treated equivalently with respect to the subtraction of collinear singularities due to the origin of the emitted parton being non-reconstructable from the experimental point of view. For this reason, the FKS pairs $(i,j)$ with $1\le i\le 2$ and $P_j=g$ or $P_j=\gamma$ induce singularities in the collinear limits $y=\pm1$. Without loss of generality and by exchanging $(1-y)\rightarrow(1-y^2)$ in Eq.~(\ref{regularisedRalpha}) and in the following formulae all collinear singularities can be regulated as in the special case of final state radiation.
	
	The integral over the phase-space $\Phi_{\text{rad}}^{(E_{\tilde{\alpha}})}$ of real squared matrix elements with all infrared singularities subtracted can be formulated as
	\begin{align}
		\int d\Phi_{\text{rad}}^{(E_{\tilde{\alpha}})}(\xi,y,\phi)\left({\mathcal{R}}_{\tilde{\alpha}}(\xi,y)- {\mathcal{R}}_{\tilde{\alpha}}(0,y)-{\mathcal{R}}_{\tilde{\alpha}}(\xi,1)+{\mathcal{R}}_{\tilde{\alpha}}(0,1)\right)
	\end{align}
	which in regularised form, by means of Eq.~(\ref{n+1PSparametr}), gives
	\begin{align}
	\begin{split}
		\int^{2\pi}_{0}d\phi \int_{-1}^{1}dy\int^{\xi_{\text{max}}}_{0}d\xi  \frac{\mathcal{J}^{(E_{\tilde{\alpha}})}(\xi,y,\phi)}{\xi^2(1-y)}\left(\hat{\mathcal{R}}_{\tilde{\alpha}}(\xi,y)- \hat{\mathcal{R}}_{\tilde{\alpha}}(0,y)-\hat{\mathcal{R}}_{\tilde{\alpha}}(\xi,1)+\hat{\mathcal{R}}_{\tilde{\alpha}}(0,1)\right)
	\end{split}
	\label{realsubtraction}
	\end{align}
	The integrand of this equation is thus well-defined in the collinear and soft limit and can be evaluated by numerical means\footnote{Note, that for a numerical Monte-Carlo evaluation of the integrals, $\xi$ obtains a rescaling according to $\xi=\xi_{\text{max}}(y)\tilde{\xi}$ with new integration variable $\tilde{\xi}\in [0,1]$.  According to Sec.~4.5 of Ref.~\cite{Alioli:2010xd}, due to this variable transformation additional terms arise with an integrand factor $\log\xi_{\text{max}}(y)$ in the soft and $\log \xi_{\text{max}}(1)$ in the soft-collinear limit. Contrary to FSR, for which $\xi_{\text{max}}(y)=1$, for the case of ISR these terms in general do not vanish.}. As explained in Sec.~(\ref{seccrosssecformulation}) the subtraction term associated with a factorising Born process $f_{b,\tilde{\alpha}}$ in the soft and/or collinear limit is uniquely defined in the correction type per region $\tilde{\alpha}$. With  emitter $E_{\tilde{\alpha}}$ and splitting $s_{\tilde{\alpha}}$ declared in Eq.~(\ref{splittingdef}) this can be expressed by the numbers $\mathcal{C}^{\text{QCD}},\mathcal{C}^{\text{QED}}\in \{0,1\}$ as
	\begin{align}
		\begin{split}
	\text{QCD splitting:  }\quad	\mathcal{C}^{\text{QCD}}(s_{\tilde{\alpha}})=1,  \mathcal{C}^{\text{QED}}(s_{\tilde{\alpha}})=0 &\quad\text{if } (P_i,P_j)\in G^{\text{QCD}} \text{ and } P_{E_{\tilde{\alpha}}}\ne \gamma\\
		\text{QED splitting:  }\quad \mathcal{C}^{\text{QCD}}(s_{\tilde{\alpha}})=0, \mathcal{C}^{\text{QED}}(s_{\tilde{\alpha}})=1 &\quad\text{if } (P_i,P_j)\in G^{\text{QED}} \text{ and } P_{E_{\tilde{\alpha}}}\ne g\\
		&\text{with }\quad P_i,P_j\in f_r \text{ and } P_{E_{\tilde{\alpha}}}\in f_b
		\end{split}
	\end{align}
	From the integrand of Eq.~(\ref{realsubtraction}) the subtraction terms $S^{(E_{\tilde{\alpha}},{s}_{\tilde{\alpha}})}$ introduced in Eq.~(\ref{realsubtractedcontr}) can be extracted
	\begin{align}
	\begin{split}
	S^{(E_{\tilde{\alpha}},{s}_{\tilde{\alpha}})}&=S^{(E_{\tilde{\alpha}},{s}_{\tilde{\alpha}})}_{\text{soft}}+S^{(E_{\tilde{\alpha}},{s}_{\tilde{\alpha}})}_{\text{coll}}+S^{(E_{\tilde{\alpha}},{s}_{\tilde{\alpha}})}_{\text{soft-coll}}\\
	&=	\frac{\mathcal{J}^{(E_{\tilde{\alpha}})}(\xi,y,\phi)}{\xi^2(1-y)}\left( \hat{\mathcal{R}}_{\tilde{\alpha}}(0,y)+\hat{\mathcal{R}}_{\tilde{\alpha}}(\xi,1)-\hat{\mathcal{R}}_{\tilde{\alpha}}(0,1)\right)\quad.
	\end{split}
	\end{align}
	The soft subtraction terms are proportional to $\hat{\mathcal{R}}_{\tilde{\alpha}}(0,y)$ which can be rewritten as
	\begin{align}
	\begin{split}
		\hat{\mathcal{R}}_{\tilde{\alpha},{(q_s,q_e)}}(0,y)=\lim_{\xi\rightarrow0}\hat{\mathcal{R}}_{\tilde{\alpha},{(q_s,q_e)}}(\xi,y)=\lim_{\xi\rightarrow0}4\pi \mathcal{S}_{\tilde{\alpha}}^{\text{soft}}\xi^2(1-y)\sum_{k,l}^{n}\frac{k_k\cdot k_l}{(k_k\cdot k_j)(k_l\cdot k_j)}\\
		\left[\alpha_s\mathcal{B}^{\text{QCD}}_{kl,{(q_s-1,q_e)}}(f_{b,\tilde{\alpha}})\mathcal{C}^{\text{QCD}}(s_{\tilde{\alpha}})+\alpha\mathcal{B}^{\text{QED}}_{kl,{(q_s,q_e-1)}}(f_{b,\tilde{\alpha}})\mathcal{C}^{\text{QED}}(s_{\tilde{\alpha}})\right]
		\label{softnonint}
	\end{split}
	\end{align}
	with the partition function $\mathcal{S}_{\tilde{\alpha}}^{\text{soft}}$ depending purely on weights $d^{\text{soft}}_{ij}$ for soft radiated gluons or photons labeled by $j$.
	
	$\mathcal{B}^{\text{QCD}}_{ij}$ represents the colour-correlated squared matrix element and $\mathcal{B}^{\text{QED}}_{ij}$ the charge-correlated, respectively. Formally, these quantities read
	\begin{align}
	\begin{split}
	\mathcal{B}^{\text{QCD}}_{kl}=-\left(\mathcal{M}_{kl}^{(n)}\right)_{\text{QCD}}J_b= - \mathcal{N}_{B}J_b\sum_{
		\text{spin, col.}}\text{Re}& \left[
	\mathcal{A}^{(n)}_{c_1,\ldots,b_k,\ldots,b_l,\ldots,c_n}\right.\\ &\times \left.T^a_{b_kd_k}T^a_{b_ld_l}\mathcal{A}^{\,*(n)}_{c_1,\ldots,d_k,\ldots,d_l,\ldots,c_n} \right]\\
	\mathcal{B}^{\text{QED}}_{kl}=-\left(\mathcal{M}_{kl}^{(n)}\right)_{\text{QED}}J_b= - \mathcal{N}_{B}J_b\sum_{
		\text{spin, col.}}\text{Re}&\left[\mathcal{A}^{( n)}Q_kQ_l\mathcal{A}^{\,*(n)}\right]
	\label{QEDchargecorrelations}
	\end{split}
	\end{align}
	$T^a_{b_id_i}$ denotes the generator of the $SU(3)$ group in the fundamental representation and $Q_i$ the charge operator, respectively. The latter is defined as
	\begin{align}
	Q_i=(-1)^{s_i}e(P_i) \quad, \qquad s_i=\left\{\begin{array}{ll}
	2, & \quad \text{outgoing (anti-)particle}\\
	1, & \quad \text{incoming (anti-)particle}
	\end{array}\right. \quad ,
	\end{align}
	with $e(P_i)$ the electric charge of the particle $P_i$ which can be read off from Table~\ref{leptonsquarks}.
	
	In the collinear limit of $k_i$ and $k_j$ of the FKS pair, first of all considering the case of ISR, we can define the energy fraction $z$ as
	\begin{align}
		&z\equiv \frac{k_{E_{\tilde{\alpha}}}^0}{{k}_i^0} &1-z=\frac{k_j^0}{{k}_{i}^0}\quad.
		\label{defzandonemz}
	\end{align}
	with $k^{\mu}_{E_{\tilde{\alpha}}}$ denoting the momentum of the initial state emitter after radiation of a parton.
	By requiring the on-shell condition of the external states before and after radiation, the momentum $k^{\mu}$ of the parton which enters the underlying Born process can be parametrised as follows.
	By fixing $j$ as the radiated parton we choose $k^{\mu}=k^{\mu}_{E_{\tilde{\alpha}}}$ with $k^2<0$ and the momentum of the incoming parton $p^{\mu}=k^{\mu}_i$ and demand $\bar{k}_{E_{\tilde{\alpha}}}^2=k_j^2=k_i^2=0$ such that $k^{\mu}$ can be described as
	\begin{align}
		&k^{\mu}=k_{E_{\tilde{\alpha}}}^{\mu}=zp^{\mu}+k^{\mu}_{\perp}-\frac{|\vec{k}_{\perp}|^{2}}{(1-z)}\frac{\eta^{\mu}}{2p\cdot \eta}
	\end{align}
	with $\eta^{\mu}$ an auxiliary light-like momentum and $k^{\mu}_{\perp}$ the transverse momentum of the radiated parton with respect to the emitter particle. In summary, for these momenta the conditions
	\begin{align}
		k_{\perp}k_j=0\qquad k_{\perp}\eta=0\qquad \eta^2=0
	\end{align}
	must be satisfied. Furthermore, we define the normalised transverse momentum
	\begin{align}
		\hat{k}_{\perp}^{\mu}=\frac{k^{\mu}_{\perp}}{|\vec{k}_{\perp}|}\quad.
	\end{align}
	In the collinear limit the splitting of a massless particle can be described by the unregularised Altarelli-Parisi splitting functions \cite{Altarelli:1977zs} and written exclusive in spin states of the splitting parton \cite{Alioli:2010xd}, i.~e.
	\begin{align}
		&\hat{P}^{ss^{\prime}}_{{E_{\tilde{\alpha}}\rightarrow (i,j)}}(z,\hat{k}_{\perp})=\biggl\langle s\biggl\lvert\hat{P}_{{E_{\tilde{\alpha}}\rightarrow (i,j)}}(z,\hat{k}_{\perp})\bigg\rvert s^{\prime}\biggr\rangle \quad \text{with } P_{E_{\tilde{\alpha}}}=f,\bar{f}\\ &\hat{P}^{\mu\nu}_{{E_{\tilde{\alpha}}\rightarrow (i,j)}}(z,\hat{k}_{\perp})=\biggl\langle \mu\biggl\lvert\hat{P}_{{E_{\tilde{\alpha}}\rightarrow (i,j)}}(z,\hat{k}_{\perp})\bigg\rvert \nu \biggr\rangle\quad \text{with } P_{E_{\tilde{\alpha}}}=\gamma,g\quad.
	\end{align}
	For simplicity, we restrict ourselves to the QED case for the collinear subtraction terms in the following. The QCD terms are analogous, replacing $\alpha$ by $\alpha_s$, $\mathcal{C}^{\text{QED}}$  by $\mathcal{C}^{\text{QCD}}$ and the QED with the corresponding QCD splitting functions.
	The polarised splitting functions enter the regularised real squared matrix element in the collinear limit, $\hat{\mathcal{R}}_{\tilde{\alpha}}(\xi,y_{\text{coll}})$ with $y_{\text{coll}}=\pm 1$, which can be written in a factorised form as
	\begin{align}
		&\hat{\mathcal{R}}_{\tilde{\alpha}}(\xi,y_{\text{coll}})=\lim_{y\rightarrow y_{\text{coll}}}\frac{8\pi \alpha}{-k^2}\xi^2(1-y^2)\hat{P}^{ss^{\prime}}_{{E_{\tilde{\alpha}}\rightarrow (i,j)}}(z,\hat{k}_{\perp})\mathcal{B}_{ss^{\prime}}(zp)\mathcal{C}^{\text{QED}}(s_{\tilde{\alpha}}) \quad \text{with }P_{E_{\tilde{\alpha}}}={f,\bar{f}} \label{regRcollf}\\
		&\hat{\mathcal{R}}_{\tilde{\alpha}}(\xi,y_{\text{coll}})=\lim_{y\rightarrow y_{\text{coll}}}\frac{8\pi \alpha}{-k^2}\xi^2(1-y^2)\hat{P}^{\mu\nu}_{{E_{\tilde{\alpha}}\rightarrow (i,j)}}(z,\hat{k}_{\perp})\mathcal{B}_{\mu\nu}(zp)\mathcal{C}^{\text{QED}}(s_{\tilde{\alpha}}) \quad \text{with } P_{E_{\tilde{\alpha}}}=\gamma\label{regRcollgam}
	\end{align}
	where $\xi=1-z$ and $k^2$ depends on $y$ as
	\begin{align}
		k^2=\left(p-k_j\right)^2=\begin{cases}
		-2p^0k^0_j(1- y)=-2(p^0)^2(1-z)(1-y) & \text{if } y\rightarrow1\\
		-2p^0k^0_j(1+ y)=-2(p^0)^2(1-z)(1+y) & \text{if } y\rightarrow-1
		\end{cases}
		\label{k2collISR}
	\end{align}
	Hence, Eq.~(\ref{regRcollf}) and (\ref{regRcollgam}) are eventually rendered finite in the collinear limit.
	The spin-correlated squared matrix elements $\mathcal{B}_{ss^{\prime}}$ and $\mathcal{B}_{\mu\nu}$ entering these equations are defined as
	\begin{align}
		&\mathcal{B}_{ss^{\prime}}=\mathcal{N}_{B}J_b\sum_{\text{colour}}\mathcal{A}_s\mathcal{A}_{s^{\prime}}^{\dagger}\\
		&\mathcal{B}_{\mu\nu}=\mathcal{N}_{B}J_b\sum_{\text{colour}}\sum_{s,s^{\prime}}\mathcal{A}_s\mathcal{A}_{s^{\prime}}^{\dagger}(\epsilon_s)^{*}_{\mu}(\epsilon_{s^{\prime}})_{\nu}\quad.
	\end{align}
	Moreover, by using the definitions above, the polarised splitting functions for ISR can be expressed as
	\begin{align}
		\hat{P}^{ss^{\prime}}_{f\rightarrow f\gamma}(z,\hat{k}_{\perp})&=\delta^{ss^{\prime}}e(f)^2\frac{1+z^2}{1-z} \label{ftofgam}\\
		\hat{P}^{ss^{\prime}}_{f\rightarrow \gamma f}(z,\hat{k}_{\perp})&=\delta^{ss^{\prime}}n_c(f)e(f)^2\left(z^2+(1-z)^2\right)\label{ftogamf}\\
		\hat{P}^{\mu\nu}_{\gamma\rightarrow f f}(z,\hat{k}_{\perp})&=e(f)^2\left(-g^{\mu\nu}z+\frac{4(1-z)}{z}\hat{k}_{\perp}^{\mu}\hat{k}_{\perp}^{\nu}\right)
		\label{gamtoff}
	\end{align}
	with $n_c(f)$ the colour degrees of freedom of $f$. The splitting functions of Eq.~(\ref{ftofgam}) and (\ref{gamtoff}) by substituting $f\leftrightarrow q$ and $\gamma\leftrightarrow g$ are identical to the analogous QCD splitting functions up to the factor $e(f)^2$ which is replaced therein by the QCD Casimir operator $C_F$. The function $\hat{P}^{ss^{\prime}}_{q\rightarrow g q}$ agrees with Eq.~(\ref{ftogamf}) if replacing $n_c(f)e(f)^2$ with $T_F$, respectively. In addition, in QCD, $g\rightarrow gg$ splittings exists and the corresponding splitting function reads
	\begin{align}
		\hat{P}^{\mu\nu}_{g\rightarrow g g}(z,\hat{k}_{\perp})=C_A\left(-2\left[\frac{z}{1-z}+z(1-z)\right]g^{\mu\nu}+\frac{4(1-z)}{z}\hat{k}_{\perp}^{\mu}\hat{k}_{\perp}^{\nu}\right)\quad.
	\end{align}
	Analogously, for the case of FSR we redefine the energy fraction $z$ as
	\begin{align}
		&z\equiv\frac{k^0_i}{\bar{k}^0_i} &1-z=\frac{k_j^0}{\bar{k}^0_i}\quad.
		\label{varFSRz}
	\end{align}
	Furthermore, $k^{\mu}$ is the momentum of the massless splitting parton. With $p^{\mu}=k^{\mu}_i$ and the on-shell conditions for the external states of the Born and real process, it can be defined as
	\begin{align}
		k^{\mu}=k_{E_{\tilde{\alpha}}}^{\mu}=\bar{k}^{\mu}_i=\frac{p^{\mu}}{z}+k^{\mu}_{\perp}-\frac{z|\vec{k}_{\perp}|^{2}}{(1-z)}\frac{\eta^{\mu}}{2p\cdot \eta}\quad.
	\end{align}
	The regularised real squared matrix elements in the collinear limit of FSR, i.~e. $\hat{\mathcal{R}}_{\tilde{\alpha}}(\xi,y_{\text{coll}})$ with $y_{\text{coll}}=1$, factorises as
	\begin{align}
				&\hat{\mathcal{R}}_{\tilde{\alpha}}(\xi,y_{\text{coll}})=\lim_{y\rightarrow y_{\text{coll}}}\frac{8\pi \alpha}{k^2}\xi^2(1-y)\hat{P}^{ss^{\prime}}_{{E_{\tilde{\alpha}}\rightarrow (i,j)}}(z,\hat{k}_{\perp})\mathcal{B}_{ss^{\prime}}\left(\frac{p}{z}\right)\mathcal{C}^{\text{QED}}(s_{\tilde{\alpha}}) \quad \text{with }P_{E_{\tilde{\alpha}}}={f,\bar{f}} \label{regRcollfFSR}\\
		&\hat{\mathcal{R}}_{\tilde{\alpha}}(\xi,y_{\text{coll}})=\lim_{y\rightarrow y_{\text{coll}}}\frac{8\pi \alpha}{k^2}\xi^2(1-y)\hat{P}^{\mu\nu}_{{E_{\tilde{\alpha}}\rightarrow (i,j)}}(z,\hat{k}_{\perp})\mathcal{B}_{\mu\nu}\left(\frac{p}{z}\right)\mathcal{C}^{\text{QED}}(s_{\tilde{\alpha}}) \quad \text{with } P_{E_{\tilde{\alpha}}}=\gamma\label{regRcollgamFSR}
	\end{align}
	with
	\begin{align}
		k^2=2k_i^0k_j^0(1-y)=2z(1-z)(\bar{k}_i^0)^2(1-y)=2\frac{z}{1-z}\xi^2s(1-y)
	\end{align}
	and the final-state polarised splitting functions for QED
	\begin{align}
				\hat{P}^{ss^{\prime}}_{f\rightarrow f\gamma}(z,\hat{k}_{\perp})&=\delta^{ss^{\prime}}e(f)^2\frac{1+z^2}{1-z} \label{ftofgamFSR}\\
		\hat{P}^{\mu\nu}_{\gamma\rightarrow f f}(z,\hat{k}_{\perp})&=n_c(f)e(f)^2\left(-g^{\mu\nu}-{4z(1-z)}\hat{k}_{\perp}^{\mu}\hat{k}_{\perp}^{\nu}\right)\quad.
		\label{gamtoffFSR}
	\end{align}
	In the same way as for ISR explained above, the analogous QCD splitting functions follow from exchanging $e(f)^2$ with $C_F$ in Eq.~(\ref{ftofgamFSR}) 
	and $n_c(f)e(f)^2$ with $T_F$ in Eq.~(\ref{gamtoffFSR}), respectively. The additional $g\rightarrow gg$ splitting function in the QCD case for FSR reads
	\begin{align}
		\hat{P}^{\mu\nu}_{g\rightarrow g g}(z,\hat{k}_{\perp})=C_A\left(-2\left[\frac{z}{1-z}+\frac{1-z}{z}\right]g^{\mu\nu}+4z{(1-z)}\hat{k}_{\perp}^{\mu}\hat{k}_{\perp}^{\nu}\right)\quad.
	\end{align}
	In the soft-collinear limit the regularised real squared matrix elements follow from taking the soft limit $\xi\rightarrow0$ of Eq.~(\ref{regRcollf}) for ISR and Eq.~(\ref{regRcollfFSR}) for FSR. In both cases $\hat{\mathcal{R}}_{\tilde{\alpha}}(0,y_{\text{coll}})$ are rendered finite by considering the fact that $1-z$ and $k^0_j$ are proportional to $\xi$.
	\subsection{Non-singular tree-level $(n+1)$-body contributions}
	\label{secnonsing}
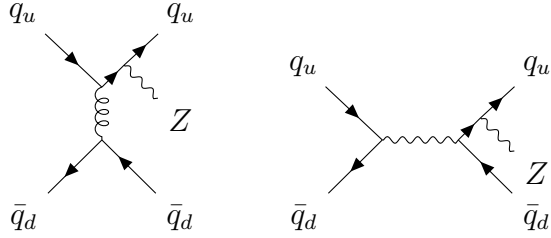
\begin{figure}
			\centering
			\begin{tikzpicture}
			\begin{feynman}
			\vertex(a);
			\vertex[above left=1cm of a] (i1){$q_u$};
			\vertex[above right=1cm of a] (f1){$q_u$};
			\vertex[below=0.7cm of a] (b);
			\vertex[below left=1cm of b] (i2){$\bar{q}_d$};
			\vertex[below right=1cm of b] (f2){$\bar{q}_d$};
			\vertex[above right=1em of a] (ii1);
			\vertex[below right=1.5em of ii1] (ii2){$Z$};
			\diagram {
				(i1) -- [fermion,small] (a) -- [fermion,small] (ii1) -- [fermion,small] (f1),
				(a) -- [gluon,small] (b),
				(i2) -- [anti fermion,small] (b) -- [anti fermion,small] (f2),
				(ii1) -- [photon,small] (ii2),
			};
			\end{feynman}
			\end{tikzpicture}
			\qquad
			\begin{tikzpicture}
			\begin{feynman}
			\vertex(a);
			\vertex[above left=1cm of a] (i1){$q_u$};
			\vertex[below left=1cm of a] (i2){$\bar{q}_d$};
			\vertex[right=1cm of a] (b);
			\vertex[above right=1cm of b] (f1){$q_u$};
			\vertex[below right=1cm of b] (f2){$\bar{q}_d$};
			\vertex[above right=1em of b] (ii1);
			\vertex[below right=1.5em of ii1] (ii2){$Z$};
			\diagram {
				(i1) -- [fermion,small] (a) -- [fermion,small] (i2),
				(a) -- [photon] (b),
				(f2) -- [fermion,small] (b) -- [fermion,small] (ii1) -- [fermion,small] (f1),
				(ii1) -- [photon,small] (ii2),
			};
			\end{feynman}
			\end{tikzpicture}
		\caption{Tree-level interfering diagrams with gluon exchange at $\mathcal{O}(g_s^2e)$ and with $W^{\pm}$ exchange at $\mathcal{O}(e^3)$ leading to $\mathcal{O}(\alpha_s\alpha^2)$ contributions.}
		\label{nonsingreal}
	\end{figure}

	For processes with at least one jet in hadron collisions and three jets in lepton collisions, considering EW corrections real-emission processes can occur to the same coupling order as tree-level LO $(n+1)$-body processes without singular regions.
	Both processes lead to same hadronic external states and thus experimentally cannot be distinguished.
	Contributions of non-singular tree-level partonic processes can occur if interferences from amplitudes to different powers in $g_s$ and $e$ are not colour-forbidden. For an example we consider the process $pp\rightarrow Zj$ at $\mathcal{O}(\alpha_s\alpha)$ which at NLO EW can get an additional jet with respect to the Born process.
	Protons and jets always involve massless quarks. Thus, the sub-processes $q_u\bar{q}_d\rightarrow Zq_u\bar{q}_d$ and its charge conjugates with $m_q=0$ at $\mathcal{O}(\alpha_s\alpha^2)$ contribute in the same way as all real emission processes  which are sub-processes of $pp\rightarrow Zjj$. These additional non-vanishing contributions are due to the interference of a diagram with $W^{\pm}$ exchange leading to $\mathcal{O}(e^3)$ amplitudes with a corresponding one with gluon exchange leading to $\mathcal{O}(g_s^2e)$ amplitudes, respectively. Exemplarily these diagrams are depicted in Fig.~\ref{nonsingreal}. Due to the absence of collinear or soft splittings for which the squared amplitudes of an underlying Born process can be factored out in the corresponding infrared limits there are effectively no phase-space regions for which the interference of these amplitudes diverges.
	
	The sum over all regions $\tilde{\alpha}$ in Eq.~(\ref{realsubtractedcontr}) implicitly contains the sum over all possible real process structures $f_r$ introduced in Sec.~\ref{secFKSphasespace} for which a singular structure of the squared amplitudes in the phase space is manifest. The last term of this equation however comprises all contributions from the non-singular processes at the same NLO coupling order, $\mathcal{O}(\alpha_s^{q_s}\alpha^{q_e})$, and can be expressed as a sum over the non-singular process particle structures $f_{r,\text{non-s.}}$,
	\begin{align}
		\mathcal{R}_{\text{non-sing},(q_s,q_e)}\left(\Phi_{n+1}\right)=\sum_{f_{r,\text{non-s.}}}\mathcal{R}_{(q_s,q_e)}\left(f_{r,\text{non-s.}},\Phi_{n+1}\right)\quad.
	\end{align}
	
	\subsection{Integrated subtraction terms of the $n$-body phase-space}
	\label{secIntegratedSubs}
	The formalism for the FKS terms presented in this section is based on the approach following Ref.~\cite{Jezo:2015aia}.

	The subtraction terms in integrated form $I_{(q_s,q_e)}(\bar{\Phi}_{n})$ from Eq.~(\ref{Isubtractionterms}) to be integrated over $n$ particle phase-space dimensions and expanded as a series in $1/\varepsilon$-poles can be achieved by making use of dimensional regularisation writing the radiated phase-space measure of Eq.~(\ref{dkn+1}) in $d=4-2\varepsilon$ dimensions,
	\begin{align}
		d\Phi_{\text{rad}}=\frac{d^{3-2\varepsilon}k_{N+1}}{2k^0_{N+1}(2\pi)^{3-2\varepsilon}}=\frac{s^{1-\varepsilon}}{(4\pi)^{3-2\varepsilon}}\xi^{1-2\varepsilon}(1-y^2)^{-\varepsilon}d\xi dyd\Omega^{(2-2\varepsilon)}\quad.
		\label{intRealdimreg}
	\end{align}
	With this phase-space measure and Eq.~(\ref{regularisedRalpha}) the diverging integral of the real emission amplitudes for a singular $\tilde{\alpha}$ FKS region over the $n+1$ phase-space can be rewritten as
	\begin{align}
	\begin{split}
		\int d\Phi_{\text{rad}}^{(E_{\tilde{\alpha}})}{\mathcal{R}}_{\tilde{\alpha}}(\xi,y)=\frac{s^{1-\varepsilon}}{(4\pi)^{3-2\varepsilon}}\int d\Omega^{(2-2\varepsilon)}\int_{-1}^{1}dy(1-y)^{-1-\varepsilon}(1+y)^{-\varepsilon}\\
		\int_{0}^{\xi_{\text{max}}}d\xi \xi^{-1-2\varepsilon}\hat{\mathcal{R}}_{\tilde{\alpha}}(\xi,y)
	\end{split}
		\label{realdimreg}
	\end{align}
	
	We define the plus-distributions
	\begin{align}
	\int_{-1}^{1}dy\left(\frac{g(y)}{1\mp y}\right)_{\delta_{o/i}} f(y)&=\int_{-1}^{1}dyg(y)\frac{f(y)-f(\pm 1)\Theta(\delta_{o/i}-1\pm y)}{1\mp y}\label{plusdistrydelta}\\
	\int_{0}^{\xi_{\text{max}}}d\xi\left(\frac{g(\xi)}{\xi}\right)_c f(\xi)&=\int_{0}^{\xi_{\text{max}}}d\xi g(\xi)\frac{f(\xi)-f(0)\Theta(\xi_c-\xi)}{\xi}\label{plusdistrxicut}
	\end{align}
	with arbitrary FKS parameters $\xi_c\in[0,\xi_{\text{max}}]$ and $\delta_{o/i} \in [0,2]$. Setting the values\footnote{The parameters $\xi_c$, $\delta_o$ and $\delta_i$ are set to these values as a default in \texttt{WHIZARD}.} $\xi_c=\xi_{\text{max}}$, $\delta_o =2$ and $\delta_i =2$, results in the simplified plus-distributions,
	\begin{align}
	\int_{-1}^{1}dy\left(\frac{g(y)}{1-y}\right)_+ f(y)&=\int_{-1}^{1}dyg(y)\frac{f(y)-f(1)}{1-y} \label{plusdistributionsy}\\
	\int_{0}^{\xi_{\text{max}}}d\xi\left(\frac{g(\xi)}{\xi}\right)_+ f(\xi)&=\int_{0}^{\xi_{\text{max}}}d\xi g(\xi)\frac{f(\xi)-f(0)}{\xi}\quad.
	\label{plusdistributions}
	\end{align}
	For simplicity, we use these plus-distributions and leave out the dependence on the arbitrary FKS parameters $\xi_c$ and $\delta_o$ ($\delta_i$ in the case of ISR) in the following. We come back to them only for the final explicit formulas.
	
	By using the general trick
	\begin{align}
		\int dx \frac{f(x)}{x^{1+\varepsilon}}=\int dx \frac{f(x)-f(0)}{x^{1+\varepsilon}}+f(0)\int dx~ {x^{-1-\varepsilon}}
	\end{align}
	and the simplified plus-distributions of Eqs.~(\ref{plusdistributionsy}) and (\ref{plusdistributions}),
	the factors $(1-y)^{-1-\varepsilon}$ and $\xi^{-1-2\varepsilon}$ in the integrand of Eq.~(\ref{intRealdimreg}) expanded in ${\varepsilon}$ can be expressed as
	\begin{align}
		&\xi^{-1-2\varepsilon}=-\frac{1}{2\varepsilon}\delta(\xi)+\left(\frac{1}{\xi}\right)_+-2\varepsilon\left(\frac{\log\xi}{\xi}\right)_++\mathcal{O}(\varepsilon^2) \label{xiexpansion}\\
		&(1-y)^{-1-\varepsilon}=-\frac{2^{-\varepsilon}}{\varepsilon}\delta(1-y)+\left(\frac{1}{1-y}\right)_+-\varepsilon\left(\frac{\log(1-y)}{1-y}\right)_++\mathcal{O}(\varepsilon^2)\quad.\label{yexpansion}
	\end{align}
	
	Inserting these identities into Eq.~(\ref{realdimreg}) the resulting term which is free of $\delta(\xi)$ and $\delta(1-y)$ functions and for which $\mathcal{O}(\varepsilon)$ and higher contributions are dropped is
	\begin{align}
		\int_{0}^{2\pi}d\phi\int_{-1}^{1}dy\int^{\xi_{\text{max}}}_{0}d\xi  {\mathcal{J}^{(E_{\tilde{\alpha}})}(\xi,y,\phi)}\left(\frac{1}{\xi}\right)_+\left(\frac{1}{1-y}\right)_+\hat{\mathcal{R}}_{\tilde{\alpha}}(\xi,y)\quad.
	\end{align}
	By the definition of plus distributions, this term is identical to the finite real-subtracted part of the cross section given in Eq.~(\ref{realsubtraction}), such that we are left with the subtraction terms containing $\delta(\xi)$ for the soft, $\delta(1-y)$ for the collinear and $\delta(\xi)\delta(1-y)$ for the soft-collinear description of the limits.
	
	The soft subtraction terms in integrated form can be obtained from Eq.~(\ref{realdimreg}) by replacing $\xi^{-1-2\varepsilon}$ with the first term of Eq.~(\ref{xiexpansion}) which yields
	\begin{align}
	\begin{split}
		I^{\text{soft}}_{\tilde{\alpha},(q_s,q_e)}(\bar{\Phi}_{n})=&-\frac{1}{2\varepsilon}\frac{s^{1-\varepsilon}}{(4\pi)^{3-2\varepsilon}}\int d\Omega^{(2-2\varepsilon)}\int_{-1}^{1}dy(1-y^2)^{-\varepsilon}\int_{0}^{\xi_{\text{max}}}d\xi \delta(\xi) \xi^2{\mathcal{R}}_{\tilde{\alpha}}(\xi,y)\\
		=&-\frac{1}{2\varepsilon}\frac{s^{1-\varepsilon}}{(4\pi)^{3-2\varepsilon}}\int d\Omega^{(2-2\varepsilon)}\int_{-1}^{1}dy(1-y^2)^{-\varepsilon}\left[\lim_{\xi\rightarrow 0}\xi^2{\mathcal{R}}_{\tilde{\alpha}}(\xi,y)\right]
	\end{split}
	\end{align}
	Performing now the sum over all regions $\tilde{\alpha}$ corresponding to one underlying Born structure $f_b$ and using the factorised form of the real squared amplitudes in the soft limit according to Eq.~(\ref{softnonint}) we get
	\begin{align}
	\begin{split}
		\sum_{\tilde{\alpha}\in \mathcal{P}_{\text{FKS}}(f_b)}I^{\text{soft}}_{\tilde{\alpha},(q_s,q_e)}(\bar{\Phi}_{n})=\sum_{k,l}^{n}\left[\frac{\alpha_s}{2\pi}\mathcal{B}^{\text{QCD}}_{kl,{(q_s-1,q_e)}}(f_b,\bar{\Phi}_{n})+\frac{\alpha}{2\pi}\mathcal{B}^{\text{QED}}_{kl,{(q_s,q_e-1)}}(f_b,\bar{\Phi}_{n})\right]\\
		\underbrace{\lim_{\xi\rightarrow0}\left(-\frac{2^{2\varepsilon}}{2\varepsilon}\frac{s^{1-\varepsilon}\mu_R^{2\varepsilon}}{(2\pi)^{1-2\varepsilon}}\right)\int d\Omega^{(2-2\varepsilon)}\int_{-1}^{1}dy(1-y^2)^{-\varepsilon}\frac{\xi^2}{4}\frac{k_k\cdot k_l}{(k_k\cdot k_j)(k_l\cdot k_j)}}_{=\sum_{\rho}\mathcal{E}^{(m_k,m_l)}_{kl,\rho}}.
		\label{softSubtractiontermint}
	\end{split}
	\end{align}
	With the eikonal integral factors $\mathcal{E}^{(m_k,m_l)}_{kl,\rho}$ which come along with poles $\varepsilon^{\rho}$ for $\rho=-2,-1,0$ we arrive at the form of the soft subtraction terms for which the finite part is represented by Eq.~(\ref{softsubint}). The divergent terms proportional to $\mathcal{E}^{(m_k,m_l)}_{kl,-1}\varepsilon^{-1}$ and $\mathcal{E}^{(m_k,m_l)}_{kl,-2}\varepsilon^{-2}$ exactly cancel with the corresponding divergent terms from virtual soft and soft-collinear emissions of massless partons. These are contained in $\mathcal{V}^{(1)}$ and $\mathcal{V}^{(2)}$ for which the regularisation scheme, here the conventional dimensional regularisation (CDR) scheme, has to be applied consistently to all terms. For the finite remainder $I^{(0)}_{(q_s,q_e)}$ the explicit form of the eikonal integral $\mathcal{E}^{(m_k,m_l)}_{kl,0}$ is given in App.~\ref{secAppendixeikonals}, where $\mathcal{V}^{(0)}$ are the finite remnants of one-loop integrals which are provided by OLPs as for example \texttt{OpenLoops}. In addition to the choice of the regularisation scheme, the same normalisation factor $\mathcal{N}$ has to be applied to the subtraction terms as well as to the virtual one-loop matrix elements. The normalisation of the subtraction terms in \texttt{WHIZARD} is chosen according to the BLHA standard\footnote{Note, that the convention for the normalisation factor can differ with respect to the OLP which is interfaced. Remarks on the convention of \texttt{RECOLA} distinct to the one of the BLHA interface and the consequential adjustments in \texttt{WHIZARD} are given in \cite{Weiss:2017qbj}.} \cite{Binoth:2010xt,Alioli:2013nda}, i.~e.
	\begin{align}
		\mathcal{N}(\varepsilon)=\frac{(4\pi)^{\varepsilon}}{\Gamma(1-\varepsilon)}\left(\frac{\mu_R^2}{Q^2}\right)^{\varepsilon}\quad,
		\label{normalisationfactor}
	\end{align}
	where $Q$ denotes the Ellis-Sexton scale \cite{Ellis:1985er}.
	
	The procedure to obtain the collinear subtraction terms in integrated form follows similarly as stated above. Since collinear singularities from ISR need a special treatment due to extra counterterms coming from the PDF evolution which will be adressed in Sec.~(\ref{secDglapremnant}), this section is restricted to the subtraction terms from collinear FSR. Note, that the soft-collinear singularities are already included by the terms proportional to $1/\varepsilon^2$ in the soft subtraction terms $I^{\text{soft}}_{\tilde{\alpha},(q_s,q_e)}$. Therefore the following definition is used, representing the expression of Eq.~(\ref{xiexpansion}) for which the $1/\varepsilon$ term is subtracted and $\mathcal{O}(\alpha^2)$ terms are dropped,
	\begin{align}
	\begin{split}
		\mathcal{P}_+(\xi)=\left(\frac{1}{\xi}\right)_+-2\varepsilon\left(\frac{\log\xi}{\xi}\right)_+
		&=\xi^{-1-2\varepsilon}+\frac{1}{2\varepsilon}\delta(\xi)\\
		&=\xi_{\text{max}}^{-1-2\varepsilon}\left((1-z)^{-1-2\varepsilon}+\frac{\xi_{\text{max}}^{2\varepsilon}}{2\varepsilon}\delta(1-z)\right)\quad.
		\label{Pplusdistr}
	\end{split}
	\end{align}
	In the r.~h.~s. of this equation the parametrisation $k_j^0=(1-z)\bar{k}^0_i$ according to Eq.~(\ref{varFSRz}) with $j$ the radiated parton and the FSR condition $\xi_{\text{max}}=2\bar{k}_i/\sqrt{s}$ is used. Replacing $(1-y)^{-1-\varepsilon}$ with the first term of Eq.~(\ref{yexpansion}) and $\xi^{-1-2\varepsilon}$ with $\mathcal{P}_+(\xi)$ in Eq.~(\ref{realdimreg}) yields
	\begin{align}
	\begin{split}
		I^{\text{coll}}_{\tilde{\alpha}}(\bar{\Phi}_{n})=\frac{s^{1-\varepsilon}}{(4\pi)^{3-2\varepsilon}}\int d\Omega^{(2-2\varepsilon)}\int_{-1}^{1}dy\int_{0}^{\xi_{\text{max}}}d\xi {\mathcal{P}_+(\xi)}\\
		\times\left[-\frac{2^{-\varepsilon}}{\varepsilon}\delta(1-y)\right](1-y)\xi^2{\mathcal{R}}_{\tilde{\alpha}}(\xi,y)
	\end{split}
	\end{align}
	which by integrating over the Born phase-space gives
	\begin{align}
	\begin{split}
		\int d\bar{\Phi}_{n}I^{\text{coll}}_{\tilde{\alpha}}(\bar{\Phi}_{n})=&\int (2\pi)^d\delta^{(d)}\left(k_{\oplus}+k_{\ominus}-\sum_{l=n_I+1}^{N+1}k_l\right)\prod_{l\neq i,j}d\Phi_l\\
		&\times\frac{\left(k^0_i\right)^{1-2\varepsilon}}{2(2\pi)^{3-2\varepsilon}}dk^0_id\Omega_i^{(3-2\varepsilon)}d\bar{k}_i^0\delta(\bar{k}_i^0-k^0_i-k^0_j)\\
		&\times\left(-\frac{2^{-\varepsilon}}{\varepsilon}\right)\frac{\left(k_j^0\right)^{1-2\varepsilon}}{2(2\pi)^{3-2\varepsilon}}dk_j^0d\Omega_j^{(2-2\varepsilon)}\frac{\mathcal{P}_+(\xi)}{\xi^{-1-2\varepsilon}}\left[\lim_{y\rightarrow1}(1-y){\mathcal{R}}_{\tilde{\alpha}}(\xi,y)\right].
		\label{collIntegral}
	\end{split}
	\end{align}
	The real squared amplitudes in the collinear limit, i.~e. the expression in brackets, can be further written in a factorised form according to Eqs.~(\ref{regRcollfFSR}) and (\ref{regRcollgamFSR}) and space-averaged due to the angular integration over $d\Omega_j^{(2-2\varepsilon)}$,
	\begin{align}
	\begin{split}
		\biggr\langle\lim_{y\rightarrow1}(1-y){\mathcal{R}}_{\tilde{\alpha}}(\xi,y)\biggr\rangle_{\Omega}=&\frac{8\pi\mu_R^{2\varepsilon}}{2(\bar{k}^0_i)^2z(1-z)}\langle \hat{P}\rangle_{\tilde{\alpha}}(z,\varepsilon)\\
		&\times \left(\alpha_s\mathcal{B}^{\text{QCD}}_{(q_s-1,q_e)}(f_{b,\tilde{\alpha}})+\alpha\mathcal{B}^{\text{QED}}_{(q_s,q_e-1)}(f_{b,\tilde{\alpha}})\right)\quad.
		\label{spaceavRalpha}
	\end{split}
	\end{align}
	Here, $\langle \hat{P}\rangle_{\tilde{\alpha}}$ denotes the space-averaged unregularised Altarelli-Parisi splitting function for which explicit formulas are given in Eqs.~(\ref{space-averagedAPsplitfgam}) to (\ref{space-averagedAPsplitgg}). Inserting Eqs.~(\ref{Pplusdistr}) and (\ref{spaceavRalpha}) into Eq.~(\ref{collIntegral}), using the relations $k_i^0=z\bar{k}_i^0$ and $k_j^0=(1-z)\bar{k}^0_i$ and performing the integration over $d\Omega^{(2-2\varepsilon)}_j$ yields
	\begin{align}
	\begin{split}
		\int d\bar{\Phi}_{n}I^{\text{coll}}_{\tilde{\alpha},(q_s,q_e)}(\bar{\Phi}_{n})=&-\frac{2^{-\varepsilon}}{\varepsilon}\frac{(4\pi)^{\varepsilon}}{\Gamma(1-\varepsilon)}\left(\frac{\mu_R}{\bar{k}^0_i}\right)^{2\varepsilon}\int d\bar{\Phi}_n\\
		&\times\left(\frac{\alpha_s}{2\pi}\mathcal{B}^{\text{QCD}}_{(q_s-1,q_e)}(f_{b,\tilde{\alpha}})+\frac{\alpha}{2\pi}\mathcal{B}^{\text{QED}}_{(q_s,q_e-1)}(f_{b,\tilde{\alpha}})\right)\\
		&\times \int_{0}^{1}dz~z^{-2\varepsilon}(1-z)\left[(1-z)^{-1-2\varepsilon}+\frac{\xi_{\text{max}}^{2\varepsilon}}{2\varepsilon}\delta(1-z)\right]\langle\hat{P}\rangle_{\tilde{\alpha}}(z,\varepsilon).
		\label{intCollsub}
	\end{split}
	\end{align}
	The prefactor on the r.~h.~s. of this equation can be further modified to
	\begin{align}
		-\frac{2^{-\varepsilon}}{\varepsilon}\frac{(4\pi)^{\varepsilon}}{\Gamma(1-\varepsilon)}\left(\frac{\mu_R}{\bar{k}^0_i}\right)^{2\varepsilon}=-\frac{\mathcal{N}(\varepsilon)}{\varepsilon}\left(\frac{2Q^2}{s}\right)^{\varepsilon}\xi_{\text{max}}^{-2\varepsilon}
	\end{align}
	such that the normalisation $\mathcal{N}(\varepsilon)$ to be used consistently for $I^{\text{soft}}$, $I^{\text{coll}}$ and $\mathcal{V}$ is factored out.
	The resulting integrals corresponding to the two terms in brackets of Eq.~(\ref{intCollsub}), which can be performed in an analytical way, are
	\begin{align}
		I^{(0)}_{z,\tilde{\alpha}}=&\int_{0}^{1}dz~z^{-2\varepsilon}(1-z)^{-2\varepsilon}\langle\hat{P}\rangle_{\tilde{\alpha}}(z,\varepsilon)\label{collinearintegrala}\\
		I^{(1)}_{z,\tilde{\alpha}}=&\int_{0}^{1}dz~z^{-2\varepsilon}(1-z)\frac{\xi_{\text{max}}^{-2\varepsilon}}{2\varepsilon}\delta(1-z)\langle\hat{P}\rangle_{\tilde{\alpha}}(z,\varepsilon)\quad.
		\label{collinearintegralb}
	\end{align}
	By evaluating the integrals in these terms, as e.~g. done in \cite{Jezo:2015aia}, and expanding
	\begin{align}
		\mathcal{T}_E=-\sum_{\tilde{\alpha}\in\mathcal{P}_{\text{FKS}}}\delta_{EE_{\tilde{\alpha}}}\frac{1}{\varepsilon}\left(\frac{2Q^2}{s}\right)^{\varepsilon}\xi_{\text{max}}^{-2\varepsilon}\left(I^{(0)}_{z,\tilde{\alpha}}+I^{(1)}_{z,\tilde{\alpha}}\right)
	\end{align}
	as a series in $\varepsilon$, the collinear integrated subtraction terms $I^{\text{coll}}_{\tilde{\alpha},(q_s,q_e)}$ summed over all singular regions $\tilde{\alpha}$ corresponding to one Born structure $f_b$ yields
	\begin{align}
	\begin{split}
		\sum_{\tilde{\alpha}\in \mathcal{P}_{\text{FKS}}(f_b)}I^{\text{coll}}_{\tilde{\alpha},(q_s,q_e)}(\bar{\Phi}_{n})=\mathcal{N}(\varepsilon)\sum_{E(f_b)}\left(\frac{\alpha_s}{2\pi}\mathcal{B}^{\text{QCD}}_{(q_s-1,q_e)}(f_{b})\mathcal{T}^{\text{QCD}}_{E(f_b)}\right.\\
		\left.+\frac{\alpha}{2\pi}\mathcal{B}^{\text{QED}}_{(q_s,q_e-1)}(f_{b})\mathcal{T}^{\text{QED}}_{E(f_b)}\right)\quad.
	\end{split}
	\end{align}
	$\mathcal{T}_E$ depends on the correction type via the corresponding splitting functions and explicitly reads
	\begin{align}
	\mathcal{T}_{E(f_b)}^{\text{QCD}}=\frac{1}{\varepsilon}\left(-2\log \xi_{\text{max}}C^{\text{QCD}}_{E(f_b)}+\gamma^{\text{QCD}}_{E(f_b)} \right)+\mathcal{Q}^{\text{QCD}}_{E(f_b)}
	\end{align}
	with
	\begin{align}
	\mathcal{Q}^{\text{QCD}}_{E(f_b)}=2\log\xi_{\text{max}}\left(\log \xi_{\text{max}}-\log \frac{Q^2}{s}\right)C_{E(f_b)}^{\text{QCD}}+\left(\log \frac{Q^2}{s}-2\log\xi_{\text{max}}\right)\gamma_{E(f_b)}^{\text{QCD}}+\gamma^{\prime \text{QCD}}_{E(f_b)}
	\label{Qcoll}
	\end{align}
	where $\mathcal{T}_{E(f_b)}^{\text{QED}}$ and $\mathcal{Q}_{E(f_b)}^{\text{QED}}$ are analogous by substituting the superscripts $\text{`QCD'}\leftrightarrow\text{`QED'}$. The Casimir operators $C_{E(f_b)}^{\text{QCD}}$ and $C_{E(f_b)}^{\text{QED}}$ and definitions $\gamma^{\text{QCD}}_{E(f_b)}$, $\gamma^{\text{QED}}_{E(f_b)}$, $\gamma^{\prime\text{QCD}}_{E(f_b)}$ and $\gamma^{\prime\text{QED}}_{E(f_b)}$ are well-defined by the particle identity $P_{E(f_b)}$ of the corresponding emitter. Their explicit values are given in App.~\ref{groupfactors}. The term proportional to $1/\varepsilon$ in $\mathcal{T}_{E(f_b)}$ in the FKS scheme is cancelled by the corresponding terms of $\mathcal{V}^{(1)}$ describing the collinear limit of virtual emitted partons. The finite remnant of the collinear subtraction terms is finally given by the general form already introduced in Eq.~(\ref{collinearsubint}). If keeping a dependence on FKS parameters $\xi_c$ and $\delta_o$ via the plus distributions of Eq.~(\ref{plusdistrydelta}) and (\ref{plusdistrxicut}) explicit, the expression of $\mathcal{Q}_{E(f_b)}$ changes by the substitution
	\begin{align}
		\frac{Q^2}{s}\rightarrow \frac{2Q^2}{s\delta_o}
	\end{align}
	as well as by adding the term
	\begin{align}
		2\left(\log \frac{2Q^2}{s\delta_o} \log \xi_{c}-\log^2\xi_c\right)C_{E(f_b)} \quad.
	\end{align}
	By setting the values $\delta_o=2$ and $\xi_c=1$, Eq.~(\ref{Qcoll}) is recovered again.
	\subsection{Initial state remnant terms}
	\label{secDglapremnant}
	The integrated subtraction terms for collinear ISR can be constructed in a similar way for collinear FSR, as discussed in the last section. However, different to final-state splittings, for which the collinear singularities of the real contribution are cancelled by counter-terms from the virtual one, the corresponding divergences of the initial state arise from collinearly splitting partons of the beam and not of the hard process. Thus, the cancellation of collinear initial-state singularities requires additional counterterms coming from the DGLAP evolution of the PDFs. The formalism presented here is based on the considerations of \cite{Frixione:1995ms}.
	
	If the beam particles $B_{\oplus}$ and $B_{\ominus}$ for the cross section introduced in Eq.~(\ref{hadronicCrossSection}) are associated with partons $d$ -- which is possible due to the universality of writing cross sections in terms of convolution products with PDFs -- the partonic densities up to NLO in $\alpha_s$ can be written as
	\begin{align}
		\Gamma^{(d)}_a(x)=\delta_{ad}\delta(1-x)-\frac{\alpha_s}{2\pi}\left(\frac{1}{\bar{\varepsilon}}P^{\text{QCD}}_{ad}(x,0)-K^{\text{QCD}}_{ad}(x)\right) +\mathcal{O}(\alpha_s^2)\quad.
		\label{PDFpartonic}
	\end{align}
	Since the expansion of the PDFs up to NLO in $\alpha$ as well as its consequences on the following formulas is completely analogous to this expansion in $\alpha_s$ for simplicity the QED case will be left out and all superscripts `QCD' dropped for the derivation of the ISR remnant.
	The function $K_{ad}(x) $ depends on the renormalisation scheme 
	where
	$K_{ad}(x)=0$ in case the $\overline{\text{MS}}$ scheme is chosen. $P_{ad}(x,0)$ are the regularised unpolarised Altarelli-Parisi splitting functions in four dimensions and depend on the unregularised ones $\hat{P}_{ad}$ as \cite{Frixione:1995ms,Kunszt:1992tn}
	\begin{align}
		P_{ad}(x,0)=\frac{(1-z)\langle\hat{P}_{ad}\rangle(z,0)}{(1-z)_+}+\gamma_a\delta_{ab}\delta(1-z)\quad.
		\label{regSplittingfun}
	\end{align}
	Related to this, the following identity for the Altarelli-Parisi kernels must be obeyed,
	\begin{align}
		2C_a\delta_{ab}\delta(1-z)=\delta(1-z)(1-z)\langle\hat{P}_{ab}\rangle\quad.
		\label{casimirsplittingfun}
	\end{align}
	The quantities $\gamma_a$ and $C_a$ as well as their explicit values are defined in \ref{groupfactors}.
	
	For partons treated as beam particles Eq.~(\ref{hadronicCrossSection}) takes the form
	\begin{align}
		d\tilde{\sigma}^{(de)}=\sum_{ab}\Gamma^{(d)}_a\star \Gamma^{(e)}_b \star d{\sigma}_{ab}
	\end{align}
	By denoting $d{\sigma}_{ab}$ as the subtracted and $d\hat{\sigma}_{ab}$ as the unsubtracted cross section these quantities on partonic level can be perturbatively expanded as
	\begin{align}
		&d\sigma_{ab}=d\sigma_{ab}^{(0)}+d\sigma_{ab}^{(1)} &d\hat{\sigma}_{ab}=d\hat{\sigma}_{ab}^{(0)}+d\hat{\sigma}_{ab}^{(1)}
		\label{partonicbeamcrosssec}
	\end{align}
	with superscript (0) for the LO and (1) for the NLO, respectively. The subtracted cross sections $d\sigma_{ab}$ to each order can be obtained by subtracting the $1/\bar{\varepsilon}$ terms according to
	\begin{align}
		&d\sigma_{ab}^{(0)}=d\hat{\sigma}_{ab}^{(0)}\\
		&d\sigma_{ab}^{(1)}=d\hat{\sigma}_{ab}^{(1)}+\underbrace{\frac{\alpha_s}{2\pi}\sum_d\left(\frac{1}{\bar{\varepsilon}}P_{da}-K_{da}\right)\star d\hat{\sigma}_{db}^{(0)}}_{d\sigma_{ab}^{(cnt,\oplus)}}+\underbrace{\frac{\alpha_s}{2\pi}\sum_e\left(\frac{1}{\bar{\varepsilon}}P_{eb}-K_{eb}\right)\star d\hat{\sigma}_{ae}^{(0)}}_{d\sigma_{ab}^{(cnt,\ominus)}}
		\label{shortdistcross1}
	\end{align}
	where the counterterms $d\sigma_{ab}^{(cnt,\oplus)}$ and $d\sigma_{ab}^{(cnt,\ominus)}$ can be extracted. The quantity $d\hat{\sigma}_{ab}^{(0)}$ is associated with the spin- and colour-averaged Born squared matrix elements summed up for a specific initial state $\{ab\}$, i.~e.
	\begin{align}
		d\hat{\sigma}_{ab}^{(0)}=\mathcal{B}_{ab}
	\end{align}
	The final aim is to compute the component $d\sigma_{ab}^{(1)}$ which contains the remnant of the collinear initial-state subtraction terms in integrated form and is introduced as $d\bar{\sigma}^{n+1}_{ab}$ with the notion `degenerate $n+1$ phase-space contribution' or `DGLAP remnant' in Sec.~(\ref{seccrosssecformulation}). The unsubtracted cross section $d\hat{\sigma}_{ab}^{(1)}$ can be constructed in the same way as the final state collinear subtraction terms $\sum_{\tilde{\alpha}}I^{\text{coll}}_{\tilde{\alpha}}$ in the previous section. For this reason, analogously to Eq.~(\ref{yexpansion}), we start by expanding
	\begin{align}
		(1-y^2)^{-1-\varepsilon}=-\frac{4^{-\varepsilon}}{2\varepsilon}\left[\delta(1-y)+\delta(1+y)\right] + \frac{1}{2}\left[\left(\frac{1}{1-y}\right)_++\left(\frac{1}{1+y}\right)_+\right]+\mathcal{O}(\varepsilon).
		\label{yISexpansion}
	\end{align}
	Using $\mathcal{P}_+(\xi)$ defined in Eq.~(\ref{Pplusdistr}) and one of the terms of Eq.~(\ref{yISexpansion}) containing a $\delta$ function, the integrated subtraction terms in the collinear limit $y=+1$ for emitter ${\oplus}$ and $y=-1$ for emitter ${\ominus}$, respectively, can be written as
	\begin{align}
	\begin{split}
		I^{\circled{\scriptsize$\pm$}\text{ coll}}_{\tilde{\alpha}}(\bar{\Phi}_{n})=\frac{s^{1-\varepsilon}}{(4\pi)^{3-2\varepsilon}}\int d\Omega^{(2-2\varepsilon)}\int_{-1}^{1}dy\int_{0}^{\xi_{\text{max}}}d\xi {\mathcal{P}_+(\xi)}\\
		\times\left[-\frac{4^{-\varepsilon}}{2\varepsilon}\delta(1\mp y)\right](1-y^2)\xi^2{\mathcal{R}}_{\tilde{\alpha}}(\xi,y)\quad.
		\end{split}
		\label{IcollIS}
	\end{align}
	Due to the angular integration measure the regularised real squared amplitudes in the collinear limit $y=+1$ can be space-averaged. By means of Eqs.~(\ref{regRcollf}) to (\ref{k2collISR}) and identifying $p^0=\sqrt{s}/2$ and $k_j^0=\xi\sqrt{s}/2$ it takes the factorised form
	\begin{align}
	\begin{split}
	\biggr\langle\lim_{y\rightarrow1}(1-y^2)\xi^2{\mathcal{R}}_{\tilde{\alpha}}(\xi,y)\biggr\rangle_{\Omega}={8\pi\alpha_s\mu_R^{2\varepsilon}}\frac{4}{s}\xi\langle \hat{P}\rangle_{\tilde{\alpha}}(z,\varepsilon)\mathcal{B}(f_{b,\tilde{\alpha}})\quad.
	\label{spaceavRalphaIS}
	\end{split}
	\end{align}
	with the unpolarised unregularised initial state splitting functions $\langle \hat{P}\rangle$ defined in eqs.~(\ref{ftofgamunpol}) to (\ref{gtogunpol}). By integrating over the angular phase-space measure $d\Omega^{(2-2\varepsilon)}$ we get the factor
	\begin{align}
		\int d\Omega^{(2-2\varepsilon)}=\frac{2\pi^{1-\varepsilon}}{\Gamma(1-\varepsilon)}\quad.
		\label{angularmeasure}
	\end{align}
	Inserting Eq.~(\ref{spaceavRalphaIS}) and (\ref{angularmeasure}) into Eq.~(\ref{IcollIS}) with the substitution $z=1-\xi$, expanding the $\varepsilon$-dependent factor as
	\begin{align}
		\frac{1}{\varepsilon}\frac{(4\pi)^{\varepsilon}}{\Gamma(1-\varepsilon)}\left(\frac{\mu_R^2}{s}\right)^{\varepsilon}=\frac{1}{\varepsilon}-\gamma_E+\log 4\pi+\log\frac{\mu_R^2}{s} +\mathcal{O}(\varepsilon)
	\end{align}
	and summing over all $\tilde{\alpha}$ regions with emitter $E_{\tilde{\alpha}}=\oplus$ leading to one initial state $\{ab\}$ yields the collinear subtraction term of the form
	\begin{align}
	\begin{split}
		\sum_{\tilde{\alpha}}I^{\oplus,\text{coll}}_{\tilde{\alpha},ab}=I^{\oplus,\text{coll}}_{ab}=-\sum_{d}\int_{0}^{\xi_{\text{max}}}d\xi~ \left[\frac{1}{\bar{\varepsilon}}+\log\frac{\mu_R^2}{s}\right]{\mathcal{P}_+(\xi)}\xi\langle \hat{P}_{da}\rangle(1-\xi,\varepsilon)\\
		\times \frac{\alpha_s}{2\pi}\mathcal{B}_{db}(1-\xi,\bar{\Phi}_n)
	\end{split}
	\end{align}
	using the definition of the $\overline{\text{MS}}$-pole
	\begin{align}
		\frac{1}{\bar{\varepsilon}}=\frac{1}{\varepsilon}-\gamma_E+\log 4\pi\quad.
	\end{align}
	Inserting Taylor-expanded splitting functions $\langle \hat{P}_{da}\rangle(1-\xi,\varepsilon)$ for $\varepsilon\rightarrow 0$ and dropping all terms of $\mathcal{O}(\varepsilon)$ results in
	\begin{align}
		\begin{split}
		I^{\oplus,\text{coll}}_{ab}=-\sum_{d}\int_{0}^{\xi_{\text{max}}}d\xi~\xi\left[\frac{1}{\bar{\varepsilon}}\left(\frac{1}{\xi}\right)_+\langle \hat{P}_{da}\rangle(1-\xi,0)+\left(\frac{1}{\xi}\right)_+\langle \hat{P}_{da}\rangle(1-\xi,0)\log \frac{\mu_R^2}{s}\right.\\
		\left.+\left(\frac{1}{\xi}\right)_+\frac{d\langle \hat{P}_{da}\rangle}{d\varepsilon}(1-\xi,\varepsilon)\biggr\rvert_{\varepsilon=0}-2\left(\frac{\log\xi}{\xi}\right)_+\langle \hat{P}_{da}\rangle(1-\xi,0) \right]\\
		\times\frac{\alpha_s}{2\pi}\mathcal{B}_{db}(1-\xi,\bar{\Phi}_n)
		\end{split}
		\label{Icollab}
	\end{align}
	Following Eqs.~(\ref{xplusxminus}) and (\ref{xle1conditions}), related to the construction of the phase-space for ISR with massless emitters in Sec.~\ref{secMasslessISem}, we observe that the maximal energy which can be radiated off one initial-state emitter in the strict collinear limit is constrained to $\xi_{\text{max}}=1-x^{\circled{\scriptsize$\pm$}}$. In the same way, this constraint leads to a lower bound for the PDF rescaling variable $z^{\circled{\scriptsize$\pm$}}$, i.~e.
	\begin{align}
	\label{zrescalingrestriction}
		x^{\circled{\scriptsize$\pm$}}\le z^{\circled{\scriptsize$\pm$}}\le 1\quad.
	\end{align}
	Due to this, the integral from the convolution product in the counterterm $d\sigma_{ab}^{(cnt,\oplus)}$, which formally reads
	\begin{align}
		d\sigma_{ab}^{(cnt,\oplus)}\propto \int^{1}_{0}dz~ \left(\frac{1}{\bar{\varepsilon}}P_{da}(z,0)-K_{da}(z)\right) \mathcal{B}_{db}(z,\bar{\Phi}_n)\quad,
	\end{align}
	has to be transformed in an integration over  $\xi=1-z^{\oplus}$.
	Using Eq.~(\ref{regSplittingfun}) expressed in terms of $\langle \hat{P}_{da}\rangle$ with the transformation of the plus distribution for $z=1-\xi$ in the collinear limit
	\begin{align}
		\left(\frac{1}{1-z}\right)_+=\left(\frac{1}{\xi}\right)_++\delta(\xi)\log \xi_{\text{max}}
	\end{align}
	as well as the identity of Eq.~(\ref{casimirsplittingfun}) we get
	\begin{align}
	\begin{split}
		d\sigma_{ab}^{(cnt,\oplus)}=\sum_{d}\int_{0}^{1-x^{\oplus}}d\xi \left[\frac{1}{\bar{\varepsilon}}\left(\frac{1}{\xi}\right)_+\langle \hat{P}_{da}\rangle(1-\xi,0)+\frac{1}{\bar{\varepsilon}}(\gamma_d+2C_d\log \xi_{\text{max}})\delta_{ad}\delta(\xi)\right.\\
		\left.-K_{da}(1-\xi)\right]
		\times \frac{\alpha_s}{2\pi}\mathcal{B}_{db}(1-\xi,\bar{\Phi}_n).
	\end{split}
	\label{counterterm+}
	\end{align}
	Since the subtracted short distance cross section $d\sigma^{(1)}_{ab}$ according to Eq.~(\ref{shortdistcross1}) depends on the terms $I^{\oplus,\text{coll}}_{ab}$ and $d\sigma_{ab}^{(cnt,\oplus)}$as
	\begin{align}
		d\sigma^{(1)}_{ab}\propto I^{\oplus,\text{coll}}_{ab}+d\sigma_{ab}^{(cnt,\oplus)}
	\end{align}
	we extract the initial-state remnant for emitter $B_{\oplus}=a$
	\begin{align}
	\begin{split}
		d\sigma_{ab}^{\oplus \text{ISR} }=&I^{\oplus,\text{coll}}_{ab}+d\sigma_{ab}^{(cnt,\oplus)}\\
		=&\frac{\alpha_s}{2\pi}\frac{1}{\bar{\varepsilon}}(\gamma_a+2C_a\log \xi_{\text{max}})\mathcal{B}_{ab}(\bar{\Phi}_n)\\
		&+\frac{\alpha_s}{2\pi}\sum_{d}\int_{0}^{1-x^{\oplus}}d\xi~\xi\left(\langle \hat{P}_{da}\rangle(1-\xi,0)\left[\left(\frac{1}{\xi}\right)_+\log \frac{s}{\mu_R^2}+2\left(\frac{\log\xi}{\xi}\right)_+\right]\right. \\
		&\left.- \left(\frac{1}{\xi}\right)_+\frac{d\langle \hat{P}_{da}\rangle}{d\varepsilon}(1-\xi,\varepsilon)\biggr\rvert_{\varepsilon=0}-K_{da}(1-\xi)\right)\mathcal{B}_{db}(1-\xi,\bar{\Phi}_n)
	\end{split}
	\label{collinearremnantISR}
	\end{align}
	with the last term, free of $1/\bar{\varepsilon}$-poles, identified as
	\begin{align}
	\begin{split}
		d\bar{\sigma}^{n+1,\oplus}_{ab}=\frac{\alpha_s}{2\pi}\sum_{d}\mathcal{K}_{da}\star \mathcal{B}_{db}.
		\end{split}
		\label{degenerate+}
	\end{align}
	Hence, the $1/\bar{\varepsilon}$ term of $I^{\oplus,\text{coll}}_{ab}$ in Eq.~(\ref{Icollab}) is cancelled by the corresponding one of $d\sigma_{ab}^{(cnt,\oplus)}$ in Eq.~(\ref{counterterm+}). Eventually, the pole term of Eq.~(\ref{collinearremnantISR}) which comes from the flavour-diagonal part of the regularised splitting functions $P_{ab}$ is cancelled exactly by the corresponding term of the virtual part $\mathcal{V}$ in the soft limit.
	Using the definition of the $\overline{\text{MS}}$-scheme pole in the exact form and an additional dependence on the arbitrary Ellis-Sexton scale $Q$ the pole $1/\bar{\varepsilon}$ can be replaced by
	\begin{align}
		\frac{1}{\bar{\varepsilon}}=\frac{(4\pi)^{\varepsilon}}{\Gamma(1-\varepsilon)}\left(\frac{\mu_R^2}{Q^2}\right)^{\varepsilon}\frac{1}{\varepsilon}-\log \frac{\mu_R^2}{Q^2}=\mathcal{N}(\varepsilon)\frac{1}{\varepsilon}-\log \frac{\mu_R^2}{Q^2}\quad.
	\end{align}
	Due to this, the finite part of the first term of Eq.~(\ref{collinearremnantISR}) can be identified as a contribution to the remainders of the soft subtraction terms,
	\begin{align}
		d\sigma_{ab,S}^{\oplus\text{ISR remn.}}=- \frac{\alpha_s}{2\pi}(\gamma_a+2C_a\log \xi_{\text{max}})\mathcal{B}_{ab}(\bar{\Phi}_n)\log \frac{\mu_R^2}{Q^2}
	\end{align}
	Varying the scale $Q$ where $\mu_R\ne Q$ should not change the complete NLO result and thus yields a sanity check on the implemented FKS scheme. 
	
	The initial-state remnant for the emitter $B_{\ominus}=b$ is almost exactly identical to $d\sigma_{\text{ISR}}^{\oplus }$
	and analogously to Eq.~(\ref{collinearremnantISR}) reads
	\begin{align}
			\begin{split}
		d\sigma_{ab}^{\ominus \text{ISR} }
		=&\frac{\alpha_s}{2\pi}\frac{1}{\bar{\varepsilon}}(\gamma_b+2C_b\xi_{\text{max}})\mathcal{B}_{ab}(\bar{\Phi}_n)\\
		&+\frac{\alpha_s}{2\pi}\sum_{d}\int_{0}^{1-x^{\ominus}}d\xi~\xi\left(\langle \hat{P}_{db}\rangle(1-\xi,0)\left[\left(\frac{1}{\xi}\right)_+\log \frac{s}{\mu_R^2}+2\left(\frac{\log\xi}{\xi}\right)_+\right]\right. \\
		&\left.- \left(\frac{1}{\xi}\right)_+\frac{d\langle \hat{P}_{db}\rangle}{d\varepsilon}(1-\xi,\varepsilon)\biggr\rvert_{\varepsilon=0}-K_{db}(1-\xi)\right)\mathcal{B}_{ad}(1-\xi,\bar{\Phi}_n)
		\end{split}
	\end{align}
	Furthermore, accounting for additional QED corrections, if required by the coupling power conditions, the analogous terms have to be added up. For these the definitions $\gamma_a^{\text{QED}}$, the underlying Born $\mathcal{B}^{\text{QED}}_{ab}$ and expressions $\mathcal{K}^{\text{QED}}_{da}$ containing the QED splitting functions must be used.
	
	Summarising the finite part of the initial-state remnant, summing over all flavour combinations $\{ab\}$ which are possible from the initial state, we have
	\begin{align}
		d\sigma^{\text{ISR remn.} }=d\sigma_{S}^{\text{ISR remn.}}+d\bar{\sigma}^{n+1}=\sum_{a,b}\left(d\sigma_{ab,S}^{\text{ISR remn.}}+d\bar{\sigma}^{n+1}_{ab}\right)
	\end{align}
	with the soft ISR remnant
	\begin{align}
	\begin{split}
		d\sigma_{S}^{\text{ISR remn.}}=-\sum_{a,b}\left[\frac{\alpha_s}{2\pi}\left(\gamma^{\text{QCD}}_a+\gamma^{\text{QCD}}_b+ 2\xi_{\text{max}}(C_a^{\text{QCD}}+C_b^{\text{QCD}})\right)\mathcal{B}^{\text{QCD}}_{ab}\right.\\
		\left.+\frac{\alpha}{2\pi}\left(\gamma^{\text{QED}}_a+\gamma^{\text{QED}}_b+2\xi_{\text{max}}(C_a^{\text{QED}}+C_b^{\text{QED}})\right)\mathcal{B}^{\text{QED}}_{ab}\right]\log \frac{\mu_R^2}{Q^2}
	\end{split}
	\end{align}
	and the $n+1$ degenerate contribution
	\begin{align}
	\begin{split}
		d\bar{\sigma}^{n+1}=&\sum_{a,b}\left(d\bar{\sigma}^{\oplus,n+1}_{ab}+d\bar{\sigma}^{\ominus,n+1}_{ab}\right)\\
		=&\sum_{a,b}\sum_{d}\left[\frac{\alpha_s}{2\pi}\left(\mathcal{K}^{\text{QCD}}_{da}\star \mathcal{B}^{\text{QCD}}_{db}+\mathcal{K}^{\text{QCD}}_{db}\star \mathcal{B}^{\text{QCD}}_{ad}\right)\right.\\
		&\left.+\frac{\alpha}{2\pi}\left(\mathcal{K}^{\text{QED}}_{da}\star \mathcal{B}^{\text{QED}}_{db}+\mathcal{K}^{\text{QED}}_{db}\star \mathcal{B}^{\text{QED}}_{ad}\right)\right]\quad.
		\end{split}
	\end{align}
	The explicit form of the ISR remnant depending on the set of arbitrary FKS parameters $\xi_c$ and $\delta_i$, the ISR counterpart to $\delta_o$, is provided by replacing the plus distributions by the modified ones of Eq.~(\ref{plusdistrydelta}) and (\ref{plusdistrxicut}), substituting
	\begin{align*}
		\frac{s}{\mu_R^2} \rightarrow\frac{s\delta_i}{2\mu^2_R}
	\end{align*}
	and by adding the term
	\begin{align*}
		-\sum_{a,b}\left[\frac{\alpha_s}{\pi}\left(C^{\text{QCD}}_a+C^{\text{QCD}}_b\right)\mathcal{B}^{\text{QCD}}_{ab}+\frac{\alpha}{\pi}\left(C^{\text{QED}}_a+C^{\text{QED}}_b\right)\mathcal{B}^{\text{QED}}_{ab}\right]\log\frac{\xi_c}{\xi_{\text{max}}}\log\frac{\mu^2_R}{Q^2}.
	\end{align*}


	\chapter{EW precision calculations for lepton collisions -- collinear ISR}
	\label{secEWprecisioncalculations}
	In order to describe SM precision physics at lepton colliders, i.~e. future colliders as for example ILC or CLIC, generally the same methods and schemes for NLO computations are applicable as for hadron collisions. However, due to the colliding leptons being colourless states of QED interactions beam effects are described fundamentally different than those of QCD initial states. These differences are manifest by the distinct perturbative nature of both of these SM theories.
	In QED, low-energy external leptons and photons which are identified as on-shell particles are fully describable by perturbative expansions due to the abelian group structure of the theory. In this section the two concepts of higher order computations treating the leptonic initial-states either in the massive or massless approach will be presented. Since these concepts essentially are equivalent for all leptons which differ only in numerical values of their masses we refer to all of these by using notions related to electrons in the following.
	
	In Sec.~\ref{secleptonicIS} it will be discussed how a resummation of large logarithms coming from ISR effects in contrast to pure fixed order corrections in $\alpha$ affects cross sections of lepton collision processes.
	The universal treatment of collinear ISR effects for MC integrated results by embedding electron PDFs into cross section computations is challenged by numerical issues of the evaluation and structure of the PDFs. Methods to solve these are suggested in Sec.~\ref{MCintegrationmethodsPDFs}.
	\section{Description of collinear radiation off the leptonic initial state}
	\label{secleptonicIS}
	Adding electroweak corrections to cross sections of $e^+e^-$ processes naively is pictured as considering matrix elements for Feynman diagrams with loops or photon radiation leading to contributions of the desired higher order in $\alpha$, which are then integrated over the phase space. However, due to collinear photon emission off the electron or positron, logarithms of the ratio $Q^2/m^2$, with $Q$ the energy scale and $m$ the electron mass, lead to large terms such that the perturbation series in $\alpha$ would be spoiled if $Q^2\gg m^2$. Analogously to QCD, a way to solve this is through the structure-function approach: By means of the DGLAP evolution equations \cite{Dokshitzer:1977sg, Gribov:1972ri,Altarelli:1977zs} the logarithms can be absorbed into universal functions, the structure functions (or PDFs), which can be factored out due to factorisation theorem. Whereas for QCD the initial conditions for the PDFs are non-perturbative functions and determined only by fits to experimental data, in QED these can be calculated perturbatively.
	
	At LO the initial conditions of the electron PDFs -- due to the description of electrons as on-shell elementary particles in the low energy limit -- are trivially given by a delta Dirac function $\delta(1-z)$, with $z$ denoting the energy fraction which is radiated away. This description however is accurate only by the  caveat that logarithms $\log Q^2/m^2$ appearing at higher orders in $\alpha$ of the cross sections are sufficiently suppressed by the coupling, i.~e. at low scales $Q$.
	For large energy scales $Q$ the Dirac delta functional behaviour of the LO PDFs is smoothened to a peak in the vicinity of $z=1$ with a tail reaching down to $z=0$ if the universal scale-dependency of terms of a perturbation series is absorbed into the PDFs which is understood in the following sense.
	A class of terms containing logarithms $\log Q^2/m^2$ appearing at each order $k$ of the cross section perturbatively expanded in $\alpha$, e.~g. all terms at the leading-logarithmic (LL) level,
	\begin{align}
		\alpha^k \log^k \frac{Q^2}{m^2}C_{\text{LL}}^{(k)}\quad,
		\label{LLres}
	\end{align}
	at the next-to-leading logarithmic (NLL) level,
	\begin{align}
		\alpha^k \log^k \frac{Q^2}{m^2}C_{\text{LL}}^{(k)}+\alpha^k \log^{k-1} \frac{Q^2}{m^2}C_{\text{NLL}}^{(k-1)}\quad,
		\label{NLLres}
	\end{align}
	and so forth, can be resummed by factorising them from the matrix elements at an arbitrary scale $\mu$. In this way the matrix elements keep a dependence on logarithms $\log Q^2/\mu^2$, whereby the factorised sum rewritten as an exponential factor contains the electron mass dependency in terms of $\log \mu^2/m^2$ and is thus included into the PDF definition of the electron. This is the basic concept of writing a cross section in the form of Eq.~(\ref{hadronicCrossSection}) which in the case of $e^+e^-$ collisions reads
	\begin{align}
	\begin{split}
		d\tilde{\sigma}^{e^+e^-}(K_{\oplus},K_{\ominus},m^2)=\sum_{i,j=e^{\pm},\gamma}\int dx_{\oplus}dx_{\ominus}\Gamma^{e^+}_i(x_{\oplus},\mu^2,m^2)\Gamma_j^{e^-}(x_{\ominus},\mu^2,m^2)\\
		\times d\sigma_{ij}(x_{\oplus}K_{\oplus},x_{\ominus}K_{\ominus},\mu^2)
		\end{split}
		\label{e+e-crosssection}
	\end{align}
	where the universal scale dependence of the PDFs is encoded in the solution of the DGLAP evolution equations
	\begin{align}
		\frac{\partial\Gamma_i(z,\mu^2)}{\partial\log \mu^2}=\frac{\alpha(\mu)}{2\pi}\left[P_{ij}\star \Gamma_j\right](z,\mu^2)\quad.
		\label{DGLAPevolutioneq}
	\end{align}
	The PDFs here are with respect to the electron, i.~e. $\Gamma_i=\Gamma_i^{e^-}$, where for convenience the superscripts are neglected.
	If choosing a scale $\mu\sim Q$ we find the partonic cross sections $d\sigma_{ij}$ being completely describable by perturbative means. In QED, this is true for initial conditions of the PDFs as well, which is just a consequence of the nature of the abelian group theory.
	
	The analytical form of the electron PDFs at LL precision has been first derived in Refs.~\cite{Cacciari:1992pz,Skrzypek:1990qs,Skrzypek:1992vk} by using the following approach.

	In general, the customary approach to solve Eq.~(\ref{DGLAPevolutioneq}) is by transforming it from the partonic basis, here $i={e^{\pm},\gamma}$, into the one of singlet (S) and non-singlet (NS) combinations such that the corresponding PDFs are of the form
	\begin{align}
		&\Gamma_{\text{S}}=\Gamma_{e^-}+\Gamma_{e^+} &\Gamma_{\text{NS}}=\Gamma_{e^-}-\Gamma_{e^+}\quad.
		\label{singlet-NSPDFs}
	\end{align}
	In this way we find the simplified equations
	\begin{align}
		\frac{\partial}{\partial\log \mu^2}\left(\begin{array}{c}
		\Gamma_{\text{S}}\\
		\Gamma_{\gamma}
		\end{array}\right)&=\frac{\alpha(\mu)}{2\pi}\mathbf{P}_{\text{S}}\star \left(\begin{array}{c}
		\Gamma_{\text{S}}\\
		\Gamma_{\gamma}
		\end{array}\right)\\
		\frac{\partial\Gamma_{\text{NS}}}{\partial\log \mu^2}&=\frac{\alpha(\mu)}{2\pi}P_{\text{NS}}\star \Gamma_{\text{NS}}
		\label{NSevolutionseq}
	\end{align}
	for which the non-singlet component decouples from the singlet-$\gamma$ system. The kernels in these equations read
	\begin{align}
		\mathbf{P}_{\text{S}}&=\left(\begin{array}{c c}
		P_{SS} &P_{S\gamma}\\
		P_{\gamma S} &P_{\gamma\gamma} 
		\end{array}\right),\qquad P_{SS}=P_{e^{\pm}e^{\pm}}+P_{e^{\pm}e^{\mp}},\quad P_{S\gamma}=2n_FP_{e^{\pm}\gamma},\quad P_{\gamma S}=P_{\gamma e^{\pm}}\\
		P_{\text{NS}}&=P_{e^{\pm}e^{\pm}}-P_{e^{\pm}e^{\mp}}
	\end{align}
	with $n_F$ the number of lepton flavour families. $n_F=1$ is assumed in the following accounting for a single (massless) lepton flavour family in the calculation, e.~g. the electron/positron in an $e^+e^-$ collision, which is a sufficiently accurate description according to Ref.~\cite{Bertone:2022ktl}.

	\subsection{LO-LL approximation}
	\label{secLLapproximation}	
	Approximately, for the LL solution of the electron PDFs a one-dimensional flavour space can be assumed, i.~e. $i=e^-$, such that the problem simplifies to solving Eq.~(\ref{NSevolutionseq}) for the non-singlet case, a scalar integro-differential equation, whereby
	\begin{align}
		\Gamma_{\text{NS}}(x,\mu^2)\simeq \Gamma_{e^-}(x,\mu^2)
	\end{align}
	As the initial condition of the electron PDF at LO is the Dirac delta function, as mentioned above, i.~e.
	\begin{align}
		\Gamma_{e}(x,\mu_0^2)=\Gamma_{e}(x,m^2)=\delta(1-x),
	\end{align}
	the non-singlet evolution equation can be written in the form
	\begin{align}
		\Gamma_{e}(x,\mu^2)=\delta(1-x)+\int_{m^2}^{\mu^2}\frac{dq^2}{q^2}\frac{\alpha(q^2)}{2\pi}\int_x^1dzP_{ee}(z)\Gamma_{e}\left(\frac{x}{z},q^2\right)
		\label{gameeEvol}
	\end{align}
	where for convenience we use `$e$' as subscript denoting the parton identity of the electron $e^-$. Using an auxiliary function $G(x,\mu^2)$ which is related to $\Gamma_e(x,\mu^2)$ as
	\begin{align}
		&G(x,\mu^2)=\int^1_xdt\Gamma_e(t,\mu^2) &\Gamma_e(x,\mu^2)=-\frac{\partial}{\partial x}G(x,\mu^2)
		\label{gammaandG}
	\end{align}
	and the form of Eq.~(\ref{gameeEvol}) we find
	\begin{align}
		G(x,\mu^2)=1+\int_{m^2}^{\mu^2}\frac{dq^2}{q^2}\frac{\alpha(q^2)}{2\pi}\int_x^1dzP_{ee}(z)G\left(\frac{x}{z},q^2\right)\quad.
		\label{Gevolution}
	\end{align}
	This equation can be solved for $G$ which provides the solution for the PDF evolution equations by the relation on the r.~h.~s. of Eq.~(\ref{gammaandG}). By means of this, for the simplest case, using the one-loop accurate regularised (unpolarised) Altarelli-Parisi kernels
	\begin{align}
		& P_{ee}(z)=\langle \hat{P}_{ee}\rangle(z)-\delta(1-z)\int_{0}^{1}dt\langle \hat{P}_{ee}\rangle(t), &\langle \hat{P}_{ee}\rangle(z)=\frac{1+z^2}{1-z}\quad,
		\label{Peeregularised}
	\end{align}
	 the solution of Gribov-Lipatov resummation to all orders in $\alpha$ in the asymptotic region $x\simeq1$ \cite{Gribov:1972ri}, i.~e. for soft-collinear radiation of photons, can be obtained,
	\begin{align}
		\Gamma_e(x,\mu^2)=\frac{e^{\eta (\frac{3}{4}-\gamma_E)}}{\Gamma(1+\eta)}\eta (1-x)^{\eta-1}
		\label{GribovLipatovSol}
	\end{align}
	with the definitions
	\begin{align}
		&\eta=\frac{\alpha}{\pi}L, &L=\log\frac{\mu^2}{m^2}
		\label{etadefinition}
	\end{align}
	and $\gamma_E=0.577...$, the Euler-Mascheroni constant. Identically, this analytic form can be achieved by transforming the integro-differential evolution equations of Eq.~(\ref{DGLAPevolutioneq}) into Mellin space by
	\begin{align}
		&M[f]\equiv f_N=\int_{0}^{1}dzz^{N-1}f(z) &M[g\star h]=M[g]M[h] \quad.
	\end{align}
	This simplifies the problem to solving ordinary differential equations, using the asymptotic limit $N\rightarrow\infty$ analogously to $z\rightarrow 1$ in $z$-space and an analytical Mellin inversion of the resulting solution according to Ref.~\cite{Bertone:2019hks}.

	The recursive solution delivers an option to find $\Gamma_e(x,\mu^2)$ for arbitrary values $x<1$. Conceptionally, the procedure is as follows.
	
	The function $G(x,\mu^2)$ as well as the PDFs $\Gamma_e(x,\mu^2)$ generally can be written as a perturbative series expressed as
	\begin{align}
		&G(x,\mu^2)=\sum_{n=0}^{\infty}G^{(n)}(x,\mu^2)=\sum_{n=0}^{\infty}\frac{\eta^n}{2^n n!}I_n(x),\label{Gseries}\\
		&\Gamma_e(x,\mu^2)=\sum_{n=0}^{\infty}\Gamma_e^{(n)}(x,\mu^2)=\sum_{n=0}^{\infty}\frac{\eta^n}{2^n n!}\frac{\partial I_n(x)}{\partial x}\quad.
		\label{GammaSeries}
	\end{align}
	Inserting $G(x,\mu^2)$ of this form in Eq.~(\ref{Gevolution}) yields the recurrence relation
	\begin{align}
		I_n(x)=\int_{x}^{1}dzP(z)I_{n-1}\left(\frac{x}{z}\right)
	\end{align}
	which with Eq.~(\ref{Peeregularised}) and the boundary conditions $G^{(0)}(x,\mu^2)=G(x,m^2)=1$ implicating $I_0(x)=1$ and $I_1(x)=\int_{x}^{1}dz P(z)$ leads to
	\begin{align}
		I_n(x)=\int_{x}^{1}dz\langle\hat{P}_{ee}\rangle(z) \left[I_{n-1}\left(\frac{x}{z}\right)-I_{n-1}(x)\right] +I_{n-1}(x)I_1(x)\quad.
	\end{align}
	The divergence of $\langle\hat{P}_{ee}\rangle(z)$ for $z\rightarrow1$ in the first term is regularised in this way by the difference $I_{n-1}\left(x/z\right)-I_{n-1}(x)$. By this relation, $I_n$ in principle can be computed for each order $n$ iteratively which, however, gets increasingly complicated with each new iteration. This leads to the solution for $G(x,\mu^2)$ and by its derivative to that for $\Gamma^e_e(x,\mu^2)$. In order to match this iterative solution for arbitrary $x$ computed to a certain order in $\alpha$ with the asymptotic $x\simeq1$ Gribov-Lipatov solution to all orders in $\alpha$ described above the following has to be considered. Adding the solution of Eq.~(\ref{GribovLipatovSol}) for $x\simeq1$ to the iterative one for all $x$ leads to a double counting of the soft terms in the $x\rightarrow 1$ regime up to the order of which the iterative solution is computed. Due to this, all terms of Eq.~(\ref{GribovLipatovSol}) expanded again as a series in $\eta$ up to that order have to be subtracted again from the sum of the asymptotic $x\simeq1$ and the `all $x$' solution.
	We thus find the standard LL accurate solution of electron PDFs with Gribov-Lipatov resummation to all orders in $\alpha$ matched with the `all $x$' solution up to $\mathcal{O}(\alpha^3)$ with the explicit form \cite{Cacciari:1992pz,Skrzypek:1990qs,Skrzypek:1992vk}
	\begin{align}
	\begin{split}
		\Gamma_e(x,\mu^2)=&\frac{e^{\eta (\frac{3}{4}-\gamma_E)}}{\Gamma(1+\eta)}\eta (1-x)^{\eta-1}-\frac{1}{2}\eta(1+x)\\
		&+\frac{1}{2^22!}\eta^2\left[-4(1+x)\log(1-x)+3(1+x)\log x-4\frac{\log x}{1-x}-5-x\right]\\
		&+\frac{1}{2^33!}\eta^3\biggr\{(1+x)[18\zeta(2)-6\text{Li}_2(x)-12\log^2(1-x)]\\
		& + \frac{1}{1-x}\left[-\frac{3}{2}(1+8x+3x^2)\log(x)-6(x+5)(1-x)\log(1-x)\right.\\
		&\left. -12(1+x^2)\log(x)\log(1-x)+\frac{1}{2}(1+7x^2)\log^2(x)\right.\\
		&\left.-\frac{1}{4}(39-24x-15x^2)\right]\biggr\}
	\end{split}
	\label{LLsoftPDF}
	\end{align}
	Note, that this is the electron PDF (or structure function) at LO-LL in $\alpha$ which matches the accuracy level of matrix elements at LO such that the overall cross section observables are LO-LL accurate.
	\subsection{NLO and NLO-NLL approximation}
	\label{secNLLapproximation}
	In order to increase the precision of observables to NLO in $\alpha$, terms up to (at least) the same order have to be taken into account coming from both the expansion of the PDFs and of the short distance partonic cross section. Schematically, by using perturbatively expanded forms of partonic cross sections and PDFs represented by terms of Eqs.~(\ref{LLres}) and (\ref{NLLres}), we symbolically rewrite Eq.~(\ref{e+e-crosssection}) for $\mu=Q$ as an expansion in $\alpha$,
	\begin{align}
	\begin{split}
		\Gamma^{e^+}(Q^2)\star \Gamma^{e^-}(Q^2)\star d\sigma(Q^2)= \left(C^{(0)}_{\text{LL}}+\frac{\alpha}{2\pi} \log \frac{Q^2}{m^2} C^{(1)}_{\text{LL}}+\frac{\alpha}{2\pi} C_{\text{NLL}}^{(0)}+\mathcal{O}(\alpha^2)\right)\\
		\star\left(d\sigma^{(0)}+\frac{\alpha}{2\pi} d\sigma^{(1)}+\mathcal{O}(\alpha^2)\right)\\
		=\underbrace{C^{(0)}_{\text{LL}}\star d\sigma^{(0)}+\frac{\alpha}{2\pi}\left[\left(\log \frac{Q^2}{m^2} C^{(1)}_{\text{LL}}+C_{\text{NLL}}^{(0)}\right)\star d\sigma^{(0)}+C^{(0)}_{\text{LL}}\star d\sigma^{(1)}\right]}_{=d\sigma_{\text{NLO}}}+\mathcal{O}(\alpha^2).
	\end{split}
	\end{align}
	We thus observe that the NLO fixed order cross section $d\sigma_{\text{NLO}}$, written symbolically in this form, implicitly contains the first $\mathcal{O}(\alpha)$ term of the expanded LL accurate PDFs by the term proportional to $\alpha/(2\pi)\log ({Q^2}/{m^2})$ as well as the initial conditions of NLL PDFs by $C_{\text{NLL}}^{(0)}$. By quantifying the factor
	\begin{align}
		\left[\frac{\alpha}{2\pi} \log \frac{Q^2}{m^2}\right]^2\sim\left[\frac{\alpha}{2\pi} \log \frac{s}{m^2}\right]^2
		\label{estimatefNLOmassiveIS}
	\end{align}
	which appears in the leading logarithmic terms at each order in $\alpha$, which are not taken into account in the NLO cross section we gain a first estimate on the reliability of a fixed-order NLO massive initial state approximation. How the electron mass can affect fixed-order calculations with higher order perturbative corrections in $\alpha$ for $e^+e^-$ annihilation is studied in detail in Refs.~\cite{Blumlein:2019srk,Blumlein:2019pqb,Blumlein:2020jrf,Ablinger:2020qvo,Blumlein:2021jdl}.
	
	Since this fixed-order approximation however is highly dependent on the physics related to that observable the recommended method for a generic approach is to resum all logarithms at NLL accuracy into universal NLO-NLL accurate PDFs by solving the DGLAP evolution equations with kernels and initial conditions at NLO in $\alpha$. Convoluted with partonic cross sections at NLO for which the initial-state mass dependencies are removed this leads to perturbatively reliable NLO-NLL accurate predictions for observables at lepton colliders. Particularly, this level of precision is desired for high centre-of-mass energies of future collider setups as $Q^2\sim s \gg m^2$ implies an enhancement of the logarithmic terms.
	
	As already mentioned above the initial conditions of NLL PDFs are of NLO accuracy in $\alpha$ and are different to the Dirac $\delta$ function as in the LO case since the leptons at NLO QED inherently mix with photons in the initial state. The solution for the NLO initial conditions of electron and photon PDFs are derived in Ref.~\cite{Frixione:2019lga} by means of an explicit short-distance cross section computation for a specific (but arbitrary) process, for which $e^+e^-\rightarrow u\bar{u}(\gamma)$ is chosen. This is achieved by computing the particle-level cross section, i.~e. the l.~h.~s. of Eq.~(\ref{e+e-crosssection}), including terms up to $\mathcal{O}(m/\sqrt{s})$, on the one hand, and the parton-level cross sections with massless electrons, on the other hand. Solving Eq.~(\ref{e+e-crosssection}) with these analytical expressions for the electron PDFs yields the NLO initial conditions which -- by applying the shorthand notation $\Gamma_i\equiv\Gamma^{e^-}_i$ -- read
	\begin{align}
		\Gamma_i=\Gamma^{(0)}_i+\frac{\alpha}{2\pi}\Gamma^{(1)}_i+\mathcal{O}(\alpha^2)
		\label{NLOIScond}
	\end{align}
	with
	\begin{align}
		\Gamma^{(0)}_i(x,\mu_0^2)&=\delta_{e^-i}\delta(1-x),\\
		\Gamma^{(1)}_{e^-}(x,\mu_0^2)&=\left[\frac{1+x^2}{1-x}\left(\log\frac{\mu_0^2}{m^2}-2\log(1-x)-1\right)\right]_++K_{ee}(x),\\
		\Gamma^{(1)}_{\gamma}(x,\mu_0^2)&=\frac{1+(1-x)^2}{x}\left(\log\frac{\mu^2_0}{m^2}-2\log x-1\right)+K_{\gamma e}(x),\label{gammaPDF}\\
		\Gamma^{(1)}_{e^+}(x,\mu_0^2)&=0 \label{NLOinitconde+}
	\end{align}
	Here, the functions $K_{ee}(x)$ and $K_{\gamma e}(x)$ are renormalisation scheme dependent and equal to zero if the $\overline{\text{MS}}$ scheme is chosen. This is analogous to the definition of $K_{ad}$ in Eq.~(\ref{PDFpartonic})).
	More details on their derivation and an alternative ansatz which is beyond the scope of this thesis are given in the reference stated above.
 
	The only difference of the NLL compared to the LL evolution of the PDFs are the QED splitting kernels being of NLO accuracy, for which the two-loop corrections are given in Ref.~\cite{deFlorian:2016gvk}. Moreover,
	the same methods as suggested for the LL evolution in Sec.~\ref{secLLapproximation} can be applied: For the asymptotic $x\simeq 1$ solution the NS evolution operator obtained by solving the non-singlet DGLAP equations of Eq.~(\ref{NSevolutionseq}) in Mellin $N$-space with Altarelli-Parisi kernels at NLO in $\alpha$ is evaluated in the limit $N\rightarrow\infty$. This evolution operator Mellin inverted into $x$-space and convoluted with the NLO initial conditions then yields the non-singlet solution for the PDFs $\Gamma_{\text{NS}}$ in the limit $x\rightarrow1$.
	Since the NLO kernel $\mathbf{P}_{\text{S}}=\mathbf{P}_{\text{S}}^{(0)}+\alpha/(2\pi)\mathbf{P}_{\text{S}}^{(1)}$ transformed into Mellin N-space according to Eq.~(5.71) in Ref.~\cite{Bertone:2019hks} gets diagonal in the limit $N\rightarrow\infty$, the singlet PDF $\Gamma_{\text{S}}$ decouples from the photon PDF $\Gamma_{\gamma}$. Furthermore, the singlet-singlet component of the $\mathbf{P}_{\text{S}}$ kernel, i.~e. $\lim_{N\to \infty}M[P_{SS}]$, yields the same expression as the corresponding one of the non-singlet case. Thus, the evolution operator for the decoupled singlet equations and -- due to Eqs.~(\ref{NLOIScond}) to (\ref{NLOinitconde+}) -- the complete solution $\Gamma_{\text{S}}(x,\mu^2)$ are identical to those resulting from the non-singlet equations. Owing to these considerations as well as Eq.~(\ref{singlet-NSPDFs}) the asymptotic NLO-NLL accurate electron PDF describing soft radiation of photons with resummation to all orders in $\alpha$ is constructed as
	\begin{align}
		\Gamma^{\text{NLL}}_{e^-,\text{asy}}(x,\mu^2)=\left(\Gamma^{\text{NLO}}_{\text{NS}}(\mu_0^2)\star M^{-1}\left[\lim_{N\to \infty}E_{\text{NS},N}(\mu_0^2,\mu^2)\right]\right) (x,\mu^2)
	\end{align}
	with the NS evolution operator $E_{\text{NS},N}$ in Mellin $N$-space and
	\begin{align}
		\Gamma^{\text{NLO}}_{\text{NS}}(\mu_0^2)=\Gamma^{\text{NLO}}_{e^-}(\mu_0^2)=\Gamma^{(0)}_{e^-}(\mu_0^2)+\frac{\alpha}{2\pi}\Gamma^{(1)}_{e^-}(\mu_0^2)\quad.
	\end{align}
	The explicit form including a running of $\alpha$ is given by \cite{Bertone:2019hks}
	\begin{align}
	\begin{split}
		\Gamma^{\text{NLL}}_{e^-,\text{asy}}(x,\mu^2)=&\frac{e^{\hat{\xi}_1-\gamma_E\xi_1}}{\Gamma(1+\xi_1)}\xi_1 (1-x)^{\xi_1-1}\\
		&\times \biggr\{1+\frac{\alpha(\mu_0)}{\pi}\left[\left(\log\frac{\mu^2_0}{m^2}-1\right)\left(A(\xi_1)+\frac{3}{4}\right)-2B(\xi_1)+\frac{7}{4}\right.\\
		&\left.+\left(\log\frac{\mu^2_0}{m^2}-1-2A(\xi_1)\right)\log(1-x)-\log^2(1-x)\right]\biggr\}
	\end{split}
	\label{NLLsoftPDF}
	\end{align}
	with quantities
	\begin{align}
		\xi_1&=2t-\frac{\alpha(\mu)}{4\pi^2b_0}\left(1-e^{-2\pi b_0t}\right)\left(\frac{20}{9}n_F+\frac{4\pi b_1}{b_0}\right),\\
		\hat{\xi}_1&=\frac{3}{2}t-\frac{\alpha(\mu)}{4\pi^2b_0}\left(1-e^{-2\pi b_0t}\right)\left(\lambda_1+\frac{3\pi b_1}{b_0}\right),\\
		t&=\frac{\alpha(\mu)}{2\pi}L-\frac{\alpha^2(\mu)}{4\pi}\left(b_0L^2-\frac{2b_1}{b_0}L\right)+\mathcal{O}(\alpha^2),\qquad L =\log\frac{\mu^2}{\mu^2_0} \label{tvariable}\\
		b_0&=\frac{n_F}{3\pi},\quad b_1=\frac{n_F}{4\pi^2},\quad \lambda_1=\frac{3}{8}-\frac{\pi^2}{2}+6\zeta(3)-\frac{n_F}{18}(3+4\pi^2)
	\end{align}
	\begin{align}
		A(\kappa)&=-\gamma_E-\psi_0(\kappa), \\
		B(\kappa)&=\frac{1}{2}\gamma_E^2+\frac{\pi^2}{12}+\gamma_E\psi_0(\kappa)+\frac{1}{2}\psi_0(\kappa)^2-\frac{1}{2}\psi_1(\kappa)
	\end{align}
	where $\psi_j(\kappa)$ is defined as:
	\begin{align}
				\psi_j(\kappa)&=\frac{d^{j+1}\log \Gamma(\kappa)}{d\kappa^{j+1}}\quad.
	\end{align}
	For small $\xi_1$, e.~g. for a collider setup with energies at the order of hundred GeV where $\xi_1\sim 0.05$, the expansions
	\begin{align}
		&A(\xi_1)=\frac{1}{\xi_1}+\mathcal{O}(\xi_1), &B(\xi_1)=-\frac{\pi^2}{6}+2\zeta(3)\xi_1+\mathcal{O}(\xi_1^2)
	\end{align}
	show that the term proportional to $A(\xi_1)\log(1-x)$ is the dominant of Eq.~(\ref{NLLsoftPDF}) even in extreme regions of $x$ in the limit $x\rightarrow1$. Due to $\xi_1\simeq \eta$ this is the essential difference of the asymptotic NLL solution compared to the LL counterpart, i.~e. the first term of Eq.~(\ref{LLsoftPDF}).
	
	For the solution of the electron PDFs at NLL accuracy valid for all $x$ values we adopt the iterative approach as suggested for the LL PDFs in Sec.~\ref{secLLapproximation}. Comparing the resummed LL accurate terms of Eq.~(\ref{LLres}) with those of NLL-accuracy in Eq.~(\ref{NLLres}), the manifest difference is the second term in the latter. Hence, the representation of the function $G$ and the PDFs $\Gamma_e$ in a perturbative series for NLL accuracy analogously to Eqs.~(\ref{Gseries}) and (\ref{GammaSeries}) read
	\begin{align}
				&G^{\text{NLL}}(x,\mu^2)=\sum_{n=0}^{\infty}G^{\text{NLL},(n)}(x,\mu^2)=\sum_{n=0}^{\infty}\frac{\eta^n}{2^n n!}\left(I^{\text{LL}}_n(x)+\frac{\alpha}{2\pi}I^{\text{NLL}}_n(x)\right),\label{GseriesNLL}\\
		&\Gamma^{\text{NLL}}_e(x,\mu^2)=\sum_{n=0}^{\infty}\Gamma_e^{{\text{NLL}},(n)}(x,\mu^2)=\sum_{n=0}^{\infty}\frac{\eta^n}{2^n n!}\left(\frac{\partial I^{\text{LL}}_n(x)}{\partial x}+\frac{\alpha}{2\pi}\frac{\partial I^{\text{NLL}}_n(x)}{\partial x}\right).
		\label{GammaSeriesNLL}
	\end{align}
	This is used for the approach of solving the equation
	\begin{align}
				G^{\text{NLL}}(x,\mu^2)=G^{\text{NLL}}(x,\mu_0^2)+\int_{m^2}^{\mu^2}\frac{dq^2}{q^2}\frac{\alpha(q^2)}{2\pi}\int_x^1dzP(z)G^{\text{NLL}}\left(\frac{x}{z},q^2\right)\quad,
		\label{GevolutionNLL}
	\end{align}
	with Altarelli-Parisi kernels at NLO in $\alpha$, $P(z)=P^{(0)}+\alpha/(2\pi)P^{(1)}$, an the relations
	\begin{align}
				&G^{\text{NLL}}(x,\mu^2)=\int^1_xdt\Gamma^{\text{NLL}}(t,\mu^2) &\Gamma^{\text{NLL}}(x,\mu^2)=-\frac{\partial}{\partial x}G^{\text{NLL}}(x,\mu^2)\quad.
		\label{gammaandGNLL}
	\end{align}
	Considering a NLL evolution operator formally it is not justified to neglect the running of $\alpha$ anymore as in the case of LL PDFs. Moreover, the running of $\alpha$ from the Thompson to the EW scale phenomenologically yields a significant difference, i.~e.
	\begin{align}
		\frac{\Delta \alpha}{\alpha(0)}=\frac{ \alpha(M_Z)-\alpha(0)}{\alpha(0)}\sim 0.07
	\end{align}
	It is thus convenient to use the variable \cite{Furmanski:1981cw}
	\begin{align}
		t=\frac{1}{2\pi b_0}\log \frac{\alpha(\mu)}{\alpha(\mu_0)}
	\end{align}
	which in an expanded form by means of
	\begin{align}
		\frac{\partial \alpha (\mu)}{\partial \log \mu^2}=\beta(\alpha)=b_0\alpha^2+b_1\alpha^3+\ldots
	\end{align}
	is already introduced in Eq.~(\ref{tvariable}), and we rewrite Eq.~(\ref{GevolutionNLL}) as
	\begin{align}
		G^{\text{NLL}}(x,t)=G^{\text{NLL}}(x,0)+\int_{0}^{t}du\frac{b_0\alpha^2(u)}{\beta(\alpha(u))}[P\star G^{\text{NLL}}]\left(x,u\right)\quad.
	\end{align}
	By this variable transformation, we find the perturbative series of Eq.~(\ref{GseriesNLL}) modified to
	\begin{align}
		G^{\text{NLL}}(x,t)=\sum_{n=0}^{\infty}\frac{t^n}{ n!}\left(J^{\text{LL}}_n(x)+\frac{\alpha(t)}{2\pi}J^{\text{NLL}}_n(x)\right)\quad.
	\end{align}
	Eventually, we find the recursive relations
	\begin{align}
		J^{\text{LL}}_n &=P^{(0)}\star J^{\text{LL}}_{n-1},\\
		\begin{split}
		J^{\text{NLL}}_n&=(-1)^n(2\pi b_0)^nG^{\text{NLL},(1)}(\mu_0^2)+\sum_{p=0}^{n-1}(-1)^p(2\pi b_0)^p\biggr\{P^{(0)}\star J^{\text{NLL}}_{n-1-p}\\
		&\qquad\qquad\qquad\qquad\qquad\qquad\quad+\left[P^{(1)}-\frac{2\pi b_1}{b_0}P^{(0)}\right]\star J^{\text{LL}}_{n-1-p}\biggr\}
		\end{split}
	\end{align}
	These equations are solved iteratively for the non-singlet and singlet-$\gamma$ evolutions up to $J^{\text{LL}}_3$ and $J^{\text{NLL}}_2$ in Ref.~\cite{Bertone:2019hks} yielding NLL accurate non-singlet, singlet and photon PDFs $\Gamma^{\text{NLL}}_{\text{rec}}$ with resummation up to the respective order by means of Eq.~(\ref{gammaandGNLL}). Since the analytical expressions are lengthy and cover several pages they are not presented here explicitly.
	
	Analogously to the LL accurate PDFs, a matching of the recursive `all $x$' solution, described by $\Gamma^{\text{NLL}}_{\text{rec}}$, to the asymptotic $x\simeq1$ one, $\Gamma^{\text{NLL}}_{\text{asy}}$ respectively, has to be applied. This can be achieved in an additive manner as proposed in \cite{Bertone:2019hks} by
	\begin{align}
		\Gamma^{\text{NLL}}_{\text{mtc}}(x)= \Gamma^{\text{NLL}}_{\text{rec}}(x)+\left[\Gamma^{\text{NLL}}_{\text{asy}}(x)-\Gamma^{\text{NLL}}_{\text{subt}}(x)\right]F(x)
		\label{PDFmatched}
	\end{align}
	with an arbitrary function $F(x)$ fulfilling the condition
	\begin{align}
		\lim_{x\rightarrow 1}F(x)=1
	\end{align}
	and terms $\Gamma^{\text{NLL}}_{\text{subt}}(x)$ which are present in both contributions, $\Gamma^{\text{NLL}}_{\text{rec}}(x)$ and $\Gamma^{\text{NLL}}_{\text{asy}}(x)$, and need to be subtracted exactly once to avoid the double counting.
	
	Another approach in order to solve the DGLAP equations with kernels at NLO in $\alpha$ is by purely numerical means. For this, the DGLAP equations are solved in Mellin space for small scale intervals into which the scale evolution range is partitioned. Multiplying up all thus obtained evolution operators per scale interval yields the complete evolution operator in Mellin space. By Mellin inverting the product of the initial conditions in $N$-space and this evolution operator finally results in the solution for the PDFs in $x$ space. Even though this procedure is beneficial in a theoretical sense, such that NLL resummation can be achieved to all orders in $\alpha$ for the whole $x$ range, it involves the drawback on the statistics on which the numerical integration over the Mellin contour depends on.
	
	A few remarks on the photon PDFs are left to be said: Formally for cross sections of $e^+e^-$ collisions -- ignoring photons from beam spectra for the moment -- photon PDFs would play a role only at the NNLO in $\alpha$: The initial conditions of Eq.~(\ref{gammaPDF}) start to contribute to $\Gamma_{\gamma}$ at $\mathcal{O}(\alpha)$ and partonic cross sections $d\sigma_{\gamma e^-}$,  $d\sigma_{e^+\gamma}$ and $d\sigma_{\gamma \gamma}$ on the r.~h.~s. of Eq.~(\ref{e+e-crosssection}) do not exist at the LO. Naively, one may conclude by this that the contributions from photon-induced channels are suppressed for an NLO calculation.
	However, photon PDFs can have a sizeable impact if soft photon radiation effects are enhanced. The photon PDFs are dominant and largely enhanced against the electron PDFs for small $x$ values \cite{vonWeizsacker:1934nji,Williams:1934ad,Budnev:1975poe,Bertone:2019hks}. It is thus phenomenologically improper to neglect them for an NLO calculation.
	
	The derivation of the solution for the NLL QED PDFs in this section is described in a rather schematic way. For more technical details and complete formulas the reader is encouraged to consider Refs.~\cite{Frixione:2019lga,Bertone:2019hks}.
	\section{MC integration methods for electron PDFs}
	\label{MCintegrationmethodsPDFs}
	Challenges of the embedding of electron PDFs into numerical cross section computations in a Monte-Carlo generator framework encompass two aspects.
	The numerical evaluation of NLL accurate electron PDFs per phase-space point requires an amount of computing time which is inappropriate for the purpose of a Monte-Carlo integration. This is facilitated by proper grid interpolation measures (similar to the case of proton PDFs) which will be covered by Sec.~\ref{gridinterpolation}. Moreover, as leptonic parton distribution functions feature a divergent but integrable structure in the radiated energy an appropriate phase-space mapping is a necessary ingredient for their inclusion in a Monte-Carlo integration. How this can be done systematically for LO-LL and LO-NLL observables and as an approach for NLO-NLL cross sections in the FKS framework which will be covered in Sec.~\ref{parametrisationPDFsing}.
	\subsection{Electron PDF grid interpolation measures}
	\label{gridinterpolation}
	The application of PDF grids and adequate interpolation measures for the electron PDFs are required for the same reason as it is commonly done for proton PDFs: The evaluation of PDFs depending on beam energy fraction $x$ and factorisation scale $\mu$ at higher orders in the coupling demands a tremendous amount of computing time due to the involved evolution procedure beyond LO. This is outlined in Sec.~\ref{secNLLapproximation} with respect to the NLL electron PDFs for which the evaluation of lengthy analytical expressions or the integration over a Mellin contour pose a time-consuming issue if used for a Monte Carlo phase-space integration. This especially concerns Monte-Carlo samplings with dynamic scales for which the evolution scale at which the PDFs are evaluated varies point by point.
	
	A grid for which PDF values for fixed points of the $x$ range and the scale $\mu$ are saved as data sets with appropriate interpolation and optional extrapolation routines offers a possibility to circumvent this problem and is the well-known method of the \texttt{LHAPDF} library~\cite{Buckley:2014ana}.
	An adequate measure in order to interpolate points on a two-dimensional regular grid thereby is represented by the application of bicubic interpolation. Regarding the $x$-binning and interpolation for proton PDFs, the $x$-bin boundaries in small $x$ regions are arranged in constant distances of $\log_{10} x$ for the grid and mapped onto this logarithmic axis for the interpolation. This is done in order to account for large gradients of PDFs for $x\rightarrow0$ as it is the case for example for gluon PDFs in $pp$ collisions.
	Conversely, this can be transferred to the binning and interpolation for the electron PDFs with singular $(1-x)^{\epsilon-1}$ behaviour close to $x=1$ as elaborated in Sec.~\ref{secleptonicIS} which is present in both, the LL and NLL approximation, and enhanced in the latter at least by factors $\log(1-x)$ and $\log^2(1-x)$ according to Eq.~(\ref{NLLsoftPDF}).
	
	\begin{figure}
		\centering
		\includegraphics[width=0.8\linewidth]{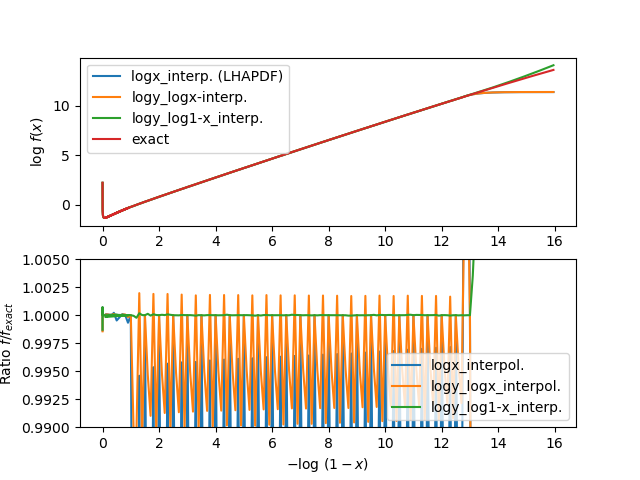}
		\caption{NLL electron PDF $f(x)$ (displayed as $\log_{10} f(x)$) as a function of $-\log_{10}{(1-x)}$ using exact values of the analytic function \cite{Frixione:2019lga,Bertone:2019hks} depicted in red (`exact'), interpolated values with grid point mapping $(x,y)\rightarrow (\log_{10} x, y)$ in blue (`logx\_interp.'), $(x,y)\rightarrow (\log_{10} x, \log_{10} y)$ in orange (`logy\_logx\_interp.') and $(x,y)\rightarrow (\log_{10} (1-x), \log_{10} y)$ in green (`logy\_log1-x\_interp.') with $y\equiv f(x)$ (upper plot); ratio $f(x)/f_{\text{exact}}(x)$ of the interpolated electron PDF function over the `exact' function depending on $-\log_{10}(1-x)$ (lower plot)}
		\label{interpolations}
	\end{figure}
\begin{figure}
		\centering
		\includegraphics[width=0.7\linewidth]{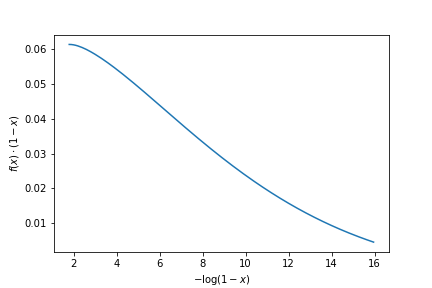}
		\caption{Function $f(x)(1-x)$ representing the electron NLL PDF times the phase-space element $\Delta x$ as an approximation of the integrand in the vicinity of $x=1$.}
		\label{xeq1limit}
	\end{figure}

	Using the code \texttt{ePDF}\footnote{The code is publicly available under \texttt{https://github.com/gstagnit/ePDF}.} \cite{Frixione:2019lga,Bertone:2019hks} for numerical PDF values resulting from the evaluation of $\Gamma^{\text{NLL}}_{\text{mtc}}(x)$ in Eq.~(\ref{PDFmatched}) the appropriate binning and interpolation for the argument $x$ can be investigated.
	Evaluating the electron PDFs for  $x$ sampled linearly within a range $0<x\le 0.9$ and logarithmically for $1\le -\log(1-x)\le 13$ and for a fixed scale $\mu$ yields the grid points between which the PDF values in the end are supposed to be obtained through interpolation. We can now probe several interpolation approaches by sampling $x$ between the fixed grid points and evaluate the PDFs for these in the two ways, once exactly and once by interpolation.
	Bicubic interpolation routines according to those used for MSTW PDFs \cite{Martin:2009iq} are applied to interpolate PDF values between the grid points. The reliability of interpolated PDF values thereby highly depends on how the grid points are mapped which is apparent from the following observation. In Fig.~\ref{interpolations} comparisons of PDF values $f(x)\equiv \Gamma^{\text{NLL}}_{\text{mtc}}(x) $ obtained by exact evaluation (in red) and by interpolation with a specific mapping of the grid points $(x,y)\equiv(x,f(x))$ are shown. Absolute values are depicted in the upper plot and the ratio $f(x)/f_{\text{exact}}(x)$ for interpolated over exact values in the lower plot, respectively.
	The interpolated PDF with mapping $(x,y)\rightarrow (\log_{10} x, y)$ which is used for example by \texttt{LHAPDF} as explained above and the corresponding ratio shown as blue curves yield a deviation from the exact values of more than $1\%$. The interpolation is improved to $\lesssim 1\%$ deviation if a mapping $(x,y)\rightarrow (\log_{10} x, \log_{10} y)$ is applied (in orange). A mapping $(x,y)\rightarrow (\log_{10} (1-x), \log_{10} y)$ (green curves) renders agreement of PDF values for an exact and an interpolated evaluation well-below $0.01\%$ in the full range of $x$ fixed by the grid points with the exception of $x$ values extremely close to zero where the deviation is still only around $0.1\%$.
	For MC predictions this accuracy can be considered as sufficiently high\footnote{Alternatively, one may interpolate using a grid point mapping $(x,y) {\rightarrow (\log_{10} (1-x), (1-x)y)}$. The accuracy of this method is not explicitly analysed here. However, due to the nearly linear shape of the function $f(x)(1-x)$ in Fig.~\ref{xeq1limit} a well performance of a bicubic interpolation is expected.}.

	Left to be discussed is the treatment of PDF values for $x$ values in the extreme region $x>x_{\text{cut}}$, where $x_{\text{cut}}=1-10^{-13}$ represents the upper boundary in $x$ for grid points of the analysis in this section. First of all, note that a fixed boundary $x_{\text{cut}}$, which naively would be assumed due to machine precision restrictions of $x$, is not required. This is because nothing prevents one from using a variable transformation, e.~g. $\bar{x}=1-x$, and evaluate the inverted region in the limit $\bar{x}\to 0$.
	This redefinition, however, must be substituted in the analytical expression for the electron PDF in order to obtain the grid points in this region. For this reason extrapolation with exemplary results for the different mappings of the grid points shown in Fig.~\ref{interpolations} is not essentially necessary since there is no restriction on the upper limit of $x$ for the grid points.
	From the behaviour of the curves which largely deviate from the exact one for that specific region it furthermore becomes apparent that if proceeded in the traditional way extrapolation is rather useless for the singular enhanced PDFs.
	
	Since the singularity of the electron PDF is integrable we can consider the following fact from the MC integration perspective.
	Approximately, for high-energy collider experiments matrix elements can be considered numerically constant varying $x$ in the region extremely close to one. Broadly speaking, the contributions to the integral in this region for one beam axis thus come from
	\begin{align}
		&\lim_{x\rightarrow1}dI(x)=dxf(x)c, &c=\text{const.},
	\end{align}
	i.~e. the PDF times the phase-space element $dx$ and a constant number $c$. We can now investigate at which $x$ these contributions eventually become negligible: Varying a technical cut in the range $x<x_{\text{cut}}<1$, above which the PDFs are set constant to $f(x_{\text{cut}})$, would not change the integrated result within a statistical error which is phenomenologically tolerable. This can be understood by estimating the contribution of the right-most bin of the $x$ range in a MC integration as negligible. Functionally, it can be approached by
	\begin{align}
		\Delta x f(x)c\simeq(1-x)f(x)c\quad.
	\end{align}
	The function $(1-x) f(x)$ depending on $-\log_{10}(1-x)$ is plotted in Fig.~\ref{xeq1limit} with respect to an exact evaluation of the electron PDFs.
	It can be seen that the area of the right-most bin is shrinking the closer $x$ approaches one which by following the trend of the `exact' curve below $1-x=10^{-16}$ yields the prerequisite for imposing a technical cut in general. In particular, it can be observed that the area $(1-x)f(x)$ for $1-x\simeq10^{-16}$ would be about $10\%$ of the contribution to the integral with respect to that for $1-x\simeq10^{-4}$ which is too large for a technical cut set to $1-x_{\text{cut}}=10^{-16}$. The investigation of the range in the vicinity of $x=1$ for which a technical cut would not affect the integrated MC cross section result within sufficiently high statistics will be deferred to future works.
	
	For the interpolation of the PDFs between grid points with distinct scale values $\mu$ and fixed $x$ we note that the scale dependence of electron PDFs is merely due to logarithms $\log (\mu^2/\mu_0^2)$, e.~g. through variables $t$ or $\xi_1$ in Eq.~(\ref{NLLsoftPDF}). Since $\mu\geq\mu_0$, a polynomial approximation of the functional dependence of the PDFs on the scale is thus accurate enough such that the standard bicubic interpolation between linearly distributed grid points along the range $\mu_0^2\leq \mu^2 \leq \mu^2_{\text{max}}$ can be considered reliable.
	\subsection{Mapping of FKS phase-space for divergent electron PDFs}
	\label{parametrisationPDFsing}
	The peaked structure of electron PDFs at both, LL and NLL accuracy, imposes the need for a revised MC sampling of phase-space points in order to account for the full area of extremely small sections of the integration variable $x$ for the beam energy fraction.
	The standard mapping applied in \texttt{WHIZARD} in the context of ISR structure functions for an improved integration efficiency for LO calculations proceeds as follows.\\
	We define mapping variables $\{p_1,p_2\}$ with $p_1,p_2\in[0,1]$ depending on the random number variables $\{r_1,r_2\}$ with $r_1,r_2\in[0,1]$ and using $\bar{r}_1=1-r_1$ as
	\begin{align}
		p_1 &= 1 - (1-r_1)^{1/\epsilon}=1- \bar{r}_1^{1/\epsilon} \label{p1 random}\\
		p_2 &=
	\begin{cases}
		 1 - {(2r_2)^{1/\epsilon}}/{2}, & u>0 \\
		{(2r_2)^{1/\epsilon}}/{2}, & u<0\\
		1/2, & u=0
	\end{cases}
	& u = 2r_2-1
	\end{align}
	where $\epsilon$ is an arbitrary small number. In \texttt{WHIZARD} this number is set by the customary choice
	\begin{align}
	\epsilon=\frac{\alpha}{\pi}q_e^2\left(\ln \frac{s}{m^2}-1\right)\quad,
	\end{align}
	with $q_e$ the charge and $m$ the mass of the initial-state particle.
	Integrating over $p_1$ and $p_2$ results in the Jacobian
	\begin{align}
		f_1 &= f_{p_1}f_{p_2} 
	\end{align}
	with
	\begin{align}
		f_{p_1} &= \frac{1}{\epsilon}(1-r_1)^{1/\epsilon-1}=\frac{1}{\epsilon}\bar{r}_1^{1/\epsilon-1}\\
		f_{p_2} &=
			\begin{cases}
		\frac{(2-2r_2)^{1/\epsilon-1}}{\epsilon}, & u>0 \\
		\frac{(2r_2)^{1/\epsilon-1}}{\epsilon}, & u<0\\
		1/\epsilon, & u=0
		\end{cases}\quad.
	\end{align}
	This mapping gives an enhanced sampling of points at the endpoints of the mapping variables, i.~e. for $p_1$ in the vicinity of $1$ and for $p_2$ approaching both, $0$ and $1$.
	
	The dependence of $\{p_1,p_2\}$ on the physical variables $\{x_1,x_2\}$ with $x_1,x_2\in[0,1]$, the fractions of the initial state momenta, is further defined as
	\begin{align}
		&p_1 = x_1 \cdot x_2  &p_2 = \frac{\log x_1}{\log (x_1\cdot x_2)}
	\end{align}
	such that the physical variables $\{x_1, x_2\}$ to be integrated over, in turn, depend on the mapping variables as
	\begin{align}
		&x_1 = p_1^{p2} &x_2 = p_1^{1-p2}.
		\label{p variables}
	\end{align}
	On the one hand, this mapping parametrisation fulfills the criterion of mapping the unit square such that the parton centre-of-mass energy is constant with respect to the centre-of-mass frame, i.~e. $x_1 \cdot x_2 $ is independent from $p_2$. On the other hand, integrating over $\{x_1,x_2\}$ results in the Jacobian
	\begin{equation}
		f_2 = |\log p_1 |
	\end{equation}
	which flattens the integrand while sampling $p_1$ close to one which follows from Eq.~(\ref{p1 random}), and thus sampling $f_2$ close to zero. The overall Jacobian for Born or LO computations then reads
	\begin{align}
		f_{\text{Born}}=f_1f_2 = f_{p_1}f_{p_2}|\log p_1|.
		\label{jacobianborn}
	\end{align}
	In this way, the integration and adaption in MC simulations is well-behaved and yield useful predictions as it is shown by simulations with \texttt{WHIZARD} \cite{Hagiwara:2005wg,Whizard:2020}. Crucially, the Jacobian in Eq.~(\ref{jacobianborn}) depends on the factor `$(1-r_1)^{1/\epsilon-1}|\log p_1|$' which flattens the integrand for $r_1\to 1$. Consequently, this factor can be used paradigmatically for similar parametrisations for NLO computations proposed in the following.
	
	The condition that the total invariant mass is conserved if a parton is radiated from the system induces a rescaling of $x_1$ and $x_2$ for initial-state emissions according to Eq.~(\ref{xplusxminus}). If applying the standard parametrisation in random number variables, this leads to a breakdown of the mapping benefits, which are described above, for non-Born initial-state momenta. Essentially, for the real-emission and DGLAP-remnant component the rescaled beam momenta (before photon emission off the initial-state) are shifted closer to $1$ with respect to the unrescaled momenta. By denoting the rescaled momenta as $x^{\prime}=x+\delta x$ one can find that even for a small variation $\delta x > 0$ which is in line with the condition $x^{\prime}\le 1$ the ratio of the rescaled over the unrescaled PDFs diverges as $x^{\prime}$ approaches $1$, i.~e.
	\begin{align}
		\lim_{x^{\prime}\rightarrow 1} \frac{\Gamma(x^{\prime})}{\Gamma(x)}=\lim_{x\rightarrow 1-\delta x}\frac{\Gamma(x+\delta x)}{\Gamma(x)}\rightarrow \infty\quad.
	\end{align}
	This means that in addition to the parametrisation of the unrescaled Born momenta $x_1$ and $x_2$ according to Eq.~(\ref{p variables}) an improved mapping parametrisation of the variation $\delta x$ is essential.
	Thus, in order to recover the previously explained integration efficiency due to the mapping we have to revise the parametrisation explicitly for rescaled initial state momenta $x_1^{\prime}$ and $x_2^{\prime}$ and thereby retaining the dependence on the FKS variables.
	
	For simplicity, we start here with the formulation of the additional NLO integration parameters for the remnant of the subtraction of collinear ISR singularities in integrated form, i.~e. the degenerate $n+1$ contribution (or DGLAP remnant) of Eq.~(\ref{degeneraten+1contr}). For this contribution one has to integrate over exactly one additional degree of freedom compared to those of the Born cross section. The momentum dependence of the PDFs rescales as
	\begin{align}
	\Gamma(x_j,\mu) \longrightarrow \Gamma(x_j/z_j,\mu)
	\end{align}
	with $x_j < x_j/z_j < 1$ and emitter $j\in\{1,2\}$.
	The standard parametrisation of ISR remnant integration variables $z_j$ following the restriction of Eq.~(\ref{zrescalingrestriction}) is defined in terms of random number variable $r_z\in [0,1]$ as
	\begin{equation}
		z_j = 1 - r_z (1-x_j)
	\end{equation}
	with Jacobian
	\begin{equation}
		f_z= 1-x_j.
	\end{equation}
	Now, we rewrite these formulas for an improved mapping adequate for using NLL electron PDFs in NLO computations.
	In order to enhance the Monte-Carlo sampling of points for $z_j$ close to $1$, it is useful to apply a bijective mapping of the random variable $r_z \in [0,1]$ defining, similar to Eq.~(\ref{p1 random}),
	\begin{align}
		p_z = 1 - (1-r_z)^{1/\epsilon}
	\end{align}
	with Jacobian factor
	\begin{align}
		f_{p_z} &= \frac{1}{\epsilon}(1-r_z)^{1/\epsilon-1}.
	\end{align}
	Regarding the condition $[0,1] \longmapsto [x_j,1]$ mapping $p_z \longrightarrow z_j$ we can find the parametrisation
	\begin{align}
    z_j = 1- p_z (1-\log p_z) (1-x_j) 
    \label{improvedz}
	\end{align}
	which results in the factor
	\begin{align}
		f_{z_j}=(1-x_j)\log p_z
	\end{align}
	for the overall Jacobian per emitter
	\begin{align}
     f_{\text{DGLAP,j}}= f_{p_z}f_{z_j}=\frac{1}{\epsilon}(1-r_z)^{1/\epsilon-1}(1-x_j)\log p_z.
     \label{DglapJacobian}
	\end{align}
	In this way, the integrand of the DGLAP remnant gets flattened which is useful with respect to the enhancement of the PDF close to $x_j=1$ for the corresponding emitter $j$.
	
	Analogously, we proceed for the real emission phase-space parametrisation and for the IS collinear subtraction terms of the real component which both require rescaled IS momenta entering the PDFs. For the first, it is less trivial to find an adequate mapping for the rescaled momenta. It requires to get Jacobian factors analogous to Eq.~(\ref{jacobianborn}) resulting in the full integration efficiency effect as stated above and, simultaneously, retaining the actual FKS parametrisation in both, $\xi$ and $y$, defined in Eq.~(\ref{xplusxminus}). According to these definitions, for the real component the momentum dependencies of the PDFs rescale as
	\begin{align}
		&\Gamma(x_1,\mu) \longrightarrow\Gamma(x_1^{\prime})= \Gamma\left(\frac{x_1}{\sqrt{1-\xi}}\sqrt{\frac{2-\xi(1-y)}{2-\xi(1+y)}}\right)\\
		&\Gamma(x_2,\mu) \longrightarrow\Gamma(x_2^{\prime})= \Gamma\left(\frac{x_2}{\sqrt{1-\xi}}\sqrt{\frac{2-\xi(1+y)}{2-\xi(1-y)}}\right)
	\end{align}
	which in the collinear limits $y=\pm 1$ have to coincide with
	\begin{align}
		&\Gamma(x_1^{\prime})\biggr\rvert_{y=1}=\Gamma\left(\frac{x_1}{1-\xi}\right) 
		&\Gamma(x_1^{\prime})\biggr\rvert_{y=-1}=\Gamma\left({x_1}\right)\\
		&\Gamma(x_2^{\prime})\biggr\rvert_{y=1}=\Gamma\left({x_2}\right)
		&\Gamma(x_2^{\prime})\biggr\rvert_{y=-1}=\Gamma\left(\frac{x_2}{1-\xi}\right)
	\end{align}
	entering the collinear subtraction terms. Because of the diverging gradient of the PDFs at $x_1\rightarrow1$ and $x_2\rightarrow1$ a proper local cancellation in the collinear limit formally guaranteed by
	\begin{align}
		&\lim_{y\rightarrow 1}\Gamma(x_1^{\prime})\Gamma(x_2^{\prime})\hat{\mathcal{R}}(\xi,y)-\Gamma\left(\frac{x_1}{1-\xi}\right)\Gamma\left({x_2}\right)\hat{\mathcal{R}}(\xi,1)\sim 0\\
		&\lim_{y\rightarrow -1}\Gamma(x_1^{\prime})\Gamma(x_2^{\prime})\hat{\mathcal{R}}(\xi,y)-\Gamma\left({x_1}\right)\Gamma\left(\frac{x_2}{1-\xi}\right)\hat{\mathcal{R}}(\xi,-1)\sim 0
	\end{align}
	gets numerically affected for a traditional MC sampling of $\xi$ and $y$ which is used for hadron collisions.
	Due to the variable $\xi$ appearing in both, the real emission amplitudes and the collinear subtraction terms, one is tempted to merely reparametrise $\xi$ analogously to the variable $z$ for the DGLAP remnant. However, one has to keep in mind that the upper boundary $\xi_{\text{max}}$, and in this turn $\xi$ itself, as well as the rescaled variables $x_1^{\prime}$ and $x_2^{\prime}$ are correlated with $y$. In this sense, it is mandatory to find a two-dimensional phase-space mapping for an improved efficiency.
	
	An idea to technically realise such a real phase-space mapping is based on the following consideration. Note that we have exactly the same number of degrees of freedom for integration variables defining the initial-state real momenta, i.~e. $\{x_1,x_2,\xi,y\}$, as the number of variables which needs to be adapted for an efficient phase-space evaluation, i.~e. $\{x_1,x_2,x_1^{\prime},x_2^{\prime}\}$. In the same way as $x_1^{\prime}$ and $x_2^{\prime}$ is defined by Eq.~(\ref{xplusxminus}) through randomly determined $x_1$, $x_2$, $\xi$ and $y$ values, we can conversely derive $\xi$ and $y$ from the variables $x_1$, $x_2$, $x_1^{\prime}$ and $x_2^{\prime}$ defined through random numbers with the condition
	\begin{align}
		&x_1\leq x_1^{\prime}<1 &x_2\leq x_2^{\prime}<1\quad.
	\end{align}
	Hence, we define the rescaled variables with mapping analogously to Eq.~(\ref{improvedz}) as
	\begin{align}
		&x^{\prime}_j=1- \hat{p}_j (1-\log \hat{p}_j) (1-x_j) &j=1,2
		\label{improvedmappingreal}
	\end{align}
	with $\hat{p}_j\in [0,1]$ which is constructed from random numbers $\hat{r}_j\in [0,1]$ as
	\begin{align}
		\hat{p}_j = 1 - (1-\hat{r}_j)^{1/\epsilon}
	\end{align}
	and leads to Jacobians
	\begin{align}
		f_{x^{\prime},j}=\frac{1}{\epsilon}(1-\hat{r}_j)^{1/\epsilon-1}(1-x_j)\log \hat{p}_j
	\end{align}
	In order to find the expressions of $\xi$ and $y$ in terms of $x_1^{\prime}$, $x_2^{\prime}$, $x_1$ and $x_2$ regarding the conditions of massless ISR construction according to Sec.~\ref{secMasslessISem} we define auxiliary quantities
	\begin{align}
		&A\equiv\frac{x_1x_2^{\prime}}{x_2x_1^{\prime}}=\frac{2-\xi(1+y)}{2-\xi(1-y)} &B\equiv\frac{x_1x_2}{x_1^{\prime}x_2^{\prime}}=1-\xi
	\end{align}
	such that $\xi$ and $y$ can be derived yielding
	\begin{align}
		&\xi = 1-B &y=\left(\frac{1+B}{1-B}\right)\left(\frac{1-A}{1+A}\right)\quad.
	\end{align}
	Considering
	\begin{align}
		d\xi dy=\mathcal{J}_{1}(A,B)~dAdB=\mathcal{J}_{1}(A,B)\mathcal{J}_{2}(x_1^{\prime},x_2^{\prime})~dx_1^{\prime}dx_1^{\prime}
	\end{align}
	with
	\begin{align}
		&\mathcal{J}_{1}(A,B)=2\left(\frac{1+B}{1-B}\right)\frac{1}{(1+A)^2} &\mathcal{J}_{2}(x_1^{\prime},x_2^{\prime})=2\frac{x_1^2}{x_1^{\prime 3}x_2^{\prime}}
	\end{align}
	we get the final Jacobian factor for $\xi$ and $y$ parametrised in random numbers $\hat{r}_{1/2}$,
	\begin{align}
		f_{\text{real},j}=\mathcal{J}_{1}(A,B)\mathcal{J}_{2}(x_1^{\prime},x_2^{\prime})f_{x^{\prime},1}f_{x^{\prime},2}\quad.
		\label{jacobianrealmapping}
	\end{align}
	This parametrisation of $\xi$ and $y$ for the radiated momentum in terms of random number variables is supposed to yield the desired improved integration efficiency due to `$(1-\hat{r}_j)^{1/\epsilon-1}\log \hat{p}_j$' factors in the Jabobian. These factors flatten the integrand if $\hat{r}_j\to 1$ which is analogous to similar factors in the Jacobians of the Born and DGLAP remnant, i.~e. in Eq.~(\ref{jacobianborn}) and (\ref{DglapJacobian}). The numerical proof is deferred to future works. However, an overview on implemented elements and first successful consistency checks achieved on the way to the complete proof is given in Sec.~\ref{secLeptonCollisions}. The desired effect for an efficient integration is per definition constrained to ISR. For FSR the radiated momentum is constructed from the same FKS variables $\xi$ and $y$ as used for the ISR singular regions. However, as the beam momentum fractions entering the PDFs receive no rescaling for FSR no additional efficiency improvement for the integration of the FKS variables is needed. In this sense, the parametrisation of $\xi$ and $y$ as shown above is redundant for FSR, but eventually has no effect on the numerical result of the MC integration.

	\chapter{Automated NLO EW calculations in the MC framework WHIZARD}
	\label{secWhizardFrameworkNLO}
For simulating physics with a predictive power at the level of exclusive data Monte-Carlo event generators in general represent indispensable tools. The multi-purpose program \texttt{WHIZARD} offers the possibility of producing cross sections and simulated event samples for generic processes. Its general capabilities are introduced in Sec.~\ref{secWHIZARD}. This framework is used for an automation of precision calculations at NLO in SM couplings for which the extension to automated NLO EW and NLO mixed corrections is the particular aim of this thesis. The technical details of the NLO automation in \texttt{WHIZARD} are presented in Sec.~\ref{secNLOmethodology}. As formally described in Sec.~\ref{mixedcouplingsSec}, NLO contributions to arbitrary coupling powers might require the subtraction of both, QCD and EW singularities. How both correction types are treated for one NLO contribution at fixed order is explained in Sec.~\ref{secMixedCoupling}. Moreover, in this section challenges for IR-safe observables imposed by EW and, in particular, mixed corrections will be discussed as well.
\section{The event generator WHIZARD}
\label{secWHIZARD}
\texttt{WHIZARD}\footnote{The program is publicly available under \texttt{https://whizard.hepforge.org/}.}, an acronym for `$W$, $H$iggs, $Z$ And Respective Decays', is a multi-purpose Monte-Carlo event generator providing the complete system of particle-physics calculations, from setting up a Lagrangian of a chosen model up to simulated unweighted event samples.

A wide range of physics models, the SM as well as simplified and extended versions of it, are implemented as internal models which is in addition to the interface to \texttt{FeynRules} \cite{Alloul:2013bka} and \texttt{SARAH} \cite{Staub:2015kfa} providing arbitrary models in the Universal Feynman Output (UFO) format \cite{Degrande:2011ua}. Scattering amplitudes for arbitrary tree-level processes of a specified model are constructed and evaluated automatically by the intrinsic matrix element generator \texttt{O'Mega} \cite{Moretti:2001zz} based on the colour-flow formalism \cite{Kilian:2012pz}.
For NLO computations virtual loop and colour-/spin-correlated matrix elements in general are accessed by one-loop providers (OLPs) \texttt{OpenLoops} \cite{Cascioli:2011va,Buccioni:2017yxi,Buccioni:2019sur}, \texttt{RECOLA} \cite{Actis:2012qn,Actis:2016mpe} and \texttt{GoSam} \cite{Cullen:2011ac,Cullen:2014yla}. The integration and event generation in \texttt{WHIZARD} happens via \texttt{VAMP} (Vegas AMPlified) \cite{Ohl:1998jn} using  grid and weight adaptive integration methods based on the \texttt{VEGAS} algorithm \cite{Lepage:1977sw,Lepage:123074} combined with multi-channel principles~\cite{Kleiss:1994qy}. The generation of multiple phase-space channels with the heuristic algorithm implemented in \texttt{WHIZARD} increases the efficiency of an adaptive integration even beyond the factorisability of effective integrands. For the phase-space evaluation on multi-core architectures the \texttt{VAMP} algorithm has been superseded by \texttt{VAMP2} \cite{Brass:2018xbv} by which the integration is parallelised via the message-passing interface (MPI). This capability is essential especially in view of high multiplicity processes requiring a large number of phase-space dimensions to be integrated over or CPU-intensive NLO computations. Arbitrary phase-space cuts can be set in the \texttt{WHIZARD} steering (\texttt{.sin}) files via commands in the scripting language \texttt{SINDARIN}.

As event samples are of usage for comparison with experimental data only if they are unweighted the phase-space generator involves a rejection algorithm for the conversion of weighted into unweighted samples. The showering and hadronisation of events which converts partonic into hadronic events proceeds via internal algorithms \cite{Kilian:2011ka} or interfaces to external tools as \texttt{PYTHIA6} \cite{Sjostrand:2006za}, by default shipped with \texttt{WHIZARD}, or \texttt{PYTHIA8} \cite{Sjostrand:2014zea}. For jet clustering algorithms in order to render observables of processes involving jets IR-safe the package \texttt{FastJet} \cite{Cacciari:2011ma} is interfaced. The output formats of event files which can be written out directly by \texttt{WHIZARD} are \texttt{StdHEP}, \texttt{LHA}, \texttt{ascii}, \texttt{LHEF2} and \texttt{LHEF3} \cite{Alwall:2006yp}. Further common formats can be used by linking \texttt{WHIZARD} to external packages such as \texttt{LCIO} \cite{Gaede:2003ip}, \texttt{HEPMC2} \cite{Dobbs:2001ck} or \texttt{HEPMC3} \cite{Buckley:2019xhk} with an interface of the latter for exporting events in \texttt{ROOT}-trees \cite{Brun:1997pa}.

Concerning NLO computations, the subtraction scheme for the regularisation of IR singularities chosen for \texttt{WHIZARD} is the FKS scheme due to its advantages mentioned already in Sec.~\ref{secRegularisationIR}. The formalism to achieve cross sections and observables at fixed NLO in couplings $\alpha$ or $\alpha_s$ is explained in detail in the previous chapter. For the generation of hadronic events at the precision level of NLO perturbative expansions the following caveat has to be considered.
In order to circumvent a double counting of terms describing the radiation of partons in both, the matrix elements and the parton shower, a proper matching scheme has to be applied. The method used in \texttt{WHIZARD} is the POWHEG matching scheme \cite{Nason:2004rx,Frixione:2007vw} which perfectly interplays with the FKS subtraction scheme. This is
because the algorithm is designed for the generation of NLO distributions relying on unique Born phase-space points due to its description with Sudakov form factors based on ratios of real emission over Born squared matrix elements. Furthermore, the POWHEG method -- related to its name which is an acronym for `Positive Weight Hardest Emission Generator' -- beats other matching schemes with respect to avoiding negative event weights which is an inherent problem in MC@NLO \cite{Frixione:2002ik} for example. The POWHEG matching scheme in \texttt{WHIZARD} first has been applied in the context of $t\bar{t}(H)$ production at an $e^+e^-$ collider \cite{ChokoufeNejad:2015kpc} and has recently been extended to $pp$ collisions \cite{StienemThesis,Bredt:2022zpn}. In order to increase the precision of event samples, merging schemes are applied which combine matrix elements with different final-state parton multiplicities and parton showers. \texttt{WHIZARD} uses the MLM scheme (of M.~L.~Mangano) \cite{Mangano:2006rw} which is based on slicing the radiation phase-space into domains for the matrix elements and the parton shower.

The framework of \texttt{WHIZARD} has been designed with a focus on physics simulation at lepton collisions. As one of the most established generators in this field it induced many studies and design reports for future colliders as ILC, CLIC, CEPC or FCC-ee \cite{Fujii:2015jha,CLIC:2018fvx,CEPCStudyGroup:2018ghi,Baer:2013cma,Behnke:2013lya,FCC:2018evy}.
This is owing to its broad range of beam structure features. It covers effects as beamstrahlung, a strong electromagnetic field induced by dense beam bunches, which is taken into account by means of the packages \texttt{CIRCE1} \cite{Ohl:1996fi} and \texttt{CIRCE2} by fitting the energy spectra obtained from \texttt{GuineaPig} \cite{Schulte:1999tx,4440556}. Additionally, there is the option for simulations to use Gaussian beam energy spreads.
ISR structure functions for lepton initial states are implemented to the LL accuracy with the formulas of Refs.~\cite{Cacciari:1992pz,Skrzypek:1990qs,Skrzypek:1992vk} as presented in Sec.~\ref{secLLapproximation} and (recently) to the NLL accuracy with the analytical form of Refs.~\cite{Bertone:2019hks,Frixione:2019lga}. Photon initial-states, e.~g. inducing background processes at lepton colliders, can be simulated with \texttt{WHIZARD} by the so called equivalent photon approximation (EPA) based on the concept of Weizs\"acker-Williams approximation \cite{vonWeizsacker:1934nji,Williams:1934ad,Budnev:1975poe}.
Another feature with respect to the lepton beams is the possibility to include polarisation of beams for arbitrary setups, e.~g. choosing polarisation beam fractions, axes or types, i. e. circular, longitudinal or transversal polarisation. Further beam structure features for lepton collisions provided by \texttt{WHIZARD} are asymmetric beams and crossing angles, for example.

For the evaluation of PDFs in hadronic collisions there is an interface to the PDF set library \texttt{LHAPDF} \cite{Buckley:2014ana} and additionally built-in grids for specific PDF sets.

From the technical side, \texttt{WHIZARD} is a modular written \texttt{Fortran} code using the modern \texttt{Fortran2008} standard.
\section{Methodology of automated NLO corrections}
\label{secNLOmethodology}
The first NLO calculations achieved with \texttt{WHIZARD} were with respect to chargino production at the ILC using special-tailored NLO-EW amplitudes \cite{Kilian:2006cj,Robens:2008sa} and the process $pp\to b\bar{b}b\bar{b}$ at NLO QCD \cite{Binoth:2009rv,Greiner:2011mp} based on CS subtraction. The work on fully-automated NLO computations using the FKS scheme in \texttt{WHIZARD} started by considering pure QCD corrections with a first proof of concept for off-shell top quark production at a lepton collider given by Refs.~\cite{Weiss:2015npa,ChokoufeNejad:2016qux,Weiss:2017qbj}. Further applications of the NLO setup related to top quark physics followed by Refs.~\cite{Bach:2017ggt,ChokoufeNejad:2017rag}. The capability of \texttt{WHIZARD} computing NLO QCD corrections to lepton and hadron collision processes with arbitrary external states in a fully-automated way has been demonstrated by Refs.~\cite{Stienemeier:2021cse,Rothe:2021sml,WhizardNLO}. The extension of this framework accounting for NLO EW corrections for generic processes is the aim of this work.

Basic ingredients of an NLO computation, the storage of matrix elements and the one-loop provision by external tools, are presented below. Furthermore, a brief overview on the workflow of an NLO calculation is given in Sec.~\ref{secworkflowNLO} and on basic EW code supplements in Sec.~\ref{secEWsupplements}.

\subsection{Bookkeeping of matrix elements}
\label{secBookkeepingME}
The bookkeeping of matrix elements is an important ingredient of an automated framework computing observables for arbitrary processes, in particular for those with external states of a multi-dimensional flavour space represented by protons or jets. In \texttt{WHIZARD}, the bookkeeping is realised by state matrices implemented as tries, i.~e. ordered tree data structures, with nodes representing sets of quantum numbers each. In this way a subprocess is fully described by the chain of nodes from the root to one of the last nodes. A set $i$ consists of the following elements: the particle identity $f_i$, also called `flavour', the colour flow indices $c_i$ which are assigned according to the colour flow formalism \cite{Kilian:2012pz}, helicity indices $h_i$ and the momentum $\{k\}_i$.
The index~$i$ thereby represents the position of a particle in Born and real flavour structures \texttt{flst\_born} and \texttt{flst\_real} with particles ordered in a code-optimising way, e.~g. with the last index of \texttt{flst\_real} referring to the radiated parton. In accordance with the FKS scheme their entries are accessed and linked in the code based on the meaning of positions of the process particle structures $f_b$ and $f_r$ defined in Eq.~(\ref{flavorstructure}).
The leaves of the trie data structure finally correspond to the stored values of the (squared) matrix elements which are assigned to each possible branch of the trie, hence to each subprocess. The way of storing the data in tries in general relies on few memory space and allows for optimised searching algorithms.

As already mentioned above, tree-level matrix elements can be obtained by the internal matrix element generator \texttt{O'Mega}. For NLO computations it is however possible to rely fully on external tools for tree-level, colour-/spin-correlated and virtual loop matrix elements. In some cases this is even required considering the status of the implementation at the time this thesis is written, e.~g. with respect to fixed coupling powers which will be commented in more detail in Sec.~\ref{secMixedCoupling}.

\subsection{One-loop providers}
\label{secOneloopProv}
For the construction and evaluation of one-loop amplitudes the OLPs which can be linked to \texttt{WHIZARD} rely on the following methods.

The tool \texttt{GoSam} uses algebraic methods generating analytical expressions for Feynman diagrams which in a next step are simplified by $D$-dimensional reduction at integrand level \cite{Ossola:2006us,Mastrolia:2008jb} using \texttt{Ninja} \cite{vanDeurzen:2013saa,Mastrolia:2012bu} or \texttt{Samurai} \cite{Mastrolia:2010nb}. Alternatively, tensor reduction \cite{Heinrich:2010ax} can be applied using \texttt{Golem95C} \cite{Cullen:2011kv,Binoth:2008uq,Guillet:2013msa}. The evaluation of scalar one-loop integrals can be performed with libraries such as \texttt{OneLOop} \cite{vanHameren:2010cp}. 

Contrary to this method, \texttt{OpenLoops} relies on a numerical algorithm for constructing one-loop amplitudes in a recursive approach. The numerical recursion based on a similar algorithm as the one presented in Ref.~\cite{vanHameren:2009vq} is thereby applied to factorised colour-stripped cut-open loop diagrams. The amplitudes are evaluated using tensorial reduction by an `on-the-fly' method \cite{Buccioni:2017yxi}, supplied with scalar one-loop integrals from \texttt{Collier} \cite{Denner:2014gla,Denner:2016kdg} or \texttt{OneLOop}. Alternatively, for the tensor reduction \texttt{Collier} or \texttt{CutTools} \cite{Ossola:2007ax} using OPP reduction can be used. Due to its numerical approach \texttt{OpenLoops} represents a fast-evaluating and efficient OLP for MC integration purposes.

\texttt{RECOLA} uses a complete recursive approach based on tree-level off-shell currents as basic building blocks. It therefore avoids the bottleneck of recomputing identical subgraphs appearing in several different Feynman diagrams, as other OLPs do, which can get severely inefficient especially for high multiplicity processes. The evaluation of one-loop scalar and tensor integrals proceeds via the \texttt{Collier} library.

While the interface to \texttt{RECOLA} is implemented as an own module, for \texttt{OpenLoops} and \texttt{GoSam} the communication with \texttt{WHIZARD} relies on the Binoth Les Houches Accord (BLHA) \cite{Binoth:2010xt,Alioli:2013nda}, a standard interface between one-loop providers and MC generators. This BLHA interface in the \texttt{WHIZARD} code is complete in the sense that the interplay with the system of storing the matrix elements values in terms of state matrices is guaranteed. Taking this and the facts described above into account, at the moment \texttt{OpenLoops} is considered as the preferred tool for NLO calculations to processes consisting of multiple subprocesses such as $pp$ induced processes. While all OLPs presented here are in general capable of computing NLO EW amplitudes, \texttt{RECOLA} is the most advanced tool considering the numerically efficient computation of EW one-loop amplitudes for arbitrary processes, including also mass dependencies of all fermions at higher orders in EW.

\subsection{Workflow of NLO calculations}
\label{secworkflowNLO}
\begin{figure}
\begin{tikzpicture}[node distance=2.cm]
\centering
\tikzstyle{startstop} = [rectangle, rounded corners, minimum width=3cm, minimum height=1cm,text centered, draw=black, fill=red!30]
\tikzstyle{process} = [rectangle, rounded corners, minimum width=3cm, minimum height=1cm, text centered, draw=black, fill=orange!30]
\tikzstyle{OLP} = [rectangle, rounded corners, minimum width=3cm, minimum height=1cm, text centered, draw=black, fill=blue!30]
\tikzstyle{kinematics} = [rectangle, rounded corners, minimum width=2cm, minimum height=1cm, text centered, draw=black, fill=green!30]
\tikzstyle{arrow} = [thick,->,>=stealth]

	\node at (0,0) ( ){};
	\node at (3,0) (start) [startstop] {$\begin{array}{c}
		\texttt{radiation\_generator}\\
		\text{Construct real flavour structures}
		\end{array}$};
	\node (pro1) [process, below of=start] {$\begin{array}{c}
		\texttt{region\_data\_init}\\
		\text{Initialise FKS regions}
		\end{array}$};
	\node (pro2) [OLP, right of=pro1, xshift=7cm] {$\begin{array}{c}
		\texttt{blha\_configuration\_write}\\
		\text{Write \texttt{.olp} file}
		\end{array}$};
	\node (pro3) [OLP, below of=pro2] {$\begin{array}{c}
			\texttt{read\_contract\_file}\\
			\text{Read \texttt{.olc} file}
			\end{array}$};
	\node (pro4) [OLP, below of=pro3] {$\begin{array}{c}
			\texttt{evaluate\_interaction}\\
			\text{Evaluate sq. matrix elements}
			\end{array}$};
		\node (pro5) [process, below of=pro4, xshift=-7.5cm] {$\begin{array}{c}
		\texttt{process\_instance\_evaluate\_trace}\\
		\text{Evaluate FKS terms}
		\end{array}$};
	\node (mci) [kinematics, below of=pro1, xshift=4cm,yshift=1.3cm] { \texttt{mci\_t}
	};
	\node (pro6) [kinematics, below of=pro1, xshift=3.5cm,yshift=-0.2cm] {$\begin{array}{c}
		 \texttt{compute\_seed\_kinematics}\\
		\text{Construct phase-space}
		\end{array}$
		};
	\node (pro7) [kinematics, below of=pro6] {$\begin{array}{c}
		\text{Compute }S_{\tilde{\alpha}},\\
		\text{cut expressions, etc.}
		\end{array}$
	};
\draw [arrow] (start) -- (pro1);
\draw [arrow] (pro1) -- (pro5);
\draw [arrow] (pro1) -- (pro2);
\draw [arrow] (pro1) -- (pro6);
\draw [arrow] (pro6) -- (pro7);
\draw [arrow] (pro7) -- (pro5);
\draw [arrow] (pro2) -- (pro3);
\draw [arrow] (pro3) -- (pro4);
\draw [arrow] (pro4) -- (pro5);
\draw [arrow] (pro6) -- (pro4);
\draw [arrow] (mci) -- (pro6);
\end{tikzpicture}
\caption{Sketched workflow for NLO calculations in \texttt{WHIZARD}}
\label{workflowNLO}
\end{figure}
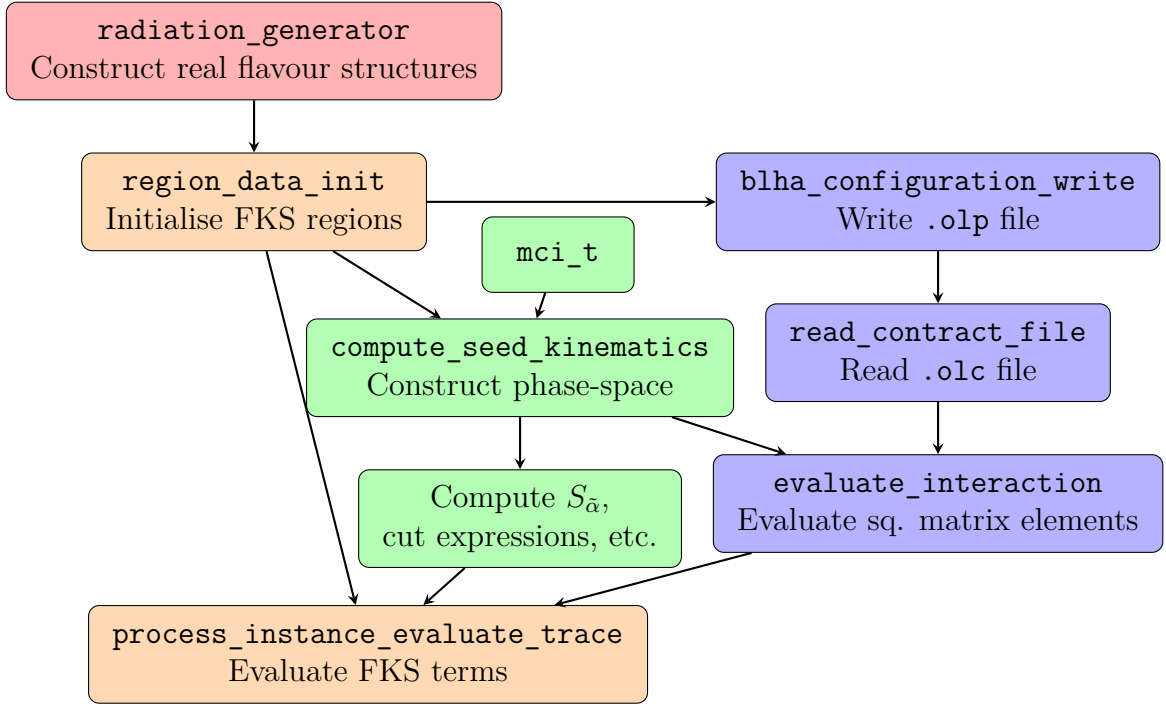

An NLO calculation in \texttt{WHIZARD} in general proceeds in the following way, sketched illustratively in Fig.~\ref{workflowNLO}.

After all process flavour structures at the Born tree-level \texttt{flst\_born} are generated for a user-defined process the module \texttt{radiation\_generator} automatically constructs all real flavour structures \texttt{flst\_real} based on possible SM vertices. This proceeds by inserting partons firstly in an arbitrary way in place of the proton and jets\footnote{The process specified by the user possibly contains particle container objects, e.~g. for multiple partons in protons and jets, defined as \texttt{alias} in the \texttt{.sin} steering file which in general increases the number of subprocesses.}, `\texttt{pr}' and `\texttt{j}', if included in the process definition, and as an additional radiated parton attached to the process. The allowed partons to be inserted thereby are represented by elements of the set $\{\texttt{pr},\texttt{j},\texttt{x}\}$, for which `\texttt{pr}' and `\texttt{j}' are otherwise excluded if not used in the process definition. Depending on the correction type of the NLO calculation, `\texttt{x}' is defined by either the gluon `\texttt{g}' in the case of QCD corrections or by the photon `\texttt{A}' in the case of EW corrections. All combinations of particle sequences thus obtained are then sorted such that only those corresponding to valid SM vertices are kept.

In the next step the FKS singular regions are initialised by applying the routine \texttt{region\_data\_init}. Each of the real flavour structures \texttt{flst\_real} is first analysed for FKS pairs possibly inducing a collinear and/or soft singularity. The singular regions are constructed for each FKS pair of \texttt{flst\_real} which are found as follows: For all pairs of legs which represent partons the emitter parton is traced back by either QCD or QED splittings such that a `test' underlying Born structure is constructed.
The pair of legs is then recognised as FKS pair if the test Born structure is contained in the list of pre-constructed Born flavour structures \texttt{flst\_born}.
The index of each thus obtained singular region is linked to the position of the emitter particle \texttt{em} in \texttt{flst\_real}, the \texttt{ftuple} representing the positions of the partons of the FKS pair and the real flavour structure \texttt{flst\_real} itself.
Subsequently, the preliminary singular regions are sorted according to all prescriptions necessary to define an FKS region as described in Sec.~\ref{secSingularregions}.
Eventually, FKS regions are initialised with index \texttt{alr} to which additionally all \texttt{ftuples} with distinct emitter but same \texttt{flst\_real} are assigned in order to compute the complete set of factors $d_{ij}$ for $S_{(i,j)}$ according to Eq.~(\ref{SalphaDef}).
Furthermore, for the later construction of subtraction terms each \texttt{alr} is associated with an underlying Born flavour structure \texttt{flst\_born} and a dynamic correction type \texttt{corr}, which will be discussed in more detail in Sec.~\ref{secMixedCoupling}. A complete set of FKS regions assembled in the so-called `FKS table' for an exemplary process is presented in App.~\ref{secFKStable}.

Describing the procedure of NLO calculations for processes with multiple subprocesses it will be referred to the tool \texttt{OpenLoops} in the following, for which the interface is the most extensively tested one in \texttt{WHIZARD}. Arrays consisting of all \texttt{flst\_real} and \texttt{flst\_born} which are part of the initialised FKS regions are the input of the routine \texttt{blha\_configuration\_write}. This routine controls all input given to OLPs using the BLHA interface by writing out \texttt{.olp} files with prefixes containing the name for a separate component of the NLO computation in \texttt{WHIZARD}.
By this, the squared matrix elements of the Born contribution are requested for in the \texttt{BORN} file. For the real part the file  \texttt{REAL} contains requests for the real squared matrix elements, and for the subtraction terms \texttt{SUB} involves those corresponding to Born, colour- and spin-correlated squared matrix elements, respectively. \texttt{LOOP} requests for all loop-tree interfering, colour- and spin-correlated as well as Born squared matrix elements for the virtual finite and integrated subtraction terms. In case PDFs are used, \texttt{DGLAP} asks for Born squared matrix elements entering the DGLAP remnant component.
An exemplary \texttt{SUB.olp} file is given in App.~\ref{secBLHAorder}. After these files are written out by \texttt{WHIZARD}, \texttt{OpenLoops} is called, registers all subprocess requests and responds with \texttt{.olc} answer files which contains a numbering of all squared matrix elements. This numbering is read in by \texttt{read\_contract\_file} and in a further step linked to \texttt{interaction\_t} objects (the branches of the trie) of the state matrix described in Sec.~\ref{secBookkeepingME}.

Outgoing from the found emitter particles by initialising the FKS regions the radiation phase-space will be constructed. From the multi-channel integration objects \texttt{mci\_t} all ingredients for the construction of the phase-space such as random numbers are provided. In these random number variables the Born phase space $\bar{\Phi}_n$ as well as the FKS variables, $\xi$ and $y$, are parametrised as \texttt{compute\_seed\_kinematics} is called. The NLO phase-space is thereby constructed for each index \texttt{i\_phs} associated with phase-space configurations distinguishing the Born-like points from those of the real emission and real phase-space configurations for distinct emitter positions \texttt{em}\footnote{If radiation is possible from both, \texttt{em=1} and \texttt{em=2}, the emitter position for an FKS region is replaced by \texttt{em=0}. This is necessary for the phase-space construction in order to treat the initial state emitter symmetrically according to Sec.~\ref{secMasslessISem} and \ref{secMassiveIS}}.

The evaluation of matrix elements proceeds in the routine \texttt{evaluate\_interaction} with the input on the phase-space points per \texttt{i\_phs} and indices of the list of squared matrix elements requests from the \texttt{.olc} files linked to the \texttt{interaction\_t} objects of the state matrix. After the \texttt{OpenLoops} routines are called for the computation of squared amplitudes, their numerical values are stored as \texttt{ME} array entries each linked to the \texttt{interaction\_t} objects. Note, that for the case that \texttt{RECOLA} is linked to \texttt{WHIZARD} the complete part of the code evaluation indicated in blue in Fig.~\ref{workflowNLO} is done in one step.

From the phase-space configurations all kinematical FKS factors entering the NLO calculation such as $S_{\tilde{\alpha}}$ as well as phase-space cut expressions are evaluated. These in addition to the previously stored squared matrix elements values are then used to evaluate the FKS terms by means of the constructed FKS singular regions. Precisely, in \texttt{process\_instance\_evaluate\_trace} all integrands represented by contributions of the Born in Eq.~(\ref{Borncomponent}), the real-subtracted in Eq.~(\ref{realsubtractedcontr}), the virtual plus integrated subtraction terms in Eq.~(\ref{finitecontrnphasespace}) and the DGLAP remnant in Eq.~(\ref{degeneraten+1contr}) are computed.
\subsection{Basic EW extensions}
\label{secEWsupplements}
An NLO EW calculation, either for integrated cross sections or event simulation, with respect to Eq.~(\ref{nloEWcalc}) can be steered with \texttt{WHIZARD} by running a SINDARIN file with commands of the usual NLO settings and additionally
\begin{align}
	\texttt{\$nlo\_correction\_type = "EW"}
	\label{SindarinEW}
\end{align}
For EW corrections all group theoretical factors entering the FKS terms have to correspond to those of QED. Due to a switch to \texttt{"EW"} as correction type these are called in place of those corresponding to QCD, which is the default for NLO calculations in \texttt{WHIZARD}. This concerns the Altarelli-Parisi splitting kernels from Eqs.~(\ref{ftofgam}) to (\ref{gamtoff}) and (\ref{ftofgamFSR}) to (\ref{gamtoffFSR}) and App.~\ref{secAppendixcollinear}, Casimir operators $C^{\text{QED}}$ and definitions $\gamma^{\text{QED}}$ and $\gamma^{\prime \text{QED}}$ defined in Table~\ref{CasimirGammaGammaP} for the collinear terms and charge correlations according to Eq.~(\ref{QEDchargecorrelations}) for the soft terms, respectively. In addition to these numerical factors the prescriptions of photons entering SM vertices have to be regarded throughout the construction of subtraction terms concerning the NLO EW automated corrections.

A technical element of the implementation considering photon-induced processes is treating the photon as either off-shell or on-shell particle in the process record. This is important with respect to renormalisation factors related to the photon at NLO EW as explained in Sec.~\ref{secExternalphotons}. The off-shell photon's technical prescriptions of the OLP \texttt{OpenLoops} which are essential for example for LHC physics are thus encoded in \texttt{WHIZARD} for automated NLO EW corrections.

For processes containing light charged fermions in the final state infrared-safety which is dictated by the KLN theorem defined in Sec.~\ref{secRegularisationIR} demands phase-space cuts to be imposed on these fermions only if they are `dressed'. This means that photons appearing at NLO EW for this class of processes are not resolved but recombined with the charged fermions if they fulfil criteria such as
\begin{align}
\label{photonreco}
	\Delta R_{f^{\pm}\gamma}\leq R_0
\end{align}
where
\begin{align}
\label{deltaR}
	\Delta R_{ij}=\sqrt{(\Delta\phi_{ij})^2+(\Delta\eta_{ij})^2}
\end{align}
and $R_0$ is a user-defined real number\footnote{$R_0$ can be set in SINDARIN by the parameter \texttt{photon\_rec\_r0} with $0.1$ as default value.}. $\Delta\phi_{ij}$ thereby denotes the difference of azimuthal angles and $\Delta\eta_{ij}$ of rapidities of particles $i$ and $j$. In case Eq.~(\ref{photonreco}) is fulfilled the sum of the momenta of the photon $\gamma$ and of the fermion $f^{\pm}$ is identified with the momentum of the dressed fermion. In any other case the non-recombined fermion is identical to the dressed one.
\section{Mixed coupling expansions}
\label{secMixedCoupling}
For the computation of NLO contributions $\delta\text{NLO}_{q_s,q_e}$ of generic coupling powers $q_s$ and $q_e$ the following technical procedures within the code have to be undertaken.

Starting from the LO coupling powers $p_s$ and $p_e$ specified in \texttt{SINDARIN} as
\begin{align}
\label{couplingsSindarin}
\begin{split}
	&\texttt{alphas\_power = <p\_s>}\\
	&\texttt{alpha\_power = <p\_e>}
\end{split}
\end{align}
and the command of Eq.~(\ref{SindarinEW}) the calculation of cross sections and fixed-order event generation at NLO in $\alpha$ can be steered. Precisely, this means a phase-space integration over the integrand given formally by the expression of Eq.~(\ref{nloEWcalc}). In this way, all contributions $\delta \text{NLO}_{p_s,p_e+1}$ can be computed. For the leading QCD coupling power contribution $\delta \text{NLO}_{n+1,m}$ (the left-most blob in the lower row in Fig.~\ref{ziehharmonika}) the NLO QCD settings such as the correction type \texttt{"QCD"} (to be replaced with \texttt{"EW"} in Eq.~(\ref{SindarinEW})) as well as the coupling powers $p_s=n$ and $p_e=m$ for the LO contribution leading in $\alpha_s$ have to be used.

A systematic approach for the inclusion of subtraction terms to any correction type of the corresponding splittings contributing to $\delta \text{NLO}_{p_s,p_e+1}$ is realised in \texttt{WHIZARD} by the following steps.

First of all, we note that the Born flavour structures \texttt{flst\_born} and those emerging from the \texttt{radiation\_generator} for the real \texttt{flst\_real} are completely independent from the coupling power. They purely depend on the process definition and the chosen model. For a fixed coupling order at NLO in $\alpha$ the raw real flavour structures \texttt{flst\_real} for this reason have to be filtered: All real flavour structures have to be rejected which do not exist for the chosen coupling order for which the NLO contributions are defined, i.~e. $q_s=p_s$ and $q_e=p_e+1$.
This can be done via a coupling power counting algorithm (CPCA) presented in Sec.~\ref{CPCA} with which coupling powers can be declared as `forbidden' having a definite flavour structure at hand.
Thus all real flavour structures are eliminated from the set of \texttt{flst\_real} objects if the fixed NLO coupling powers $\{p_s,p_e+1\}$ corresponds to one coupling power combination which is forbidden from the flavour structure. Likewise, all Born flavour structures are vetoed if neither the coupling powers $\{p_s,p_e\}$ nor $\{p_s-1,p_e+1\}$ are allowed from the particle content of the corresponding objects \texttt{flst\_born} with the CPCA.

The resulting reduced set of \texttt{flst\_born} and \texttt{flst\_real} are used in a second step for the initialisation of FKS singular regions as described in Sec.~\ref{secworkflowNLO}.
A crucial ingredient for NLO contributions in mixed coupling expansions is the introduction of a dynamical correction type which varies between QCD and EW depending on the FKS region. A singular FKS region $\tilde{\alpha}$ is well-defined in the correction type due to a definite emitter $E_{\tilde{\alpha}}$ and FKS pair $(i,j)$ determining the splitting in Eq.~(\ref{splittingdef}), and thus the SM vertex associated with it.
For the construction of singular regions in \texttt{region\_data\_init} as explained in Sec.~\ref{secworkflowNLO} this induces a new degree of freedom, i.~e. considering possible splittings for both correction types. Hence, FKS regions are created from classifying external legs $(i,j)$ of real flavour structures into the sets $G^{\text{QCD}}$ and $G^{\text{QED}}$ defined in Eqs.~(\ref{GroupG1}) to (\ref{GroupG3}) and constructing splittings from the associated correction types.
In the special case of FKS pairs $(q,\bar{q})$ (if $i$ and $j$ are final-states), $(q,{q})$ and $(\bar{q},\bar{q})$ (if $i$ is an initial-state), one is tempted to assume that this must imply two FKS regions associated with the two correction types.
However this is only the case if the underlying Born structure, i.~e. tracing this FKS pair back to a photon or a gluon, is allowed for coupling powers $\{p_s,p_e\}$ or $\{p_s-1,p_e+1\}$ which can be found out via the CPCA. An example here would be the process $pp\to t\bar{t}j$: The real-emission process $q\bar{q}\to t\bar{t} q'\bar{q}'$ at $\mathcal{O}(\alpha_s^2\alpha^2)$ with the FKS pair $(q',\bar{q}')$ either is due to a $\gamma\to q'\bar{q}'$ splitting with the Born process $q\bar{q}\to t\bar{t} \gamma$ at $\mathcal{O}(\alpha_s^2\alpha)$ or a $g\to q'\bar{q}'$ splitting with the Born process $q\bar{q}\to t\bar{t} g$ at $\mathcal{O}(\alpha_s\alpha^2)$.

Simultaneously, assigning a correction type to an FKS region the underlying Born flavour structure is attributed to it. The dynamical treatment of the correction type is visualised by assignments \texttt{"qcd"} and \texttt{"ew"} to each region \texttt{alr} in the FKS table shown for the exemplary process $pp\to t\bar{t}$ with $p=\{b,\bar{b},g,\gamma\}$ in App.~\ref{secFKStable}.
If no underlying Born structure from the reduced set of \texttt{flst\_born} can be found to be attributed to either QCD or QED splittings of legs of a real flavour structure \texttt{flst\_real}, a pseudo-FKS region is created. This means that for this real process no singularity from collinear or soft splittings exist, but the tree-level $n+1$ process for this flavour structure is still contributing at NLO which is explained in detail in Sec.~\ref{secnonsing}.
Concerning the technical bookkeeping, for the dynamical correction type the string \texttt{"none"} is assigned to these pseudo-FKS regions. Due to this labeling the subtraction terms are omitted for these regions.

According to the labels \texttt{"qcd"} or \texttt{"ew"} assigned to the other FKS regions, the corresponding group factors entering the real subtraction terms are chosen. This labelling is reused for the subtraction terms in integrated form due to the linking of the underlying Born flavour structures to the dynamical correction types in each FKS region.
If a Born flavour structure is linked to both correction types, e.~g. for the process $b\bar{b}\to t\bar{t}$ once at $\mathcal{O}(\alpha_s^2)$ with correction type \texttt{"ew"} and once at $\mathcal{O}(\alpha_s \alpha)$ with correction type \texttt{"qcd"}, the integrated subtraction terms have to be evaluated for both of these cases. Note that for this special case the opposite applies for the virtual finite loop terms entering NLO contributions at coupling powers $\{p_s,p_e+1\}$, for which QCD and EW corrections overlap in the loop-tree interfering amplitudes. Due to this, virtual loop contributions from QCD and EW corrections can not be evaluated separately but only by a single OLP request for loop-tree interfering matrix elements. This request per default contains the Born flavour structure, the coupling powers $\{p_s,p_e\}$ and the correction type \texttt{EW} as OLP input parameters.

In the context of steering \texttt{OpenLoops} via the BLHA interface the input of coupling powers is required not only for the loop-tree interfering matrix elements but for squared matrix elements for each element of the NLO calculation. Subtraction terms for NLO contributions of order $\{p_s,p_e+1\}$ can be constructed with respect to either QCD or QED splittings and thus need squared factorised Born tree amplitudes at the corresponding coupling order, i.~e. either at $\{p_s,p_e\}$ or $\{p_s-1,p_e+1\}$. Applying the CPCA to the set of \texttt{flst\_born} the allowed coupling powers from the particle content of the flavour structures are checked for compatibility with either $\{p_s,p_e\}$ or $\{p_s-1,p_e+1\}$. Note that for processes containing two or more $q\bar{q}$ pairs the coupling powers are not uniquely defined due to interfering diagrams with two possibilities for intermediate states, gluons and EW bosons.
Tracking these Born-level process structures and requesting the two different tree-level squared amplitudes is treated as a special case in \texttt{WHIZARD} throughout all parts of the NLO calculation.\footnote{The code implementation for the extraction of squared tree-level matrix elements of coupling powers $\{p_s-1,p_e+1\}$ for this special case is based on sums over off-diagonal entries of the colour-correlated squared matrix elements for this coupling order.}

For Born-level processes at the subleading QCD coupling powers $\{p_s-1,p_e+1\}$, colour correlated squared matrix elements must be requested from \texttt{OpenLoops} which enter the subtraction terms corresponding to QCD soft splittings. These need the correction type input `\texttt{QCD}'. Due to this, for each \texttt{OpenLoops} BLHA request associated with Born flavour structures -- except for those entering the \texttt{BORN.olp} file -- the correction type, either \texttt{QCD} or \texttt{EW}, as well as the coupling powers have to be set.
An example for squared amplitudes requests in the \texttt{SUB.olp} file for $pp\to t\bar{t}$ with $p=\{b,\bar{b},g,\gamma\}$ is shown in App.~\ref{secBLHAorder}. The \texttt{Born.olp} file contains only tree-level squared matrix elements requests for Born processes at the coupling powers $\{p_s,p_e\}$ which are selected by means of the CPCA. The squared matrix elements accessed by this file enter the $\text{LO}_{p_s,p_e}$ contribution of the complete NLO calculation.
\subsection{Coupling power counting algorithm}
\label{CPCA}
An algorithm determining coupling powers from the particle content of the flavour structures is useful as it flexibly provides the information of contributing subprocesses throughout an NLO calculation with mixed QCD-EW corrections. The algorithm presented in this section provides a mapping from flavour structures to potentially allowed coupling powers.

First of all, we note that for any $2\to n$ tree-level process the total number of coupling powers of either $\alpha_s$ or $\alpha$ is given by
\begin{align}
	n_{\text{tot}}\equiv p_s+p_e=n_{\text{legs}}-2=n
\end{align}
with $n_{\text{legs}}$ denoting the number of external legs. From this we can derive the complete set of possibilities for powers $l_s$ of the QCD coupling $\alpha_s$ and $l_e$ of the electromagnetic coupling $\alpha$, i.~e.
\begin{align}
	&\{l_s,l_e\}=\{n-k,k\}, & 0\leq k\leq n\quad.
\end{align}
In order to constrain this range of possibilities for arbitrary processes we can make the following naive observations.

Gluons as external states always couple to processes by the coupling $\alpha_s$ to the power of one. Counting the number of gluon external states $n_{g}$ gives a first constraint on the upper boundary of the range of $k$:
\begin{align}
	0\leq k \leq n -n_{g}\quad.
\end{align}
Similarly, external legs associated with EW bosons $\gamma$, $W$, $Z$ and $H$ couple to the process each by $\alpha$ of order one which further leads to
\begin{align}
	n_{W}\leq k \leq n -n_{g}\quad.
\end{align}
with $n_{W}$ the number of external EW bosons. Furthermore, leptons (including neutrinos) always couple as pairs with $\alpha$ to the power of one. The number of external leptons $n_{l}$ thus gives another constraint yielding
\begin{align}
	n_{W}+\frac{n_l}{2}\leq k \leq n -n_{g}\quad.
	\label{generallyallowedcouplings}
\end{align}
As quarks are particles of both theories, QCD and EW, these are the only SM particles which as external states may couple to a process in pairs by both $g_s$ and $e$ on the amplitude-level. At the level of interfering amplitudes this would lead to ill-defined coupling powers of $\alpha_s$ and $\alpha$. 
As the coupling $g_s$ or $e$ is either due to a gluon or an EW boson identifiable intermediate and external states beyond the quarks further constrain the range of $k$. This concerns the following considerations:
\begin{itemize}
	\item For the case of exactly one $q\bar{q}$ pair and only gluons as external states, i.~e. $n_q+n_{g}=n_{\text{legs}}$, we find
	\begin{align}
		k=0\quad.
	\end{align}
	\item For the case of exactly one $q\bar{q}$ pair and only EW bosons or leptons as external states, i.~e. $n_q+n_{W}+n_{l}=n_{\text{legs}}$, we find
	\begin{align}
		k=n\quad.
	\end{align}
	\item If the quarks as external states are all of different flavours which implies pure EW couplings to $W^{\pm}$ of the quarks, we find
	\begin{align}
		k=n -n_{g}=n_{W}+\frac{n_{l}}{2}+\frac{n_{q}}{2}\quad.
		\label{interferenceCP}
	\end{align}
\end{itemize}
Note that if none of the conditions of these bullet points are fulfilled the ambiguity in the coupling powers merely is due to two or more $q\bar{q}$ pairs as external particle states of the process. Thus, for the allowed coupling powers due to Eq.~(\ref{generallyallowedcouplings}), the flavour structure is still uniquely defined.

In the implemented algorithm a chosen input coupling power is checked on equality to any of those cases represented by the allowed $k$ according to the rules above for an arbitrary input flavour structure. In this way, for example a flavour structure can be rejected from further evaluations in the code if the corresponding tree-level process does not exist for a certain coupling order. Selecting flavour structures in this way for each part of the NLO calculation in addition to the user-defined coupling powers $\{p_s,p_e\}$ for the LO process yields a well-defined input for the OLP and the evaluation of FKS terms. This also includes subprocesses with two or more $q\bar{q}$ pairs as the ambiguity in the coupling powers purely influences the actual numerical value of the squared or interfering matrix elements. By means of the user-defined coupling powers $\{p_s,p_e\}$ and the associated ones with respect to the different terms and factors entering the NLO calculation the evaluation of OLP requests for these flavour structures is eventually well-defined.
\subsection{IR-safe observables}
\label{secIRsafeobservables}
The criterion for observables being infrared-safe is enforced by the KLN theorem. It demands that adding a collinear or soft particle to a process does not change the value of the observable of that process. In practice, if phase-space cuts are imposed on partonic events a jet clustering algorithm which classifies partons into jets guarantees IR-safety of observables.

In \texttt{WHIZARD}, jet clustering algorithms such as the $k_T$ and anti-$k_T$ algorithm \cite{Catani:1993hr,Ellis:1993tq,Cacciari:2008gp} are provided by an interface to \texttt{FastJet} \cite{Cacciari:2011ma}. These are inevitable concerning QCD corrections to processes with final-state jets. In order to guarantee IR-safety for QED corrections photon recombination algorithms as explained in Sec.~\ref{secEWsupplements} have to be applied which act on all charged fermions.

For NLO QCD-EW mixed corrections the jet definition of processes has to be extended including particles beyond QCD partons. There are different prescriptions for the jet definition to be applied for NLO contributions $\delta \text{NLO}_{q_s,q_e}$ represented by the lower row of Fig.~\ref{ziehharmonika}. In order to ensure IR-safe observables for QCD corrections to the process at leading $\alpha_s$ order, i.~e the contribution $\delta \text{NLO}_{n+1,m}$, the definition of jets must include all coloured (light) partons, quarks and gluons.

For NLO EW corrections to the process at leading $\alpha_s$ order, i.~e. inducing the contribution $\delta \text{NLO}_{n,m+1}$, photons must be recombined with quarks to fulfil the IR-safety criterion. This can be achieved by so-called `democratic' jets \cite{Frederix:2016ost} which means to extend the particle set forming a jet object to account for photons in the same way as for gluons. Crucially, for real emissions of photons off quarks the jet clustering algorithm acts on both of these partons and according to a jet resolution criterion recombines them. 
Furthermore, having both a gluon and a photon as final-state particles in the real emission event each of them possibly induces IR singularities. If the photon is included democratically in the jet definition, IR divergences of real squared matrix elements from emissions of soft gluons are rendered finite as the counter-event contains a Born configuration with the photon consistently treated as the jet fulfilling phase-space cut criteria.

For contributions $\delta\text{NLO}_{n-i,m+1+i}$ with $i>0$ the jet definition must be extended additionally to account for leptons. This is due to the fact that photons already occur at the LO-level for processes with jets at coupling powers $\{n-1,m+1\}$. For higher orders in $\alpha$ photons can split not only into $q\bar{q}$ pairs which yield hadronic jets but also into lepton-antilepton pairs, $l^{\pm}l^{\mp}$. Observables sensitive to the jet-energy for this reason would not be IR-safe if the leptons are not taken into account in the jet definition as light lepton pairs emerging from a splitting photon can induce collinear singularities.

As protons are similar hadronic particle containers as jets analogous particle set extensions can be made for them provided that for the used PDF set the additional parton is included in the fit. However, there are no IR-safety criteria which have to be imposed on protons representing initial-state particles. For that reason the proton particle set extension to account in addition for photons (and leptons) merely changes the numerical result due to the additional contributions. 

QED PDFs in general are suppressed compared to quark and gluon PDFs as protons constitute a hadronic system with predominant QCD interactions. In fact, approximately $0.5\%$ of the proton's momentum is carried by the photon \cite{Bertone:2017bme}. The actual contributions of photon-induced processes however highly depend on the process and the observable one is looking at. There are certain exceptions for which these can get very large such as contributions of about $5$-$10\%$ for high invariant masses $M_{ll}$ in Drell-Yan mass distributions \cite{Bertone:2017bme}. Crucially, the contribution of the $q\gamma$ PDF channel can grow up to $40\%$ for $W^+W^+W^-x$ production in $W^-$ transverse momentum distributions \cite{Dittmaier:2017bnh}. In particular single photon-induced processes are represented by PDF channels opened up in a mixed coupling expansion of LO and NLO contributions such as $g\gamma\to t\bar{t}$ in top pair production. This in general implies that photon-induced processes for this purpose should not be neglected.

There are PDF sets including lepton PDFs \cite{Bertone:2015lqa,Bauer:2017isx,Buonocore:2020nai} which in principle yield non-zero contributions. However, they are mostly numerically irrelevant for applications related to LHC processes as for example $l\gamma$ luminosities with respect to $\gamma\gamma$ ones are suppressed by a factor of about $200$-$300$ \cite{Buonocore:2020nai}.

Recommendations how to set \texttt{alias} definitions \texttt{pr} and \texttt{jet} in \texttt{SINDARIN} denoting the particle content of protons and jets for each NLO contribution indicated by blobs of the lower row in Fig.~\ref{ziehharmonika} are shown in Table~\ref{jetdefinitions}. The coupling order of an NLO contribution thereby is set by the second to fourth columns with the input \texttt{<p\_s>} and \texttt{<p\_e>} given through the coupling powers related to Eq.~(\ref{couplingsSindarin}). The suggestions for the proton and jet definitions in the last two columns follow the considerations above on IR-safety conditions and the contributions of photon (and lepton) PDFs. The following specifications for the SINDARIN files are used for the expressions \texttt{q} and \texttt{l} in Table~\ref{jetdefinitions}:

\begin{align}
	\begin{split}
	&\texttt{alias q = u:d:s:c:b:U:D:S:C:B}\\
	&\texttt{alias l = e1:e2:e3:E1:E2:E3}
	\end{split}
	\label{aliasdefql}
\end{align}
The definition of \texttt{pr/j} for $pp \to X$ processes, where $X$ consists of leptons and/or massive particles, is different compared to that for $pp \to X+\text{jets}$ only with respect to EW corrections to subleading QCD processes (last two rows in Table \ref{jetdefinitions}). This is mandated by the IR-safety condition for photons existing at LO-level and splitting into lepton pairs in real emission diagrams as explained above. Furthermore, the definition of the proton \texttt{pr}, which for consistency reasons have to be set equal to \texttt{j}, for some cases require PDF sets to include photons (or even leptons) where necessary from a phenomenological point of view. Technically, by adding photons and/or leptons to the particle sets \texttt{WHIZARD} sets up corresponding PDFs with zero values if the PDF set itself does not contain PDFs for these added particles.

	\begin{table}
	\centering
	{	\onehalfspacing
		\begin{tabularx}{1.\linewidth}{l|c|c|c|l|l}
		$\text{LO}_{p_s,p_e}+\delta \text{NLO}_{q_s,q_e}$& \texttt{\small p\_s}& \texttt{\small p\_e}&\texttt{\small \$nlo\_} & \multicolumn{2}{l}{\texttt{alias pr/j}} \\
		(with $0<i<n$) &&&\texttt{\small correction\_type} & $pp\to X$& $pp\to X~\text{jets}$\\
		\hline
		$\text{LO}_{n,m}+\delta\text{NLO}_{n+1,m}$&\texttt{n} &\texttt{m}& \texttt{"QCD"}& \texttt{q:g}&\texttt{q:g}\\
		$\text{LO}_{n,m}+\delta\text{NLO}_{n,m+1}$&\texttt{n} &\texttt{m}& \texttt{"EW"}& \texttt{q:g:A}&\texttt{q:g:A}\\
		$\text{LO}_{n-i,m+i}+\delta\text{NLO}_{n-i,m+1+i}$&\texttt{n-i} &\texttt{m+i}& \texttt{"EW"}& \texttt{q:g:A}&\texttt{q:l:g:A}\\
		$\text{LO}_{0,m+n}+\delta\text{NLO}_{0,m+n+1}$&\texttt{0} &\texttt{m+n}& \texttt{"EW"}& \texttt{q:A}&\texttt{q:l:A}\\
	\end{tabularx}}
	\caption{Setting recommendations steering NLO computations in \texttt{WHIZARD} with NLO contributions $\delta \text{NLO}_{q_s,q_e}$ corresponding to the blobs of Fig.~\ref{ziehharmonika} in addition to contributions $\text{LO}_{p_s,p_e}$, where \texttt{n} denotes the $\alpha_s$ power and \texttt{m} the $\alpha$ power with respect to the leading QCD order contribution (upper left-most blob in Fig.~\ref{ziehharmonika}). While the \texttt{alias} definitions \texttt{q} and \texttt{l} for quarks and leptons are given in Eq.~(\ref{aliasdefql}), \texttt{g} denotes the gluon and \texttt{A} the photon, respectively.}
	\label{jetdefinitions}
\end{table}

If the final state of processes consists of jets as well as charged leptons for NLO EW calculations both measures to ensure IR-safe observables, jet clustering and dressing of leptons, must be undertaken.
This yields a non-trivial phase-space cut evaluation since photons have to be recombined with all charged fermions. On the one hand, photons render the charged leptons dressed. However, as they are part of the jet definition, they enter the jet clustering algorithm simultaneously to be recombined with QCD partons. This implies the condition to treat fermion dressing and jet clustering in a systematic way such that photons are always recombined with the closest charged fermion. This can be achieved in two ways \cite{Frederix:2018nkq}. One possibility is to recombine the photon with all charged fermions, i.~e. light quarks and leptons, in the first step since the photon-recombination cone is typically defined more narrow as the jet clustering cone. If the photon fulfils the recombination criterion, i.~e. it can be found within the cone with radius $R_0$ according to Eq.~(\ref{photonreco}), the photon is discarded for the subsequent step of clustering. For the jet clustering eventually only gluons, dressed quarks and non-recombined photons are taken into account. Alternatively, a jet-lepton separation criterion can be applied.

	\part{Validation and Results}

	\chapter{Hadron collisions in NLO mixed coupling expansions}
	\label{secHadronCollisions}
	In this chapter the validation as well as illustrative results of NLO calculations in mixed coupling expansions for several benchmark processes at the LHC are shown. Total cross sections and differential distributions are obtained with the automated MC framework of \texttt{WHIZARD} which has been extended to account for EW corrections. The methods employed to this end are explained in detail in the previous chapters: the phase-space construction for the case of massless initial-state radiation presented in Sec.~\ref{secMasslessISem}, the FKS subtraction scheme in mixed coupling expansions described in Sec.~\ref{mixedcouplingsSec} and the technical background of an NLO computation performed with \texttt{WHIZARD} given in Sec.~\ref{secWhizardFrameworkNLO}.
	
	Input parameters and further details on the setup used for the calculations of the results are given in Sec.~\ref{secSetupLHC}. In Sec.~\ref{secPureEWLHC} NLO EW cross sections and differential distributions for pure electroweak LHC processes are presented. NLO mixed corrections are computed for processes for which the leading QCD coupling contributions are of $\mathcal{O}(\alpha_s)$ or higher, with results shown in Sec.~\ref{secMixedCorrLHC}. Dedicated cross-checks with other MC tools are provided for both of these process classes.
	\section{Setup of the calculation}
	\label{secSetupLHC}
	For the results of NLO calculations related to $pp$ processes in this work, the following settings and input parameters are used.
	
	Because $pp$-induced processes come along with large sets of subprocesses \texttt{OpenLoops} is used for the matrix element computations. As explained in Sec.~\ref{secOneloopProv} this OLP represents the most efficient tool concerning the interplay with \texttt{WHIZARD} for this process class. The renormalisation of SM parameters due to UV divergences is performed by applying the complex-mass scheme (CMS) which is the default\footnote{For any other case the flag \texttt{?openloops\_use\_cms = false} can be set in \texttt{SINDARIN}.} in \texttt{OpenLoops} and explained in detail in Sec.~\ref{secOnshellCMS}.
	
	Furthermore, for the calculations presented here the $G_{\mu}$ input scheme is used for the computation of the electromagnetic coupling $\alpha$ for which the running of the coupling up to the EW scale is implicitly contained. This choice in \texttt{SINDARIN} corresponds to the setting
	\begin{align*}
		\texttt{\$blha\_ew\_scheme = "GF"}\quad.
	\end{align*}
	The coupling $\alpha$ is defined by the input for the $G_{\mu}$ constant
	\begin{equation}
	G_{\mu}= 1.16639\cdot 10^{-5} \; \text{GeV}^{-2}
	\end{equation}
	and the pole masses and widths of $W$ and $Z$ bosons according to Eq.~(\ref{alphaGmu}). These are set through the transformation formulas
	\begin{align}
		&M_V={M_V^{\text{BW}}}/{\sqrt{1+\left(\Gamma_V^{\text{BW}}/M_V^{\text{BW}}\right)^2}} &\Gamma_V={\Gamma_V^{\text{BW}}}/{\sqrt{1+\left(\Gamma_V^{\text{BW}}/M_V^{\text{BW}}\right)^2}}
	\end{align}
	using the Breit-Wigner masses and widths
	\begin{eqnarray}
	\begin{split}
		M_W^{\text{BW}} &=\; \phantom{1}80.385\phantom{0} \;\text{GeV} \quad\qquad
		\Gamma_W^{\text{BW}} &=2.0897\;    \;\text{GeV} \quad~ \\
		M_Z^{\text{BW}} &=\; \phantom{1}91.1876 \;\text{GeV} \quad\qquad
		\Gamma_Z^{\text{BW}} &=2.4955\;    \;\text{GeV} \quad.
	\end{split}
	\label{WZmasseswidths}
	\end{eqnarray}
	 For reasons of comparisons with reference MC tools, these are chosen according to those of Ref.~\cite{Frederix:2018nkq}. The masses and widths in this study correspond to those of PDG \cite{ParticleDataGroup:2016lqr} using a Breit-Wigner (OS-like) lineshape approach.
	Other heavy particle masses and widths are set by the numerical values
	\begin{eqnarray}
	\begin{split}
		M_H &=\; 125.0\phantom{000} \;\text{GeV} \quad\qquad
		\Gamma_H &=0\;     \;\\
		m_t &=\; 173.34\phantom{00} \;\text{GeV} \quad\qquad
		\Gamma_t &=1.36918\;\text{GeV}    \;
	\end{split}
	\label{Htmasseswidths}
	\end{eqnarray}
	where light quarks and leptons are considered as massless, i.~e.
	\begin{eqnarray}
		m_{q\neq t}  &=\;0 \quad\qquad m_l&=\;0 \quad.
	\end{eqnarray}
	Note that for processes with on-shell bosons or top quarks corresponding particle widths of Eqs.~(\ref{WZmasseswidths}) and (\ref{Htmasseswidths}) are set to zero. The explicit settings for these cases are indicated in the discussion part below, where necessary.

	The centre-of-mass energy for the $pp$ collisions is chosen according to the collider energy of LHC Run II,
	\begin{align}
		\sqrt{s} = 13 \text{ TeV}\quad.
	\end{align}

	For the PDFs the \texttt{LUXqed\_plus\_PDF4LHC15\_nnlo\_100} PDF set \cite{Butterworth:2015oua,Manohar:2016nzj}, a fit including photon PDFs, from the library
	\texttt{LHAPDF} (version 6.2.3) \cite{Buckley:2014ana} is used. The running of the the QCD coupling constant $\alpha_s$ at three-loop accuracy is performed with \texttt{LHAPDF} with reference value
	\begin{align}
		\alpha_s(M_Z)=0.118
	\end{align}
	associated with the chosen PDF set.
	
	The renormalisation and factorisation scale corresponds to
	\begin{align}
		&\mu_R=\mu_F=\frac{H_T}{2} &H_T=\sum_i\sqrt{p_{T,i}^2+m_i^2}
	\end{align}
	where $H_T$ denotes the transverse mass with $p_{T,i}$ and $m_i$ the transverse momentum and masses of the final-state particles. By defining \texttt{scale} in this way $\mu_R$ and $\mu_F$ are set simultaneously in \texttt{WHIZARD}.
	
	For the photon recombination with massless fermions the parameter $R_0$ according to Eq.~(\ref{photonreco}) is set by
	\begin{align*}
		\texttt{photon\_rec\_r0 = 0.1}
	\end{align*}
	in the SINDARIN file. For processes with jets the anti-$k_T$ clustering algorithm \cite{Cacciari:2008gp} with jet radius $R=0.4$ is applied by specifying
	\begin{align*}
		\begin{split}
		&\texttt{jet\_algorithm = antikt\_algorithm}\\
			&\texttt{jet\_r = 0.4}
		\end{split}
	\end{align*}
	which is used as an input to \texttt{FastJet} \cite{Cacciari:2011ma}. Phase-space cut expressions acting on dressed fermions and clustered jets follow the conditions
	\begin{itemize}
		\item $p_{T,l^{\pm}}> 10$ GeV and $\lvert \eta_{l^{\pm}}\rvert<2.5$ on charged dressed leptons
		\item $\Delta R_{l^{+}l^{-}}>0.4$ and $M_{l^{+}l^{-}}> 30$ GeV on pairs of charged dressed leptons with same flavour and opposite charge
		\item $p_{T,j}> 30$ GeV and $\lvert \eta_{j}\rvert<4.5$ on clustered jets
	\end{itemize}
	with distance $\Delta R_{ij}$ given by Eq.~(\ref{deltaR}).
	
	The process-specific settings for an NLO computation, i.~e. the coupling powers defined by Eq.~(\ref{couplingsSindarin}), the correction type as well as \texttt{alias} definitions of proton and jet objects, are consistently applied as prescribed by Table \ref{jetdefinitions}. By this photon-induced processes are taken into account automatically in the NLO computations where relevant with results presented in the subsequent sections.
	An exemplary \texttt{.sin} file containing all settings for steering an NLO EW cross section computation for the process $pp\to e^+e^-j$ is attached in App. \ref{secSindarin}. All the NLO EW calculations are performed by using \texttt{WHIZARD}'s MPI parallelised setup. The computing time highly depends on the considered process. However, the estimated average time for cross section integrations is a couple of hours on around $\mathcal{O}(100)$ cores.
	
	\section{NLO EW corrections to pure electroweak processes at the LHC}
	\label{secPureEWLHC}
	In this section numerical results on NLO EW total cross sections and differential distributions to pure electroweak processes at the LHC are presented. These processes can be further classified into those involving on-shell and off-shell massive vector-bosons. Results and validation checks on the former are discussed in Sec.~\ref{secOnShellLHC} and those on the latter in Sec.~\ref{secOffshellLHC}, respectively.

	In order to quantify the relative correction of the NLO EW with respect to the LO cross section in the following we define the quantity
	\begin{align}
	\delta\equiv\frac{\sigma^{\text{tot}}_{\text{NLO}}-\sigma^{\text{tot}}_{\text{LO}}}{\sigma^{\text{tot}}_{\text{LO}}}
	\label{delta}
	\end{align}
	with $\sigma^{\text{tot}}_{\text{LO}}$ denoting the LO and $\sigma^{\text{tot}}_{\text{NLO}}$ the NLO total cross section, respectively.
	
	\subsection{On-shell boson production}
	\label{secOnShellLHC}
	In this section the production of EW massive bosons for $pp$ processes with two and three final states at NLO EW is studied. This involves neutral- and charged-current processes with $VV$, $VH$, $VVV$, $VVH$ and $VHH$ production for all possible combinations of $V=W^{\pm},Z$.
	For the calculation of cross sections and fixed-order differential distributions for this class of processes the QCD coupling power $p_s$ is set to zero, and the electromagnetic coupling power $p_e$ to the number of final states, respectively.
	Furthermore, the LHC setup of Sec.~\ref{secSetupLHC} is adjusted with the additional requirement to set the vector-boson widths of Eq.~(\ref{WZmasseswidths}) to zero.
	\subsubsection{Total cross sections}
	In order to validate the NLO EW calculation for these processes the total cross sections are compared with twofold reference results of the MC framework \texttt{MUNICH/MATRIX} \cite{Kallweit:2014xda,Buonocore:2021rxx,Bonciani:2021zzf} applying both schemes, CS and $q_T$ subtraction. In Table \ref{EWonshell} results of the cross-checks\footnote{For technical reasons of the comparison the top-quark width is set to zero for all processes except for $W^+W^-$, $W^+W^-Z$ and $W^+W^-H$ production, for which top resonances can occur in real-emission diagrams. The width for these cases is set to $\Gamma_t=1.44262$ GeV.} with reference values using CS subtraction are shown. The checks can be assessed by evaluating the parameters
		\begin{sidewaystable}
		\centering
		\begin{tabularx}{0.85\textwidth}{l|r|r|r|r|r|r}
			process   &$\texttt{MUNICH}_{\text{(CS)}}\texttt{+OpenLoops}$  &\multicolumn{3}{c|}{\texttt{WHIZARD+OpenLoops}}   & \textit{dev} [\%] & $\sigma^{\text{sig}}_{\text{NLO}}$\\
			$pp \rightarrow X$ &$\sigma_{\text{NLO}}^{\text{tot}}$  [fb] &$\sigma_{\text{LO}}^{\text{tot}}$  [fb] &$\sigma_{\text{NLO}}^{\text{tot}}$ [fb] &$\delta$ [\%]&&\\
			\hline\hline
			$ZZ$       &        $ 1.05729(1) \cdot 10^4 $      &  $1.10367(1)\cdot 10^4 $ & $ 1.0573(1) \cdot 10^4 $    & $ -4.20 $ & $ 0.0001 $ & $ 0.01 $\\
			$W^+Z$       &        $ 1.71505(2) \cdot 10^4$      & $ 1.71760(2)\cdot 10^4 $  & $ 1.71507(2) \cdot 10^4 $    & $-0.15 $ & $ 0.001 $ & $ 0.88 $\\
			$W^-Z$  &      $ 1.08576(1) \cdot 10^4 $   & $  1.084976(8)\cdot 10^4 $ & $ 1.08574(1) \cdot 10^4 $ & $ +0.07 $  & $ 0.001 $ & $ 0.90 $\\
			$W^+W^-$ &           $ 7.93106(7) \cdot 10^4 $    & $ 7.58598(4)\cdot 10^4 $    &   $ 7.9309(2)\cdot 10^4 $     & $ +4.55 $ & $ 0.002 $ & $ 0.89 $\\
			$ZH$   &       $ 6.18523(6) \cdot 10^2 $       &  $  6.53082(2)\cdot 10^2 $   & $ 6.18533(6) \cdot 10^2 $   &   $ -5.29 $ & $ 0.002 $ &$ 1.17 $  \\
			$W^+H$  &   $ 7.18070(7) \cdot 10^2 $         & $ 7.35035(4)\cdot 10^2 $  &  $ 7.18072(9) \cdot 10^2 $   &   $ -2.31 $ & $ 0.0003 $ &$ 0.18 $ \\ 
			$W^-H$     &      $ 4.59289(4) \cdot 10^2 $     & $  4.69379(2)\cdot 10^2 $  &    $ 4.59299(5) \cdot 10^2 $   &  $ -2.15 $  & $ 0.002 $ &$ 1.62 $ \\
			$ZZZ$      &     $ 9.7429(2) \cdot 10^0 $       & $  1.07605(8)\cdot 10^1 $  & $ 9.742(1) \cdot 10^0 $   &  $-9.47 $ & $ 0.012 $ & $ 1.01 $ \\
		    {$W^+W^-Z$}      &     $ 1.08288(2) \cdot 10^2 $  & $ 1.00587(2)\cdot 10^2 $  &   $1.0829(1) \cdot 10^2 $    & { $ +7.67 $} & $ 0.004 $ & $ 0.45 $ \\
			$W^+ZZ$   &   $ 2.0188(4) \cdot 10^1 $ & $  1.9875(2) \cdot 10^1 $ &$  2.019(2) \cdot 10^1 $ & $ +1.58 $ & $ 0.0001 $ & $ 0.01 $ \\
			$W^-ZZ$    &      $ 1.09844(2) \cdot 10^1 $ &  $  1.06546(8)\cdot 10^1 $   &   $ 1.0984(1) \cdot 10^1 $  &  $ +3.09 $ & $ 0.006 $ & $ 0.51 $\\
			$W^+W^-W^+$   &     $ 8.7979(2) \cdot 10^1 $  &  $ 8.2867(7)\cdot 10^1 $  &   $ 8.7991(15) \cdot 10^1 $  &   $ +6.18 $ & $ 0.014 $ & $ 0.79 $\\
			$W^+W^-W^-$   &     $ 4.9447(1) \cdot 10^1 $   &   $  4.6149(1)\cdot 10^1 $ &   $ 4.9441(2) \cdot 10^1 $  &   $ +7.13 $ & $ 0.013 $ & $ 2.52 $\\
			$ZZH$     &    $ 1.91607(2) \cdot 10^0 $ & $ 2.1006(1)\cdot 10^0 $ &$ 1.91614(18) \cdot 10^0 $ &  $ -8.78 $ & $ 0.004 $ & $0.39$ \\
			$W^+ZH$      &         $  2.48068(2) \cdot 10^0 $&  $ 2.4409(2) \cdot 10^0 $ & $ 2.48095(28) \cdot 10^0 $ & $ +1.64 $ & $ 0.011 $ & $ 0.96 $ \\
			$W^-ZH$     &     $ 1.34001(1) \cdot 10^0 $    &   $ 1.30731(9) \cdot 10^0 $  &    $ 1.3402(1) \cdot 10^0 $   &  $ +2.51 $ & $ 0.011 $ & $ 1.02 $\\
			$W^+W^-H$     &    $ 9.7012(2) \cdot 10^0 $    &   $ 8.8313(2) \cdot 10^0 $  &    $ 9.700(2) \cdot 10^0 $   &  $ +9.83 $ & $ 0.014 $ & $ 0.75 $\\
			$ZHH$     &    $ 2.39350(2) \cdot 10^{-1} $    &   $ 2.6909(2) \cdot 10^{-1} $   &    $ 2.3934(3) \cdot 10^{-1} $   &  $ -11.06 $ & $ 0.005 $ & $ 0.41 $\\
			{$W^+HH$}     &    $ 2.44794(2) \cdot 10^{-1} $   &   $  2.7827(2) \cdot 10^{-1} $   &    $  2.4478(2) \cdot 10^{-1} $   & {$ -12.04 $} & $ 0.007 $ & $ 0.74 $\\
			$W^-HH$     &     $ 1.33525(1) \cdot 10^{-1} $     &    $ 1.5087(1)\cdot 10^{-1} $ &    $ 1.33471(19) \cdot 10^{-1} $   &  $ -11.53 $ & $ 0.041 $ & $ 2.80 $\\
		\end{tabularx}
				\caption[pure EW onshell]{Comparison of NLO EW total cross sections for processes with two and three on-shell bosons at $\sqrt{s}=13$ TeV obtained with \texttt{MUNICH} and \texttt{WHIZARD} using the OLP \texttt{OpenLoops} with numbers $\delta$ defined in Eq.~(\ref{delta}) and \textit{dev} and $\sigma^{\text{sig}}$ in Eq.~(\ref{MunWZparams}), respectively.}
				\label{EWonshell}
\end{sidewaystable}
		\begin{align}
	 &\textit{dev}\equiv  \frac{|\sigma^{\text{tot}}_{\texttt{WHIZARD}}-\sigma^{\text{tot}}_{\texttt{MUNICH}}|}{\sigma^{\text{tot}}_{\texttt{WHIZARD}}} & \sigma^{\text{sig}}\equiv\frac{|\sigma^{\text{tot}}_{\texttt{WHIZARD}}-\sigma^{\text{tot}}_{\texttt{MUNICH}}|}{\sqrt{\Delta_{\texttt{WHIZARD}}^2+\Delta_{\texttt{MUNICH}}^2}}
	\label{MunWZparams}
	\end{align}
	for the NLO EW cross sections $\sigma^{\text{tot}}$ obtained with \texttt{MUNICH} and \texttt{WHIZARD} and statistical MC uncertainties $\Delta_{\text{err}}$. For all processes in Table~\ref{EWonshell}, agreement is found at a numerical accuracy level of relative MC errors well below sub-per-mille level. This can be concluded from $\textit{dev}\lesssim 0.01\%$ in the sixth column together with $\sigma^{\text{sig}}\lesssim3$ in the last column. Thus, the possibility for even minor bugs on the implemented subtraction schemes is ruled out.
	
	Apart from these checks, the numerical values of the relative correction $\delta$ of the NLO EW with respect to the LO result, defined in Eq.~(\ref{delta}), can be understood by the following physical effects playing a role for the inclusion of EW corrections.
	
	An inherent effect playing a role for all shown processes is the EW virtual massive vector-boson exchange between external state legs of a process which yields a negative contribution to the NLO correction. This effect is  enhanced by double logarithms of the form $\alpha \log^2 (s/M_W^2)$ for soft-collinear emission of EW vector bosons in the high-energy limit $s\gg M_W$. In general these can be approximated by EW Sudakov correction factors \cite{Ciafaloni:1998xg,Kuhn:1999de,Kuhn:1999nn,Denner:2000jv,Denner:2001gw}. For di-boson processes this effect is dominant in particular for $ZZ$ and $ZH$ with suppressions of about $-5\%$. NLO EW cross sections to $ZZZ$, $ZZH$, $ZHH$ and $W^{\pm}HH$ with $\delta\sim- 10\%$ yield the largest suppressions for triple-boson production.
	
	There are multiple effects from real-emission diagrams counteracting the EW Sudakov logarithmic suppression with a different impact depending on the process. The most prominent one is the enhancement due to single-top resonances in real-emission diagrams to the processes $pp\to W^+W^-$, $W^+W^-Z$ and $W^+W^-H$. The positive NLO contributions due to this overcompensate the EW Sudakov effects yielding enhancements up to $\delta \sim +10\%$ for $W^{+}W^-H$. 
	A positive relative correction of the NLO EW cross sections with respect to the LO ones can be caused as well by quasi-collinear splittings such as $q^*\to q'W^{\pm}$ or $q^*\to qZ$ in high-energy phase-space regions inducing large positive amplitudes \cite{Frixione:1992pj}. This effect is enhanced especially if $q\gamma$ channels open up by the corresponding real configuration and the photon couples through the vertex $\gamma W^+W^-$. In addition to spin-1 exchange in the $t$-channel this vertex leads to large amplitudes for $W$ bosons propagating in forward direction \cite{Bredenstein:2005pk,Denner:1996wm}. These amplifying effects are present both in real-emission diagrams $q\gamma\to W^{\pm} Z q'~(+H/Z)$ with $q^*\to Zq$ and in $q\gamma \to W^+W^-q'~(+W^{\pm}/H/Z)$ with $q^*\to W^{\pm}q'$ quasi-collinear splittings. This twofold enhancement effect for the corresponding processes may be one of the reasons for their (over-) compensation of the EW Sudakov suppression to be seen from their values $\delta \gtrsim 0$ in Table~\ref{EWonshell}. Note furthermore, that this NLO effect plays neither a role for the production of purely neutral bosons, where there is no $W$ boson, nor the Higgsstrahlung processes, $W^{\pm}H$ and $W^{\pm}HH$, with no significant enhancements due to quasi-collinear splittings at NLO. Dominant negative EW virtual contributions thus are reasonable for these processes explaining their negative relative correction $\delta$.
	
	Another effect which can be relevant for the processes exhibiting $\delta \gtrsim 0$ is the so called `radiation zero' \cite{Mikaelian:1979nr,Brown:1995pya,Stirling:1999sj}. In general, this effect means that for $q\bar{q}^{\prime}\to W^{\pm}\gamma$ squared amplitudes for certain phase-space configurations, due to roots of the cross section differential in the photon-beam angle, vanish. The condition for this effect is based on the fact that the photon can couple to the $W^{\pm}$ inducing an $s$-channel in the same way as to the quarks inducing a $t$-channel.
	In this way, LO contributions for this process become suppressed with respect to real-emission ones, leading to large K~factors. Due to the similarity with respect to $W^{\pm}\gamma$ production the `radiation zero' effect and the consequential enhancement of NLO with respect to LO contributions may be also transferred in a mitigated form to $W^{\pm}Z$ production. In addition, it could play a role as well for the production of $W^{\pm}ZZ$, $W^{\pm}ZH$ and $W^+W^-W^{\pm}$ (due to $qq^{\prime}\to W^{\pm}\gamma^*(\to W^+W^-)$ diagrams). As contributions for the processes of $W^{\pm}H$ and $W^{\pm}HH$ merely come from $s$-channel diagrams at LO the `radiation zero' effect explicitly plays no role there.
	
	\subsubsection{Fixed-order differential distributions}
\begin{figure}
		\centering
		\includegraphics[width=0.49\linewidth]{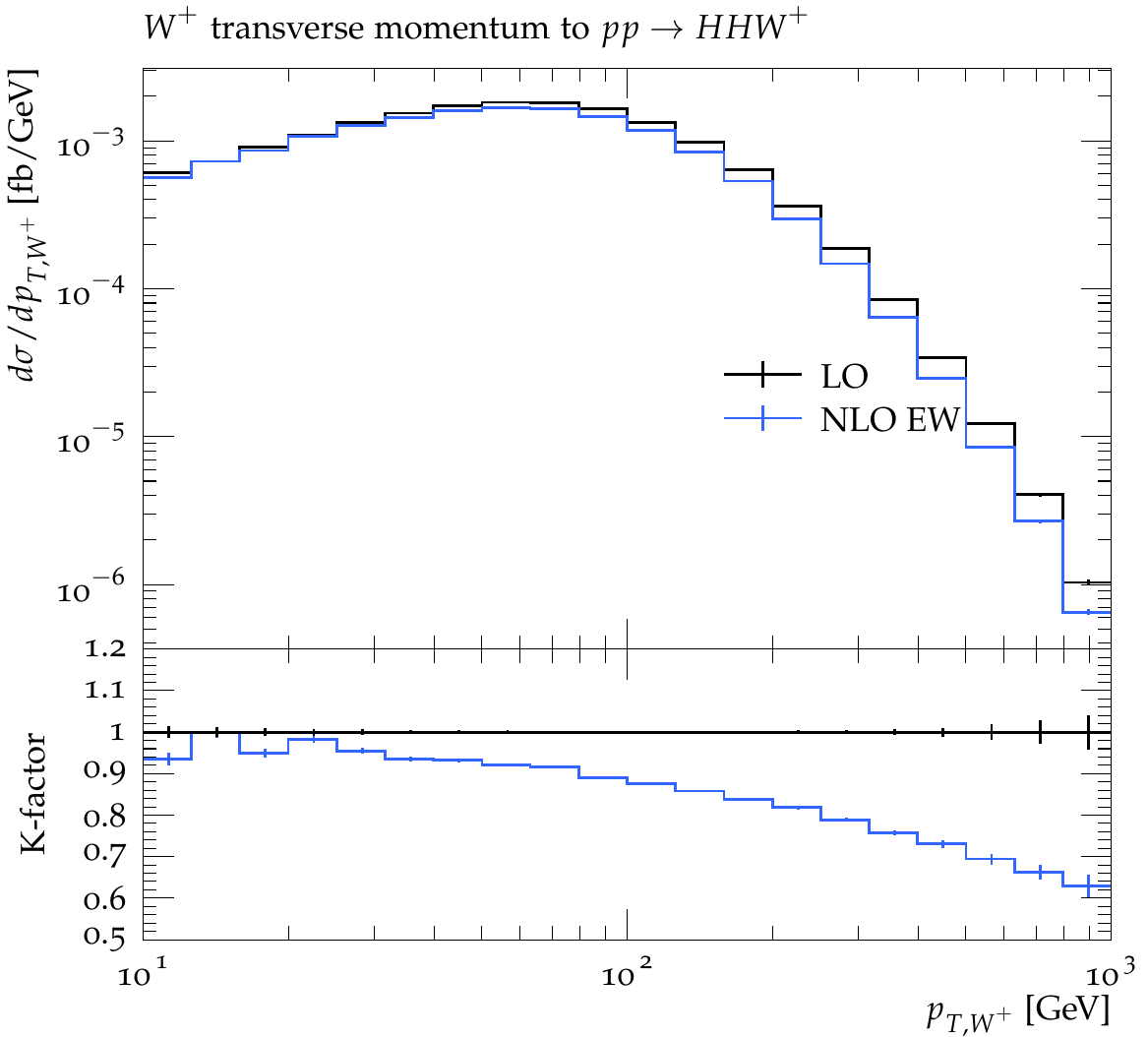}
		\includegraphics[width=0.49\linewidth]{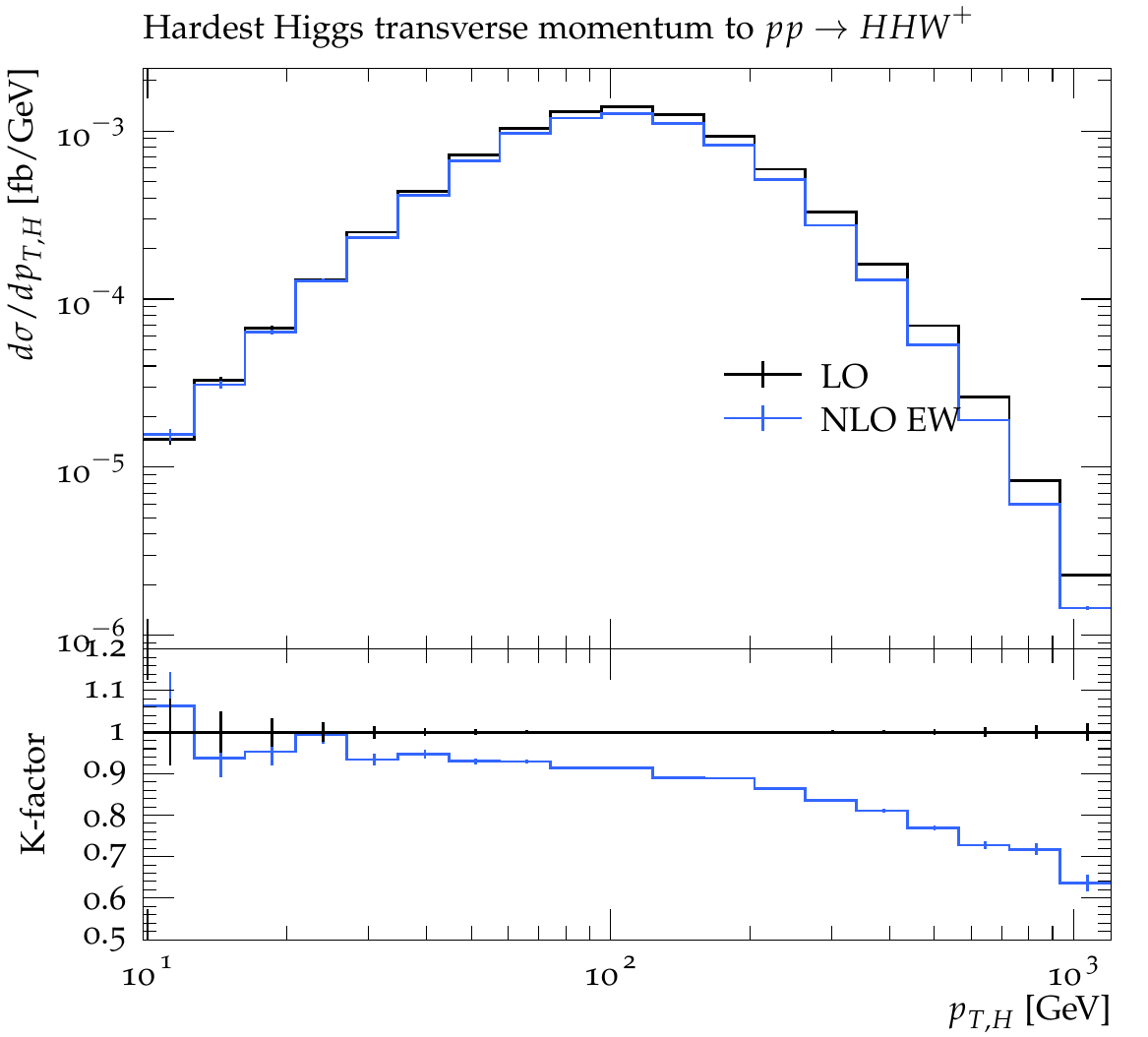}
		\caption{Differential distributions of  the transverse momentum of the $W^+$ boson (left) and the hardest Higgs (right) for the process $pp\to HHW^+$ at $13$~TeV}
		\label{HHWtransverseM}
	\end{figure}

\begin{figure}
	\centering
	\includegraphics[width=0.49\linewidth]{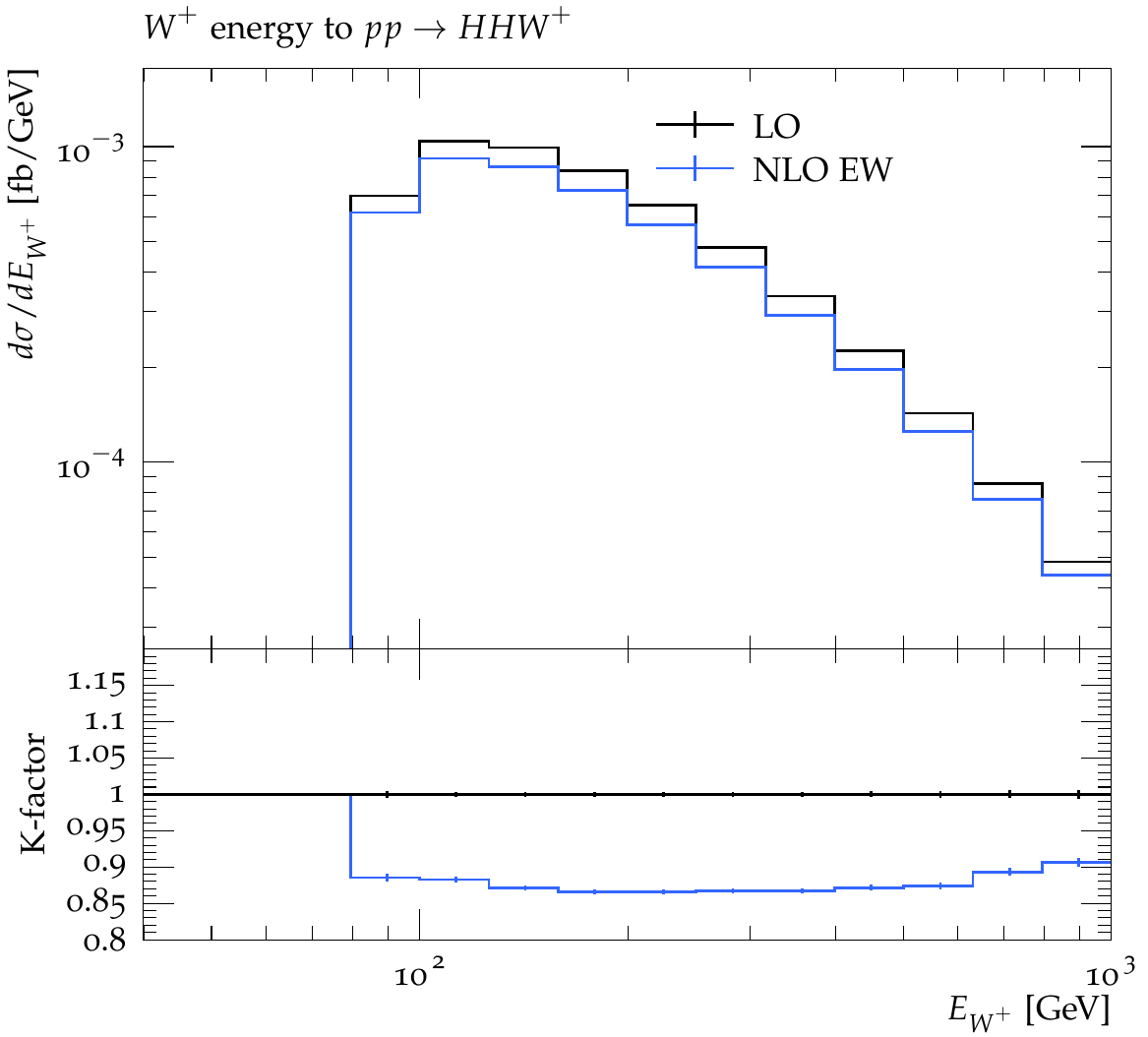}
	\includegraphics[width=0.49\linewidth]{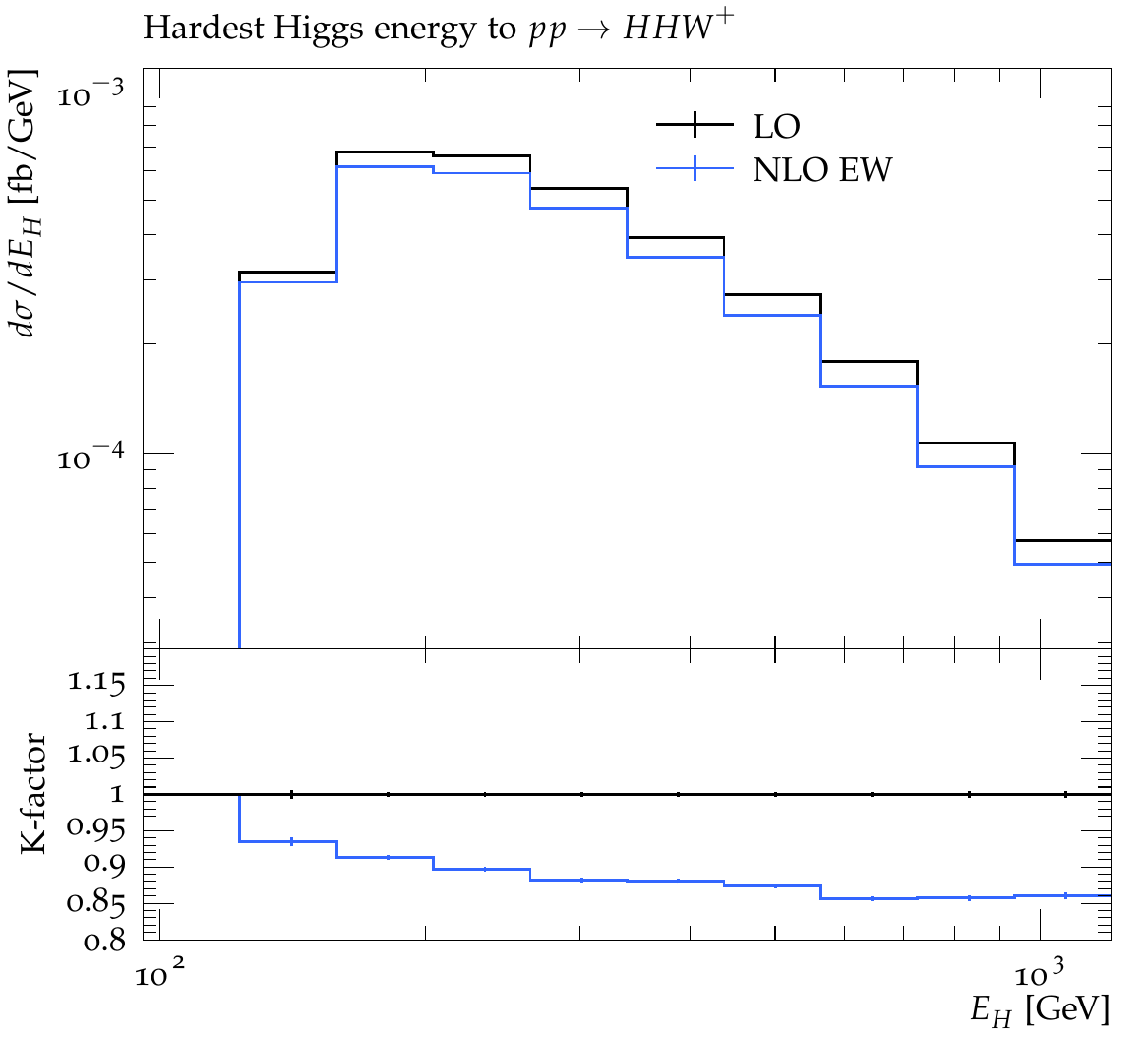}
	\caption{Differential distributions of  the energy of the $W^+$ boson (left) and the hardest Higgs (right), respectively, for the process $pp\to HHW^+$ at $13$~TeV}
	\label{HHWenergy}
	\end{figure}

\begin{figure}
	\centering
	\includegraphics[width=0.49\linewidth]{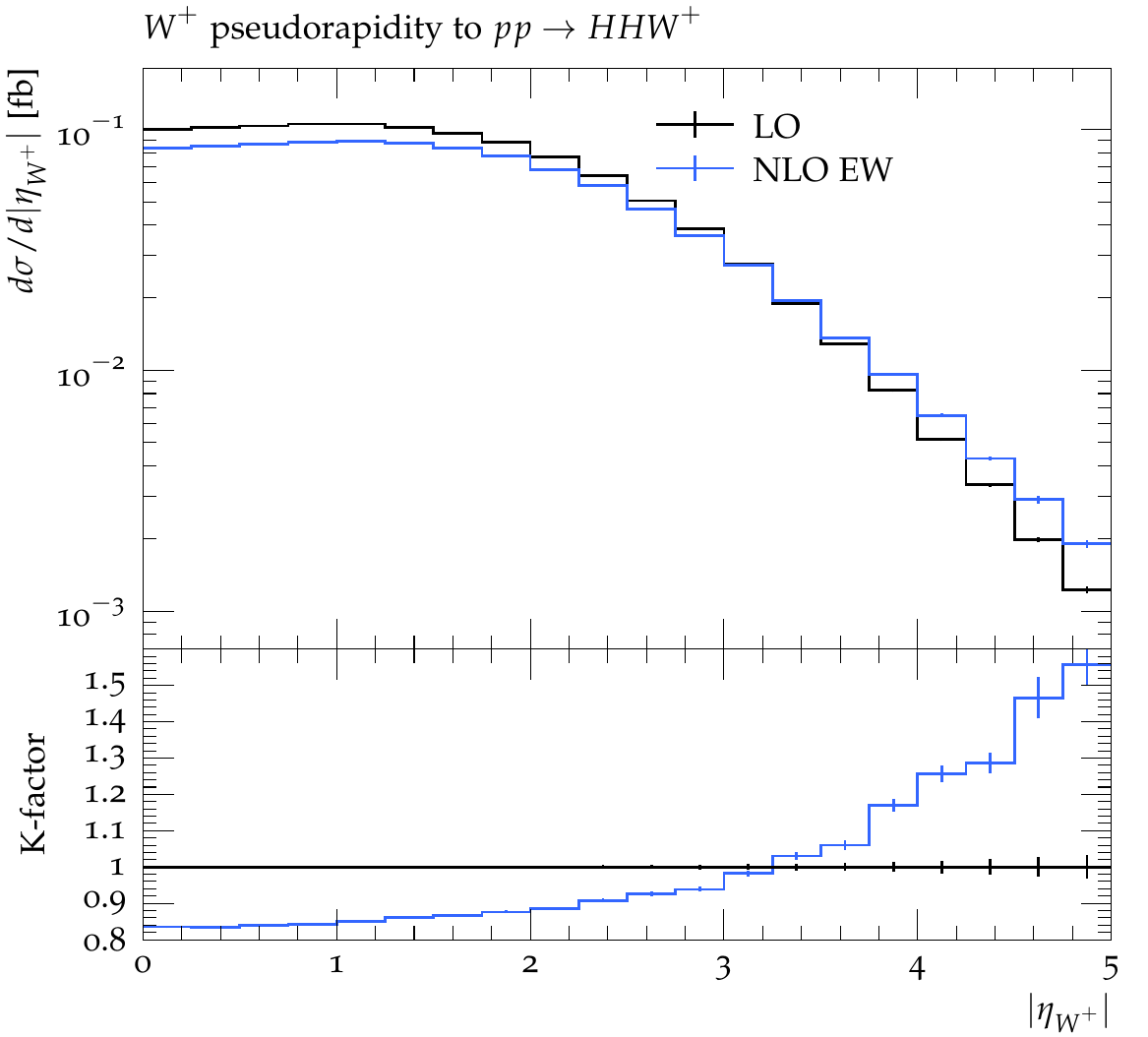}
	\includegraphics[width=0.49\linewidth]{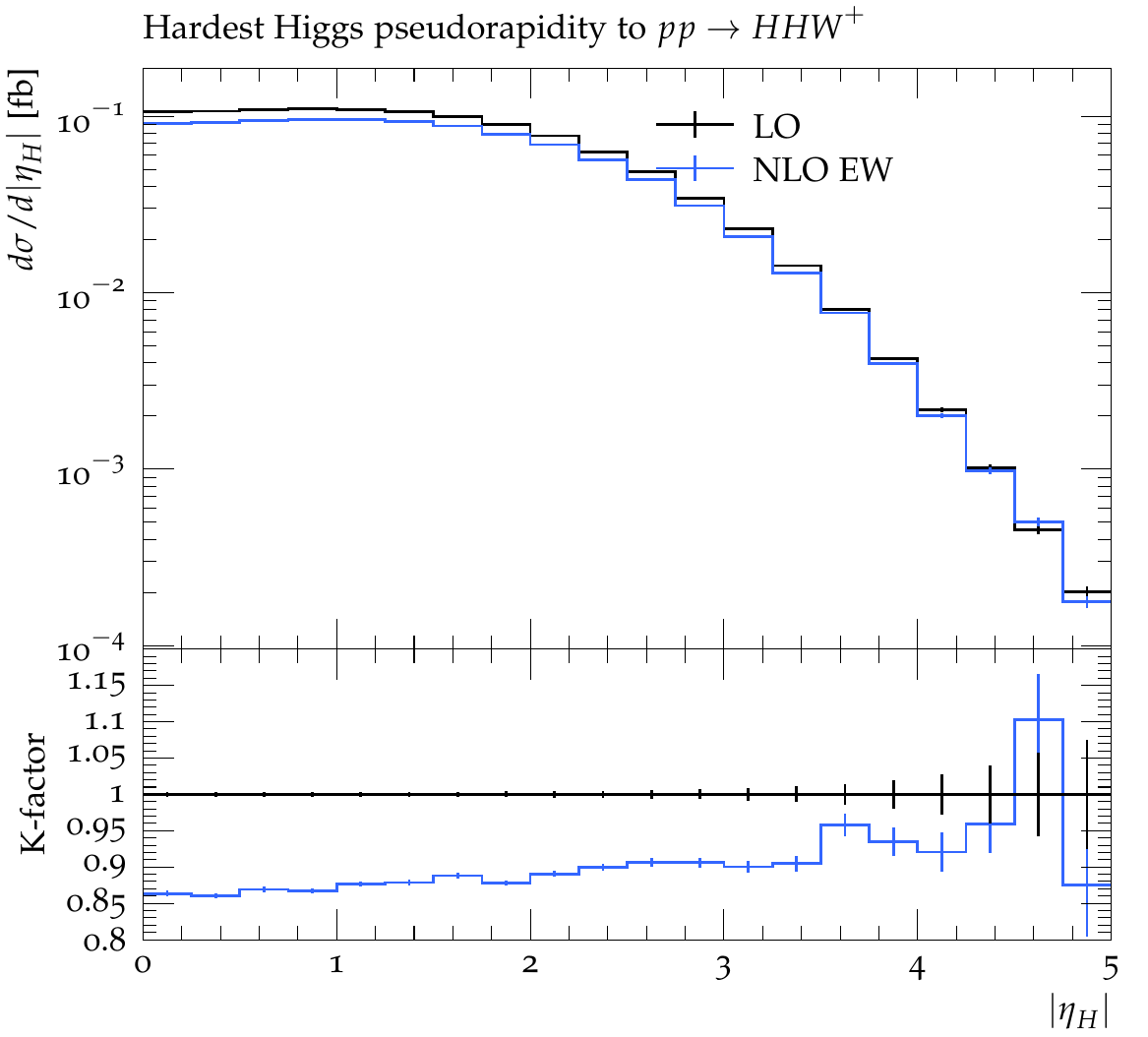}
	\caption{Differential distributions of the pseudorapidity of the $W^+$ boson (left) and the hardest Higgs (right), respectively, for the process $pp\to HHW^+$ at $13$~TeV}
	\label{HHWpseudorap}
\end{figure}

\begin{figure}
	\centering
	\includegraphics[width=0.7\linewidth]{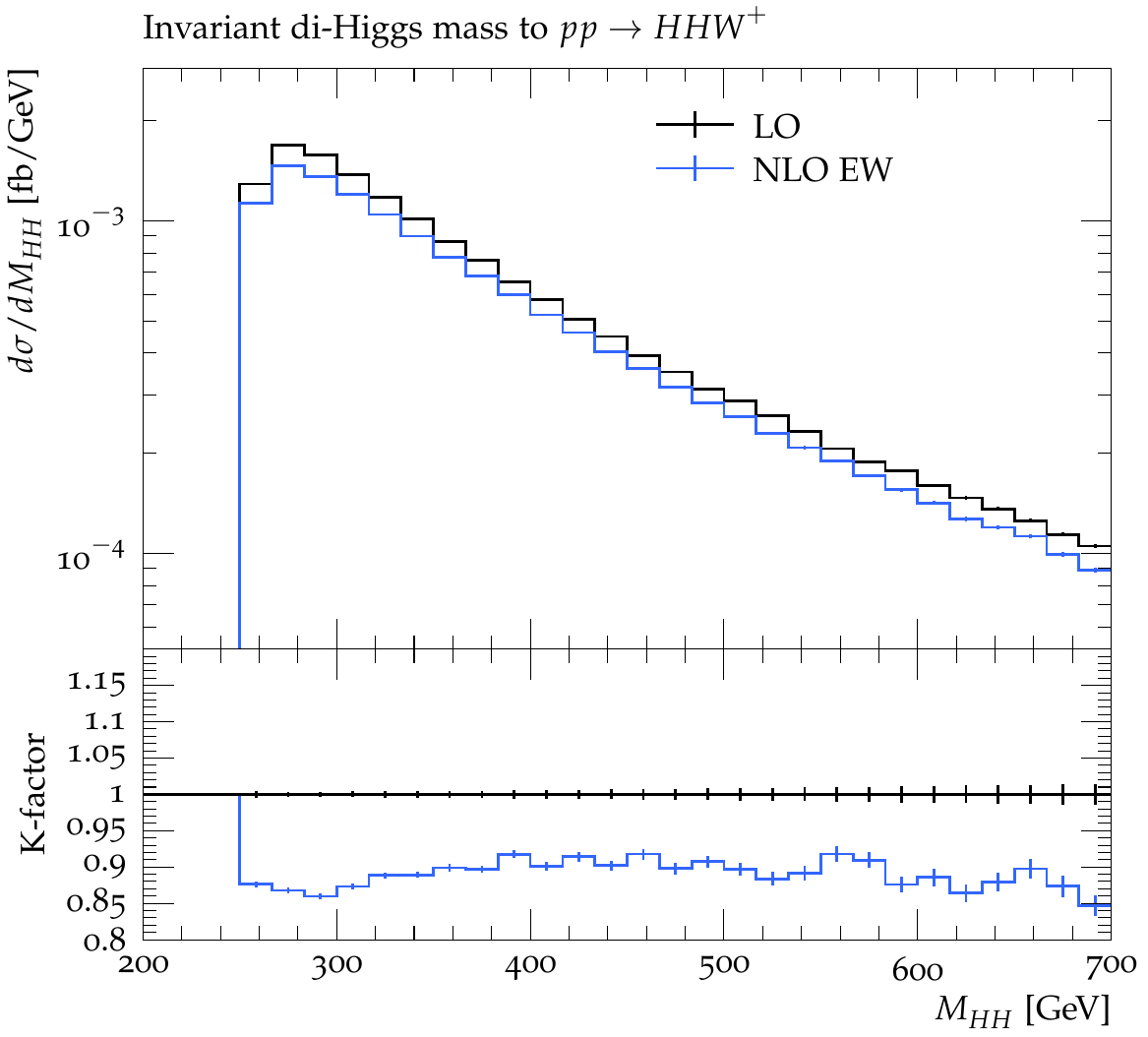}
	\caption{Differential distribution of the invariant mass of the di-Higgs system for the process $pp\to HHW^+$ at $13$~TeV}
	\label{HHWInvMass}
	\end{figure}

	Illustratively, for the process $pp\to HHW^+$ fixed-order differential distributions at LO and NLO EW are presented in Figs.~\ref{HHWtransverseM} to \ref{HHWInvMass}. The histograms are produced with the analysis tool \texttt{Rivet} \cite{Buckley:2010ar} by means of simulated \texttt{HepMC} event samples using \texttt{WHIZARD}.
	Note that the distributions shown here represent fixed-order accurate results of the NLO calculation. For a realistic event-simulation of differential cross sections a proper matching to QED parton showers must be applied. As this is a topic beyond the scope of this thesis, in the following the focus is on understanding the physics effects occurring at NLO EW for the considered fixed-order observables. These apply in the same way for the charge-conjugated process, $pp\to HHW^-$.
	
	In Fig.~\ref{HHWtransverseM} differential distributions of the transverse momentum of the $W^+$ boson and the Higgs boson with the largest transverse momentum, i.~e. the hardest Higgs boson, is shown. First of all, the global behaviour of the two distributions at LO relative to each other can be interpreted in the following way. The distribution $d\sigma/dp_{T,W^+}$ for transverse momenta $p_{T,W^+}$ in the range $10$ - $100$ GeV only slightly increases, with a plateau around $40$ - $100$ GeV with a size of about $10^{-3}$ fb/GeV. In contrast, the distribution $d\sigma/dp_{T,H}$ has a much steeper increase for $p_{T,H}$ in this range of about two orders of magnitudes starting from about $10^{-5}$ fb/GeV. This behaviour can be traced back to the fact that there are much less events for the hardest Higgs with low $p_T$'s as for low $p_{T,W}$ as the $W$ recoils against both of the two Higgs bosons.

The behaviour of the distributions at NLO EW with respect to the corresponding LO ones for the $p_T$ of the two different final states is similar: The differential K factor, in fact, appears nearly identical for both of the histograms with a suppression from $1$ to about $0.65$ for the complete $p_T$ ranges shown in Fig.~\ref{HHWtransverseM}. This suppression can be directly related to negative EW Sudakov logarithms of the form $\sim-\alpha \log^2 p_T^2 / M_W^2$. In general, these double logarithms for an EW Sudakov approximation are defined with arguments $r_{kl}/M_W^2$ with $r_{kl}\gg M_W^2$ where $r_{kl}$ denotes the kinematic invariant $(p_k+p_l)^2$ for external states $k$ and $l$ \cite{Denner:2000jv}.
Large suppression factors of virtual loop with respect to Born amplitudes due to this effect can be related to large invariants of a process. For $pp\to HHW^+$ the largest suppressions can be found for high-$p_{T,W}$ regions for which all invariants $r_{kl}$ of the process are maximal. These can be quantified approximately by the invariant corresponding to the $W$ boson and the hardest Higgs. As this quantity grows symmetrically in both $p_{T,W^+}$ and $p_{T,H}$ the similarity of both differential K~factors, in particular for high $W$ transverse momenta, is manifest.

In Fig.~\ref{HHWenergy}, the energy-distributions at LO and NLO EW for each, $W$ boson and hardest Higgs, are depicted. The K~factor for the distribution differential in the $W$ energy can be considered as constant. This behaviour underlines the fact that the actual energy of the $W$ -- which may vary at NLO EW in addition due to the real radiation of a photon -- is not directly in correlation with the EW Sudakov suppression factor for which the double-logarithms depend purely on invariants of two external state legs of the (Born) process. As explained with respect to the total cross sections, this EW virtual effect is eventually overcompensating the effect of QED radiation.
The K~factor of the differential distribution in  $E_H$, the energy of Higgs identified with the hardest $p_T$, is decreasing starting from the threshold until it stays constant at around $500$ GeV.
As explained above, the $p_T$ of the hardest Higgs can be seen as a measure for the kinematical invariant $r_{kl}$ on which the EW logarithmic Sudakov suppression factors depend on. This has an impact also on the variable $E_H$:
for small numbers of $E_H$ the correlation to the variable $p_{T,H}$ is enhanced as in general for very low energies also low transverse momenta are expected. The second-$p_{T}$-hardest Higgs due to its definition always receives transverse momenta below that of the $p_T$-hardest Higgs. Conversely, large energies $E_H$ do not necessarily induce large $p_{T,H}$ values.
To that end, EW Sudakov suppression factors indirectly depend on the variable $E_H$, in particular for small energies of the $p_T$-hardest Higgs.
%

The cross sections differential in the pseudorapidity of the $W^+$ and the hardest Higgs are shown in Fig.~\ref{HHWpseudorap}. From the distributions at LO as well as at NLO EW it can be concluded that the bulk of the total cross section comes from events with momenta of the final states distributed close to the plane perpendicular to the beam axis. This is due to the fact that contributions to the Born amplitudes purely come from the $s$-channel. At NLO EW, hard-photon ISR in the real-emission diagrams due to radiative return causes large real amplitudes if the photons are radiated in forward direction. This induces a boost of the $HHW$ system along the beam axis. This effect is visible in the distribution differential in $\lvert \eta_{W^+}\rvert$ where for large numbers of this variable the K factor amounts up to $1.5$. For pseudorapidities close to zero the K~factor is around $0.8$ which can be related to large EW Sudakov logarithms which play a role purely for the virtual diagrams. These loop diagrams have Born-like phase-space configurations in the same way as the Born tree-level diagrams with enhanced amplitudes for $W^+$ scattered perpendicular to the beam axis.
For this reason and due to the shift of real events to phase-space regions with forward radiated $W^+$ bosons the suppression due to EW virtual logarithms is mostly visible for small pseudorapidities of the $W^+$ boson.
In contrast, the hardest Higgs pseudorapidity distribution, shown on the right hand side of Fig.~\ref{HHWpseudorap}, exhibits a relatively constant large suppression of the NLO EW with respect to the LO curves. By definition the `hardest Higgs'  corresponds to that Higgs boson final state with the largest $p_{T}$. Due to this, the latter explained effect of the boost of the Born final-state system in the case of ISR is less pronounced for the hardest Higgs than for the $W$ boson. In other words, the boost-energy is mainly deposited to the $W^+$ and second-$p_T$-hardest Higgs which per definition is the Higgs boson scattered closest to the beam axis.

Fig.~\ref{HHWInvMass} shows the distribution differential in the invariant mass of the di-Higgs system at LO and NLO EW. The general behaviour of the curves can be understood as follows: The sharp edge at $250$ GeV indicates the threshold of the di-Higgs system which can be produced starting from an invariant mass of twice the Higgs mass.
From that mass the distributions decrease with $M_{HH}$ which is the typical behaviour as the centre-of-mass energy $\sqrt{\hat{s}}$ increases with this variable leading to an overall suppression of the amplitudes. The K~factor related to these distributions can be considered as constant in $M_{HH}$ with the statistical precision. This can be related to the fact that real-emission and Born-like events in general are not kinematically separated in the di-Higgs mass, as e.~g. in the case of the $W^+$ pseudorapidity. Photon radiation at NLO EW is merely due to initial-state or $W^+$ emitters but does not concern the di-Higgs system. The overall suppression of the LO distribution can be associated with  approximately $\delta\sim -12\%$ which corresponds to the value of the relative correction of the total cross section of this process in Table~\ref{EWonshell}.

	\subsection{Off-shell vector boson production}
	\label{secOffshellLHC}
	Electroweak processes associated with off-shell vector bosons impose new technical challenges for a complete NLO EW computation. This is due to generally higher final-state multiplicities with respect to on-shell boson processes involving massless QCD partons and leptons. These massless final states require phase-space cut criteria as well as IR safety conditions. In this section the NLO EW corrections to several LHC benchmark processes are computed by using the setup defined in Sec.~\ref{secSetupLHC}.
	\subsubsection{Total cross sections}
	The total cross sections at NLO EW for neutral- and charged-current processes (with and without associated Higgs) as well as vector-boson fusion (VBF) and single-top plus jet processes are computed with \texttt{WHIZARD+Openloops} with numerical results presented in Table~\ref{pureEWoffshell}. Corresponding reference results computed with \texttt{MG5\_aMC@NLO} (\texttt{MG5}) \cite{Frederix:2018nkq} are added which serve as checks in order to validate the automated NLO EW framework of \texttt{WHIZARD} for this class of processes.
	For comparison we again define the quantity
	\begin{equation}
	\sigma^{\text{sig}}\equiv \frac{|\sigma^{\text{tot}}_{\texttt{WHIZARD}}-\sigma^{\text{tot}}_{\texttt{MG5}}|}{\sqrt{\Delta_{\texttt{WHIZARD}}^2+\Delta_{\texttt{MG5}}^2}}\quad.
	\label{MG5WZsigma}
	\end{equation}
	with cross sections $\sigma^{\text{tot}}$ obtained with \texttt{MG5} and \texttt{WHIZARD} and corresponding statistical MC uncertainties $\Delta$. In the same way as \texttt{WHIZARD}, the MC generator \texttt{MG5} uses the FKS subtraction scheme.
		\begin{sidewaystable}
		\centering
		\begin{tabularx}{0.78\textwidth}{l|r|r|r|r|r|r}
			process  & $\alpha^m$ &\texttt{MG5\_aMC@NLO}  &\multicolumn{3}{c|}{\texttt{WHIZARD+OpenLoops}}  
			 & $\sigma^{\text{sig}}_{\text{NLO}}$\\
			$pp \rightarrow X~ (+\text{jets})$ && $\sigma_{\text{NLO}}^{\text{tot}}$ [pb]& $\sigma_{\text{LO}}^{\text{tot}}$ [pb] &$\sigma_{\text{NLO}}^{\text{tot}}$ [pb] & $\delta$ [\%]&\\
			\hline\hline
			$e^+\nu_e$          & $\alpha^2$ &      $ 5.2005(8)\cdot10^3 $      & $5.2377(4)\cdot10^3$ &       $ 5.1994(4)\cdot10^3 $     & $- 0.73 $ 
			 & $ 1.24 $\\
			$e^+e^-$       & $\alpha^2$ &       $ 7.498(1)\cdot10^2 $    & $7.535(1)\cdot10^2$ &    $ 7.498(1)\cdot10^2 $    & $ -0.50 $
			 & $ 0.004 $\\
			$e^+\nu_e\mu^-\bar\nu_\mu$  & $\alpha^4$ &     $ 5.2794(9)\cdot10^{-1} $ & $5.0938(7)\cdot10^{-1}$ &   $ 5.2816(9)\cdot10^{-1} $ & $ +3.69 $  
			& $ 1.69 $\\
			$e^+e^-\mu^+\mu^-$ & $\alpha^4$ &          $ 1.2083(3)\cdot10^{-2} $ &  $1.2747(2) \cdot10^{-2}$ &     $ 1.2078(3) \cdot10^{-2}$     & $- 5.25 $
			& $ 1.26 $\\
			$H e^+\nu_e$   &     $\alpha^3$    &  $ 6.4740(17)\cdot10^{-2} $  &   $6.7488(6)\cdot10^{-2}$  &     $ 6.4763(6)\cdot10^{-2} $   &   $ -4.04 $ 
			&$ 1.24 $  \\
			$ H e^+e^-$  & $\alpha^3$  &   $ 1.3699(2)\cdot10^{-2} $    &  $1.4552(1)\cdot10^{-2}$   &     $ 1.3699(1)\cdot10^{-2} $   &   $- 5.86 $ 
			&$ 0.32 $ \\ 
			$ H jj$     &  $\alpha^3$      &     $ 2.7058(4)\cdot10^{0} $  &  $2.8251(4)\cdot10^{0} $ &      $ 2.7056(6)\cdot10^{0} $   &  $ -4.23 $ 
			&$ 0.27 $ \\
			$tj$     &   $\alpha^2$     &   $ 1.0540(1)\cdot10^2 $  &  $1.0615(1)\cdot10^2$  &    $ 1.0538(1)\cdot10^2 $   &  $ -0.72 $ 
			& $ 0.74 $\\
		\end{tabularx}
\caption[pure EW offshell]{Comparison of NLO EW total cross sections for LHC benchmark processes associated with off-shell vector-bosons at $\sqrt{s}=13$ TeV obtained with \texttt{MG5\_aMC@NLO} and \texttt{WHIZARD} using the OLP \texttt{OpenLoops} with numbers $\delta$ defined in Eq.~(\ref{delta}) and $\sigma^{\text{sig}}$ in Eq.~(\ref{MG5WZsigma}), respectively.}
\label{pureEWoffshell}
\end{sidewaystable}

	The NLO EW cross section results of both MC generators in Table~\ref{pureEWoffshell} agree for relative MC uncertainties of $\mathcal{O}(0.01\%)$ with respect to the corresponding absolute values at the level $\sigma_{\text{NLO}}^{\text{sig}}< 2$. In view of the relative correction $\delta$ with magnitudes $\geq0.5\%$ this yields a convincing check. It confirms the correctness of the implementation of NLO EW automated corrections in \texttt{WHIZARD} for this class of processes requiring cut criteria on charged dressed fermions.
	
	Towards the discussion of the results, the first observation which can be made with respect to relative corrections is that the only positive value appears for the charged-current process $pp\to e^+\nu_e\mu^-\bar{\nu}_{\mu}$. This is due to the underlying process $pp\to W^{+}W^{-}$ which already in the case of on-shell $W$ bosons gets positive corrections to be seen in Table~\ref{EWonshell}. The same argument for the enhancement of the NLO EW with respect to the LO cross sections holds for the off-shell process: Large real-emission amplitudes for the configuration $b\gamma\to e^+\nu_e\mu^-\bar{\nu}_{\mu}b$ (identically with $b\leftrightarrow\bar{b}$) are expected due to single-top resonances. This effect in addition to others causing large real amplitudes overcompensate the respective suppression effect due to negative EW virtual Sudakov logarithms.
	
	Similar as for the production of on-shell bosons $ZZ$, $W^+H$ and $ZH$ in Table~\ref{EWonshell} correction factors $\delta$ of a few negative percent can be observed as well for those containing diagrams with corresponding off-shell vector-bosons\footnote{Note that for the neutral-current processes the gauge-bosons $Z$ and $\gamma$ mix with each other. However, for high energies such as those at the LHC the photonic channels typically are suppressed.}, i.~e. $e^+e^-\mu^+\mu^-$, $He^+\nu_e$ and $He^+e^-$ production. As explained above in the context of the vector-boson on-shell processes for these diagrams the EW Sudakov suppression effect is dominant.
	
	As the process of $Hjj$ production contains similar sub-graphs with additional interfering VBF diagrams a suppression factor of similar size as for the latter processes is evident.
	
	That the magnitude of the relative correction $\delta$ for $e^+\nu_e$, $e^+e^-$ and $tj$ final states is below $1\%$ can be explained by the fact that these processes involve only two final states. EW Sudakov factors in general add up for all possible kinematical invariants $r_{kl}=(p_k+p_l)^2$ of the process. Thus, a small number of final states, and in this turn possible invariants, leads to a reduced size of the virtual suppression factor with respect to higher particle multiplicity processes.
	\subsubsection{Fixed-order differential distributions}
	For the charged-current process $pp\to e^+\nu_e\mu^-\bar{\nu}_{\mu}$ cross section distributions differential in properties of  $e^+\nu_e$ and $e^+\mu^-$ systems at LO and NLO EW are shown in Figs.~\ref{Menu} - \ref{Memu}. Apart from their qualitative discussion given below, for the validation of the event generation with \texttt{WHIZARD} checks with \texttt{MG5\_aMC@NLO} with respect to several differential distributions at NLO EW are provided in App.~\ref{secChecksDiffDistr}.
	\begin{figure}
		\centering
		\includegraphics[width=0.7\linewidth]{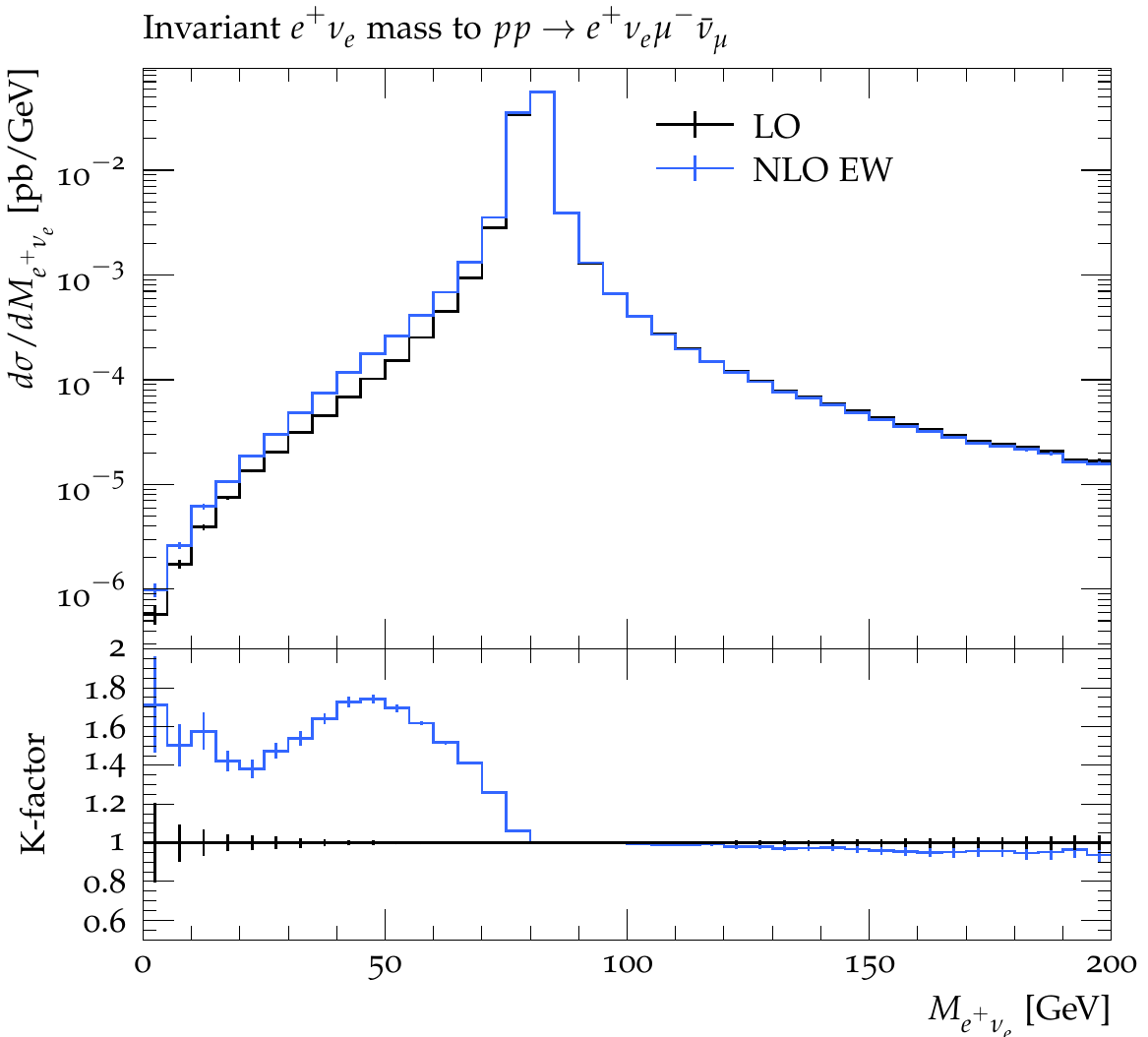}
		\caption{Differential distribution of the invariant mass of the $e^+\nu_e$ system for the process $pp\to e^+\nu_e\mu^-\bar{\nu}_{\mu}$ at $13$ TeV}
		\label{Menu}
	\end{figure}

	The distributions differential in the invariant mass of the $e^+\nu_e$ system, $M_{e^+\nu_e}$, at LO and NLO EW are presented in Fig.~\ref{Menu}. The global behaviour of both curves corresponds to the typical shape of the invariant mass of final-states emerging from the decay of a $W$ boson with a peak at the $W$ mass. While for $M_{e^+\nu_e}$ above this mass the K~factor is close to $1$, for values below this mass it amounts up to about $1.8$. This large K~factor can be traced back to the effect which is called `radiative tail':
	The bulk of the events contributing to the cross section at LO comes from phase-space configurations with invariant masses $M_{e^+\nu_e}$ at the peak around $M_W$. At NLO due to radiation of photons these events are shifted from the threshold to lower values of $M_{e^+\nu_e}$.
	\begin{figure}
		\centering
		\includegraphics[width=0.7\linewidth]{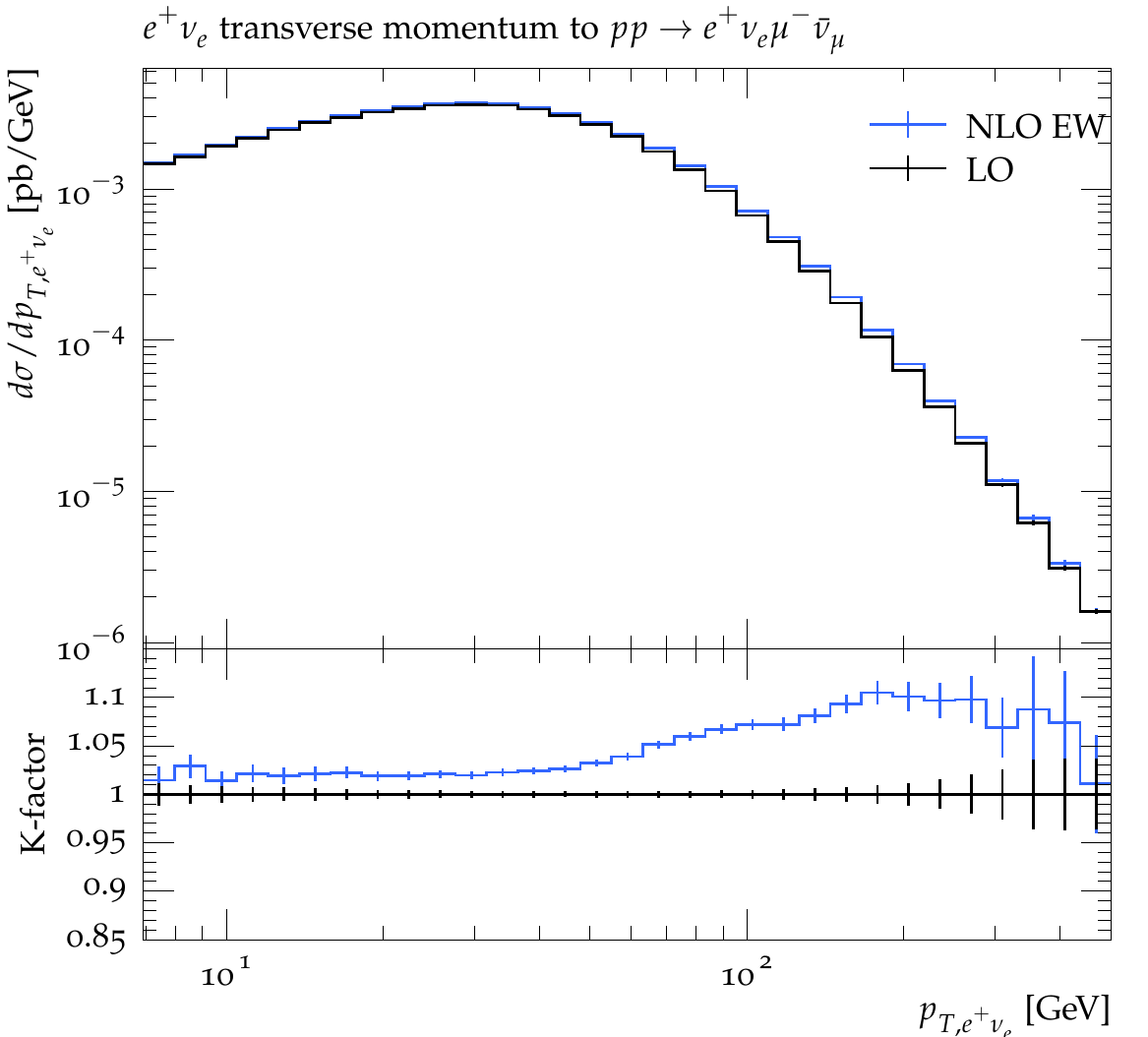}
		\caption{Differential distribution of the transverse momentum of the $e^+\nu_e$ system for the process $pp\to e^+\nu_e\mu^-\bar{\nu}_{\mu}$ at $13$ TeV}
		\label{Ptenu}
	\end{figure}

	In Fig.~\ref{Ptenu} the distribution differential in $p_{T,e^+\nu_e}$, the transverse momentum of the $e^+\nu_e$ system, is shown. The behaviour of the K~factor can be related to the superposition of the following physical effects.
	
	First of all, we observe a constant K~factor of about $1.03$ up to $p_{T,e^+\nu_e}\sim 50$ GeV. The positive relative correction associated with it can be understood by the fact that in general contributions from the LO graphs $\gamma\gamma\to W^{+}W^{-}$ are enhanced for large momentum transfers. This is due to the momentum enhancement of the splitting $\gamma^*\to W^{+}W^{-}$, which e.~g. is amplified compared to the $\gamma^*\to f\bar{f}$ vertex, and the additional $t$-channel nature of the process \cite{Bredenstein:2005pk,Denner:1996wm,Dittmaier:2021}. At NLO $\gamma q \to W^{+}W^{-}q$ channels open up containing this $\gamma\gamma\to WW$ sub-graph. These yield much larger contributions as the LO $\gamma\gamma\to W^{+}W^{-}$ channels since the photon PDFs in general are suppressed with respect to quark PDFs. In contrast to the $\gamma\gamma$ induced processes, the $\gamma q$ (and $\gamma\bar{q}$) channels also cover a large number of flavour and charge degrees of freedom for which all contributions add up for the complete real-emission part of the differential cross section. From the fact that this $\gamma\gamma\to WW$ sub-graph exists in the LO as well as in the NLO diagrams a constant behaviour of the K~factor can be understood.
	
	For $p_{T,e^+\nu_e}\gtrsim50$ GeV the K~factor increases with the transverse momentum up to about $1.1$. The bulk of this effect can be explained by the fact that contributions due to single-top resonances in the real-emission diagrams $\gamma b\to W^{+}W^{-}b$ increase with $W^{\pm}b$ energies approaching the top mass. This explanation is in line with the observation that the largest K~factor is reached for $p_{T,e^+\nu_e}$ values around the top mass apart from the low statistics for bins with large $p_{T,e^+\nu_e}$.
	In general, this positive contribution in addition to enhancements of real amplitudes due to  $q^*\to Wq'$ quasi-collinear splittings in the high-$p_T$ regions overcompensate the suppressive effect of EW Sudakov virtual corrections. 
	
	\begin{figure}
		\centering
		\includegraphics[width=0.7\linewidth]{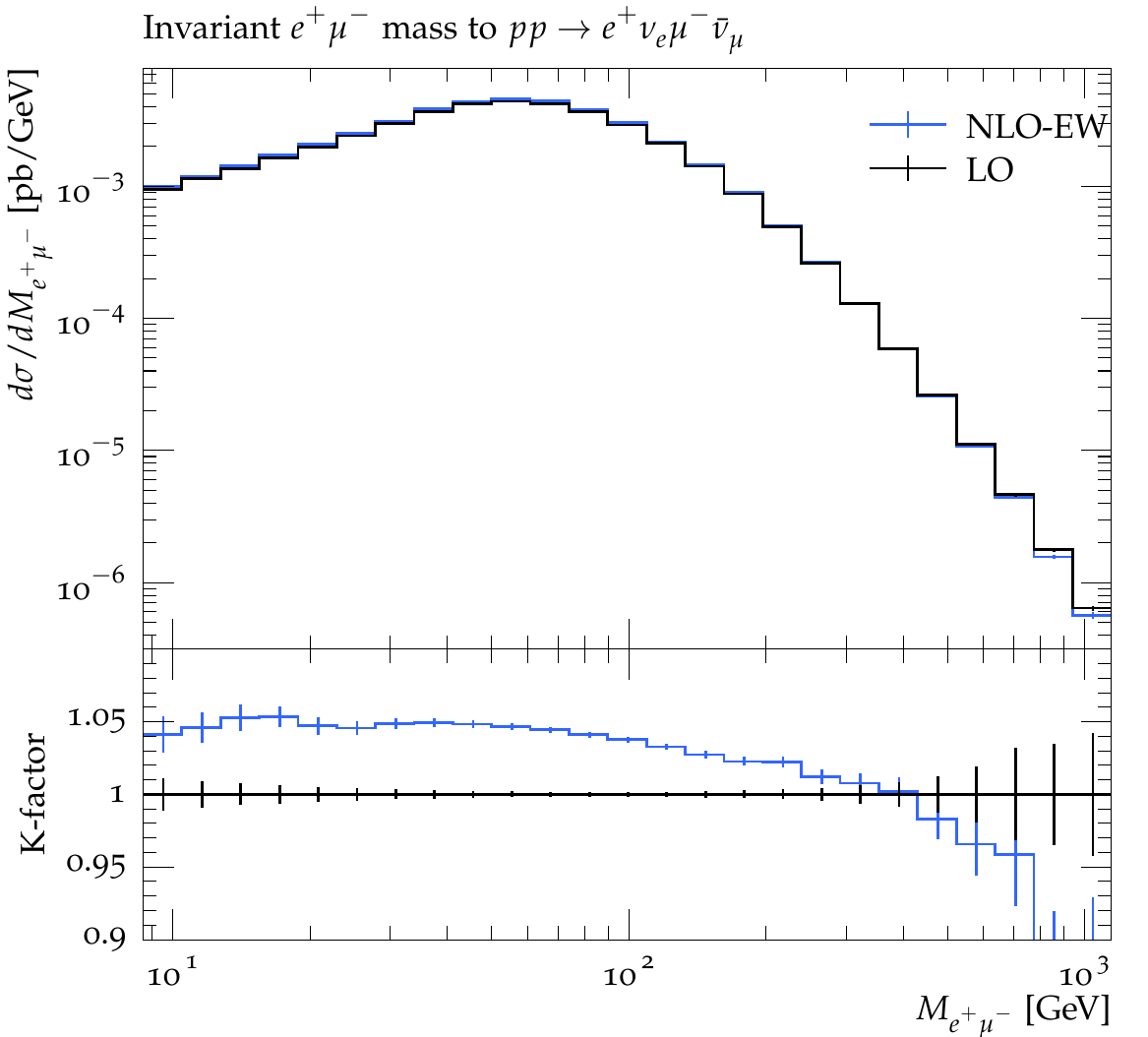}
		\caption{Differential distribution of the invariant mass of the $e^+\mu^-$ system for the process $pp\to e^+\nu_e\mu^-\bar{\nu}_{\mu}$ at $13$ TeV}
		\label{Memu}
	\end{figure}
	The LO and NLO EW distribution differential in the invariant mass of the $e^+\mu^-$ system, $M_{e^+\mu^-}$, is depicted in Fig.~\ref{Memu}. The shape of the K~factor for these distributions originates from two effects which are directly related to $M_{e^+\mu^-}$. The first one is represented by large contributions of real amplitudes for $q\gamma$ PDF channels which are opened up with respect to the underlying Born diagrams $\gamma\gamma\to WW$, described in detail above. This effect unambiguously plays a role in the distribution differential in $M_{e^+\mu^-}$ as this variable crucially depends on the invariant mass of the intermediate $WW$ system. This property in turn is independent from momentum transfers in the $t$-channel diagrams which for large values, corresponding to small angles of the $W$ bosons to the beam axis, induce large amplitudes for $\gamma\gamma\to WW$ subgraphs. Events with these phase-space configurations thus contribute to any $M_{e^+\mu^-}$. This in particular leads to a constant high enhancement of the real with respect to the Born contributions.
	It can be seen from the constant shift of the K~factor of about $+5\%$ for low invariant masses where EW virtual Sudakov logarithmic effects can be neglected.
	In fact, it can be seen that the K~factor in Fig.~\ref{Memu} decreases starting from masses $M_{e^+\mu^-}$ at $\mathcal{O}(100~\text{GeV})$. The most plausible explanation is the suppression due to EW Sudakov virtual corrections
	which scale indeed with logarithms of the invariant $r_{kl}=M_{e^+\mu^-}$.

	\section{NLO mixed corrections to processes at the LHC}
	\label{secMixedCorrLHC}
	For processes at the LHC with non-zero $\alpha_s$ power at Born level, EW corrections are less trivial to compute as for pure EW processes. Cancellations of IR-singularities of both gauge groups, QCD and QED, must be taken into account simultaneously. In order to achieve this for arbitrary processes and LO coupling powers, the formal description of the FKS scheme in mixed coupling expansions in Sec.~\ref{mixedcouplingsSec} and technical methods of Sec.~\ref{secMixedCoupling} are implemented in the framework of \texttt{WHIZARD}.
	The checks which validate the automated computation of NLO mixed corrections are presented in the following. In Sec.~\ref{secToppairLHC} the validation of LO and NLO contributions to all possible coupling powers for $t\bar{t}$, $t\bar{t}W$, $t\bar{t}Z$ and $t\bar{t}H$ production at the LHC is exhibited. Sec.~\ref{secWjZj} includes the check for representative processes with massless final states which additionally require intricate criteria ensuring IR-safety as outlined in Sec.~\ref{secIRsafeobservables}.
	\subsection{Top-quark pair production with associated $W/Z/H$}
	\label{secToppairLHC}
	In general, regarding all contributions from the mixed coupling expansion, an NLO calculation yields an increased precision for theoretical predictions. For processes with LO contributions from more than one coupling order the interference of correction types for observables at NLO plays a role.
	Top-quark pair production at the LHC is the simplest representative process of this class. Considering Fig.~\ref{ziehharmonika} observables for this process entail contributions from three different coupling orders at LO, i.~e. $\text{LO}_{20}$, $\text{LO}_{11}$ and $\text{LO}_{02}$, and from four different ones at NLO, i.~e. $\delta\text{NLO}_{30}$, $\delta\text{NLO}_{21}$, $\delta\text{NLO}_{12}$ and $\delta\text{NLO}_{03}$. The contributions $\delta\text{NLO}_{21}$ and $\delta\text{NLO}_{12}$ thus require the consistent subtraction of IR-singularities of both QCD and QED splittings.
	The whole tower of coupling power contributions are computed with \texttt{WHIZARD} using the methods described in Sec.~\ref{secMixedCoupling} and the LHC setup of Sec.~\ref{secSetupLHC}. Needless to say, as the top quark is identified in the final-state it is considered as on-shell with the top width set to zero.
	
	All results for contributions to the cross sections per coupling power combinations are shown in Table~\ref{ttbarmixed}. In addition, twofold reference results of \texttt{MUNICH/MATRIX} obtained with its NLO EW extension using CS \cite{Kallweit:2014xda} and $q_T$ subtraction \cite{Buonocore:2021rxx,Bonciani:2021zzf} are included. The parameters $\sigma^{\text{sig}}$ and \textit{dev} correspond to those defined in Eq.~(\ref{MunWZparams}). The relative MC uncertainties are chosen for each LO contribution at $\mathcal{O}(0.001\%)$ and for the NLO contribution at $\mathcal{O}(0.01\%)$ - $\mathcal{O}(0.1\%)$ with respect to the corresponding absolute result.
	
	In view of these relative MC uncertainties the cross-checks can be considered as convincing. Furthermore, relative deviations \textit{dev} of the results are found at the order of magnitude of the relative MC errors and the parameter $\sigma^{\text{sig}}$ takes values below $2$ for all contributions. This yields the proof of validity for the implemented FKS scheme in \texttt{WHIZARD} and, simultaneously, for the two implemented subtraction schemes, CS and $q_T$, of \texttt{MUNICH/MATRIX} in the context of NLO mixed corrections. In order to prove the automation for computations of NLO corrections of this kind similar cross-checks are performed for the processes with associated Higgs, $Z$ and $W$ boson, i.~e. $t\bar{t}H$, $t\bar{t}W^+$, $t\bar{t}W^-$ and $t\bar{t}Z$ production. For these processes, similarly to $t\bar{t}$ production, particle widths of the top quark and massive bosons are set to zero. Results on these checks are presented in Tables \ref{ttHmixed} - \ref{ttZmixed}. Likewise, agreement of all contributions of the different subtraction schemes with respect to each other are found using sufficiently high statistics.
	
	The discussion of the underlying physics of the NLO cross section entailing all contributions from the mixed coupling expansion is as follows. For the process $pp\to t\bar{t}$, the largest contribution at LO comes from $\text{LO}_{20}$ at highest order in $\alpha_s$. The smallest, $\text{LO}_{02}$, is about 2 orders of magnitudes suppressed with respect to $\text{LO}_{20}$ which is clear from the approximate factor $\alpha^2/\alpha_s^2 \sim 1/100$ in the squared amplitudes. The contribution $\text{LO}_{11}$ is smaller than $\text{LO}_{02}$. This can be explained because of vanishing interfering amplitudes of diagrams  $q\bar{q}\to t\bar{t}$ with $q\neq b$ due to colour. Explicitly for these processes, the $s$-channel diagram at $\mathcal{O}(g_s^2)$ with internal gluon propagator interfering with that at $\mathcal{O}(g_e^2)$ with internal $\gamma/Z$ propagator is colour-forbidden. This fact clearly can be seen from Table~\ref{pdfchannelsLO} containing the LO contributions split up into PDF channels, where the contribution of the $q\bar{q}$ channel (with $q\neq b$) vanishes. Non-zero contributions to $pp\to t\bar{t}$ at $\mathcal{O}(\alpha_s\alpha)$ hence merely come from $b\bar{b}$ and $g\gamma$ channels. Due to the relative size of bottom and photon PDFs with respect to those of other partons these channels yield rather small contributions.
		\begin{sidewaystable}
		\centering
		\small
		{\onehalfspacing
		\begin{tabularx}{\textwidth}{c|c|c|c|c|c|c|c}
			 \multicolumn{1}{c}{ }&& \multicolumn{3}{c|}{$\sigma^{\text{tot}}$ [fb]}   & \multicolumn{3}{c}{$\sigma^{\text{sig}}$ / \textit{dev}}\\
			 $pp \rightarrow t\bar{t}$ & $\alpha_s^n\alpha^m$& \multicolumn{2}{c|}{\texttt{MUNICH}}
			 	 &\texttt{WHIZARD} & \multicolumn{3}{c}{\texttt{MUNICH}-\texttt{WHIZARD}}\\
			\hline\hline
			$\text{LO}_{20}$       & $\alpha_s^2$ & \multicolumn{2}{c|}{      $ 4.37969(3) \cdot 10^5 $ }      &    $ 4.37972(6) \cdot 10^5 $    &  \multicolumn{3}{c}{$ 0.54$ / $ 0.001\%$} \\
			$\text{LO}_{11}$       & $\alpha_s\alpha$ &        \multicolumn{2}{c|}{ $ 1.77303(4) \cdot 10^3 $  }      &  $ 1.77300(5) \cdot 10^3 $  &     \multicolumn{3}{c}{$ 0.50 $ / $0.002\%$} \\
			$\text{LO}_{02}$       & $\alpha^2$ & \multicolumn{2}{c|}{ $ 2.76254(2) \cdot 10^3 $   }   &    $ 2.76253(4) \cdot 10^3 $    & \multicolumn{3}{c}{$ 0.20 $ / $0.000\%$} \\
						\hline
			$\sum_{\sigma^{\text{tot}}_{\text{LO}}}$&& \multicolumn{2}{c|}{$4.42504(3) \cdot 10^5$ }&$4.42507(6) \cdot 10^5$ &\multicolumn{3}{c}{ }\\
			&&qT&CS&FKS&${\texttt{M}_{\text{qT}}}$-${\texttt{M}_{\text{CS}}}$&$\texttt{M}_{\text{qT}}$-$\texttt{W}_{\text{FKS}}$ & ${\texttt{M}_{\text{CS}}}$-$\texttt{W}_{\text{FKS}}$\\
			\hline\hline
			$\delta\text{NLO}_{30}$       & $\alpha_s^3$ & $ 2.0226(9) \cdot 10^5 $  &      $ 2.02154(3) \cdot 10^5 $      &    $ 2.0218(2) \cdot 10^5 $   & $ 1.13 $ / $0.053\%$ & $0.79$ / $0.038\%$ &$ 1.20$ / $0.015\%$\\
			$\delta\text{NLO}_{21}$       & $\alpha_s^2\alpha$ & -$ 4.7005(6) \cdot 10^3 $ &      -$ 4.7007(2) \cdot 10^3 $     &    -$ 4.7003(3) \cdot 10^3 $    & $ 0.29 $ / $0.004\%$ & $0.31$ / $0.004\%$ &$ 1.27$ / $0.008\%$\\
			$\delta\text{NLO}_{12}$       & $\alpha_s\alpha^2$ & $ 2.4058(4) \cdot 10^3 $  &      $ 2.40514(6) \cdot 10^3 $      &    $ 2.4053(3) \cdot 10^3 $    & $ 1.80 $ / $0.027\%$ & $1.09$ / $0.021\%$ &$ 0.53$ / $0.006\%$\\
			$\delta\text{NLO}_{03}$       & $\alpha^3$ &$ 2.0870(9) \cdot 10^1 $ &      $ 2.0874(5) \cdot 10^1 $     &    $ 2.090(1) \cdot 10^1 $    & $ 0.41 $ / $0.019\%$ & $1.95$ / $0.154\%$ &$ 1.90$ / $0.135\%$\\
			\hline
			$\sum_{\sigma^{\text{tot}}_{\text{LO}}}+\sum_{\sigma^{\text{tot}}_{\text{NLO}}}$& \multicolumn{1}{c|}{ }&$6.4249(9) \cdot 10^5$ 
			&$6.42384(4) \cdot 10^5$ & $6.4242(3) \cdot 10^5$ &$1.14$ /  $0.017\%$&$0.76$ / $0.012\%$&$1.32$ / $0.005\%$
		\end{tabularx}}
				\caption{ Results of cross-checks of \texttt{WHIZARD} and \texttt{MUNICH} for cross section contributions to the process $pp\to t\bar{t}$ at $\sqrt{s}=13$~TeV for each coupling power combination at LO and NLO}
				\label{ttbarmixed}
\end{sidewaystable}
		\begin{sidewaystable}
		\centering
		\small
		{\onehalfspacing
		\begin{tabularx}{\textwidth}{c|c|c|c|c|c|c|c}
			 \multicolumn{1}{c}{ }&& \multicolumn{3}{c|}{$\sigma^{\text{tot}}$ [fb]}   & \multicolumn{3}{c}{$\sigma^{\text{sig}}$ / \textit{dev}}\\
			 $pp \rightarrow t\bar{t}H$ & $\alpha_s^n\alpha^m$& \multicolumn{2}{c|}{\texttt{MUNICH}}
			 	 &\texttt{WHIZARD} & \multicolumn{3}{c}{\texttt{MUNICH}-\texttt{WHIZARD}}\\
			\hline\hline
			$\text{LO}_{21}$       & $\alpha_s^2\alpha$ & \multicolumn{2}{c|}{      $  3.44865(1)\cdot 10^2 $ }      &    $  3.4487(1)\cdot 10^2 $    &  \multicolumn{3}{c}{$ 0.76$ / $ 0.003\%$} \\
			$\text{LO}_{12}$       & $\alpha_s\alpha^2$ &        \multicolumn{2}{c|}{ $  1.40208(2)\cdot 10^0 $ }      &  $  1.4022(1)\cdot 10^0 $  &     \multicolumn{3}{c}{$ 1.44 $ / $0.011\%$} \\
			$\text{LO}_{03}$       & $\alpha^3$ & \multicolumn{2}{c|}{ $  2.42709(1)\cdot 10^0 $   }   &    $ 2.4274(2) \cdot 10^0 $    & \multicolumn{3}{c}{$ 2.07 $ / $0.011\%$} \\
						\hline
			$\sum_{\sigma^{\text{tot}}_{\text{LO}}}$&& \multicolumn{2}{c|}{$ 3.48694(1)\cdot 10^2$ }&$ 3.4870(1)\cdot 10^2$ &\multicolumn{3}{c}{ }\\
			&&qT&CS&FKS&${\texttt{M}_{\text{qT}}}$-${\texttt{M}_{\text{CS}}}$&$\texttt{M}_{\text{qT}}$-$\texttt{W}_{\text{FKS}}$ & ${\texttt{M}_{\text{CS}}}$-$\texttt{W}_{\text{FKS}}$\\
			\hline\hline
			$\delta\text{NLO}_{31}$       & $\alpha_s^3\alpha$ & $ 9.98(1) \cdot 10^1 $  &      $  9.9656(4)\cdot 10^{1} $      &    $ 9.968(4) \cdot 10^1 $   & $ 1.05 $ / $0.108\%$ & $ 0.79 $ / $ 0.086\%$ &$ 0.62$ / $ 0.023\%$\\
			$\delta\text{NLO}_{22}$       & $\alpha_s^2\alpha^2$ & $  6.210(4)\cdot 10^0 $  &      $ 6.209(1) \cdot 10^0 $     &    $  6.208(2)\cdot 10^0 $    & $0.37 $ / $ 0.024\%$ & $0.46 $ / $0.033 \%$ &$ 0.20$ / $ 0.009 \%$\\
			$\delta\text{NLO}_{13}$       & $\alpha_s\alpha^3$ & $ 1.7248(7) \cdot 10^0 $  &      $  1.7238(2)\cdot 10^0 $      &    $ 1.7232(5) \cdot 10^0 $    & $ 1.33 $ / $0.058\%$ & $1.92$ / $0.097\%$ &$1.24 $ / $0.040\%$\\
			$\delta\text{NLO}_{04}$       & $\alpha^4$ &$  1.5051(4)\cdot 10^{-1} $ &      $  1.5053(3)\cdot 10^{-1} $     &    $  1.5060(7)\cdot 10^{-1}  $    & $ 0.28 $ / $0.010\%$ & $1.10$ / $0.058\%$ &$1.00 $ / $0.048\%$\\
			\hline
			$\sum_{\sigma^{\text{tot}}_{\text{LO}}}+\sum_{\sigma^{\text{tot}}_{\text{NLO}}}$& \multicolumn{1}{c|}{ }&$4.565(1)\cdot 10^2$ 
			&$ 4.56433(4)\cdot 10^2 $ & $ 4.5646(4)\cdot 10^2$ &$1.07$ / $0.024\%$ & $0.72$ / $0.017\%$ & $0.82$ / $0.007\%$
		\end{tabularx}}
				\caption[pure EW onshell]{Results of cross-checks of \texttt{WHIZARD} and \texttt{MUNICH} for cross section contributions to the process $pp\to t\bar{t}H$  at $\sqrt{s}=13$~TeV for each coupling power combination at LO and NLO}
				\label{ttHmixed}
\end{sidewaystable}
		\begin{table}
		\centering
		\small
		{\onehalfspacing
		\begin{tabularx}{0.95\textwidth}{c|c|c|c|c}
			 \multicolumn{1}{c}{ }&& \multicolumn{2}{c|}{$\sigma^{\text{tot}}$ [fb]}   & \multicolumn{1}{c}{$\sigma^{\text{sig}}$ / \textit{dev}}\\
			 $pp \rightarrow t\bar{t}W^+$ & $\alpha_s^n\alpha^m$& {$\texttt{MUNICH}_{\text{(CS)}}$}
			 	 &\texttt{WHIZARD} & \multicolumn{1}{c}{$\texttt{MUNICH}_{\text{(CS)}}$-\texttt{WHIZARD}}\\
			\hline\hline
			$\text{LO}_{21}$       & $\alpha_s^2\alpha$ & \multicolumn{1}{c|}{      $  2.411403(1)\cdot 10^2 $ }      &    $  2.4114(1)\cdot 10^2 $    &  \multicolumn{1}{c}{$ 0.72 $ / $ 0.003\%$} \\
			$\text{LO}_{12}$       & $\alpha_s\alpha^2$ &        \multicolumn{1}{c|}{ $ 0.000 $ }      &  $ 0.000 $  &     \multicolumn{1}{c}{$ 0.00 $ / $0.000\%$} \\
			$\text{LO}_{03}$       & $\alpha^3$ & \multicolumn{1}{c|}{ $ 2.31909(1) \cdot 10^0 $   }   &    $ 2.3193(1) \cdot 10^0 $    & \multicolumn{1}{c}{$ 1.76$ / $0.009\%$} \\
						\hline
			$\sum_{\sigma^{\text{tot}}_{\text{LO}}}$&& \multicolumn{1}{c|}{$ 2.434594(9)\cdot 10^2$ }&$ 2.4347(1)\cdot 10^2$ &\multicolumn{1}{c}{$0.74$ / $0.003\%$ }\\
			&&&& \\
			\hline\hline
			$\delta\text{NLO}_{31}$       & $\alpha_s^3\alpha$ &       $  1.18993(2)\cdot 10^2 $      &    $  1.1905(5)\cdot 10^2 $   & $ 1.06 $ / $0.048\%$  \\
			$\delta\text{NLO}_{22}$       & $\alpha_s^2\alpha^2$ &      $- 1.09511(9) \cdot 10^1 $     &    $-  1.0947(3)\cdot 10^1 $    & $ 1.13 $ / $ 0.035\%$  \\
			$\delta\text{NLO}_{13}$       & $\alpha_s\alpha^3$ &      $  2.93251(3)\cdot 10^1 $      &    $2.9334(8)  \cdot 10^1 $    & $ 1.14 $ / $0.030\%$ \\
			$\delta\text{NLO}_{04}$       & $\alpha^4$ &       $ 5.759(3) \cdot 10^{-2} $     &    $  5.756(4)\cdot 10^{-2} $    & $ 0.58 $ / $0.049\%$ \\
			\hline
			$\sum_{\sigma^{\text{tot}}_{\text{LO}}}+\sum_{\sigma^{\text{tot}}_{\text{NLO}}}$& \multicolumn{1}{c|}{ }&$ 3.80884(2)\cdot 10^2$ & $ 3.8096(6)\cdot 10^2$ &\multicolumn{1}{c}{ $1.40$ / $0.020\%$ }
		\end{tabularx}}
				\caption[pure EW onshell]{Results of cross-checks of \texttt{WHIZARD} and \texttt{MUNICH} for cross section contributions to the process $pp\to t\bar{t}W^+$ at $\sqrt{s}=13$ TeV for each coupling power combination at LO and NLO}
				\label{ttWpmixed}
\end{table}
		\begin{table}
		\centering
		\small
		{\onehalfspacing
		\begin{tabularx}{0.95\textwidth}{c|c|c|c|c}
			 \multicolumn{1}{c}{ }&& \multicolumn{2}{c|}{$\sigma^{\text{tot}}$ [fb]}   & \multicolumn{1}{c}{$\sigma^{\text{sig}}$ / \textit{dev}}\\
			 $pp \rightarrow t\bar{t}W^-$ & $\alpha_s^n\alpha^m$& {$\texttt{MUNICH}_{\text{(CS)}}$}
			 	 &\texttt{WHIZARD} & \multicolumn{1}{c}{$\texttt{MUNICH}_{\text{(CS)}}$-\texttt{WHIZARD}}\\
			\hline\hline
			$\text{LO}_{21}$       & $\alpha_s^2\alpha$ & \multicolumn{1}{c|}{      $ 1.217532(4) \cdot 10^2 $ }      &    $ 1.21756(5) \cdot 10^2 $    &  \multicolumn{1}{c}{$ 0.44 $ / $ 0.002\%$} \\
			$\text{LO}_{12}$       & $\alpha_s\alpha^2$ &        \multicolumn{1}{c|}{ $  0.000$ }      &  $  0.000 $  &     \multicolumn{1}{c}{$ 0.00 $ / $0.000\%$} \\
			$\text{LO}_{03}$       & $\alpha^3$ & \multicolumn{1}{c|}{ $ 1.115922(4) \cdot 10^{0} $   }   &    $  1.11596(6)\cdot 10^0 $    & \multicolumn{1}{c}{$ 0.62 $ / $0.003\%$} \\
						\hline
			$\sum_{\sigma^{\text{tot}}_{\text{LO}}}$&& \multicolumn{1}{c|}{$ 1.228691(4)\cdot 10^2$ }&$ 1.22871(5)\cdot 10^2$ &\multicolumn{1}{c}{ $0.45$ / $0.002\%$}\\
			&&&& \\
			\hline\hline
			$\delta\text{NLO}_{31}$       & $\alpha_s^3\alpha$ &       $ 6.18022(9) \cdot 10^1 $      &    $  6.178(2)\cdot 10^1 $   & $ 0.72 $ / $0.029\%$  \\
			$\delta\text{NLO}_{22}$       & $\alpha_s^2\alpha^2$ &      $ - 4.3881(4)\cdot 10^0 $     &    $-4.388(2)  \cdot 10^0 $    & $ 0.10 $ / $ 0.004\%$  \\
			$\delta\text{NLO}_{13}$       & $\alpha_s\alpha^3$ &      $  1.48300(2)\cdot 10^1 $      &    $  1.4822(4)\cdot 10^1 $    & $ 1.74 $ / $0.052\%$ \\
			$\delta\text{NLO}_{04}$       & $\alpha^4$ &       $  3.949(1)\cdot 10^{-2} $     &    $ 3.951(2) \cdot 10^{-2} $    & $ 0.83 $ / $0.046\%$ \\
			\hline
			$\sum_{\sigma^{\text{tot}}_{\text{LO}}}+\sum_{\sigma^{\text{tot}}_{\text{NLO}}}$& \multicolumn{1}{c|}{ }&$ 1.95153(1)\cdot 10^2$ & $ 1.9513(3)\cdot 10^2$ &\multicolumn{1}{c}{ $0.90$ / $0.012\%$ }
		\end{tabularx}}
				\caption[pure EW onshell]{Results of cross-checks of \texttt{WHIZARD} and \texttt{MUNICH} for cross section contributions to the process $pp\to t\bar{t}W^-$ at $\sqrt{s}=13$ TeV for each coupling power combination at LO and NLO}
				\label{ttWmmixed}
\end{table}
		\begin{table}
		\centering
		\small
		{\onehalfspacing
		\begin{tabularx}{0.95\textwidth}{c|c|c|c|c}
			 \multicolumn{1}{c}{ }&& \multicolumn{2}{c|}{$\sigma^{\text{tot}}$ [fb]}   & \multicolumn{1}{c}{$\sigma^{\text{sig}}$ / \textit{dev}}\\
			 $pp \rightarrow t\bar{t}Z$ & $\alpha_s^n\alpha^m$& {$\texttt{MUNICH}_{\text{(CS)}}$}
			 	 &\texttt{WHIZARD} & \multicolumn{1}{c}{$\texttt{MUNICH}_{\text{(CS)}}$-\texttt{WHIZARD}}\\
			\hline\hline
			$\text{LO}_{21}$       & $\alpha_s^2\alpha$ & \multicolumn{1}{c|}{      $ 5.04648(2) \cdot 10^2 $ }      &    $  5.0463(2)\cdot 10^2 $    &  \multicolumn{1}{c}{$ 0.82$ / $ 0.003\%$} \\
			$\text{LO}_{12}$       & $\alpha_s\alpha^2$ &        \multicolumn{1}{c|}{ $  -3.48777(7)\cdot 10^0 $ }      &  $ -3.4876(2) \cdot 10^0 $  &     \multicolumn{1}{c}{$ 0.71 $ / $0.004\%$} \\
			$\text{LO}_{03}$       & $\alpha^3$ & \multicolumn{1}{c|}{ $ 1.141785(6) \cdot 10^1 $   }   &    $  1.14179(4)\cdot 10^1 $    & \multicolumn{1}{c}{$ 0.008 $ / $0.000\%$} \\
						\hline
			$\sum_{\sigma^{\text{tot}}_{\text{LO}}}$&& \multicolumn{1}{c|}{$ 5.16066(2)\cdot 10^2$ }&$ 5.1605(2)\cdot 10^2$ &\multicolumn{1}{c}{$0.82$ / $0.003\%$}\\
			&&&& \\
			\hline\hline
			$\delta\text{NLO}_{31}$       & $\alpha_s^3\alpha$ &       $  2.26389(6)\cdot 10^2 $      &    $  2.264(1)\cdot 10^2 $   & $ 0.26 $ / $0.017\%$  \\
			$\delta\text{NLO}_{22}$       & $\alpha_s^2\alpha^2$ &    $ -4.231(2) \cdot 10^0 $      &    $  -4.228(3)\cdot 10^0 $    & $ 1.02 $ / $0.084\%$ \\
			$\delta\text{NLO}_{13}$       & $\alpha_s\alpha^3$ &        $ 4.2428(7) \cdot 10^0 $     &    $  4.245(3)\cdot 10^0 $    & $ 0.76 $ / $ 0.062\%$  \\
			$\delta\text{NLO}_{04}$       & $\alpha^4$ &       $  -4.111(1)\cdot 10^{-1} $     &    $ -4.111(2) \cdot 10^{-1} $    & $ 0.21 $ / $0.012\%$ \\
			\hline
			$\sum_{\sigma^{\text{tot}}_{\text{LO}}}+\sum_{\sigma^{\text{tot}}_{\text{NLO}}}$& \multicolumn{1}{c|}{ }&$ 7.42056(7)\cdot 10^2$ & $ 7.420(1)\cdot 10^2$ &\multicolumn{1}{c}{ $0.31$ / $0.006\%$}
		\end{tabularx}}
				\caption[pure EW onshell]{Results of cross-checks of \texttt{WHIZARD} and \texttt{MUNICH} for cross section contributions to the process $pp\to t\bar{t}Z$ at $\sqrt{s}=13$~TeV for each coupling power combination at LO and NLO}
				\label{ttZmixed}
\end{table}

	For NLO contributions $\delta \text{NLO}_{q_s,q_e}$ to $pp\to t\bar{t}$, apart from $\delta \text{NLO}_{30}$ which is -- about a factor of $100$ -- enhanced with respect to the others due to the coupling powers, the origin of the numerical size and sign of a contribution is not obvious anymore. With respect to contributions $\delta\text{NLO}_{21}$, $\delta\text{NLO}_{12}$ and $\delta\text{NLO}_{03}$ this can be seen as well from the splitting of each into PDF channel contributions presented in Table~\ref{pdfchannelsNLO} which vary severely in the size and in the sign per channel. One observation can be deduced nevertheless: For $\delta \text{NLO}_{21}$ the largest size of PDF channel contributions comes from $gg$ initiated processes which is negative and about one order of magnitude enhanced with respect to the others. This can be concluded in first line from the amplification due to the PDFs. Secondly, as this contribution has a negative sign large EW Sudakov virtual corrections considered as dominant effect can be found as reasonable. This is underlined by the fact that EW corrections for this channel induce positive real-emission contributions only by photon radiation off the final-state. Virtual loop diagrams with EW boson exchange for which amplitudes can grow up to large values in high-energy phase-space regions in this way overcompensate the real contributions.
	
\begin{figure}
		\centering
			\includegraphics[width=0.4\textwidth]{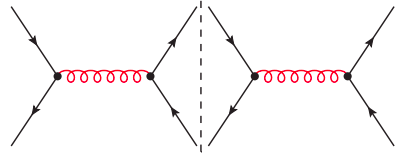}\qquad
			\includegraphics[width=0.4\textwidth]{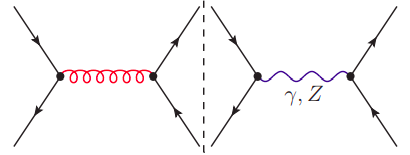}\\
			\includegraphics[width=0.4\textwidth]{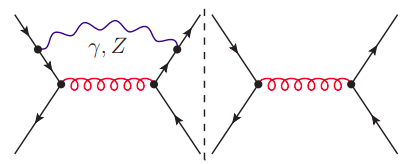}\qquad
			\includegraphics[width=0.4\textwidth]{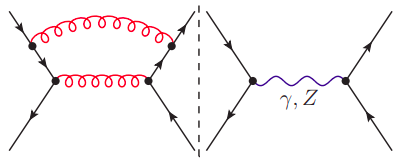}
			\caption{Interfering diagrams to the process $q\bar{q}\to q^{\prime}\bar{q}^{\prime}$ at the coupling orders $\mathcal{O}(\alpha_s^2)$ and $\mathcal{O}(\alpha_s \alpha)$ at LO (upper row) and $\mathcal{O}(\alpha_s^2\alpha)$ (lower row) at NLO; cf.~\cite{Kallweit:2014xda}}
			\label{mixedfeynman}
	\end{figure}

	For this particular channel, EW virtual effects thus can be clearly detected as the coupling power of $\alpha_s$, which is two, at LO and NLO is associated with the coupling of two gluons to the hard process. This is not any more unambiguous considering quark-pair initiated processes, i.~e. $q\bar{q}$ and $b\bar{b}$ channels, since QCD and EW corrections overlap in the virtual diagrams. This is depicted schematically in Fig.~\ref{mixedfeynman} for exemplary $q\bar{q}\to q^{\prime}\bar{q}^{\prime}$ Feynman graphs. The diagrams on the left-hand side represent the EW corrections to the Born tree-level interfering graphs with pure QCD couplings. On the right-hand side the QCD corrections to the Born tree-level QCD-EW interfering graphs are shown.
	Since the lower left loop-tree interfering diagrams can be considered equally as QCD corrections to the upper right interfering graphs at Born-level, the following fact becomes clear: In general, QCD and EW corrections cannot be distinguished from each other for a fixed coupling order, here $\mathcal{O}(\alpha_s^2\alpha)$, to which the process would be considered at NLO in either $\alpha$ or $\alpha_s$ with respect to a well-defined LO coupling order. This concerns the quark-pair channels of $pp\to t\bar{t}$ for both coupling order contributions, $\delta\text{NLO}_{21}$ and $\delta\text{NLO}_{12}$. Equally, this concerns contributions $\delta\text{NLO}_{22}$ and $\delta\text{NLO}_{13}$ of $t\bar{t}$ production with associated $W/Z/H$ boson.
	For $\delta\text{NLO}_{03}$ with respect to $\delta\text{NLO}_{30}$ for $pp\to t\bar{t}$ in the same way as $\delta\text{NLO}_{04}$ with respect to $\delta\text{NLO}_{31}$, we observe a suppression of about 3 orders of magnitudes. This obviously can be traced back to the difference in the coupling powers corresponding to an approximate factor of $(\alpha/\alpha_s)^3\sim 1/1000$.
	
	Considering the numerical size, two contributions of the processes $pp\to t\bar{t}W^+$ in Table~\ref{ttWpmixed} and of $pp\to t\bar{t}W^-$ in Table~\ref{ttWmmixed} are prominent. First of all, the $\text{LO}_{12}$ contributions in both processes vanish. This can be traced back to the colour-forbiddance, as described above, since only up-, down-, strange- and charm-quarks can occur in the initial-state at LO for the charged-current processes with one $W^{\pm}$ in the final-state. Obviously, at $\mathcal{O}(\alpha_s\alpha)$ contributions originate purely from the $\mathcal{O}(g_s^2)$ diagram with gluon $s$-channel propagator interfering with the one at $\mathcal{O}(g_e^2)$ involving a $\gamma/Z$ propagator which couples to the $t\bar{t}$ pair. From the colour structure -- the interfering amplitudes lead to a trace over a single Gell-Mann matrix -- these contributions vanish identically.
	
	A second contribution to $pp\to t\bar{t}W^{\pm}$ which stands out corresponds to $\delta \text{NLO}_{13}$ leading to a relative correction with respect to $\text{LO}_{21}$ of about $+12\%$. This comes from real-emission diagrams involving subgraphs with $tW\to tW$ scattering with Higgs boson exchange. This leads to an enhancement of the amplitudes with these real configurations as outlined in Refs.~\cite{Frederix:2017wme,Dror:2015nkp}.
	\subsection{Jets and charged leptons in the final-state}
	\label{secWjZj}
	For LHC processes with final-state jets additional challenges to achieve theoretical predictions with EW corrections arise. Similar as for top-pair production, these processes require the subtraction of both QCD and QED IR-singularities for NLO contributions apart from $\delta \text{NLO}_{n+1,m}$ and $\delta \text{NLO}_{0,m+n+1}$ (left- and right-most blob in Fig.~\ref{ziehharmonika}). For contributions $\delta \text{NLO}_{n,m+1}$ overlapping QCD and EW corrections do not only occur in the virtual loop-tree interfering amplitudes, as it is the case e.~g. for $t\bar{t}$ production, but also in real-emission diagrams. Treating the photon on a democratic basis with the gluon in the jet definition, as it is dictated by IR-safety conditions, IR divergences can occur from either the photon or gluon. For example, squared real-emission amplitudes to the process $pp\to Zj$ at $\mathcal{O}(\alpha_s\alpha)$ corresponding to flavour structures $q\bar{q}\to Z \gamma g$ for the subtraction of IR-divergences must be considered as both, EW corrections to $q\bar{q}\to Zg$ and QCD corrections to $q\bar{q}\to Z \gamma$ Born contributions.
	
	Additional intricacies due to IR-safety conditions arise for the evaluation of phase-space cuts to massless final-states. Requirements to be met especially concern processes involving both massless charged leptons and jets. For these, photons must be included in the jet definition and recombined with leptons simultaneously in order to render observables IR-safe. In \texttt{WHIZARD} this can be technically achieved for example, as suggested in Sec.~\ref{secIRsafeobservables}, by first applying photon recombination to all charged fermions. This is followed by a jet clustering on dressed quarks, gluons and non-recombined photons. Finally, the cuts defined in Sec.~\ref{secSetupLHC} are imposed on dressed leptons and clustered jets. An explicit example for \texttt{SINDARIN} commands in order to compute NLO EW corrections to the process $pp\to e^+e^-j$ at $\mathcal{O}(\alpha_s\alpha^2)$ is given in App.~\ref{secSindarin}.

	\begin{table}
	\centering
	\begin{tabularx}{\textwidth}{l|r|r|r|r|r|r|c}
		process  & $\alpha_s^n\alpha^m$ &\multicolumn{2}{c|}{\texttt{MG5\_aMC@NLO}}  &\multicolumn{3}{c|}{\texttt{WHIZARD+OpenLoops}}  
		& $\sigma^{\text{sig}}$\\
		$pp \rightarrow X j$ &&$\sigma_{\text{LO}}^{\text{tot}}$ [pb]& $\sigma_{\text{NLO}}^{\text{tot}}$ [pb]&$\sigma_{\text{LO}}^{\text{tot}}$ [pb]& $\sigma_{\text{NLO}}^{\text{tot}}$ [pb] & $\delta$ [\%]& {\small LO/NLO}\\
		\hline\hline
		$e^+\nu_ej$          & $\alpha_s\alpha^2$ &  $914.81(6)$ &   $ 904.75(8) $        &  $914.74(7)$  &   $ 904.59(7) $     & $ -1.11 $ &  $0.8$/$1.5 $\\
		$e^+e^-j$       & $\alpha_s\alpha^2$ &  $150.59(1)$  &   $ 149.09(2) $      & $150.59(1)$ &  $ 149.08(2) $    & $ -1.00 $ & $0.05$/$ 0.4 $
	\end{tabularx}
	\caption{Checks for cross sections at LO and NLO EW, $\sigma^{\text{tot}}_{\text{NLO}}=\text{LO}_{12}$ and $\sigma^{\text{tot}}_{\text{NLO}}=\text{LO}_{12}+\delta\text{NLO}_{13}$, for representative LHC processes with jets and leptons in the final state at $\sqrt{s}=13$ TeV; the parameters $\delta$ and $\sigma^{\text{sig}}$ are defined according to Eqs.~(\ref{delta}) and (\ref{MG5WZsigma})}
	\label{ppllj}
	\end{table}
	Note that if jets are involved in the final-state, non-singular tree-level $(n+1)$-body contributions as explained in Sec.~\ref{secnonsing} play a role. These result from the interference of amplitudes contributing to the NLO coupling order for which no common propagator exists from which two external states emerge and could induce an IR-divergence. Processes of this kind contributing to $pp\to e^+\nu_ej$ at NLO are interfering amplitudes at $\mathcal{O}(\alpha_s\alpha^3)$ for example of $uu\to e^+\nu_e ud$ and $\bar{c}\bar{s}\to e^+\nu_e \bar{c}\bar{c}$. Independently from the fact that some processes at this coupling order are suppressed due to colour such as $u\bar{d}\to e^+\nu_e b\bar{b}$ all real flavour structures possible from the coupling powers are taken into account in \texttt{WHIZARD}. This yields a consistent treatment of matrix elements to the desired coupling powers in the automated framework.
	However, it causes a loss in computation efficiency due to the calling of the OLP \texttt{OpenLoops} and matrix-element evaluation for zero-number contributions for a large number of real flavour structures. Even though this efficiency loss can be avoided in a technical sense a more inherent efficiency issue remains: The FKS phase-space is constructed such that there is merely one set of FKS variables, $\xi$ and $y$, to be integrated over for all FKS regions. As the flavour-space for the radiated parton for this class of processes is extended accounting for the photon, in addition, the cut evaluation acting on jets and dressed leptons becomes highly complicated. The adaption of the integration over merely two FKS variables thus is affected by a complicated topology of non-vetoed phase-space configurations.
	
	In Table~\ref{ppllj} results for checks\footnote{For these checks, slightly different $W$ and $Z$ boson widths as stated in Sec.~\ref{secSetupLHC} are used, i.~e. $\Gamma_W=2.4952$ GeV and $\Gamma_Z=2.0850$ GeV. These correspond to the PDG widths  \cite{ParticleDataGroup:2016lqr}.} with \texttt{MG5\_aMC@NLO} \cite{Frederix:2018nkq} on the total cross section at LO and NLO EW for the processes $pp\to e^+\nu_ej$ and $pp\to e^+e^-j$ at $\mathcal{O}(\alpha_s\alpha^2)$, i.~e. including $\text{LO}_{12}$ and $\delta\text{NLO}_{13}$ contributions, are shown. The MC uncertainty of each calculation is chosen at $\mathcal{O}(0.01\%)$ of the absolute cross section result which makes the check convincing with respect to a relative NLO correction $\delta$ parametrically of $\mathcal{O}(1\%)$. As the parameter $\sigma^{\text{sig}}$ is below $2$ for both processes the NLO EW computation performed with \texttt{WHIZARD+OpenLoops} can be considered as reliable for this class of processes.

	\chapter{Lepton collisions at NLO EW}
	\label{secLeptonCollisions}
	For making precise predictions for lepton collider physics \texttt{WHIZARD} represents a suitable tool which provides a broad set of beam structure features as outlined in Sec.~\ref{secWHIZARD}. The extension of this framework accounting for NLO EW corrections in arbitrary lepton collision processes requires the following considerations.
	
	For a universal treatment of collinear ISR effects for the precision level of an NLO EW calculation NLO-NLL lepton PDFs must be applied. By this, logarithms of the ratio $\mu^2/m^2$, where $\mu$ denotes the factorisation scale and $m$ the initial-state mass, are resummed and factorised into the PDFs as explained in Sec.~\ref{secNLLapproximation}. For $\mu^2\sim s$ the perturbative series of the expansion of partonic cross sections in $\alpha$ is thus valid for arbitrary lepton collider processes with respect to QED corrections.
	
	Formally, this would yield the most suitable approach for precise predictions in this context. However, the technical achievement of results at this precision level in a MC setup highly depends on the applied integration methods as the PDFs contain a singular structure. This is the case already for LO calculations applying LL PDFs for which the ISR structure function mapping presented in Sec.~\ref{parametrisationPDFsing} is implemented in \texttt{WHIZARD} \cite{Hagiwara:2005wg,Whizard:2020} and has proven itself sufficiently stable. An extension of this phase-space mapping with respect to NLO calculations applying NLL electron PDFs is suggested in the same section. At this time, the analytical solution for the NLL PDF's functions have been implemented and validated in \texttt{WHIZARD}. Numerically exact same results as obtained with the code \texttt{ePDF} \cite{Frixione:2019lga,Bertone:2019hks}, using its default settings, have been reproduced.
	
	The implementation of the  phase-space mappings adjusted for NLO computations overcoming the extremely peaked behaviour of the electron PDFs is completed for separate components, i.~e. the Born and virtual part with Born-like phase-space configurations as well as the DGLAP remnant. For the latter, the PDFs entering FKS terms with ISR in the collinear limit are rescaled with a parametrisation of the respective variable according to Eq.~(\ref{improvedz}).
	This extended mapping is probed together with the grid interpolation method proposed in Sec.~\ref{gridinterpolation} using numerical NLL PDF values obtained with \texttt{ePDF} as grid points. A sufficiently high integration efficiency and adaption quality for these separate components is achieved. In addition, internal consistency checks are performed showing the correctness of the new mapping procedure. For the purpose of these checks, the electron PDFs are set to constant value~$1$, and the integration is performed once with the original and once with the adjusted parametrisation of the NLO phase-space variables. Both integrations yield the same numerical results.
	The component of the real-emission and counterterms requires a twofold mapping as the rescaling of PDFs follows the FKS requirements of Sec.~\ref{secMasslessISem}. The proposed parametrisation according to Eqs.~(\ref{improvedmappingreal}) - (\ref{jacobianrealmapping}) is based on a similar mapping as for the DGLAP remnant component. Their implementation in \texttt{WHIZARD} is under development.
	
	Another approach to achieve a sufficiently good performance of a MC integration applying electron PDFs within the \texttt{MG5\_aMC@NLO} framework is presented in Refs.~\cite{Frixione:2021zdp,Bertone:2022ktl}. This however will not be discussed in the following as the preferred choice of methods always depends on the fundamental structure of the MC tool itself. The checks in \texttt{WHIZARD} applying NLL electron PDFs to NLO calculations, described above, are merely on partial terms of an NLO observable which separately have no physical meaning. For this reason, the complete validation of the setup in \texttt{WHIZARD}, including also cross-checks with \texttt{MG5\_aMC@NLO}, will be deferred to a future study.
	
	Apart from this approach, treating collinear ISR effects in a resummed way, NLO EW corrections to lepton collider observables can be computed at fixed order by keeping the initial-state mass-dependencies explicit in the matrix elements. This ansatz inevitably depends on the collider setup as for some cases the ratios $s/m^2$ might lead to large logarithms such that the NLO QED fixed-order perturbative expansion would give an unreliably imprecise result. However, by considering kinematic cases where terms involving collinear logarithms of a perturbative order beyond NLO are sufficiently suppressed by $\alpha$, e.~g. by means of Eq.~(\ref{estimatefNLOmassiveIS}), fixed NLO EW corrections can be considered as reliable for certain collider setups.
	The FKS phase-space construction for ISR in this approach follows the considerations on radiation off massive initial-state emitters according to Sec.~\ref{secMassiveIS}.
	
	With the implementation of this phase-space construction in \texttt{WHIZARD} NLO EW calculations for lepton collisions can be performed by treating the leptons as massive. By this, collinear radiation off the initial state is regularised by the mass itself and neither PDFs nor collinear counterterms are needed. The OLP which can be used for this purpose is \texttt{RECOLA} as it can account for full mass dependence of leptons in the amplitudes.
	
	While in Sec.~\ref{secValidationee} the validation of this framework is presented, Sec.~\ref{scans} contains NLO EW results on cross sections and differential distributions on multi-boson processes at a future muon collider obtained with it.
	
	For the results presented in this chapter the parameter
	\begin{align}
		\delta_{\text{EW}}=\frac{\sigma_{\text{NLO}}^{\text{tot}}}{\sigma_{\text{LO}}^{\text{tot}}}-1
	\end{align}
	is used describing the relative NLO EW correction of the cross section at NLO, $\sigma_{\text{NLO}}^{\text{tot}}$, with respect to that at LO, $\sigma_{\text{LO}}^{\text{tot}}$, for a lepton collision process.
	\section{Validation for $e^+e^-$ processes}
	\label{secValidationee}
	
	In this section explicit checks on total cross sections at LO and NLO EW are shown which validate the framework of \texttt{WHIZARD+RECOLA}. These are performed besides internal FKS sanity checks. Automated NLO EW corrections to lepton collider processes are thus proven reliable for the massive initial-state fixed-order approximation. 
	
	\begin{table}
		\centering
		\small
		\begin{tabularx}{\linewidth}{l|r|r|r|r|r|c}
			$e^+e^-\to HZ$&\multicolumn{2}{c|}{\texttt{MCSANCee} \cite{Sadykov:2020dgm}}&\multicolumn{3}{c|}{\texttt{WHIZARD+RECOLA}}&\multicolumn{1}{c}{$\sigma^{\text{sig}}$}\\
			$\sqrt{s}$ [GeV]& $\sigma_{\text{LO}}^{\text{tot}}$ [fb] & $\sigma_{\text{NLO}}^{\text{tot}}$ [fb] & $\sigma_{\text{LO}}^{\text{tot}}$ [fb] & $\sigma_{\text{NLO}}^{\text{tot}}$ [fb] & $\delta_{\text{EW}}$ [\%]& (LO/NLO)\\
			\hline\hline
			$\phantom{0}250$       & $225.59(1)$ &       $ 206.77(1)$      &    $225.60(1)$    & $207.0(1)$& $-8.25$ &$0.4$/$2.1$\\
			$\phantom{0}500$       & $53.74(1)$ &       $62.42(1) $      &    $53.74(3)$    & $62.41(2)$& $+16.14$ &$0.2$/$0.3$\\
			$1000$  & $12.05(1)$ &     $14.56(1)$   &   $12.0549(6)$ & $14.57(1)$&$+20.84$&$0.5$/$0.5$
		\end{tabularx}
		\caption[muon collider NLO EW]{Comparison of LO and NLO total inclusive cross sections results of the \texttt{MCSANCee} and \texttt{WHIZARD+RECOLA} setup for the process $e^+e^-\rightarrow HZ$ with unpolarised beams}
		\label{eeHZtable}
	\end{table}
	
	The cross-checks are performed using reference results of \texttt{MCSANCee} \cite{Sadykov:2020dgm,Arbuzov:2022mij}. In Table~\ref{eeHZtable} cross section results at LO and NLO EW for both tools with respect to the process $e^+e^-\to HZ$ for unpolarised beams are shown. In addition, the relative correction $\delta_{\text{EW}}$ and the parameter
	\begin{align}
		\sigma^{\text{sig}}\equiv \frac{|\sigma^{\text{tot}}_{\texttt{WHIZARD}}-\sigma^{\text{tot}}_{\texttt{MCSANCee}}|}{\sqrt{\Delta_{\texttt{WHIZARD}}^2+\Delta_{\texttt{MCSANCee}}^2}}
	\end{align}
	using the MC uncertainties $\Delta_{\texttt{WHIZARD}}$ and $\Delta_{\texttt{MCSANCee}}$ are included. The cross sections are obtained using the input parameters of the reference results in Ref.~\cite{Sadykov:2020dgm} which corresponds to an ILC-like setup with centre-of-mass energies of $\sqrt{s}=250$, $500$ and $1000$ GeV, respectively. Moreover, it involves the application of the $\alpha(0)$ input and on-shell renormalisation scheme for which all particle widths are set to zero.
	
	From the $\sigma^{\text{sig}}$ values in Table~\ref{eeHZtable} we conclude that the results of both MC frameworks agree very well for MC uncertainties below the permille-level of the absolute cross section values. For relative corrections $\delta_{\text{EW}}$ at $\mathcal{O}(10\%)$ this is sufficient to say that the framework \texttt{WHIZARD+RECOLA} yields reliable predictions including NLO EW corrections for such processes.

	\begin{table}
	\centering
	\small
	\begin{tabularx}{\linewidth}{l|r|r|r|r|r|c}
		$e^+e^-\to \mu^+\mu^-$&\multicolumn{2}{c|}{\texttt{MCSANCee} \cite{Arbuzov:2022mij} }&\multicolumn{3}{c|}{\texttt{WHIZARD+RECOLA}}&\multicolumn{1}{c}{$\sigma^{\text{sig}}$}\\
		$\sqrt{s}$ [GeV]& $\sigma_{\text{LO}}^{\text{tot}}$ [pb] & $\sigma_{\text{NLO}}^{\text{tot}}$ [pb] & $\sigma_{\text{LO}}^{\text{tot}}$ [pb] & $\sigma_{\text{NLO}}^{\text{tot}}$ [pb] & $\delta_{\text{EW}}$ [\%]& (LO/NLO)\\
		\hline\hline
		$\phantom{0}5$       & $2978.6(1)$ &       $3434.2(1) $      &    $2978.7(1)$    & $ 3433.5(3)$& $+15.27$ &$0.3$/$2.2$\\
		$\phantom{0}7$       & $1519.6(1)$ &       $ 1773.8(1) $      &    $1519.605(4)$    & $1773.1(2) $& $+16.68$ &$0.05$/$3.0$
	\end{tabularx}
	\caption[muon collider NLO EW]{Comparison of LO and NLO total cross sections results of the \texttt{MCSANCee} and \texttt{WHIZARD+RECOLA} setup for the process $e^+e^-\to \mu^+\mu^-$ with unpolarised beams}
	\label{eemmtable}
	\end{table}
	\begin{table}
	\centering
	\small
	\begin{tabularx}{\linewidth}{l|r|r|r|r|r|c}
		$e^+e^-\to \tau^+\tau^-$&\multicolumn{2}{c|}{\texttt{MCSANCee} \cite{Arbuzov:2022mij} }&\multicolumn{3}{c|}{\texttt{WHIZARD+RECOLA}}&\multicolumn{1}{c}{$\sigma^{\text{sig}}$ }\\
		$\sqrt{s}$ [GeV]& $\sigma_{\text{LO}}^{\text{tot}}$ [pb] & $\sigma_{\text{NLO}}^{\text{tot}}$ [pb] & $\sigma_{\text{LO}}^{\text{tot}}$ [pb] & $\sigma_{\text{NLO}}^{\text{tot}}$ [pb] & $\delta_{\text{EW}}$ [\%]&(LO/NLO)\\
		\hline\hline
		$\phantom{0}5$       & $2703.3(1)$ &       $2816.7(1) $      &    $2703.353(5)$    & $ 2816.7(2)$& $+4.19$ &$0.5$/$0.2$\\
		$\phantom{0}7$       & $1503.0(1)$ &       $1648.8(1) $      &    $1503.066(4)$    & $1648.5(1)$& $+9.68$ &$0.7$/$1.6$
	\end{tabularx}
	\caption[muon collider NLO EW]{Comparison of LO and NLO total cross sections results of the \texttt{MCSANCee} and \texttt{WHIZARD+RECOLA} setup for the process $e^+e^-\to \tau^+\tau^-$ with unpolarised beams}
	\label{eetautautable}
	\end{table}

	A similar check can be done for off-shell neutral-current processes including final-state radiation.
	For the processes $e^+e^-\to \mu^+\mu^-$ and $e^+e^-\to \tau^+\tau^-$ at low collider energies, $\sqrt{s}=5$ and $7$ GeV, LO and NLO EW cross sections computed with \texttt{WHIZARD+RECOLA} are compared with those of \texttt{MCSANCee} from Ref.~\cite{Arbuzov:2022mij}. Results of these checks, for which settings and parameters are described in the latter reference, are displayed in Tables~\ref{eemmtable} and \ref{eetautautable}.
	In both of these, agreement between the shown cross sections from the two generators can be found for MC uncertainties parametrically below $\mathcal{O}(0.01\%)$ of the absolute results.
	
	Due to this, the framework of \texttt{WHIZARD+RECOLA} can be considered validated for NLO EW precision predictions for these kind of processes. This is useful with respect to the collider energies which are of the typical order of magnitude of those at contemporary $e^+e^-$ collider experiments such as Belle II at SuperKEKB or BES III at IHEP in Beijing. Note that for these low collider energies collinear ISR effects inducing the logarithmic terms as described above are much less pronounced. However, the effects of pure soft photon radiation beyond the NLO in general may play an increased role depending highly on the process and the collider setup. For example, $e^+e^-\to HZ$ at $250$ GeV or $e^+e^-\to \tau^+\tau^-$ at $5$ GeV, due to the final-state mass thresholds, yield the phase-space for the radiated photon at NLO much more restricted compared to higher collider energies. This could imply the need for soft photon resummation which is captured for example by the YFS approach \cite{Yennie:1961ad}. As this is beyond the scope of this thesis it will not be discussed here.
	
	\section{Multi-boson processes at a muon collider}
	In this section NLO EW cross sections and distributions for multi-boson
	processes  $\mu^+\mu^- \to V^n H^m$ with $V \in \{W^\pm,Z\}$
	and $n+m \leq 4$ are computed with results and discussions presented in Ref.~\cite{Bredt:2022dmm}.
		
	For this purpose, one-loop EW virtual contributions are provided by \texttt{RECOLA} accounting for the full mass dependence of fermions and bosons.  Additionally, for NLO QED corrections to $HZ$ and $ZZ$ production the interface to the OLP \texttt{OpenLoops} is used.
	The FKS phase-space construction with massive particles in the initial-state, i.~e. the mapping between the Born and real-radiation phase-space parameterisations, follows the considerations of Sec.~\ref{secMassiveIS}.
	The integration proceeds via numerical phase-space sampling with
	multi-channel adaption using \texttt{WHIZARD}'s MPI-based
	parallelisation~\cite{Brass:2018xbv}.
	
	Further details on the setup for the numerical results achieved in this section are as follows:
	The $G_{\mu}$ scheme is applied for the computation of the electromagnetic coupling $\alpha$ for which light fermion mass logarithms are resummed in the
	incorporated running of $\alpha$.
	Regarding massive vector bosons, on-shell renormalisation conditions are imposed and particle widths set to zero. 
	Throughout the calculation, non-zero masses for all particles except for photon and neutrinos are applied, and the corresponding Yukawa couplings are included. For light fermions in loops this is merely a technical detail without phenomenological significance.
	However, keeping the muon mass non-zero regulates IR singularities from collinearly radiated photons off the initial-state.  In fact, QED corrections beyond NLO due to ISR at a muon collider are parametrically of order $(\alpha/\pi)^2\log^2(s/m_\mu^2) \sim 0.1\;\%$ according to the considerations with respect to Eq.~(\ref{estimatefNLOmassiveIS}). Hence, they can be considered as sufficiently suppressed in the particular context of this study. This allows us to treat the colliding $\mu^+\mu^-$ system perturbatively without the need for higher-order resummation by applying leptonic PDFs.
	
	The following numerical input parameters are used for the calculations with results presented in this section:
	\begin{equation*}
	G_{\mu}= 1.166379\cdot 10^{-5} \; \text{GeV}^{-2} \\
	\end{equation*}
	\begin{eqnarray*}
		m_u &=\; \phantom{17}0.062           \;\text{GeV}\qquad\qquad
		m_d &=\; 0.083           \;\text{GeV} \\
		m_c &=\; \phantom{17}1.67\phantom{0} \;\text{GeV}\qquad\qquad
		m_s &=\; 0.215           \;\text{GeV} \\
		m_t &=\;          172.76\phantom{0}  \;\text{GeV}\qquad\qquad
		m_b &=\; 4.78\phantom{0} \;\text{GeV}
	\end{eqnarray*}
	\begin{eqnarray*}
		M_W &=\; \phantom{1}80.379\phantom{0} \;\text{GeV} \quad\qquad
		m_e &=\;  0.0005109989461  \;\text{GeV}  \\
		M_Z &=\; \phantom{1}91.1876 \;\text{GeV}  \quad\qquad
		m_{\mu} &=\; 0.1056583745\phantom{000} \;\text{GeV} \\
		M_H &=\; 125.1\phantom{000} \;\text{GeV} \quad\qquad
		m_{\tau} &=\; 1.77686\phantom{00000000} \;\text{GeV}\qquad.
	\end{eqnarray*}
	
	For ensuring the correctness of the calculations, in addition to the checks of Sec.~\ref{secValidationee} with an analogous beam setup, technical sanity checks on the implemented FKS subtraction scheme for muon collider processes are performed. These include numerical checks for the IR cancellation in the soft limit and consistency checks. For the latter, integrated results using the FKS real phase-space parameterisation to an underlying Born process $\mu^+\mu^- \to X$ and the LO parameterisation of $\mu^+\mu^- \to X + \gamma$ with a well-defined photon, i.~e. using a (technical) energy cut $E_{\gamma}>10$ GeV, are compared.
	The derived $\mu^+\mu^- \to HZ$ EW Sudakov correction factor of App.~\ref{HZapprox} as well as reference results of \texttt{SHERPA}'s automated evaluation of EW Sudakov factors \cite{Bothmann:2020sxm} served as additional cross-checks on the presented differential distributions.
	
	In Sec.~\ref{scans} cross section scans in the collider energy for the multi-TeV range for $HZ$ and $ZZ$ production are shown. Sec.~\ref{secmultibosonMuon} contains total cross section results of two-, three- and four-boson processes at a muon collider for proposed collider energies, $\sqrt{s}= 3$, $10$ and $14$ TeV. For the process $\mu^+\mu^- \to HZ$ distributions differential in Higgs properties are presented and discussed in Sec.~\ref{differentialresults}.
		\subsection{Collider energy scans for cross sections of $HZ$ and $ZZ$ production}
	\label{scans}
\begin{figure}
	\centering
		\includegraphics[width=.3\textwidth]{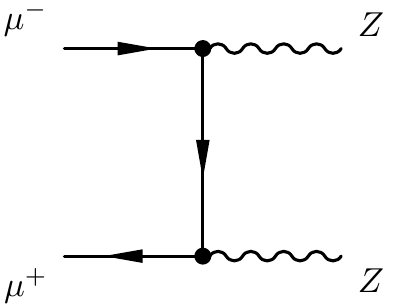}
		\includegraphics[width=.3\textwidth]{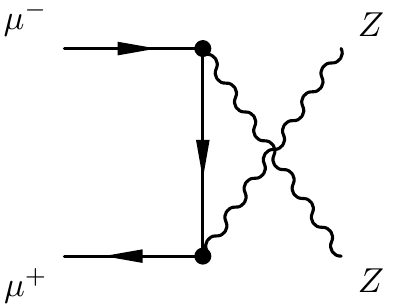}
		\includegraphics[width=.3\textwidth]{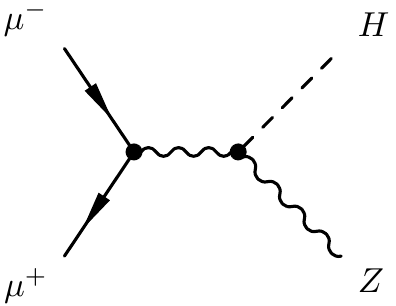}
	\\
		\includegraphics[width=.3\textwidth]{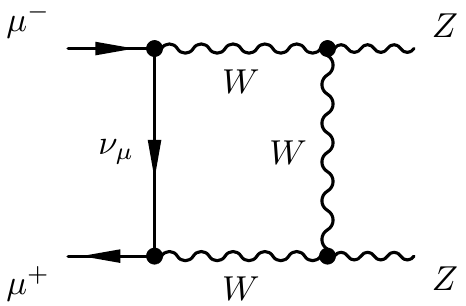}
	\qquad
		\includegraphics[width=.3\textwidth]{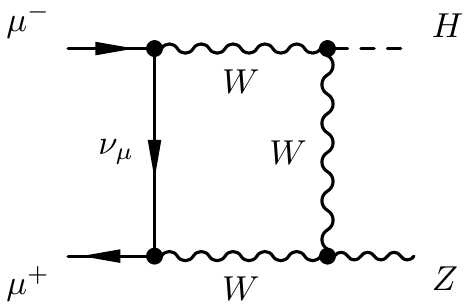}
	\caption{Upper row: tree-level diagrams for the processes
		$\mu^+\mu^-\rightarrow ZZ$ and to $\mu^+\mu^-\rightarrow HZ$
		(omitting contributions from the muon-Yukawa coupling). Lower row: representative one-loop
		diagrams for the virtual contribution to $\mu^+\mu^-\rightarrow
		ZZ$ and to $\mu^+\mu^-\rightarrow HZ$; cf.~\cite{Bredt:2022dmm}}
	\label{HZdiagram}
\end{figure}

	The simplest process class from which fundamental EW higher order
	perturbative effects can be understood is neutral di-boson production. These processes, by neglecting contributions associated with muon-Higgs Yukawa couplings, have a clear kinematical structure at both LO and NLO EW. The LO dominant contributions come from the $s$-channel diagram for $HZ$ and from $t/u$-channel diagrams for $ZZ$ production which are depicted in the upper row in Fig.~\ref{HZdiagram}. This classification of the two kind of processes is useful in order to distinguish different kinds of effects which are enhanced at NLO EW as explained below. Furthermore, for the neutral di-boson processes QED radiation in the corresponding real-emission diagrams exclusively comes from the initial state at NLO in $\alpha$.
	The lower row of Fig.~\ref{HZdiagram} depicts typical one-loop diagrams, underlining the
	fact that with respect to the virtual one-loop diagrams these processes are closely related.

	\begin{figure}
		\centering
		\includegraphics[width=0.9\textwidth]{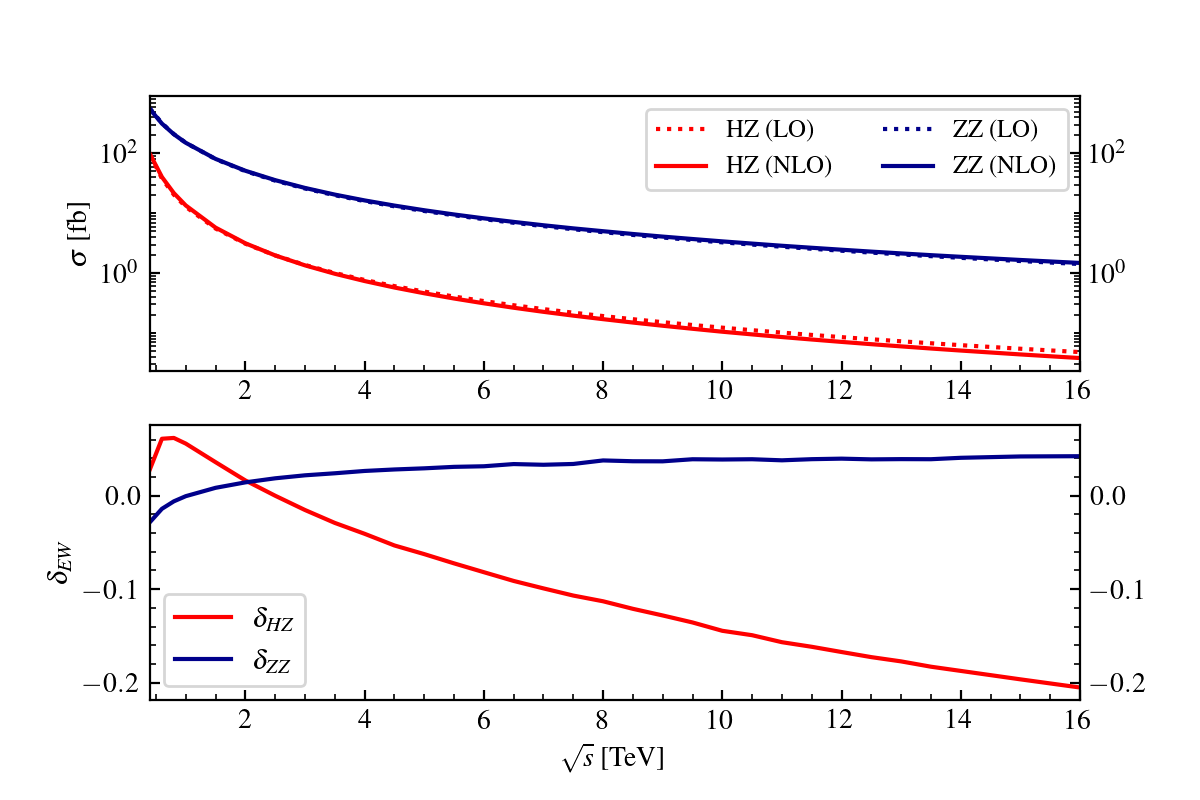}
		\caption{LO and NLO inclusive cross section scans in
			$\sqrt{s}$ for $HZ$ and $ZZ$ production in the upper plot;
			relative NLO correction $\delta_{\text{EW}}$ in the lower
			plot}
		\label{HZaZZ}
	\end{figure}
	The relative corrections $\delta_{\text{EW}}$ (lower plot) for both processes can be interpreted by scans over $\sqrt{s}$ for LO and NLO inclusive cross sections of $HZ$ and $ZZ$ production at the muon collider, shown in
	Fig.~\ref{HZaZZ} (upper plot).
	
	First, consider the Higgsstrahlung process $\mu^+\mu^-\to HZ$.  A large suppression in Fig.~\ref{HZaZZ} which increases in size starting from the peak of the cross section at $\sqrt{s}\sim 0.8$ TeV can be observed which can be attributed to large virtual effects as will be shown in the following.
	
	In order to understand the behaviour in the regime of high centre-of-mass energies, in general the approximation of EW Sudakov logarithmic correction factors can be applied for which pioneering works have been done in
	\cite{Kuhn:1999nn,Denner:2000jv,Denner:2001gw,Bell:2010gi}.
	In the Sudakov limit 
	\begin{align}
		r_{kl}= (p_k+p_l)^2 \sim s \gg M_W^2 
		\label{sudakovlimit}
	\end{align}
	with external states $k$ and $l$ and by using a fictitious photon mass $\lambda=M_W$ these correction factors effectively correspond to the EW virtual contributions in addition to real corrections with photon transverse momenta smaller than a cut-off scale at the order of $M_W$. They represent form factors in terms of double and single logarithms of the ratio $r_{kl}/M_W^2$ which are factorised in the soft and/or collinear limit.
	
	The exact treatment of QED IR subtraction, which is mandated for FKS subtraction, naturally demands photons to be massless. The virtual loop contributions in this way contain diagrams with massless photon exchange regulated by the IR scheme. In contrast, virtual contributions with massive weak vector boson exchange are still regularised by their masses which are at the EW scale $M_W$. Therefore, for large $\sqrt{s}$, these contributions are implicitly contained in the EW next-to-leading logarithmic (NLL) Sudakov factors.
	Combining FKS QED subtraction and these NLL EW Sudakov factors in order to obtain inclusive results comparable with those shown in Fig.~\ref{HZaZZ} in general requires a consistent matching of these for an NLO calculation:
	The subtraction of QED IR poles in the FKS virtual terms is defined via $\overline{\text{MS}}$ subtraction at the scale $\mu=\mu_R$ such that remnant finite terms from  virtual loop contributions receive a logarithmic dependence on this scale. More details on this are given within the description of virtual subtraction terms in the FKS formalism presented in Sec.~\ref{secIntegratedSubs}.
	Using an EW Sudakov approximation with the cut-off scale of the effective photon mass $\lambda=M_W$ in place of these virtual (finite) loop contributions however yields missing terms of order $\alpha\log(\mu_R/M_W)$ for the complete NLO computation \cite{Granata:2017iod}.
	Because of the technical challenges of this combined treatment of FKS QED subtraction and virtual EW Sudakov approximation factors, it is considered as beyond the scope of this work. For a completely automated treatment of EW Sudakov correction factors in the FKS framework see Ref.~\cite{Pagani:2021vyk}. The EW Sudakov factor to $\mu^+\mu^-\to HZ$, discussed in the following, thus merely serves for purposes of explanations.
	
	In the following, it will be shown that the effective EW virtual loop contributions of $HZ$, implicitly contained in the respective results of Fig.~\ref{HZaZZ}, are quantitatively approximated well enough by the Sudakov approach.
	\begin{figure}
		\centering
		\includegraphics[width=0.9\textwidth]{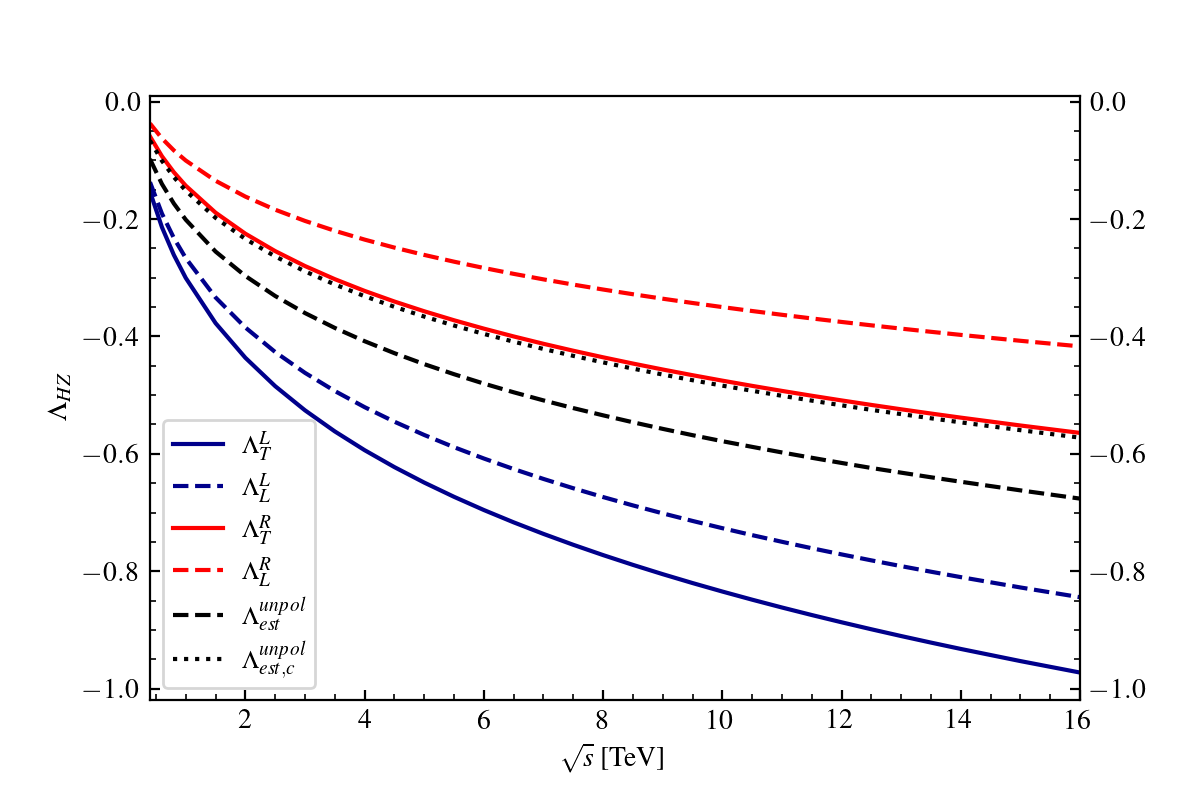}
		\caption{Sudakov factors $\Lambda^{\kappa}_{\lambda}$ for muon chiralities $\kappa=L,R$ and $Z$ polarisations $\lambda=T,L$ and
			estimated unpolarised correction factor
			$\Lambda_{\text{est}}^{\text{unpol}}$ at
			$\theta_{H}=90^{\circ}$ as well as
			$\Lambda^{\text{unpol}}_{\text{est,c}}$ (without angular dependent terms) for $HZ$ production
			at the muon collider as a function of the collider
			energy, $\sqrt{s}$.}
		\label{sudpic}
	\end{figure}
	To this end, the NLL Sudakov form factor to $\mu^+\mu^- \to HZ$ is derived in
	App.~\ref{HZapprox} by using analytical results to the process $q\bar{q}\rightarrow HZ$~\cite{Granata:2017iod} in analogy. For the final result we arrive at the estimate of
	Eq.~(\ref{sudoverall}) with all technical details and considerations explained in App.~\ref{HZapprox}.
	This Sudakov factor depending on
	$\sqrt{s}$ for a fixed polar angle $\theta_H=90^{\circ}$ of the Higgs as
	shown in Figure~\ref{sudpic} exhibit large suppressions in particular
	for left-handed muons in the initial state due to the larger
	coupling to $SU(2)$ interactions. Note, that the results in Fig.~\ref{sudpic} are not representing Sudakov approximations for inclusive cross sections (integrated over $\theta_{H}$). This would either require the technical implementation of these factors in \texttt{WHIZARD} in order to multiply them with the corresponding Born contributions on amplitude-level or a completely analytical approach. In addition, results according to this would also miss terms related to the matching with QED FKS subtraction, as explained above, in order to give a completely accurate description of the virtual EW corrections approximated for results of Fig.~\ref{HZaZZ}. As the factors of Fig.~\ref{sudpic} are presented merely for purposes of clarifications the angular integration is left out here and the following considerations are made.
	
	In a further approach we approximated the unpolarised correction factor $\Lambda^{\text{unpol}}_{\text{est,c}}$ for which the angular dependent part of Eq.~(\ref{sudangle}) is excluded, shown as black dotted curve. This correction factor corresponds to the amount to which at least the inclusive result due to the virtual loop corrections is suppressed.
	According to Eq.~(\ref{sudangle}), the angular-dependent part of the Sudakov factor is negative, of subleading logarithmic type and amounts to about $-17\%$ at $\sqrt{s}=16
	\text{ TeV}$ for angles $\theta_H$ in the perpendicular plane,
	i.~e. for $\theta_H\sim90^{\circ}$.
	For these angles also the Born process is
	enhanced according to
	\cite{Chanowitz:1985hj,Bohm:2001yx}
	\begin{align}
	\left(\frac{d\sigma}{d\Omega}\right)_{\text{Born}}\propto
	\frac{\beta
		M_Z^2}{(s-M_Z^2)^2}\left(\frac{s\beta^2}{8M_Z^2}\sin^2\theta_{H}+1\right)
	\quad.
	\label{bornHZcrosssec}
	\end{align}
	which can be observed as well in the differential
	cross sections for the Born case presented in
	Sec.~\ref{differentialresults} within Fig.~\ref{thetadist}.
	
	The	estimated unpolarised correction factor, $\Lambda^{\text{unpol}}_{\text{est}}$,  given by Eq.~(\ref{finaldeltaunpol}) including the angular-dependent terms with a chosen polar angle $\theta_{H}=90^{\circ}$, is shown as black dashed curve in Fig.~\ref{sudpic}. This decreases the cross section with a largest suppression of about $-65\%$ for high energies.
	Concerning the magnitude of $\Lambda^{\text{unpol}}_{\text{est}}$ at
	this angle relatively to $\Lambda^{\text{unpol}}_{\text{est},c}$ as
	well as the similar enhancement behaviour at angles around $\theta_{H}=90^{\circ}$ for the Born differential in $\theta_{H}$ the relative inclusive virtual corrections to $HZ$ production can be estimated to be of the order of $\Lambda^{\text{unpol}}_{\text{est}}$. The counteracting effects of QED radiation in an inclusive calculation  will be discussed in the following.
	
	\begin{figure}
		\centering
		\includegraphics[width=0.7\textwidth]{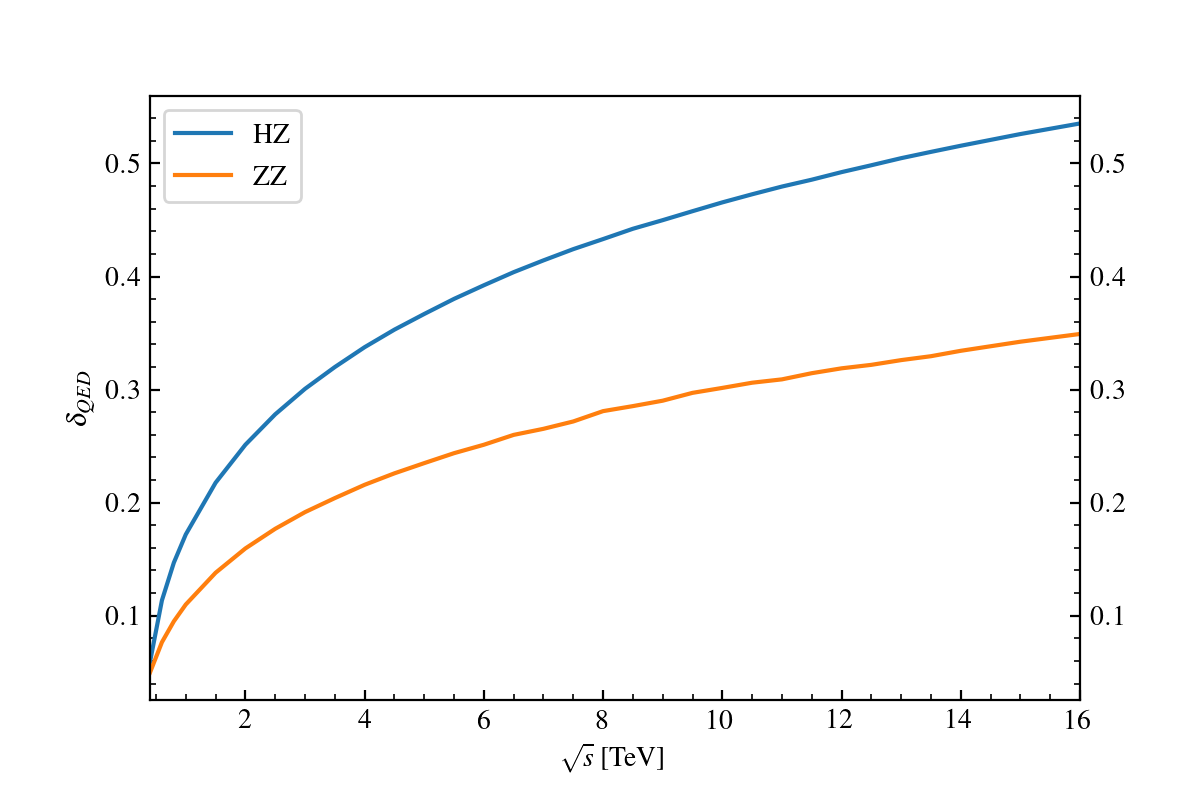}
		\caption{Relative QED corrections
			$\delta_{QED}=\sigma^{\text{incl}}_{\text{NLO,QED}}/\sigma^{\text{incl}}_{\text{LO}}-1$
			to $HZ$ and $ZZ$ production at the muon collider as a function of
			the collider energy, $\sqrt{s}$.}
		\label{HZQED}
	\end{figure}

	For including only pure NLO QED corrections in the calculation the resulting relative correction $\delta_{QED}$ as depicted in Fig.~\ref{HZQED} is positive and growing with $\sqrt{s}$. About the
	contributing NLO parts we can make the following qualitative statements:
	In general, virtual loop-tree interfering amplitudes are supposed to give negative contributions such that a positive overall NLO correction factor can be explained by dominating real corrections. In particular, since the main contribution for $HZ$ production at Born-level is $s$-channel induced, large QED real corrections are expected due to large amplitudes for hard photons radiated in forward
	direction, the effect of which is enhanced with growing $\sqrt{s}$. This explains the bulk of the correction factor to be seen in Figure~\ref{HZQED} as the blue curve.
	From the superposition of these two effects, the negative EW Sudakov logarithms overcompensating the QED real corrections, a reduction of $\delta_{{HZ}}$ down to a range of about $ -20~\%$ to $-10~\%$ at high energies in Figure~\ref{HZaZZ} is found to be reasonable.
	
	For $\mu^+\mu^-\to ZZ$ a detailed discussion about the composition of pure weak and QED parts of the NLO inclusive corrections is left out here since the process compared to $HZ$ production is more intricate from its angular dependency at LO as well as at NLO. Explicitly, the result for the Sudakov
	correction factors of the analogous process $e^+e^-\rightarrow ZZ$ given in \cite{Denner:2000jv} cannot be related straightforwardly to an estimation for the Sudakov suppression of the inclusive cross section.
	This is due to the completely different
	angular dependence of the Sudakov factor depicted in Fig.~7 within Ref.~\cite{Denner:2000jv} which is minimal for $Z$ polar angles around $90^{\circ}$ compared to the Born amplitudes with the largest contributions at angles around the beam axis.
	Moreover, considering the suppression  $\delta_{QED,ZZ}$ relative to $\delta_{QED,HZ}$ in Fig.~\ref{HZQED} leads to the conclusion that the impact of real emission amplitudes with hard photon radiation on the relative NLO correction is reduced compared to that of $HZ$ production.
	This can be traced back to the effect that for $ZZ$ production tree-level amplitudes due to the forward scattering of the $Z$ bosons at high energies are enhanced already at Born-level due to the dominant $t/u$-channel process. The positively contributing QED real corrections due to hard photon radiation thus yield a mitigated enhancement of the cross section compared to the $s$-channel Higgsstrahlung process.
	In Fig.~\ref{HZaZZ} an overall positive correction $\delta_{ZZ}$ can be seen. This can be explained by an overcompensation of the EW Sudakov effects by the QED real corrections in the inclusive results.
	
	\subsection{Total cross sections for multi-boson processes}
	\label{secmultibosonMuon}
	In the same way as for $\mu^+\mu^-\to HZ$ and $\mu^+\mu^-\to ZZ$ NLO EW corrections can be computed for all other possible combinations of final-state bosons. Numerical results of the LO and NLO inclusive cross sections for all these combinations in two-, three- and four-boson production processes at $\sqrt{s}= 3$ TeV are presented in Table \ref{table3TeV}. Table \ref{table10TeV} and \ref{table14TeV} contain the corresponding results for two- and three-boson production at $\sqrt{s}= 10$ and $14$ TeV.
	For very high centre-of-mass energies as well as high EW boson multiplicities, the fixed-order perturbative approximation breaks down.
	To this end, the computations for four-boson production at $\sqrt{s}= 10$ and $14$ TeV are omitted which render meaningful results only by taking appropriate EW higher order resummation approaches, e.~g. by soft-collinear effective field theory, into account.
	
	In order to investigate collinear ISR effects influencing the presented NLO EW results from a kinematical point of view, the LL lepton PDF with its analytical form given in Eq.~(\ref{LLsoftPDF}) is applied to the LO cross sections.
	Note, that for the purpose of computing inclusive fixed-order cross sections, higher-order QED contributions beyond NLO are parametrically subleading, as explained above.  It can be seen that the PDF convolution captures the dominant QED-radiation effect within the NLO cross section, to be compared with the complete NLO EW results.
	
	For each collider energy, the cross sections computed by using ISR structure functions are displayed in
	Table~\ref{table3TeVISR}, \ref{table10TeVISR} and \ref{table14TeVISR}. The relative correction $\delta_{\text{ISR}}$ is
	defined as
	\begin{align}
	\delta_{\text{ISR}} =
	\frac{\sigma_{\text{LO+ISR}}^{\text{incl}}}{\sigma_{\text{LO}}^{\text{incl}}}-1
	\qquad .
	\end{align}
	
	As mentioned above, for inclusive observables QED effects beyond NLO are suppressed compared to those of NLO EW corrections to the hard scattering process. However, resummation methods in general improve the numerical accuracy of predicted cross sections. They become essential, in particular, when computing exclusive observables. Beyond collinear resummation via the application of lepton PDFs, higher-order soft-photon radiation effects are captured by YFS resummation methods \cite{Yennie:1961ad}.
	In addition, for high-energy muon colliders EW PDFs might play a phenomenological role~\cite{Han:2020uid}. Their effects are not taken into account here.
	
	A few general remarks on the discussion of the LO and NLO EW results:
	First of all, we observe that for all processes the absolute value of the cross section decreases with $\sqrt{s}$ and with the number of bosons in the final state. Apart from the processes with pure Higgs final-state, these range from $\sim10^{-4}$ to $\sim10^2$ fb. As the
	cross sections for the tree-level processes of $ HH$ and $HHH$ production are numerically below $\mathcal{O}(10^{-7}\text{ fb})$, owing to the mass suppressed muon-Higgs couplings at the collider energies used for this simulation, a detailed discussion on the respective results is left out here.
	The abnormally  large corrections to cross sections labeled with `$*$' in in Table~\ref{table3TeV}, \ref{table10TeV} and \ref{table14TeV} at this fixed order can be related to missing squared one-loop amplitudes with e.~g. $W^{\pm}$
	exchange. These phenomenologically have to be taken into account but require cross section calculations beyond fixed coupling powers if considered in the predictions. However, although not
	relevant with respect to the small size of the cross sections at tree-level, the LO and formal NLO $\mu^+\mu^-\to
	HH$ and $\mu^+\mu^-\to HHH$ numerical cross section results are included in these tables for completeness.
	
	\begin{table}
		\centering
		\small
		\begin{tabularx}{0.85\textwidth}{l|r|r|r}
			$\mu^+\mu^-\to X, \sqrt{s}=3$ TeV  &
			$\sigma_{\text{LO}}^{\text{incl}}$ [fb]  &
			$\sigma_{\text{NLO}}^{\text{incl}}$ [fb] &
			$\delta_{\text{EW}} $ [\%]\\
			&&&
			\\\hline\hline
			$W^+W^-$       &        $  4.6591(2)\cdot 10^2 $      &   $ 4.847(7) \cdot 10^2 $    & $ +4.0(2) $ \\
			$ZZ$      &       $ 2.5988(1) \cdot 10^1$      &  $ 2.656(2) \cdot 10^1 $    & $ +2.19(6) $ \\
			$HZ$   &     $ 1.3719(1) \cdot 10^0 $   &   $ 1.3512(5) \cdot 10^0 $ & $  -1.51(4)$ \\
			$HH$  &    $ 1.60216(7) \cdot 10^{-7} $
			&  $ 5.66(1)\cdot 10^{-7}~^* $     & $  $
			\\\hline
			
			$ W^+W^-Z$  &   $ 3.330(2) \cdot 10^1 $       &     $ 2.568(8) \cdot 10^1 $   &   $ -22.9(2) $ \\
			$W^+W^-H$      &     $ 1.1253(5) \cdot 10^0 $    &      $ 0.895(2)\cdot 10^{0} $   &  $ -20.5(2) $ \\
			$ZZZ$       &   $ 3.598(2) \cdot 10^{-1} $     &  $ 2.68(1) \cdot 10^{-1} $   &  $ -25.5(3)$\\
			$HZZ$      &    $  8.199(4)\cdot 10^{-2} $  &   $ 6.60(3)\cdot 10^{-2} $    & { $ -19.6(3) $} \\
			$ HHZ$   & $ 3.277(1) \cdot 10^{-2} $ &  $ 2.451(5) \cdot 10^{-2} $ & $ -25.2(1)$\\
			$ HHH$     &   $ 2.9699(6) \cdot
			10^{-8} $  &   $ 0.86(7) \cdot 10^{-8}~^* $  &  $  $
			\\\hline
			
			$W^+W^-W^+W^-$   &   $ 1.484(1) \cdot 10^0 $  &    $  0.993(6)\cdot 10^0 $  &   $ -33.1(4) $ \\
			$W^+W^-ZZ$     &   $  1.209(1)\cdot 10^0 $  & $ 0.699(7) \cdot 10^0 $ &  $ -42.2(6) $ \\
			$W^+W^-HZ$      &       $ 8.754(8)  \cdot 10^{-2} $ & $ 6.05(4) \cdot 10^{-2} $ & $ -30.9(5) $\\
			$W^+W^-HH$    &   $ 1.058(1) \cdot 10^{-2} $      &   $ 0.655(5)\cdot 10^{-2} $   &  $ -38.1(4) $\\
			$ZZZZ$     &   $ 3.114(2) \cdot 10^{-3} $    &     $ 1.799(7)\cdot 10^{-3} $   &  $ -42.2(2) $\\
			$HZZZ$     &   $ 2.693(2)\cdot 10^{-3} $    &     $ 1.766(6)\cdot 10^{-3} $   &  $ -34.4(2) $\\
			{$HHZZ$}     &   $ 9.828(7) \cdot 10^{-4} $    &     $ 6.24(2)  \cdot 10^{-4} $   & {$ -36.5(2)$}\\
			$HHHZ$     &   $ 1.568(1) \cdot 10^{-4} $   &      $ 1.165(4) \cdot 10^{-4} $   &  $ -25.7(2) $
		\end{tabularx}
		\caption[muon collider NLO EW]{Total inclusive cross sections
			at LO and NLO EW with corresponding relative corrections
			$\delta_{\text{EW}}$, for two-, three- and four-boson
			production at $\sqrt{s}= 3$ TeV. For (*), see remarks in the text.}
		\label{table3TeV}
	\end{table}

	\begin{table}
		\centering
		\small
		\begin{tabularx}{0.85\textwidth}{l|r|r|r}
			$\mu^+\mu^-\to X, \sqrt{s}=10$ TeV  &
			$\sigma_{\text{LO}}^{\text{incl}}$ [fb]  &
			$\sigma_{\text{NLO}}^{\text{incl}}$ [fb] &
			$\delta_{\text{EW}}$ [\%]\\
			&&&\\
			\hline\hline
			$W^+W^-$        &       $ 5.8820(2) \cdot 10^1 $  &    $ 6.11(1) \cdot 10^1 $    & $ +3.9(2) $ \\
			$ZZ$        &       $ 3.2730(4) \cdot 10^0$     &    $ 3.401(4)\cdot 10^0 $    & $+3.9(1) $ \\
			$HZ$   &     $ 1.22929(8) \cdot 10^{-1} $  &   $  1.0557(8)\cdot 10^{-1} $ & $ -14.12(7) $ \\
			$HH$ &          $ 1.31569(5) \cdot
			10^{-9} $  &     $ 42.9(4)\cdot 10^{-9}~^{*} $     & 
			\\\hline
			
			$ W^+W^-Z$   &   $ 9.609(5) \cdot 10^{0} $      &     $  5.86(4)\cdot 10^0 $   &   $ -39.0(2) $ \\
			$W^+W^-H$           &     $ 2.1263(9)\cdot 10^{-1} $    &      $ 1.31(1)\cdot 10^{-1} $   &  $ -38.4(5)  $ \\
			$ZZZ$        &   $ 8.565(4) \cdot 10^{-2} $  &  $ 5.27(8) \cdot 10^{-2} $   &  $ -38.5(9)$\\
			{$HZZ$}       &    $  1.4631(6)\cdot 10^{-2} $  &  $0.952(6)\cdot 10^{-2} $    & { $ -34.9(4) $} \\
			$ HHZ$   & $ 6.083(2) \cdot 10^{-3} $  &$  2.95(3)\cdot 10^{-3} $ & $ -51.6(5) $\\
			$ HHH$     &   $ 2.3202(4) \cdot 10^{-9} $  &   $ -1.0(2) \cdot 10^{-9}~^{*}  $  & 
			\end{tabularx}
			\caption[muon collider NLO EW]{Total inclusive cross sections
				at LO and NLO with corresponding relative correction
				$\delta_{\text{EW}}$ for di- and tri-boson production at
				$\sqrt{s}= 10$ TeV. For (*), see remarks in the text.}
			\label{table10TeV}
			\end{table}

			\begin{table}
			\centering
			\small
			\begin{tabularx}{0.87\linewidth}{l|r|r|r}
			$\mu^+\mu^-\to X, \sqrt{s}=14$ TeV  &
			$\sigma_{\text{LO}}^{\text{incl}}$ [fb]  &
			$\sigma_{\text{NLO}}^{\text{incl}}$ [fb] &
			$\delta_{\text{EW}}$ [\%]\\
			&&&\\
			\hline\hline
			$W^+W^-$        &       $  3.2423(1)\cdot 10^1 $   &     $  3.358(8)\cdot 10^1 $    & $ +3.6(2) $ \\
			$ZZ$       &       $ 1.80357(9) \cdot 10^0$   &     $ 1.872(4)\cdot 10^0 $    & $ +3.8(2)$ \\
			$HZ$   &     $  6.2702(4)\cdot 10^{-2} $  &   $  5.097(6)\cdot 10^{-2} $ & $ -18.7(1) $ \\
			$HH$  &          $ 3.4815(1) \cdot 10^{-10} $  &     $
			217(2)\cdot 10^{-10}~^{*} $     & $ $
			\\\hline
			
			$ W^+W^-Z$  &    $ 6.369(3) \cdot 10^{0} $    &        $  3.51(3)\cdot 10^0 $   &   $ -45.0(4) $ \\
			$W^+W^-H$           &     $ 1.2846(6)\cdot 10^{-1} $   &      $ 0.73(1)\cdot 10^{-1} $   &  $ -43.3(9) $ \\
			$ZZZ$        &   $  5.475(3)\cdot 10^{-2} $   &   $  3.06(3)\cdot 10^{-2} $   &  $ -44.2(6)$\\
			{$HZZ$}      &    $  8.754(4)\cdot 10^{-3} $ &    $5.28(3)\cdot 10^{-3} $    & { $ -39.7(4) $} \\
			$ HHZ$    & $  3.668(1)\cdot 10^{-3} $  & $  1.49(1)\cdot 10^{-3} $ & $ -59.4(3)$\\
			$ HHH$     &   $  1.1701(2)\cdot 10^{-9} $  &   $  -0.739(8)\cdot 10^{-9}~^{*} $  &  $  $\\
			\end{tabularx}
			\caption[muon collider NLO EW]{Total inclusive cross sections
				at LO and NLO with corresponding relative correction
				$\delta_{\text{EW}}$ for di- and tri-boson production at
				$\sqrt{s}= 14$ TeV. For (*), see remarks in the text.}
			\label{table14TeV}
			\end{table}
			
			For the di-boson processes $ZZ$ and $HZ$ production a discussion on inclusive cross sections is given already in detail in the analysis of the previous chapter. However, the different kinematical effects of these processes are highlighted by several aspects given below.
			
			$W^+W^-$ production with dominant contributions from $t$-channel diagrams at Born level and high energies is similar to $ZZ$ exept for the possibility for photon radiation off the final-state at NLO. This can induce quasi-collinear effects in $W^{\pm}\to W^{\pm}\gamma$ splittings which increase with the energy scale of the process.
			However, this is a minor effect compared to large real-emission amplitudes due to hard photon quasi-collinear radiation off the light massive muons. Another difference to $ZZ$ is that the $W$ bosons can have longitudinal gauge boson polarisations corresponding to two charged Goldstone bosons in the final-state for which the $s$-channel process is dominant. This contribution is suppressed with $1/s$ \cite{Beenakker:1993tt,Beenakker:1994vn}.
			These considerations may explain the similarity of the relative corrections $\delta_{\text{EW}}$ of the two gauge boson pair production processes at high energies $10$ and $14$ TeV and their deviation at $3$ TeV.
			
			Quantitatively, the influence of the kinematics which causes the main difference in $\delta_{\text{EW}}$ of gauge boson pair production $WW$ and $ZZ$ compared to Higgsstrahlung $HZ$ at the considered collider energies can be seen as well in their different $\delta_{\text{ISR}}$ factors listed in Table \ref{table3TeVISR}, \ref{table10TeVISR} and \ref{table14TeVISR}. According to these the correction due to ISR resummation for $HZ$ is approximately twice as big as for $ZZ$ and $WW$ and grows with the energy. This originates from the same effect as explained for the comparison of NLO QED cross sections with respect to Fig.~\ref{HZQED}:
			We note, that NLO QED as well as NLO EW corrections implicitly involve terms at $\mathcal{O}(\alpha)$ of the LL PDF expansions containing the leading logarithms which are dominant with respect to ISR effects.
			These positive $\mathcal{O}(\alpha\ln (Q^2/m_{\mu}^2))$ contributions get large due to hard collinear photon ISR inducing a boost of the photon recoil-system along the beam axis which is enhanced with the collider energy.
			For $HZ$ production this induces a radiative return back to the threshold. In this way, the cross section at NLO gets largely enhanced as the Higgsstrahlung cross sections at Born-level receive dominant contributions from the $s$-channel diagrams which fall off with $1/s$. For $VV$ production however this $1/s$ fall-off at Born-level gets damped by large contributions from amplitudes with small scattering angles of the gauge bosons. Hence, at NLO the radiative return leads to a less pronounced enhancement of the cross sections for gauge-boson pair production.
			
			\begin{figure}
			\centering
			\includegraphics[width=.33\textwidth]{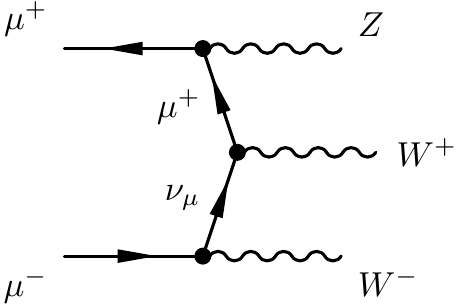}
			\qquad\qquad
			\includegraphics[width=.33\textwidth]{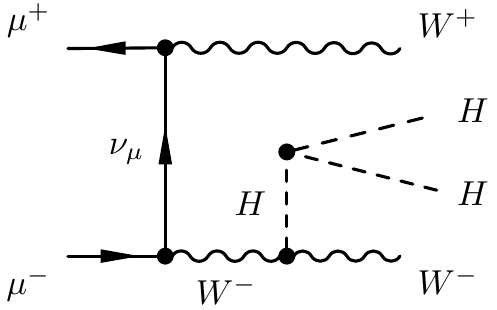}
			\\
			\quad\\
			\includegraphics[width=.37\textwidth]{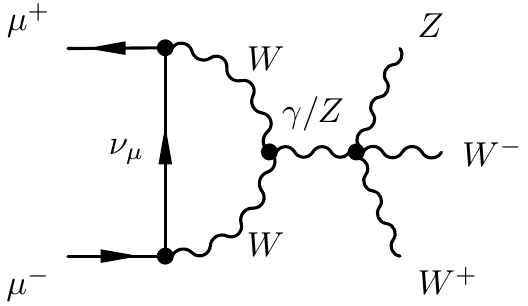}
			\qquad\qquad
			\includegraphics[width=.37\textwidth]{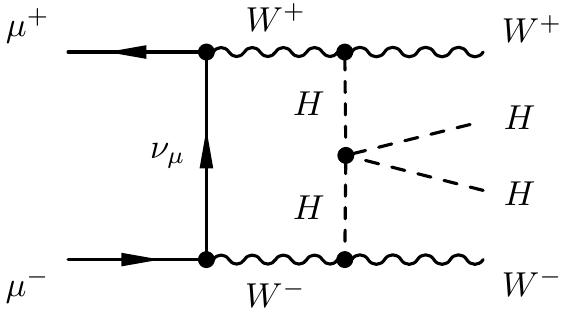}
			\caption{Upper row: exemplary tree-level diagrams for the processes
				$\mu^+\mu^-\rightarrow W^+W^-Z$ (left) and to
				$\mu^+\mu^-\rightarrow W^+W^-HH$ (right). Lower row: one-loop diagrams for the virtual contribution to $\mu^+\mu^-\rightarrow
				W^+W^-Z$ (left) and to $\mu^+\mu^-\rightarrow W^+W^-HH$ (right); cf.~\cite{Bredt:2022dmm}}
			\label{feyndiag_2to34}
			\end{figure}
			
			For the results of three- and four-boson production processes, for which exemplary Feynman diagrams at tree- and one-loop level are depicted in Fig.~\ref{feyndiag_2to34}, the following observations can be made.
			For all triple-boson processes at $3$ TeV except for pure Higgs final-state production, the NLO EW cross sections are suppressed by $-25\%$ to $-20\%$ relatively to the LO results, to be seen in Table~\ref{table3TeV}.
			That the absolute values of $\delta_{\text{EW}}$ of the di-boson processes in general are smaller than those of the triple-boson processes can be explained by the fact that negative Sudakov logarithm factors add up for all kinematic invariants of external states \cite{Denner:2001gw}.
			For four-boson final-state processes, this effect is enhanced, seen by $\delta_{\text{EW}}$ ranging from $-26~\%$ to $ - 42~\%$. In addition, Table \ref{table3TeVISR}, \ref{table10TeVISR} and 	\ref{table14TeVISR} show that with increasing number of bosons in the final-state the enhancement due to ISR effects gets damped.
			This originates from the multi-peripheral kinematics of these hard processes. Similar to the discussion above on the di-boson processes: the radiative return is not so much pronounced as there is not a single dominating threshold to return to, except for the Higgs- and multi-Higgsstrahlung processes. For the latter class of processes a larger enhancement can be seen from their $\delta_{\text{ISR}}$ values in comparison with those of other processes at fixed number of bosons and beam energies.
			Consequently, concerning the NLO EW results the dominant contribution to $\delta_{\text{EW}}$ of the three-boson and nearly the complete size of $\delta_{\text{EW}}$ of the four-boson processes can be led back to negative EW virtual final-state correction factors.
			
			\begin{table}
			\centering
			\small
			\begin{tabularx}{0.8\textwidth}{l|r|r|r}
			$\mu^+\mu^-\to X, \sqrt{s}=3$ TeV  & $\sigma_{\text{LO}}^{\text{incl}}$ [fb] & $\sigma_{\text{LO+ISR}}^{\text{incl}}$ [fb]  & $\delta_{\text{ISR}} $ [\%]\\
			&&&\\
			\hline\hline
			$W^+W^-$       &       $  4.6591(2)\cdot 10^2 $      & $5.303(2)\cdot 10^2$    & $ +13.82(4) $ \\
			$ZZ$        &       $ 2.5988(1) \cdot 10^1$      &  $3.007(1) \cdot 10^1 $    & $ +15.71(4) $ \\
			$HZ$  &     $ 1.3719(1) \cdot 10^0 $   & $1.7868(4)
			\cdot 10^0 $ &  $ +30.24(3) $ \\
			\hline
			%
			$ W^+W^-Z$  &   $ 3.330(2) \cdot 10^1 $     &  $ 3.427(2)\cdot 10^1 $     &   $ +2.90(9) $ \\
			$W^+W^-H$     &     $ 1.1253(5) \cdot 10^0 $    & $1.2052(7)\cdot 10^0$ &        $ +7.10(8) $ \\
			$ZZZ$      &   $ 3.598(2) \cdot 10^{-1} $    &  $3.786(2)\cdot 10^{-1} $  &   $+5.24(8) $\\
			$HZZ$      &    $  8.199(4)\cdot 10^{-2} $  & $ 8.887(5)\cdot 10^{-2} $ &   { $ +8.39(8) $} \\
			$ HHZ$   & $ 3.277(1) \cdot 10^{-2} $ & $3.525(2)\cdot 10^{-2}$ &  $ +7.58(7)$\\
			\hline
			$W^+W^-W^+W^-$   &   $ 1.484(1) \cdot 10^0 $  & $1.465(1)\cdot 10^0$ &     $ -1.3(1) $ \\
			$W^+W^-ZZ$     &   $  1.209(1)\cdot 10^0 $ & $1.187(1)\cdot 10^0$ &  $ -1.8(1) $ \\
			$W^+W^-HZ$     &       $ 8.754(8)  \cdot 10^{-2} $ & $8.742(8)\cdot 10^{-2}$ &  $ -0.1(1) $\\
			$W^+W^-HH$    &   $ 1.058(1) \cdot 10^{-2} $      & $1.076(1)\cdot 10^{-2}$  &   $ +1.7(1) $\\
			$ZZZZ$     &   $ 3.114(2) \cdot 10^{-3} $    & $ 3.139(3) \cdot 10^{-3} $ &     $ +0.8(1) $\\
			$HZZZ$      &   $ 2.693(2)\cdot 10^{-3} $    & $2.730(2)\cdot 10^{-3}$ &    $ +1.4(1) $\\
			{$HHZZ$}     &   $ 9.828(7) \cdot 10^{-4} $    & $10.042(8)\cdot 10^{-4}$   & {$ +2.2(1) $}\\
			$HHHZ$      &   $ 1.568(1) \cdot 10^{-4} $   &  $1.657(1)\cdot 10^{-4}$    &  $ +5.7(1) $\\
			\end{tabularx}
			\caption[muon collider NLO EW]{Total inclusive cross sections
				at LO with LL ISR photon resummation, and including relative correction
				$\delta_{\text{ISR}}$ for two-, three- and four-boson production at
				$\sqrt{s}= 3$ TeV.}
			\label{table3TeVISR}
			\end{table}
			
			\begin{table}
			\centering
			\small
			\begin{tabularx}{0.83\textwidth}{l|r|r|r}
			$\mu^+\mu^-\to X, \sqrt{s}=10$ TeV  & $\sigma_{\text{LO}}^{\text{incl}}$ [fb] &  $\sigma_{\text{LO+ISR}}^{\text{incl}}$ [fb]  & $\delta_{\text{ISR}}$ [\%]\\
			&&&\\
			\hline\hline
			$W^+W^-$       &       $ 5.8820(2) \cdot 10^1 $  & $7.295(7)\cdot 10^1$      & $ +24.0(1) $ \\
			$ZZ$      &       $ 3.2730(4) \cdot 10^0$   & $4.119(4)\cdot 10^0$      & $+25.8(1) $ \\
			$HZ$  &     $ 1.22929(8) \cdot 10^{-1} $ & $1.8278(5)\cdot 10^{-1}$  & $ +48.69(4) $ \\
			\hline
			$ W^+W^-Z$  &   $ 9.609(5) \cdot 10^{0} $      &  $10.367(8)\cdot 10^{0}$    &   $ +7.9(1) $ \\
			$W^+W^-H$     &     $ 2.1263(9)\cdot 10^{-1} $    & $2.410(2)\cdot 10^{-1}$   &  $ +13.3(1)  $ \\
			$ZZZ$     &   $ 8.565(4) \cdot 10^{-2} $  &  $9.431(7)\cdot 10^{-2}$     &  $+10.1(1) $\\
			{$HZZ$}      &    $  1.4631(6)\cdot 10^{-2} $  & $1.677(1)\cdot 10^{-2}$      & { $ +14.62(8) $} \\
			$ HHZ$  & $ 6.083(2) \cdot 10^{-3} $ & $6.916(3)\cdot 10^{-3}$  & $ +13.68(6) $
			\end{tabularx}
			\caption[muon collider NLO EW]{Total inclusive cross sections
				at LO with LL ISR photon resummation and relative correction
				$\delta_{\text{ISR}}$ for two- and three-boson production at
				$\sqrt{s}= 10$ TeV.}
			\label{table10TeVISR}
			\end{table}
			\begin{table}
			\centering
			\small
			\begin{tabularx}{0.82\linewidth}{l|r|r|r}
			$\mu^+\mu^-\to X, \sqrt{s}=14$ TeV & $\sigma_{\text{LO}}^{\text{incl}}$ [fb] & $\sigma_{\text{LO+ISR}}^{\text{incl}}$ [fb] &  $\delta_{\text{ISR}}$ [\%]\\
			&&&\\
			\hline\hline
			$W^+W^-$       &       $  3.2423(1)\cdot 10^1 $   & $4.162(4)\cdot 10^1$  &   $ +28.4(1) $ \\
			$ZZ$       &       $ 1.80357(9) \cdot 10^0$   &  $2.288(1)\cdot 10^0$    & $ +26.86(6) $ \\
			$HZ$  &     $  6.2702(4)\cdot 10^{-2} $ & $9.692(3)\cdot 10^{-2}$  & $ +54.57(5) $ \\
			\hline
			$ W^+W^-Z$  &   $ 6.369(3) \cdot 10^{0} $    &   $6.961(6)\cdot 10^{0}$   &   $ +9.3(1) $ \\
			$W^+W^-H$     &     $ 1.2846(6)\cdot 10^{-1} $   & $1.477(1)\cdot 10^{-1}$   &  $ +14.98(9) $ \\
			$ZZZ$      &   $  5.475(3)\cdot 10^{-2} $   &  $6.110(5)\cdot 10^{-2}$  &    $ +11.6(1) $\\
			{$HZZ$}      &    $  8.754(4)\cdot 10^{-3} $ & $10.197(7)\cdot 10^{-3}$ &    { $ +16.49(9) $} \\
			$ HHZ$   & $  3.668(1)\cdot 10^{-3} $ & $4.237(2)\cdot 10^{-3}$ &  $ +15.51(7)$\\
			\end{tabularx}
			\caption[muon collider NLO EW]{Total inclusive cross sections
				at LO with LL ISR photon resummation and relative correction
				$\delta_{\text{ISR}}$ for two- and three-boson production at
				$\sqrt{s}= 14$ TeV.}
			\label{table14TeVISR}
			\end{table}
			
			After all, a pattern in the relative NLO correction $\delta_{\text{EW}}$ of the $s$- and $t$-channel dominant processes for all the three collider energies, $3$, $10$ and $14$ TeV, becomes apparent which is directly related to their kinematical structure:
			The same reasons explaining similar suppression factors at high energies of the processes $WW$ and $ZZ$ can be attributed to those with approximately the same size for $WWZ$, $WWH$, $ZZZ$ and $HZZ$ in all tables.
			The bulk of their contributions at LO and NLO comes from $t$-channel diagrams which induce large amplitudes if these bosons are scattered in the forward direction.
			These suppressions are significantly different to those of the Drell-Yan-like Higgs- and di-Higgsstrahlung processes, i.~e. $HZ$ and $HHZ$ production The former compared to $WW/ZZ$ and the latter compared to $WWH/WWZ/ZZZ/ZZH$ have distinct $\delta_{\text{EW}}$ of $-20~\%$ to $-15~\%$ in Table \ref{table10TeV} and \ref{table14TeV}. 
			This observation is essentially the same comparing $\delta_{{HZ}}$ with $\delta_{{ZZ}}$ depicted in Fig.~\ref{HZaZZ} which are increasingly different with $\sqrt{s}$. Conclusively, the kinematical structure, either $t$- or $s$-channel induced, of the dominant Born process has a decisive impact on the relative size of NLO EW corrections to inclusive cross sections.
			
	\subsection{Differential distributions for $HZ$ production}
	\label{differentialresults}
\begin{figure}
		\centering
		\includegraphics[width=0.53\textwidth]{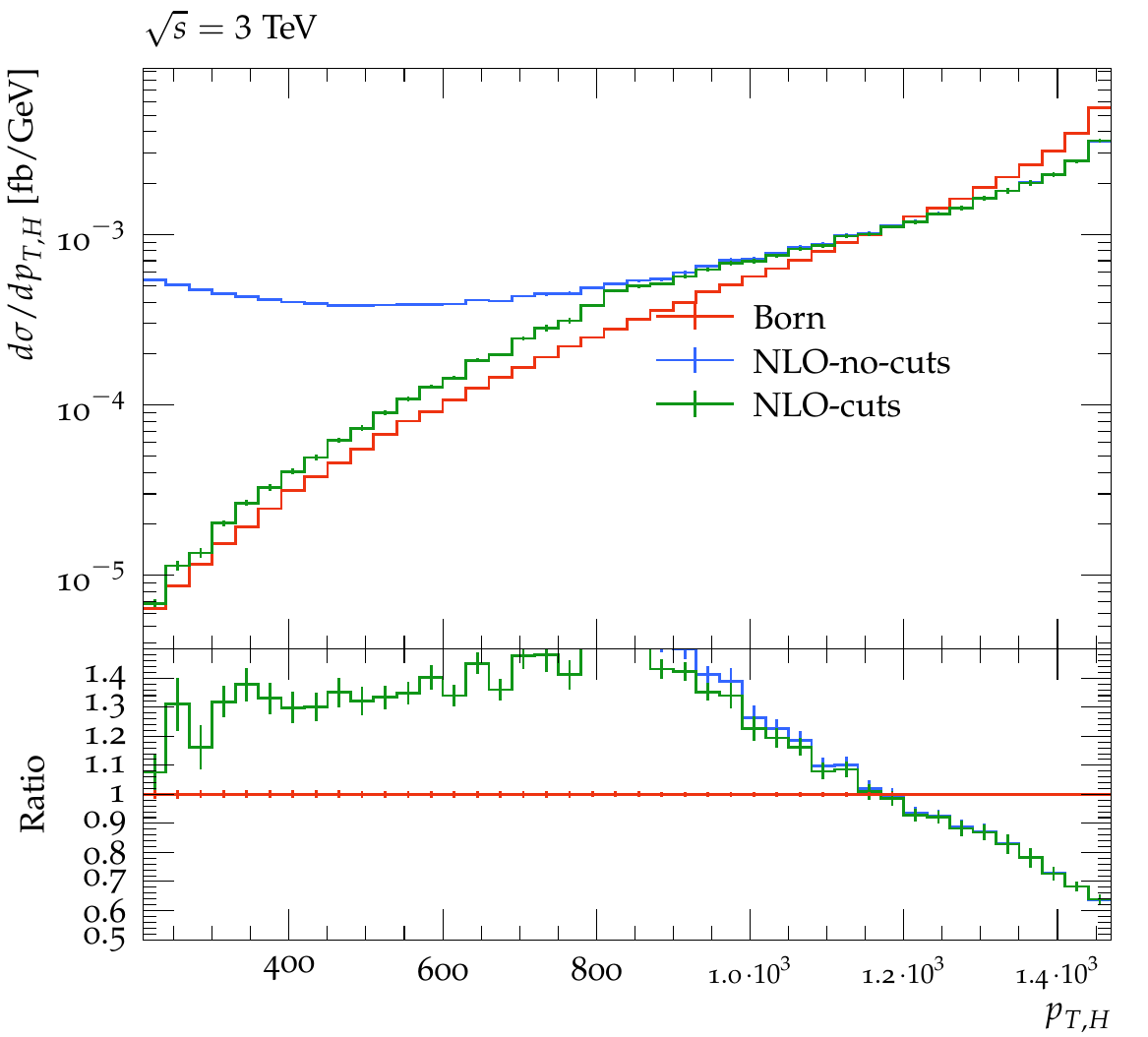}\\
		\includegraphics[width=0.53\textwidth]{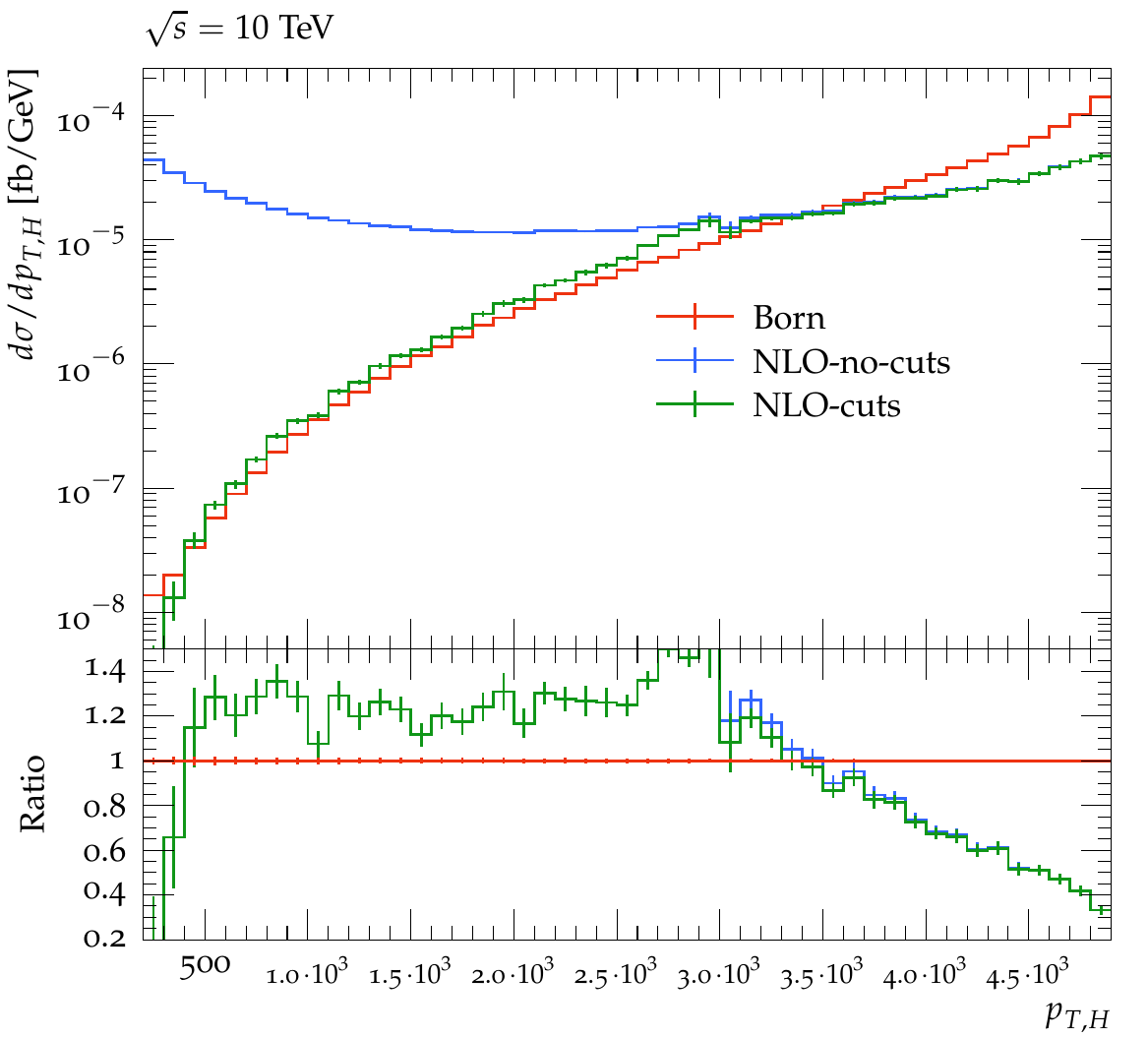}\\
		\includegraphics[width=0.53\textwidth]{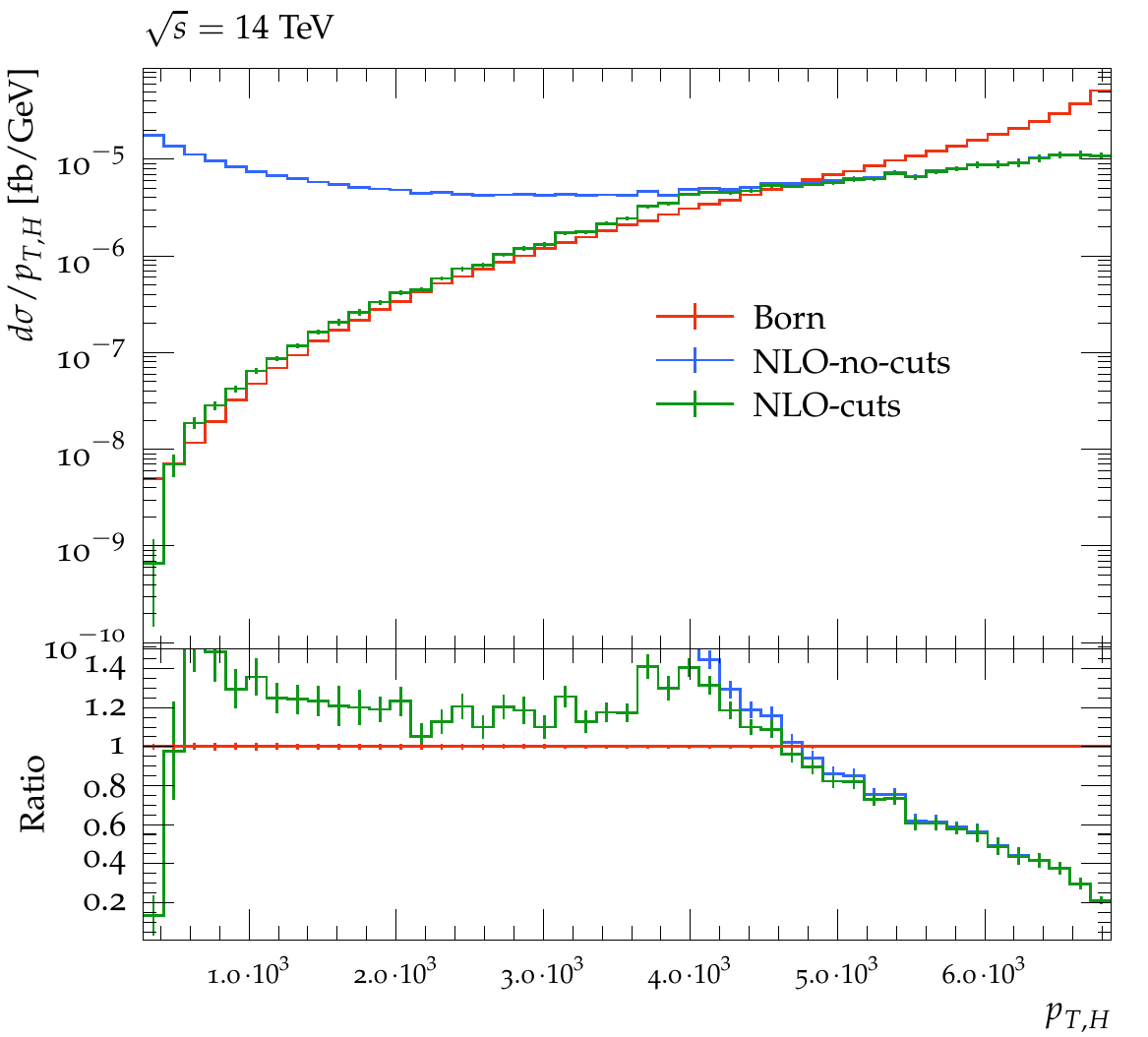}
		\caption{Higgs transverse
			momentum distributions, $d\sigma/dp_{T,H} (\mu^+\mu^-
			\rightarrow HZ)$, at $\sqrt{s} = 3$, $10$ and $14$
			TeV, respectively.}
		\label{ptdist}
	\end{figure}

\begin{figure}
		\centering
		\includegraphics[width=0.53\textwidth]{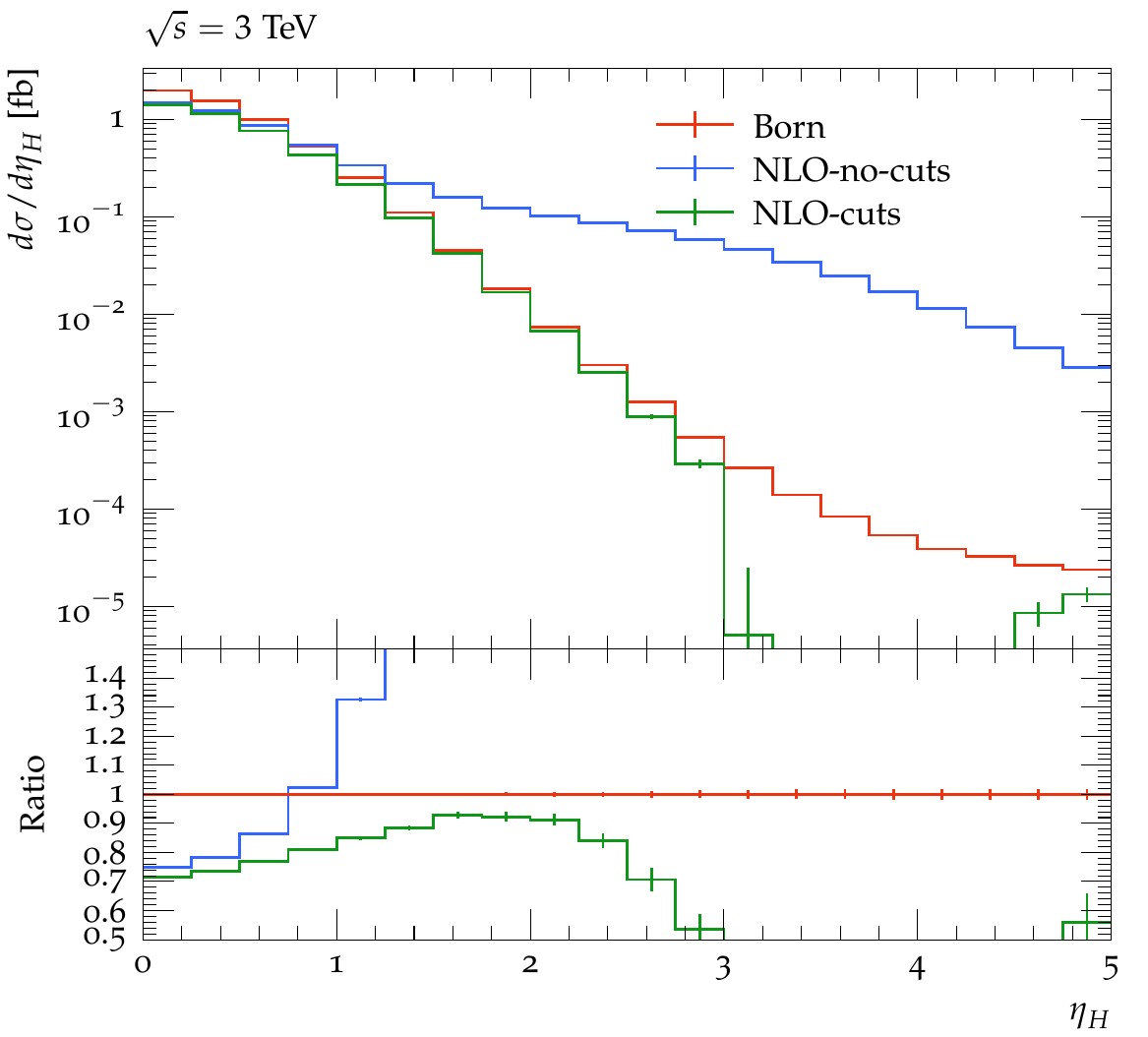}\\
		\includegraphics[width=0.53\textwidth]{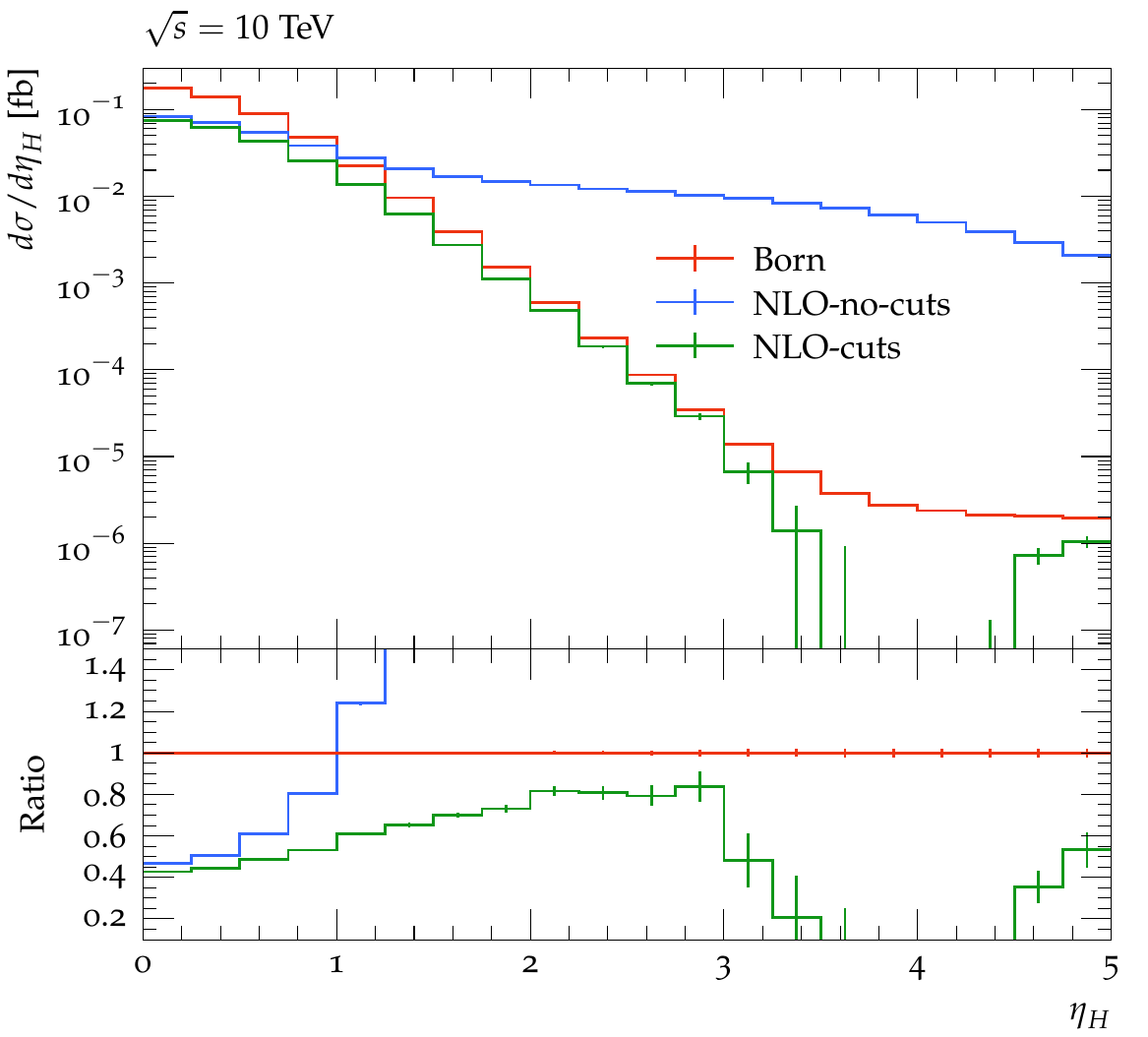}\\
		\includegraphics[width=0.53\textwidth]{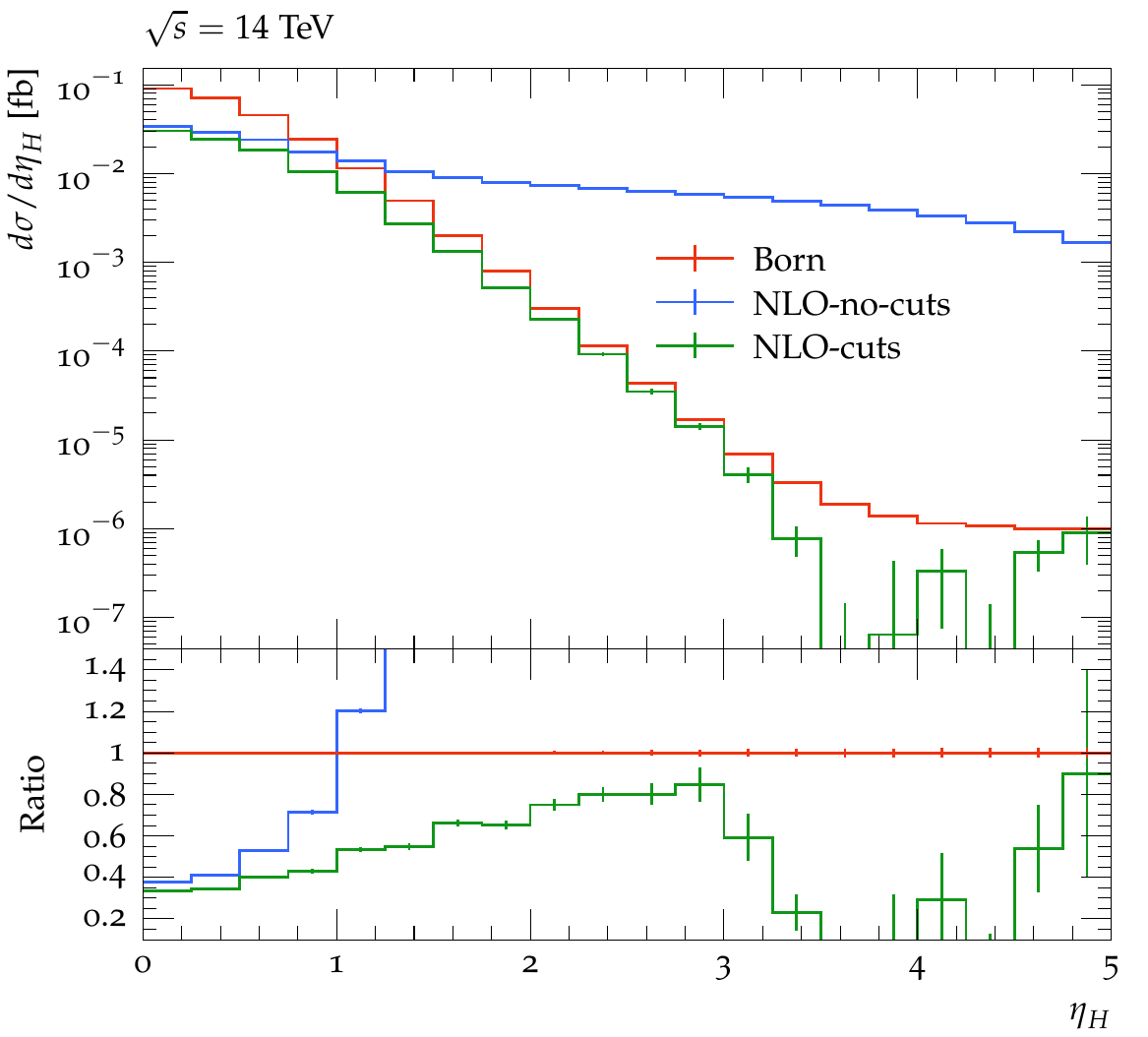}
		\caption{Higgs pseudorapidity distributions,
			$d\sigma/d\eta_{H} (\mu^+\mu^- \rightarrow HZ)$,
			at $\sqrt{s} = 3$, $10$ and $14$ TeV, respectively.
			The distributions are symmetric in $\eta_H$.}
		\label{etadist}
	\end{figure}

\begin{figure}
		\centering
		\includegraphics[width=0.53\textwidth]{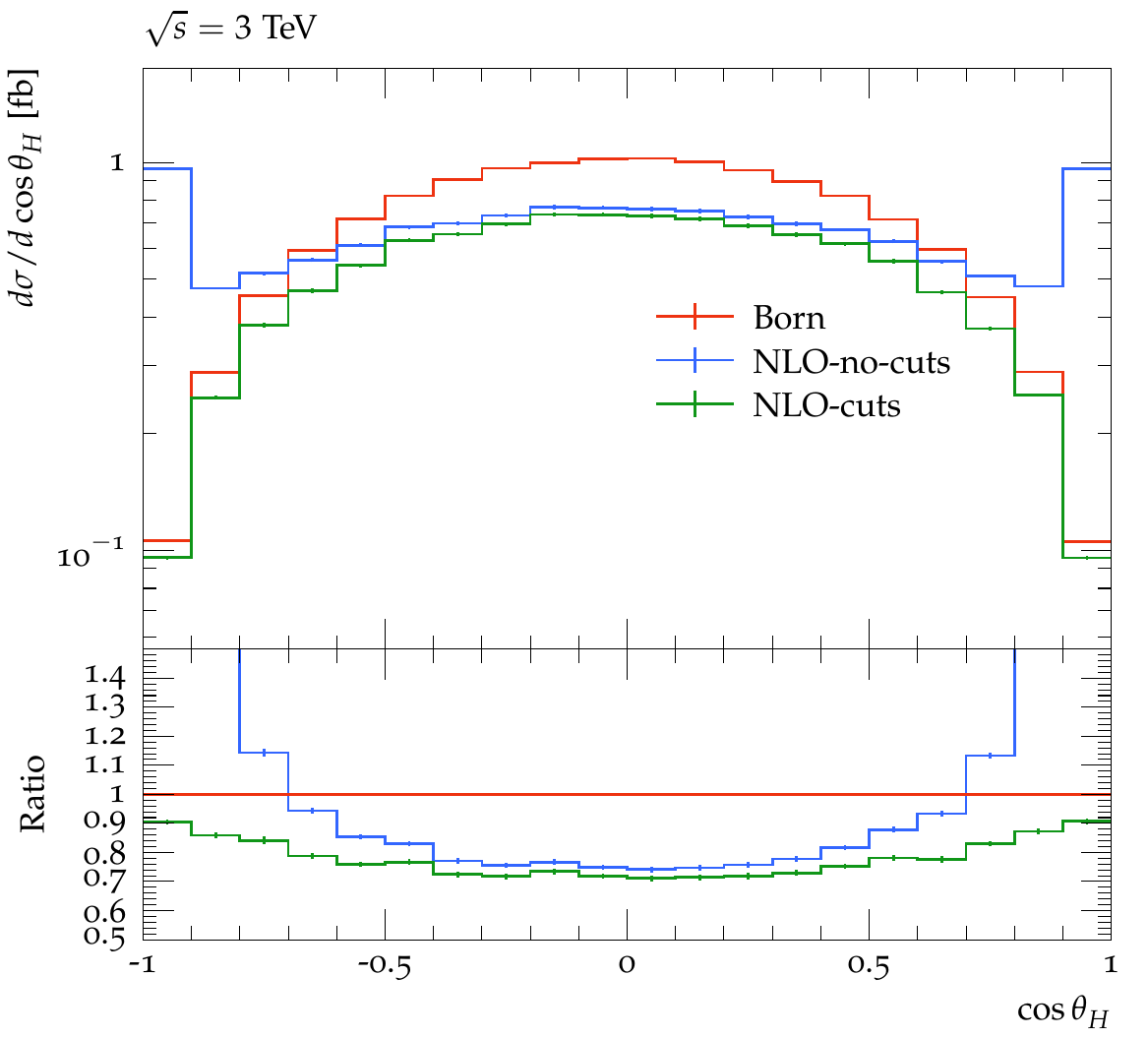}\\
		\includegraphics[width=0.53\textwidth]{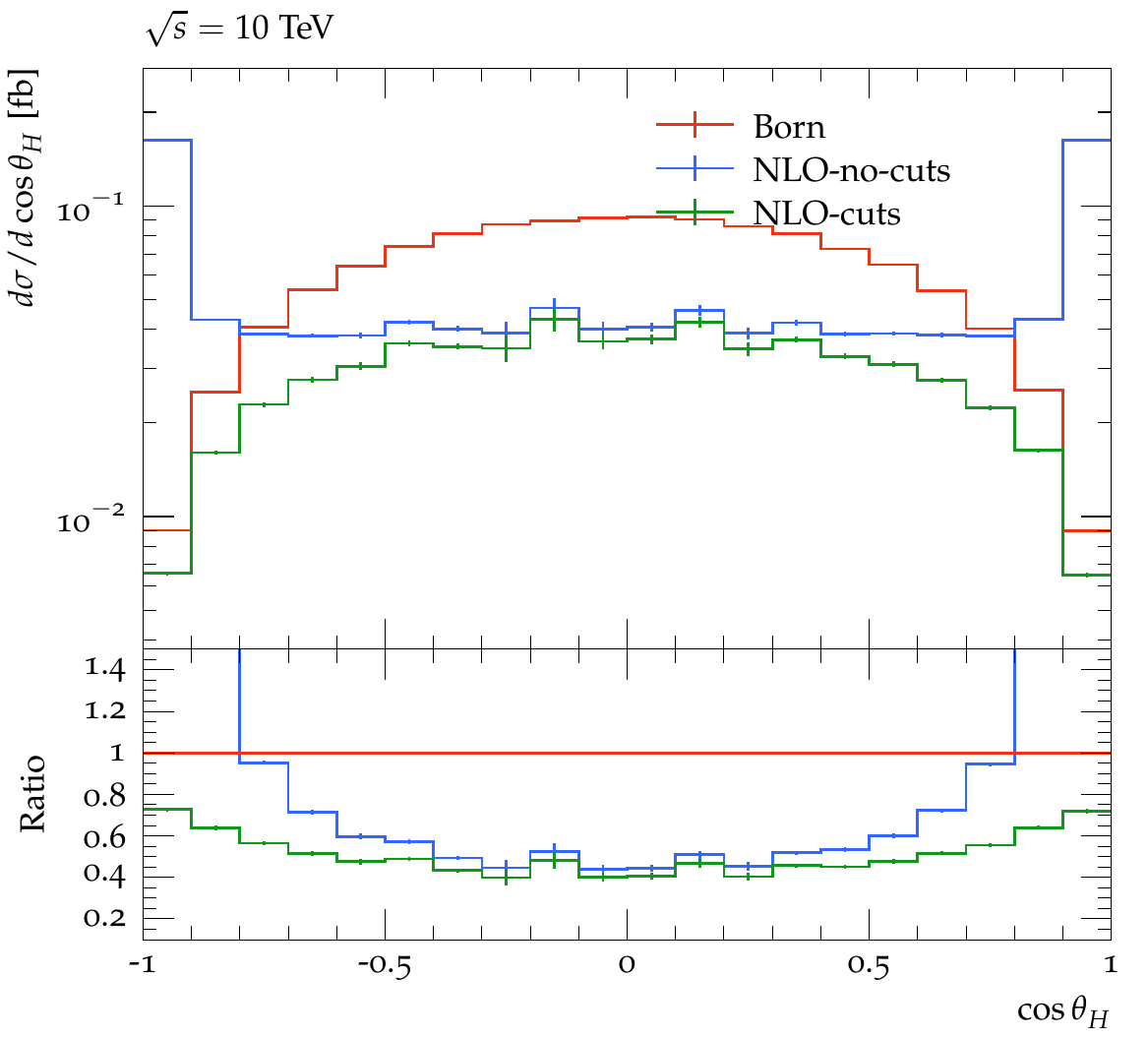}\\
		\includegraphics[width=0.53\textwidth]{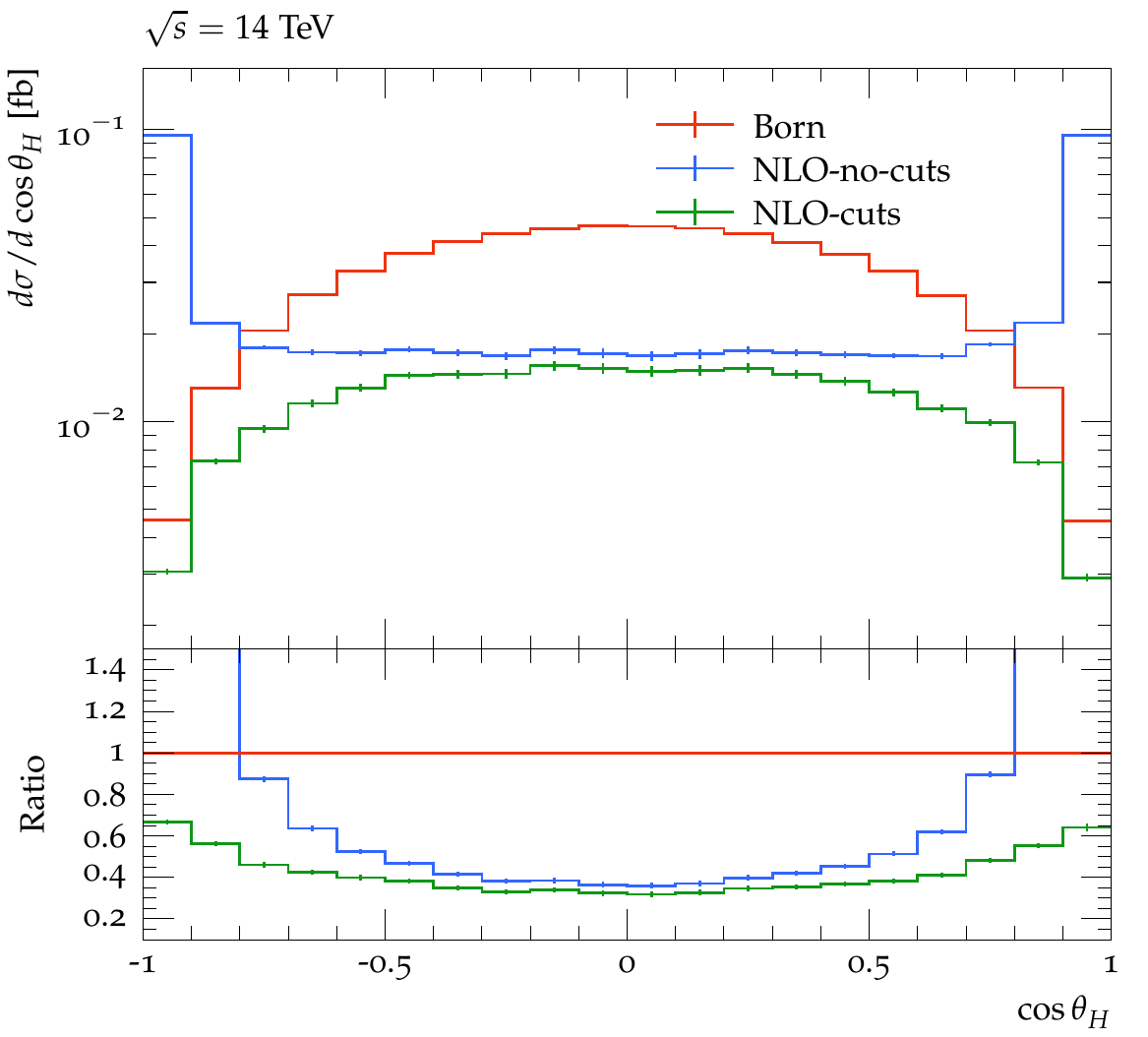}
		\caption{Differential Higgs polar angle distributions,
			$d\sigma/d\cos\theta_{H}(\mu^+\mu^- \rightarrow HZ)$ at
			$\sqrt{s} = 3$, $10$ and $14$ TeV, respectively.}
		\label{thetadist}
	\end{figure}

	In order to give an overview on the impact of NLO EW corrections on differential cross sections, distributions for the process $\mu^+\mu^-\rightarrow HZ$ at $\sqrt{s} = 3$, $10$ and $14$ TeV  differential in Higgs properties are displayed in Fig.~\ref{ptdist}, \ref{etadist} and \ref{thetadist}, respectively. These are fixed-order differential distributions. A realistic physics simulation would require a proper matching to QED parton showers combined with YFS soft-photon resummation in addition. As this is beyond the scope of this work it will not be covered in the results presented here.
	However, the effects of EW corrections on observables, for which fiducial cuts on the  phase space are imposed, will be discussed in this section. In particular, for phase space points with photon energies exceeding a certain value photons are considered as `observable' and hence these events are discarded from the analysis.
	In order to visualise the impact
	of this phase space cut, in addition to the curve representing the
	Born observable two curves for the NLO observables are included: One for
	the case that no cuts are imposed, called `NLO-no-cuts', and one with
	application of the cut
	\begin{align}
	E_{\gamma}<0.7\cdot \sqrt{s}/2,
	\label{photoncut}
	\end{align}
	which is dubbed `NLO-cuts'. This cut is based on the typical photon-veto which is imposed for experimental analyses at high-energy lepton colliders like ILC or CLIC \cite{Berggren:2022}.
	
	Distributions for cross sections differential in the Higgs transverse
	momentum $p_{T,H}$ in three plots, each for one center-of-mass energy
	value, are shown in Figure \ref{ptdist}. In these plots the K factor, i.e the ratio of NLO over Born differential cross sections, decreases for $\sqrt{s}/4 \lesssim p_{T,H} \lesssim \sqrt{s}/2$.
	In this part of the $p_{T,H}$ range, the curves `NLO-no-cuts' and `NLO-cuts' almost coincide, and the decrease of the ratio is the steeper the larger the collider energy: It drops to about $0.6$ for $\sqrt{s}=3$ TeV, $0.4$ for $\sqrt{s}=10$ TeV and $0.2$ for $\sqrt{s}=14$ TeV.
	
	The origin of this suppression can be traced back to EW Sudakov logarithmic factors in terms of $\log^2\left[(p_H+p_Z)^2/M_W^2\right]$ which grow with the invariant mass of the Born $HZ$ system. Obviously, this behaviour is enhanced with the centre-of-mass energy of the process.
	
	The cut on the photon energy influences the differential distributions
	in regions which are kinematically not accessible at Born level and which
	hence receive so-called `huge' K factors without cut. This happens in phase-space regions with small transverse momenta of the Higgs, i.~e. $p_{T,H} \lesssim p_{T,H}^{\text{max}}/2$.
	Due to the cut the NLO differential cross section shrinks to a constant ratio to the Born differential cross section of the order
	$\sim1.2$. Comparing both NLO distributions this means a difference in
	the differential cross section of two orders of magnitude for the lowest
	$\sqrt{s}$ and more than three for $\sqrt{s}=14$ TeV, respectively, for the first bin in the histograms. The cut has the most effect on the first bins since small Higgs transverse momenta due to small Higgs energies correspond
	to a recoiling system with large invariant mass which consists of the
	$Z$ and the photon for the real emission process. Due to momentum
	conservation this in turn leads to high energetic photons likely to
	not pass the cut criterium of Eq.~(\ref{photoncut}).
	
	In Figure \ref{etadist} distributions differential in the Higgs pseudorapidity $|\eta_{H}|$ for the proposed collider energies are depicted.
	Note, that these distributions are symmetric with respect to the central axis of the detector. For this reason the shortcut $\eta_{H}\equiv |\eta_{H}|$ is used for the differential distributions.
	First of all the most significant deviation of the two NLO distributions in each of the plots is at large pseudorapidities which is directly related to the fact that hard photons in this process induce large real amplitudes if they are radiated close to the beam axis. These phase-space configurations correspond to momenta of the recoiling system including
	those of the Higgs boosted into forward direction.
	On the other hand for small $|\eta_{H}|$ photon radiation for the NLO distributions is suppressed and virtual effects play a much more significant role.
	For $|\eta_{H}|=0$ the Higgs is radiated in the plane perpendicular to the beam axis such that the differential K factor reduced by one is directly comparable with the Sudakov supression factor $\Lambda^{\text{unpol}}_{\text{est}}$ of Eq.~(\ref{finaldeltaunpol}) which is shown in Fig.~\ref{sudpic}.
	In fact, the relative deviation of the NLO differential cross section for which cuts are applied with respect to the Born distribution agrees with $\Lambda^{\text{unpol}}_{\text{est}}$ in the first bin for
	all the shown plots at the level of a few percent.
	
	The same physics effects as in the pseudorapiditiy distributions are encoded in the differential distributions for the Higgs polar angle, $d\sigma/d\cos \theta_{H} $ shown in Fig.~\ref{thetadist} for all three collider energies. That the bulk of the Born contribution is located at the central part of the detector around $\theta_{H}=90^{\circ}$ can be understood from the $\sin^2
	\theta_{H}$ dependence of Born squared amplitudes with longitudinally polarised $Z$ bosons. These are enhanced by $s/M_Z^2$ compared to those with transversal polarisations of the $Z$ boson \cite{Bohm:2001yx}. According to this, the angular dependence of the differential cross sections at Born-level is functionally described by Eq.~(\ref{bornHZcrosssec}).
	By comparing the curves labelled `NLO-no-cuts' and `NLO-cuts' it becomes clear again that the cut on the photon's energy has the most significant impact on the NLO distributions for Higgs momenta boosted along the beam axis due to collinearly emitted hard photons. At angles in the plane perpendicular to the beam axis the cut has only a minor effect and the curves deviate at a few percent.
	As in the case of the central description of distributions differential in the pseudorapidity, i.~e. at $\eta_{H} \sim 0$, the Sudakov factor $\Lambda^{\text{unpol}}_{\text{est}}$ can be found as an accurate approximation for $d\sigma/d\cos\theta_{H}$ at $\theta_{H}= 90^{\circ}$, especially for high collider energies and when applying a hard photon veto.
	
	\subsection{Concluding remarks}
		
	In general, two physical effects are counteracting each other for all considered processes: On the one hand, positive contributions due to collinear initial-state photon radiation emerge at NLO EW which are enhanced with the centre-of-mass energy of the process. On the other hand, large negative virtual corrections due to EW Sudakov logarithmic terms contribute to the observables at NLO EW.
	For all processes, the only exception being gauge-boson pair production, $\mu^+\mu^-\to VV$, the relative corrections for the NLO EW with respect to the LO results are negative and increase in size with $\sqrt{s}$. This can be traced back to an overcompensation of the effects due to EW Sudakov logarithms with respect to QED radiation effects.
	The suppression of NLO EW with respect to LO inclusive cross sections amounts up to about $-40\%$ for four-boson production at $3$ TeV and $-60\%$ for three-boson production at $14$ TeV. The results for Higgs- and multi-Higgsstrahlung processes stand out in a way that they exhibit much smaller relative corrections, i.~e. NLO EW cross sections are much more suppressed compared to the LO ones, with respect to the other processes. As explained in this work, this is due to the $s$-channel kinematical structure of their dominant Born contribution.
	
	Furthermore, distributions differential in Higgs properties for the process $\mu^+\mu^-\to HZ$ have been studied. For the differential distributions $d\sigma/d\cos \theta_{H}$ and $d\sigma/d\eta_{H}$ it can be observed that the suppression factor of the NLO EW  with respect to the LO curve for Higgs polar angles perpendicular to the beam axis is described very well by the NLL EW Sudakov approximation factor to $\mu^+\mu^-\to HZ$. It is shown, that this approximative factor agrees  with the calculated relative NLO EW correction even at the percent-level for these phase-space configurations if a cut criterion on hard photons appearing at NLO EW, i.~e. $E_{\gamma}<0.7\sqrt{s}$, is imposed.

	\chapter{Summary, Conclusions and Outlook}
	Precise predictions of the SM provided by MC event generators play a key role in the search for new physics at current and future collider experiments, which have become increasingly difficult in the last decade since the discovery of the Higgs boson. An automated MC framework accounting for higher order QCD and EW corrections in simulations is thus of high importance for scrutinising the broad range of LHC or future hadron-collider processes for direct or indirect final-state signals. Likewise, simulating lepton-collider processes with higher order EW corrections by an automated setup yields precise predictions for a large set of processes relevant for example for future-collider design studies.
	
	As an extension to the automation of NLO QCD corrections in the MC generator of \texttt{WHIZARD}, this thesis provides the full validation for automated NLO EW corrections to processes at hadron and lepton colliders. The FKS scheme for the subtraction of IR singularities implemented in \texttt{WHIZARD} has been generalised accounting for NLO QCD, EW and mixed corrections.
	
	In particular, for hadron collisions, this required the consideration of the interference of correction types for the construction of subtraction terms with respect to overlapping QCD-EW NLO contributions in the mixed coupling expansion. Non-negligible contributions from photon-induced processes and non-singular $n+1$ tree-level processes as well as IR-safety conditions impose additional technical intricacies for automated NLO EW calculations for processes at the LHC.
	
	Cross-checks with the MC tools \texttt{MG5\_aMC@NLO} and \texttt{MUNICH/MATRIX} yield agreement for fixed-order cross sections with relative MC uncertainties at the level of $\mathcal{O}(0.01\%)$ for a large set of benchmark processes at the LHC. This includes not only pure EW processes with on-shell and off-shell vector-bosons, but also processes with Born-level contributions at $\mathcal{O}(\alpha_s^n\alpha^m)$ with $n>0$. Explicitly, all LO and NLO contributions of mixed coupling expansions to $pp\to t\bar{t}(W/Z/H)$ have been validated. Additionally, cross sections at LO and NLO EW for the processes $pp\to e^+\nu_e j$ and $pp\to e^+e^- j$ computed with \texttt{WHIZARD+OpenLoops} agree at the sub-per-mille level -- a factor $\sim 0.01$ below the relative NLO correction -- with the corresponding reference results.
	
	For lepton-collider processes, tiny masses of the initial-state impose a great challenge for the automated computation of observables including NLO EW corrections. There are two approaches for lepton-collider observables at higher orders in $\alpha$ taking corrections due to collinear ISR into account.
	
	In the first, logarithmic terms induced by collinear radiation off the initial-state leptons are resummed and factorised in QED PDFs. The convolution of the lepton PDFs with partonic cross sections due to the divergent PDF structure at $x=1$ requires an adequate parametrisation of the fractions of the beam energy in terms of random number variables in order to ensure an efficient adaption of the MC integration. The formal description of such a mapping extended to NLO calculations is included in this thesis.
	
	The second approach assumes terms containing collinear logarithms, i.~e. $\log(Q^2/m^2)$ with $Q$ the energy scale of the process and $m$ the lepton mass, as sufficiently suppressed by the electromagnetic coupling $\alpha$. This allows for reliable fixed-order expansions of lepton collision observables including $\mathcal{O}(\alpha)$ corrections. Under certain conditions of lepton colliders, i.~e. the collider energy is sufficiently small with respect to the the initial-state mass, this approach can be considered as useful. The construction of the real radiation phase-space in the approach of massive initial-state emitters follows on-shell conditions of external particles and momentum conservation rules. The respective formalism for this construction and the validation of the \texttt{WHIZARD+RECOLA} framework for simulating NLO EW cross sections to $e^+e^-$ collision processes has been shown in this work.
	
	As an application of this approach, results for cross sections and differential distributions at NLO EW in multi-boson processes at a future muon collider,  $\mu^+\mu^- \to V^n H^m$ with $V \in \{W^\pm,Z\}$, $n+m \leq 4$ and $n\neq0$, at collider energies of $\sqrt{s}= 3$, $10$ and $14$ TeV have been computed for the first time. These processes constitute one of the most important cornerstones both of the EW precision and the discovery program of a future muon collider.
	In conclusion, the results show the significant impact of EW corrections on observables for large boson multiplicities and high energies in the multi-TeV range of a future muon collider. This is relevant for precise predictions for such a collider in order to investigate physics of the EW sector.
	
	The framework of automated NLO EW corrections in \texttt{WHIZARD} provides the basis for new features and capabilities of the MC generator which can be applied in many-faceted simulations of collider physics.
	
	For realistic predictions comparable with experimental data, simulated event samples must be showered and hadronised. For NLO QCD calculations the matching to parton showers, which in \texttt{WHIZARD} is based on the POWHEG scheme, has recently been proven to be functional for hadron and lepton collision processes \cite{StienemThesis}. The next aim would be to extend this matching for NLO EW calculations using QED and QCD-QED mixed parton showers.
	
	Also useful for future studies would be the application of EW corrections to decays of massive particles as well as the combination of production and decay, i.~e. factorised processes. As the simulation of observables for this class of processes is well developed in \texttt{WHIZARD}, e.~g. including NLO QCD corrections to the factorised off-shell $e^+e^-\to t\bar{t}$ process, the NLO EW corrections are supplementary to increase the EW precision. With the infrastructure provided by this work, the technical incorporation of this process class in the NLO EW framework in \texttt{WHIZARD} should require only minor efforts.
	
	Similarly, progress have been made with respect to the interface to one-loop-providers in \texttt{WHIZARD} in the context of this thesis. This can be exploited for simulating loop-induced processes with \texttt{WHIZARD}. For certain hadron- and lepton-collider processes they yield significant contributions at the accuracy level comparable with the current precision frontier of theoretical predictions. Relevant in particular for studying the trilinear Higgs coupling are loop-induced processes with di-Higgs production.
	Furthermore, for studies related to BSM and SM effective field theories (SMEFT) including perturbative corrections, the IR-cancellation mechanism provided by the automated NLO framework in \texttt{WHIZARD} can be used also in a broader context.
	
	In general \texttt{WHIZARD} provides a setup simulating cross sections and distributions to completely arbitrary processes. This imposes the potential for an extension of the automated NLO SM corrections to processes at any collider, including also $ep$ or photon colliders, with presumably few technical efforts.
	
	For lepton colliders, the next steps to pursue are towards a higher-order QED/EW theoretical description of observables. This will include the combination of NLL QED PDFs for lepton collisions with NLO EW computations. To achieve this, MC integration methods, which involve an adjusted phase-space parametrisation of NLO integration variables regarding the singular integrable structure of electron PDFs, will have to be implemented within the MC. Such a parametrisation has been presented in this thesis.
	In a further step, this new NLO ISR module can be supplemented with the simulation of beamstrahlung via the beam spectra generator \texttt{CIRCE}.
	
	Furthermore, the feature of \texttt{WHIZARD} to apply beam polarisation to lepton collision processes in the nearest future will be extended to NLO EW calculations using the fixed-order approximation with massive initial-state emitters. This, in particular, yields the potential of precise spin-dependent SM predictions for future muon collider studies possibly helping to unravel the mystery towards $(g-2)_{\mu}$.
	
	QED resummation of soft photons radiated off initial- and final-states can be achieved by YFS methods. Paradigmatic for an implementation of these in \texttt{WHIZARD} is the framework of \texttt{SHERPA} which is the first tool accounting for YFS resummation in an automated way \cite{Krauss:2022ajk,Price:2021qnc}. 
	Concerning higher order QED corrections, a precision level of up to NNLO QED by now can be reached using the FKS subtraction scheme as basis. This so far has been realised by the generator \texttt{McMule} \cite{Banerjee:2020rww}. In this framework, the massive treatment of leptons simplifying the FKS subtraction at that precision level is combined with a YFS resummation approach rendering photons exclusive. This yields precise predictions which are relevant in particular for low-energy lepton-collider studies. The working FKS subtraction using massive initial-state leptons in \texttt{WHIZARD}, demonstrated in this thesis, allows for an extension to the NNLO QED precision level of computations as a future project.
	
	Finally, the resummation of EW Sudakov logarithms in addition to an automated evaluation of EW Sudakov approximation factors as it is realised for example by \texttt{SHERPA} \cite{Bothmann:2020sxm} and \texttt{MG5\_aMC@NLO} \cite{Pagani:2021vyk} will be future goals to pursue for the NLO EW automated framework of \texttt{WHIZARD}.
	
	\chapter*{Acknowledgements}
	\addcontentsline{toc}{chapter}{Acknowledgements}
	Many people accompanied me on the journey of working on and towards this thesis. I want to use the lines of this chapter to express my acknowledgements to some explicitly.

First of all, I would like to express my deepest gratitude to my supervisor J\"urgen Reuter for giving me the opportunity to become a part of his group and to work on this PhD project. I appreciate his encouragement, invaluable scientific advice, providing help also in detailed technical aspects of this work and for his way being accessible to his PhD students. I am indebted to him by the plentiful helpful remarks he provided while finalising this manuscript and for promoting me throughout the years of this project.

I thank my co-supervisor Gudrid Moortgat-Pick for great support and promoting advice on scientific and career aspects.

For agreeing to be the chair of my examination commission I would like to express my appreciation to Sven-Olaf Moch. I thank Elisabetta Gallo and Markus Diehl for agreeing to be further members of this committee.

Furthermore, I want to thank my mentor Elli Pomoni for coaching and advice on a good work-life balance (especially during the Corona pandemic period).

I am grateful to be part of the \texttt{WHIZARD} collaboration which besides my supervisor consists of Wolfgang Kilian, Thorsten Ohl, Nils Kreher, Krzysztof M{\c e}ka{\l}a, Pascal Stienemeier and Tobias Striegl. In particular, I want to thank Wolfgang for his constant support and indispensable scientific and computational advice. I am thankful for many helpful details to the code and scientific aspects of this PhD project provided by Thorsten.
From the bottom of my heart, I want to thank Pascal for introducing me into all technical details of \texttt{WHIZARD}, for interesting and fruitful discussions and a very productive collaboration. Furthermore, he is one of my role models with respect to diligent working and endurance.
Also, I want to highlight the help and support of the former members, Simon Bra\ss~ and Vincent Rothe. Simon, Vincent and Pascal essentially contributed to this work with their expertise on parallel computing, event generation and automation of NLO corrections in the \texttt{WHIZARD} framework.
In addition, their code work paved the way for the project of this PhD thesis.

For the extensive cross-checks with \texttt{MUNICH/MATRIX} presented in this thesis as well as a lot of scientific advice and useful explanations on EW corrections I owe deepest thanks to Stefan Kallweit. I feel very honoured to have worked with such a great scientist.
I appreciate the collaboration with Jonas Lindert who in addition to \texttt{OpenLoops} process libraries and BLHA support provided important aspects concerning EW corrections. His technical advice on \texttt{OpenLoops} prescriptions contributed decisively to the code development of the \texttt{WHIZARD}-\texttt{OpenLoops} interplay.
For helpful scientific and technical details on automated NLO EW corrections in the first years of this PhD project I want to thank Davide Pagani.
Moreover, I want to thank Alan Price for performing dedicated cross-checks with \texttt{SHERPA}, his invaluable advice on EW precision calculations for lepton collisions, proofreading the complete thesis and kind supporting efforts.
For sharing his expertise on PDFs related to EW corrections I would like to thank Christopher Schwan.
I thank Matthew Lim and Peter Pl\"o\ss l for proofreading parts of this thesis and providing important advice on special theoretical aspects of it.
Also, I appreciate Florian Fabry's advice on technical methods concerning PDFs and that he always had an open ear.

During my time at DESY, it was a pleasure to share an office with Aron Bodor, Ivan Novikov and Krzysztof M{\c e}ka{\l}a.
For the lovely and heart-warming atmosphere of floor 1b at DESY I furthermore thank Georgios Billis, Pedro Cal, Bahman Dehnadi, Juhi Dutta, Oskar Grocholski, Rebecca von Kuk, Daniel Meuser, Johannes Michel and M. Olalla Olea Romancho in addition to people named already above. I very much enjoyed all our breaks with coffee, cake and good talks.

I thank my family and friends for mental support as well as my dear husband for constantly empowering me.
Also, I want to thank god for countless blessings during the time of this PhD project.
	\appendix
	\chapter{The infrared limit}
	\section{Explicit formulas for FKS subtraction terms}
	\subsection{Eikonal integrals}
	\label{secAppendixeikonals}
	Based on the approach used in \cite{Alioli:2010xd} the evaluation of the eikonal integral of Eq.~(\ref{softSubtractiontermint}) yields the following results. Depending on the FKS parameter $\xi_c$ the complete eikonal integral takes the general form
	\begin{align}
		\sum_{\rho}\mathcal{E}^{(m_k,m_l)}_{kl,\rho}=-\frac{2^{2\varepsilon}\xi_c^{-2\varepsilon}}{2\varepsilon}\frac{s^{-\varepsilon}\mu_R^{2\varepsilon}}{(2\pi)^{1-2\varepsilon}}\underbrace{\int d\Omega^{(2-2\varepsilon)}\int_{-1}^{1}dy(1-y^2)^{-\varepsilon}\frac{s\xi^2}{4}\frac{k_k\cdot k_l}{(k_k\cdot k_j)(k_l\cdot k_j)}}_{=I(k_k,k_l)}
		\label{eikonals}
	\end{align}
	The prefactor of this equation can be rewritten in terms of the normalisation factor $\mathcal{N}(\varepsilon)$ defined in Eq.~(\ref{normalisationfactor}), expanding the rest as a series in $\varepsilon$,
	\begin{align}
		-\frac{2^{2\varepsilon}}{2\varepsilon}\frac{s^{-\varepsilon}\mu_R^{2\varepsilon}}{(2\pi)^{1-2\varepsilon}}=-\frac{1}{2\varepsilon}\mathcal{N}(\varepsilon)\left[1+\varepsilon\log\frac{Q^2}{\xi^2_cs}+\left(\frac{1}{2}\log^2\frac{Q^2}{\xi^2_cs}-\frac{\pi^2}{6}\right)\varepsilon^2+\mathcal{O}(\varepsilon^3)\right].
		\label{prefactorexpans}
	\end{align}
	Note, that by setting the value $\xi_c=1$ we recover the eikonal integrals to be used without adding the factor of the Heaviside functions of Eq.~(\ref{plusdistrxicut}) to the integrand of Eq.~(\ref{realdimreg})\footnote{In this context the $\xi_c$ parameter is set to this value as a default in \texttt{WHIZARD}.}.
	The integral $I(k_k,k_l)$ written as a series in $\varepsilon$ takes the basic form
	\begin{align}
	\begin{split}
		I(k_k,k_l)&=\int d \cos\theta\frac{d\phi}{\pi}(\sin \theta\sin \phi)^{-2\varepsilon}\left[\frac{s\xi^2}{4}\frac{k_k\cdot k_l}{k_k\cdot k_j~k_l\cdot k_j}\right]\\
		&=\frac{1}{\varepsilon}I_{-1}(k_k,k_l)+I_0(k_k,k_l)+\varepsilon I_{\varepsilon}(k_k,k_l)\quad.
	\end{split}
	\end{align}
	For the evaluation of the integrals $I_{-1}$, $I_0$ and $I_{\varepsilon}$ three different kinematical cases have to be distinguished for the momenta $k_k$ and $k_l$, i.~e. the purely massless case with $m_k=0$ and $m_l=0$, the massless-massive case with $m_k=0$ and $m_l\ne 0$ and the massive-massive case with $m_k\ne 0$ and $m_l\ne0$. Eventually, the eikonal integral $\sum_{\rho}\mathcal{E}^{(m_k,m_l)}_{kl,\rho}$ of the form
	\begin{align}
		\sum_{\rho}\mathcal{E}^{(m_k,m_l)}_{kl,\rho}=\frac{1}{\varepsilon^2}A+\frac{1}{\varepsilon}B+C
	\end{align}
	can be obtained for all three cases. By identifying $C=\mathcal{E}^{(m_k,m_l)}_{kl,0}$ we get the following functions for the eikonals contributing to the finite part of the soft subtraction terms.
	
	\begin{itemize}
		\item \textbf{The massless-massive case}\\
		For the case that particle $k$ is massless, $k_k^2=m_k^2=0$, particle $l$ massive,~$k_l^2=m_l^2\ne 0$, $k\ne l$ and the definitions
		\begin{align}
			\hat{k}_k=\frac{k_k}{k^0_k}, \qquad \hat{k}_l=\frac{k_l}{k^0_l}, \qquad \beta = \frac{\lvert \vec{k}_l\rvert}{k_l^0}
		\end{align}
		the resulting expressions to be extracted from $I(k_k,k_l)$ are \cite{Alioli:2010xd}
		\begin{align}
			&I_{-1}(k_k,k_l)=-1\\
			&I_{0}(k_k,k_l)=\log \frac{(\hat{k}_k\cdot \hat{k}_l)^2}{\hat{k}_l^2}
			\label{Izero}
			\end{align}
		\begin{align}
		\begin{split}
			I_{\varepsilon}(k_k,k_l)=-2\left[\frac{1}{4}\log^2\frac{1-\beta}{1+\beta}+\log\frac{\hat{k}_k\cdot{k}_l}{1+\beta}\log\frac{\hat{k}_k\cdot{k}_l}{1-\beta}+\text{Li}_2\left(1-\frac{\hat{k}_k\cdot{k}_l}{1+\beta}\right)\right.\\
			\left.+\text{Li}_2\left(1-\frac{\hat{k}_k\cdot{k}_l}{1+\beta}\right)\right]
			\label{Ieps}
			\end{split}
		\end{align}
		Inserting these expressions into Eq.~(\ref{eikonals}) and using the expansion of Eq.~(\ref{prefactorexpans}) yields the finite part of the eikonal integral
		\begin{align}
			C=\mathcal{E}^{(0,m_l)}_{kl,0}=\frac{1}{2}\left[\log^2\frac{Q^2}{s\xi^2_c}-\frac{\pi^2}{6}\right]-\frac{1}{2}I_{0}(k_k,k_l)\log \frac{Q^2}{s\xi_c^2}-\frac{1}{2}I_{\varepsilon}(k_k,k_l) .
		\end{align}
		\item \textbf{The massless-massless case}\\
		For two massless particles with $k_k^2=k_l^2=0$ and $k\ne l$ using the identity
		\begin{align}
			\frac{k_k\cdot k_l}{k_k\cdot k_j~k_l\cdot k_j}=\frac{k_k\cdot (k_k+k_l)}{k_k\cdot k_j~(k_k+k_l)\cdot k_j}+\frac{k_l\cdot (k_k+k_l)}{k_l\cdot k_j~(k_k+k_l)\cdot k_j}
		\end{align}
		the integral $I$ can be rewritten in the more practical form
		\begin{align}
			I(k_k,k_l)=I(k_k,k_k+k_l)+I(k_l,k_k+k_l)
		\end{align}
		such that Eqs.~(\ref{Izero}) and (\ref{Ieps}) can be used for the finite part
		\begin{align}
		\begin{split}
			C=\mathcal{E}^{(0,0)}_{kl,0}=&\left[\frac{1}{2}\log^2\frac{Q^2}{s\xi^2_c}-\frac{\pi^2}{6}\right]-\frac{1}{2}\log\frac{Q^2}{s\xi^2_c}\left[I_0(k_k,k_k+k_l)+I_0(k_l,k_k+k_l)\right]\\
			&-\frac{1}{2}\left[I_{\varepsilon}(k_k,k_k+k_l)+I_{\varepsilon}(k_l,k_k+k_l)\right]
			\end{split}
		\end{align}
		The massless self-eikonals, i.~e. the eikonals for the case $k_k^2=k_l^2=0$ and $k = l$, are simply given by
		\begin{align}
			\mathcal{E}^{(0,0)}_{kk,0}=0\quad.
		\end{align}
		\item \textbf{The massive-massive case}\\
		For the case that two particles are massive, i.~e. $k_k^2=m_k^2\ne 0$, $k_l^2=m_l^2\ne 0$ and $k\ne l$, the $1/\varepsilon$ term vanishes for $I$ such that
		\begin{align}
			I(k_k,k_l)&=I_0(k_k,k_l)+\varepsilon I_{\varepsilon}(k_k,k_l)\\
			C=\mathcal{E}^{(m_k,m_l)}_{kl,0}&=-\frac{1}{2}I_0(k_k,k_l)\log\frac{Q^2}{s\xi_c^2}-\frac{1}{2}I_{\varepsilon}(k_k,k_l)
		\end{align}
		with
		\begin{align}
			&I_0(k_k,k_l)=\frac{1}{\beta}\log \frac{1+\beta}{1-\beta}, &\beta=\sqrt{1-\frac{k_k^2k_l^2}{(k_k\cdot k_l)^2}}\quad.\label{Izeromm}
		\end{align}
		The part $I_{\varepsilon}$ is constructed in a non-trivial way by using the following variable definitions
		\begin{align}
			&\vec{\beta}_1=\frac{\vec{k}_k}{k_k^0},\qquad \vec{\beta}_2=\frac{\vec{k}_l}{k_l^0}\\
			&a=\beta_1^2+\beta_2^2-2\vec{\beta}_1\cdot \vec{\beta}_2,\qquad b=\frac{\beta_1^2\beta_2^2-(\vec{\beta}_1\cdot\vec{\beta}_2)^2}{a}, \quad c=\sqrt{\frac{b}{4a}} \label{abc}\\
			&x_1=\frac{\beta_1^2-\vec{\beta}_1\cdot\vec{\beta}_2}{a}, \qquad x_2=\frac{\beta_2^2-\vec{\beta}_1\cdot\vec{\beta}_2}{a}=1-x_1\\
			&z_+=\frac{1+\sqrt{1-b}}{\sqrt{b}},\qquad z_-=\frac{1-\sqrt{1-b}}{\sqrt{b}}\\
			&z_1=\frac{\sqrt{x_1^2+4c^2}-x_1}{2c}, \qquad z_2=\frac{\sqrt{x_2^2+4c^2}-x_2}{2c} \label{z1z2}
		\end{align}
		\begin{align}
			\begin{split}
			K(z)=&-\frac{1}{2}\log^2\frac{(z-z_-)(z_+-z)}{(z_++z)(z_-+z)}-2\text{Li}_2\left(\frac{2z_-(z_+-z)}{(z_+-z_-)(z_-+z)}\right)\\
			&-2\text{Li}_2\left(-\frac{2z_+(z_-+z)}{(z_+-z_-)(z_+-z)}\right).\label{Kz}
			\end{split}
		\end{align}
		We get
		\begin{align}
			I_{\varepsilon}(k_k,k_l)=\left[K(z_2)-K(z_1)\right]\frac{1-\vec{\beta}_1\cdot\vec{\beta}_2}{\sqrt{a(1-b)}}\quad.
			\label{Ieps_mm}
		\end{align}
		For masses of the particles $k$ and  $l$ in extreme limits, e.~g. $m_k\gg m_l$, the numerical evaluation of Eq.~(\ref{Ieps_mm}) might get affected by the limitations of the machine precision. For that reason, $1-\beta$ used in Eq.~(\ref{Izeromm}) is expanded for small $k_k^2k_l^2/(k_k\cdot k_l)^2$. Furthermore, it is useful to redefine variables of Eqs. (\ref{abc}) to (\ref{z1z2}) as
		\begin{align}
			\tilde{b}=ab, \qquad \tilde{x}_{1/2}=ax_{1/2}, \qquad  \tilde{z}_{+/-}=\sqrt{\tilde{b}}z_{+/-}, \qquad \tilde{z}_{1/2}=\sqrt{\tilde{b}}z_{1/2},
		\end{align}
		make corresponding substitutions in Eq.~(\ref{Kz}) and expand the fractions of $K(z_2)$ in $\tilde{b}$ \cite{WhizardNLO}.
	\end{itemize}
	\subsection{Space-averaged Altarelli-Parisi kernels}
	\label{secAppendixcollinear}
	The space-averaged Altarelli-Parisi splitting kernels for FSR and ISR originate from the results of \cite{Altarelli:1977zs}. For QED final-state splittings they read
	\begin{align}
		\langle\hat{P}_{f\rightarrow f\gamma}\rangle(z,\varepsilon)&=e(f)^2\left(\frac{1+z^2}{1-z}-\varepsilon(1-z)\right) \label{space-averagedAPsplitfgam}\\
		\langle\hat{P}_{\gamma\rightarrow f f}\rangle(z,\varepsilon)&=n_c(f)e(f)^2\frac{(1-z)^2+z^2-\varepsilon}{1-\varepsilon}\quad.\label{space-averagedAPsplitff}
	\end{align}
	which are identical to the analogous QCD formulas for $q\rightarrow qg$ and $g\rightarrow q\bar{q}$ splitting by exchanging $e(f)^2\leftrightarrow C_F$ in Eq.~(\ref{space-averagedAPsplitfgam}) and $n_c(f)e(f)^2\leftrightarrow T_F$ in Eq.~(\ref{space-averagedAPsplitff}). For $g\rightarrow gg$ splittings the corresponding function is
	\begin{align}
		\langle\hat{P}_{g\rightarrow gg}\rangle(z,\varepsilon)=2C_A\left(\frac{z}{1-z}+\frac{1-z}{z}+z(1-z)\right)
		\label{space-averagedAPsplitgg}
	\end{align}
	
	Using the shorthand notation $\hat{P}_{E_{\tilde{\alpha}}\rightarrow(i,j)}=\hat{P}_{iE_{\tilde{\alpha}}}$ the analogous unpolarised QED initial-state splitting functions are given by
	\begin{align}
				\langle\hat{P}_{ff}\rangle(z,\varepsilon)&=e(f)^2\left[\frac{1+z^2}{1-z}-\varepsilon(1-z)\right]  \label{ftofgamunpol}\\
		\langle\hat{P}_{\gamma f}\rangle(z,\varepsilon)&=n_c(f)e(f)^2\frac{(1-z)^2+z^2-\varepsilon}{1-\varepsilon}\label{ftogamfunpol}\\
		\langle\hat{P}_{f\gamma }\rangle(z,\varepsilon)&=e(f)^2\left[\frac{1+(1-z)^2}{z}-\varepsilon z\right]
		\label{gamtoffunpol}
	\end{align}
	with the QCD counterparts obtained by the same substitutions as stated above for the FSR case and the additional QCD function for $g\to gg$ splittings
	\begin{align}
		\langle\hat{P}_{gg}\rangle(z,\varepsilon)= 2C_A\left[\frac{z}{1-z}+\frac{1-z}{z}+z(1-z)\right]
		\label{gtogunpol}
	\end{align}
	\section{Group factors for the infrared limits}
	\label{groupfactors}
	While in the soft and soft-collinear limit the Casimir operators $C$ are important ingredients for the subtraction terms in IR schemes, in the collinear limit all group theoretical factors $C$, $\gamma$ and $\gamma^{\prime}$ play an essential role as outlined in Sec.~\ref{secIntegratedSubs} and \ref{secDglapremnant}. As the definitions $\gamma$ and $\gamma^{\prime}$ for the description of the collinear limit address massless partons they are relevant primarily for QCD and QED interactions concerning energy scales of contemporary collider experiments.
	
	In general the Casimirs $C$ are defined for a particle identity $P$ by
	\begin{align}
	\vec{Q}(P)\vec{Q}(P)\equiv Q^2(P)=C(P)
	\end{align}
	with $\vec{Q}(P)$ the representations (or charges) of the group. In the following we use the shorthand notation $C_a\equiv C(P_a)$. 
	
	\subsection{QCD and QED group factors}
		The definitions $\gamma_a\equiv \gamma(P_a)$ come from the virtual contributions to the flavour-diagonal Altarelli-Parisi kernels $P_{aa}$. They can be extracted as a constant contribution of the following expansion for small $\xi_{\text{max}}$ \cite{Kunszt:1992tn}
	\begin{align}
	\begin{split}
	\int_{0}^{\xi_{\text{max}}}d\xi~P_{aa}(\xi)=&\int_{0}^{\xi_{\text{max}}}d\xi\left[\left(\frac{1}{\xi}\right)_++\delta(\xi)\log\xi_{\text{max}}\right]\xi\langle\hat{P}_{aa}\rangle(\xi)+\gamma_a\\
	\approx&2 C_a\log \xi_{\text{max}}+\gamma_a+\mathcal{O}(\xi_{\text{max}})
	\end{split}
	\end{align}
	for which the identities of Eqs.~(\ref{regSplittingfun}) and (\ref{casimirsplittingfun}) have been used.
	The definitions $\gamma^{\prime}_a\equiv\gamma^{\prime}(P_a)$ are constant contributions to the collinear integrals defined in Eqs.~(\ref{collinearintegrala}) and (\ref{collinearintegralb}).
	Numerical values of $\gamma$ and $\gamma^{\prime}$ for QCD and QED interactions are given in Table~\ref{CasimirGammaGammaP}.

	For the $SU(3)$ symmetry group we find the Casimir operators
	\begin{align}
		C_a^{\text{QCD}}=C_F=\sum_{a=1}^{N_c^2-1}t^at^a=\frac{N_c^2-1}{2N_c},\qquad\text{for }P_a\in\{q,\bar{q}\}
	\end{align}
	for generators $t^a$ of the fundamental representation of the colour algebra and
	\begin{align}
		C_a^{\text{QCD}}=C_A=\sum_{a,b}f^{abc}f^{abc}=C_A=N_c,\qquad\text{for }P_a=g
	\end{align}
	for generators $(T^a)_{bc}=-if^{abc}$ of the adjoint representations of the Lie algebra.
	
	In the case of QED interactions we need the representations of the $U(1)$ symmetry group, i.~e. the electic charge $e(P)$, which yield the trivial Casimir operators
	\begin{align}
		C^{\text{QED}}_a=e^2(P_a)
	\end{align}
	for charged fermions and bosons $P_a$. Explicit values of $C$ for QCD and QED interactions are collected in Table~\ref{CasimirGammaGammaP}.

	\begin{table}
		\centering
		{	\onehalfspacing
			\small
			\begin{tabularx}{\linewidth}{c|c|c|c|c|c|c}
				$P$&$C^{\text{QCD}}$&$\gamma^{\text{QCD}}$&$\gamma^{\prime\text{QCD}}$&$C^{\text{QED}}$&$\gamma^{\text{QED}}$&$\gamma^{\prime\text{QED}}$\\
				\hline
				$g$&$C_A$&$\begin{array}{r}
				\frac{11}{6}C_A\\
				-\frac{2}{3}T_Fn_F
				\end{array}$&$\begin{array}{r}
				\left(\frac{67}{9}-\frac{2\pi^2}{3}\right)C_A\\
				-\frac{23}{9}T_Fn_F
				\end{array}$&$0$&$0$&$0$\\
				$\gamma$&$0$&$-$&$-$&$0$&$-\frac{2}{3}f$&$-\frac{23}{9}f$\\
				$W^{\pm}$&$0$&$-$&$-$&$e(W^{\pm})^2$&$-$&$-$\\
				$Z/H$&$0$&$-$&$-$&$0$&$-$&$-$\\
				$q$&$C_F$&$\frac{3}{2}C_F$&$\left(\frac{13}{2}-\frac{2\pi^2}{3}\right)C_F$&$e(q)^2$&$\frac{3}{2}e(q)^2$&$\left(\frac{13}{2}-\frac{2\pi^2}{3}\right)e(q)^2$\\
				$l^{\pm}$&$0$&$-$&$-$&$e(l^{\pm})^2$&$\frac{3}{2}e(l^{\pm})^2$&$\left(\frac{13}{2}-\frac{2\pi^2}{3}\right)e(l^{\pm})^2$\\
				$\nu_l$&$0$&$-$&$-$&$0$&$-$&$-$\\
		\end{tabularx}}
		\caption{Explicit numbers of QCD and QED Casimirs and constants $\gamma$ and $\gamma^{\prime}$ for each SM particle using the definition $f=n_c\sum_{i=1}^{n_F}e^2(q_i)+\sum_{i=1}^{n_l}e^2(l_i)$}
		\label{CasimirGammaGammaP}
	\end{table}
	
	\subsection{EW Casimir operators}
	\label{secEWCasimirs}
	The formulas and considerations of this section correspond to those given in Refs.~\cite{Pozzorini:2001rs,Denner:2000jv}.
	For representations of the $SU(2)\times U(1)$ semi-simple group of EW interactions we have to consider both, diagonal and non-diagonal Casimir operators. In general, for generators $I^{V^a}$ the electroweak Casimir reads
	\begin{align}
		C^{\text{EW}}_{\varphi\varphi^{\prime}}=\sum_{V^a=A,Z,W^{\pm}}\left(I^{V^a}I^{\bar{V}^a}\right)_{\varphi\varphi^{\prime}}=\frac{1}{c_W^2}\left(\frac{Y_W}{2}\right)_{\varphi\varphi^{\prime}}^2+\frac{1}{s_W^2}C^{SU(2)}_{\varphi\varphi^{\prime}}
	\end{align}
	with
	\begin{align}
		C^{SU(2)}_{\varphi\varphi^{\prime}}=\sum_{a=1}^{3}(T^a)^2\quad.
	\end{align}
	For generators $T^a$ of irreducible representations of the $SU(2)$ group, i.~e. all SM particles $P_a$ except for gauge bosons, the Casimir operators are diagonal,
	\begin{align}
		C^{\text{EW}}_{\varphi,\text{irr}}=\delta_{\varphi\varphi^{\prime}}\left[\frac{Y_W^2}{4c_W^2}+\frac{I_W(I_W+1)}{s_W^2}\right]
	\end{align}
	where $Y_W$ denotes the hypercharge and $I_W$ the weak isospin with explicit values given in Table~\ref{leptonsquarks}. Note, that scalar fields of the Higgs doublet $\Phi =(\phi^+,\phi_0)^\top $ are fundamental representations of $SU(2)$ such that their Casimir operators correspond to those of left-handed leptons by
	\begin{align}
		\phi^+\leftrightarrow\bar{l}^{L},\qquad\phi^0\leftrightarrow\bar{\nu}^{L},\qquad\phi^-\leftrightarrow{l}^{L},\qquad\phi^*\leftrightarrow{\nu}^{L}\quad.
	\end{align}
	For EW gauge bosons, the adjoint representations, the Casimir operators read
	\begin{align}
		C^{\text{EW}}_{V^aV^b}=\frac{2}{s_W^2}\delta^{SU(2)}_{V^aV^b}
	\end{align}
	with the matrix $\delta^{SU(2)}_{V^aV^b}$ defined in the physical basis $V_a=(I^A,I^Z,I^{W^+},I^{W^-})$ as
	\begin{align}
		\delta^{SU(2)}_{V^aV^b}=\left( \begin{array}{rrrr}
		s_W^2&-s_Wc_W & 0 & 0 \\
		-s_Wc_W&c_W^2 & 0 & 0 \\ 
		0&0 & 1 & 0 \\
		0&0 & 0 & 1 \\ 
		\end{array}\right)\quad.
	\end{align}
		\section{The NLL EW Sudakov approximation factor to $\mu^+\mu^-\to HZ$}
	\label{HZapprox}
	Formulas and considerations of this chapter are published in Ref.~\cite{Bredt:2022dmm}.
	
	The analytic form of the NLL EW Sudakov correction factor to
	$\mu^+\mu^-\rightarrow HZ$ can be derived by applying the general factorisation
	formalism of \cite{Denner:2000jv,Denner:2001gw}. Due to the
	same EW coupling behaviour in
	$f\bar{f}\rightarrow HZ$ with fermions $f\neq t$ at one-loop level, neglecting all masses
	$m_f$ and apart from QED couplings with electric
	charges $Q_f$, this is in analogy to the results of the flavour and chirality generic formulas to $q\bar{q}\rightarrow HZ$
	in \cite{Granata:2017iod}.  For $s\gg M_{W}$, using the abbreviations for double and single logarithmic factors
	\begin{align}
	&L(s,M^2_W)=\frac{\alpha}{4\pi}\log^2\frac{s}{M^2_W}
	&l(s,M^2_W)=\frac{\alpha}{4\pi}\log\frac{s}{M^2_W} \qquad ,
	\end{align}
	we can approximate the leading logarithmic, angular-independent, terms
	coming from exchange of soft-collinear gauge bosons between pairs of
	external legs by
	\begin{align}
	\Lambda^{\kappa}_{l,\lambda}=A^{\kappa}_{\lambda} \; L(s,M_W^2) +
	B^{\kappa}_{\lambda} \; \log\frac{M_Z^2}{M_W^2}l(s,M_W^2)+C_{\lambda}
	\qquad .
	\label{Sud}
	\end{align}
	Here, $\lambda=T,L$ denote the transverse and longitudinal
	polarisation of the $Z$ boson, and $\kappa=L,R$ the muon initial state
	chirality, respectively. The constant factors read
	\begin{subequations}
		\begin{align}
		A^{\kappa}_{T}&=\;-\frac{1}{2}\left[2C^{\text{EW}}_{\mu^{\kappa}}+C^{\text{EW}}_{\Phi}+C^{\text{EW}}_{ZZ}\right]
		&
		A^{\kappa}_{L}&=\;-\left[C^{\text{EW}}_{\mu^{\kappa}}+C^{\text{EW}}_{\Phi}\right]
		\\
		B^{\kappa}_{T}&=\;2(I^Z_{\mu_{\kappa}})^2+(I^Z_H)^2
		&
		B^{\kappa}_{L}&=\;2\left[(I^Z_{\mu_{\kappa}})^2+(I^Z_H)^2\right]
		\\
		C_{T}&=\;\delta^{LSC,h}_H
		&
		C_L&=\;\delta^{LSC,h}_H+\delta^{LSC,h}_{\chi} \qquad.
		\end{align}
	\end{subequations}
	EW Casimir operators $C^{\text{EW}}$ calculated from the formulas of App.~\ref{secEWCasimirs} in addition to
	explicit values for $(I^Z)^2$ and $\delta^{LSC,h}$ extracted from
	\cite{Pozzorini:2001rs} are given by
	\begin{subequations}
		\begin{align}
		C^{\text{EW}}_{\mu^L}&=\;C^{\text{EW}}_{\Phi}=\frac{1+2c_W^2}{4s_W^2c_W^2} &
		C^{\text{EW}}_{\mu^R}&=\;\frac{1}{c_W^2} &
		C^{\text{EW}}_{ZZ}&=\;2\frac{c_W^2}{s_W^2}\\
		(I^Z_{\mu_L})^2&=\;\frac{(c_W^2-s_w^2)^2}{4s_W^2c_W^2} &
		(I^Z_{\mu_R})^2&=\;\frac{s_W^2}{c_W^2} &
		(I^Z_{H})^2&=\;\frac{1}{4s_W^2c_W^2}
		\end{align}
		\begin{align}
		\delta^{LSC,h}_H&=\;\frac{\alpha}{8\pi s_W^2}\left[\frac{1}{2c_W^2}\ln^2 \left(\frac{M_H^2}{M_Z^2}\right)+\ln^2 \left(\frac{M_H^2}{M_W^2}\right)\right] &
		\delta^{LSC,h}_{\chi}&=\;\frac{\alpha}{16\pi s_W^2c_W^2}\ln^2 \left(\frac{M_H^2}{M_Z^2}\right)
		\end{align}
	\end{subequations}
	In the same context, subleading, angular-dependent, terms
	proportional to $l(s,M_W)\log(|t|/s)$ and $l(s,M_W^2)\log(|u|/s)$ due
	to $W^{\pm}$ boson exchange between initial- and final-state legs arise. For the
	considered process they take the form
	\begin{align}
	\label{sudangle}
	\Lambda^{\kappa}_{\theta,\lambda}=- \delta_{\kappa L}
	\frac{D_{\lambda}}{I^Z_{\mu_{\kappa}}} \;
	l(s,M_W^2)\left[\log\frac{|t|}{s} +
	\log\frac{|u|}{s}\right]
	\end{align}
	with constants
	\begin{equation}
	D_T =\; - \frac{c_W(1+c_W^2)}{2s_W^2} \qquad
	D_L =\; -\frac{c_W}{s_W} \qquad ,
	\end{equation}
	using the shortcuts $s_W=\sin \theta_W$ and $c_W=\cos
	\theta_W$. Considering the Mandelstam variables $t$ and $u$ approximated in the
	high-energy limit
	\begin{equation}
	t=\;(p_{\mu^+}-p_H)^2\sim -\frac{s}{2}(1-\cos \theta_{H}) \qquad
	u=\;(p_{\mu^+}-p_Z)^2\sim -\frac{s}{2}(1+\cos \theta_{H}) \quad,
	\end{equation}
	Eq.~(\ref{sudangle}) can be written in terms of the Higgs polar angle
	$\theta_{H}$.
	Single-logarithmic terms $\Lambda^{\kappa}_{s,\lambda}$ originating from virtual
	soft/collinear gauge bosons emitted from single external legs and $\Lambda^{\kappa}_{\text{PR},\lambda}$ from  the renormalisation of coupling parameters can be expressed as
	\begin{equation}
	\Lambda^{\kappa}_{s,\lambda} =\; E^{\kappa}_{s,\lambda} \;
	l(s,M_W^2) + \frac{\alpha}{4\pi} F^{\kappa}_{s,\lambda}
	\qquad
	\Lambda^{\kappa}_{\text{PR},\lambda} =\;
	E^{\kappa}_{\text{PR},\lambda} \; l(s,M_W^2) +
	\frac{\alpha}{4\pi}F^{\kappa}_{\text{PR},\lambda} \quad .
	\label{singlePR}
	\end{equation}
	Explicitly, the quantities $E^{\kappa}_{s,\lambda}$ and $F^{\kappa}_{s,\lambda}$ for $\Lambda^{\kappa}_{s,\lambda}$ read
	\begin{subequations}
		\begin{align}
		&E^{\kappa}_{s,T} = 3C^{\text{EW}}_{\mu^{\kappa}}+2C^{\text{EW}}_{\Phi}+\frac{1}{2}b^{\text{EW}}_{ZZ}
		- \frac{3}{4s_W^2}\frac{m_t^2}{M_W^2}
		\\
		&E^{\kappa}_{s,L} = 3C^{\text{EW}}_{\mu^{\kappa}}+4C^{\text{EW}}_{\Phi}-\frac{3}{2s_W^2}
		\frac{m_t^2}{M_W^2}
		\\
		&F^{\kappa}_{s,T} = \left(\frac{3}{4s_W^2}\frac{m_t^2}{M_W^2}+T_{ZZ}\right)\log
		\frac{m_t^2}{M_W^2}+\left(\frac{M_Z^2}{24s_W^2M_W^2}-2C^{\text{EW}}_{\Phi}\right)\log
		\frac{M_H^2}{M_W^2}
		\\
		&F^{\kappa}_{s,L} = \frac{3}{2s_W^2}\frac{m_t^2}{M_W^2}\log
		\frac{m_t^2}{M_W^2}+\left(\frac{M_Z^2}{8s_W^2M_W^2}-2C^{\text{EW}}_{\Phi}\right)\log
		\frac{M_H^2}{M_W^2}
		\end{align}
	\end{subequations}
	while 
	$E^{\kappa}_{\text{PR},\lambda}$ and $F^{\kappa}_{\text{PR},\lambda}$ for $\Lambda^{\kappa}_{\text{PR},\lambda}$ are given by
	\begin{subequations}
		\begin{align}
		&E^{\kappa}_{\text{PR},T} =
		-b^{\text{EW}}_{WW}+\rho_{\mu_{\kappa}}\frac{s_W}{c_W}b^{\text{EW}}_{AZ}+2C^{\text{EW}}_{\Phi}
		- \frac{1}{2}b^{\text{EW}}_{ZZ}-\frac{3}{4s_W^2}\frac{m_t^2}{M_W^2}
		\\
		&E^{\kappa}_{\text{PR},L}=-b^{\text{EW}}_{WW}+\rho_{\mu_{\kappa}}\frac{s_W}{c_W}b^{\text{EW}}_{AZ}
		\\
		\begin{split}
		&F^{\kappa}_{\text{PR},T}=\frac{5}{6}\left(\frac{1}{s_W^2} +
		\frac{\rho_{\mu_{\kappa}}}{c_W^2}-\frac{M_Z^2}{2s_W^2M_W^2}\right)\log
		\frac{M_H^2}{M_W^2}
		\\
		&\qquad\qquad-\left[\frac{9+6s_W^2-32s^4_w}{18s_W^2}\left(\frac{1}{s_W^2}
		+ \frac{\rho_{\mu_{\kappa}}}{c_W^2}\right)+T_{ZZ} -
		\frac{3}{4s_W^2}\frac{m_t^2}{M_W^2}\right]\log
		\frac{m_t^2}{M_W^2}
		\end{split}
		\\
		&F^{\kappa}_{\text{PR},L}=\left(\frac{1}{s_W^2} +
		\frac{\rho_{\mu_{\kappa}}}{c_W^2}\right)\left[\frac{5}{6}\log
		\frac{M_H^2}{M_W^2} -\frac{9+6s_W^2-32s_W^4}{18s_W^2}\log
		\frac{m_t^2}{M_W^2} \right]
		\end{align}
	\end{subequations}
	The explicit values of the $\beta$-function coefficients
	$b^{\text{EW}}_{ZZ}$, $b^{\text{EW}}_{AZ}$ and $b^{\text{EW}}_{WW}$ and the
	coefficient $T_{ZZ}$ used here are \cite{Pozzorini:2001rs}
	\begin{subequations}
		\begin{align}
		&b^{\text{EW}}_{ZZ}=\frac{19-38s_W^2-22s_W^4}{6s_W^2c_W^2} &
		b^{\text{EW}}_{AZ}=-\frac{19+22s_W^2}{6s_Wc_W}&
		&b^{\text{EW}}_{WW}=\frac{19}{6s_W^2}\\
		&T_{ZZ}=\frac{9-24s_W^2+32s_W^4}{36s_W^2c_W^2} \qquad.
		\end{align}
	\end{subequations}
	The constants $E^{\kappa}_{\text{PR},\lambda}$ and
	$F^{\kappa}_{\text{PR},\lambda}$ depend on the parameter
	$\rho_{\mu_{\kappa}}$ which is defined by the EW quantum numbers of
	the muon as \cite{Granata:2017iod}
	\begin{align}
	\rho_{\mu_{\kappa}}=\frac{Q_{\mu}-T^3_{\mu_{\kappa}}}{T^3_{\mu_{\kappa}}-Q_{\mu}s_W^2}
	\qquad.
	\end{align}
	Summing the contributions from Eqs.~(\ref{Sud}), (\ref{sudangle}) and
	(\ref{singlePR}) the overall Sudakov correction factor depending on the
	muon chirality $\kappa$ and $Z$ boson polarisation $\lambda$ can be
	formulated as
	\begin{equation}
	\label{sudoverall}
	\Lambda^{\kappa}_{\lambda} = \Lambda^{\kappa}_{l,\lambda} +
	\Lambda^{\kappa}_{\theta,\lambda} + \Lambda^{\kappa}_{s,\lambda} +
	\Lambda^{\kappa}_{\text{PR},\lambda} \qquad .
	\end{equation}
	This factor can now be used for an approximation of the relative NLO
	correction to the unpolarised process for which the photon radiation
	effects are subtracted, i.~e. emulating the virtual EW effects in the
	high-energy limit. For the definition of this correction factor two different approximations are applied. For the first one consider
	\begin{equation}
	\label{longitudinalestimation}
	\Lambda^{\kappa}_{\lambda}\,\mathcal{M}^{\mu^+_{\kappa}\mu^-_{\kappa}\rightarrow
		HZ_{\lambda}}_0 \quad \xrightarrow{s\gg M_W^2} \quad
	\delta_{\lambda L}\Lambda^{\kappa}_{\lambda} \,
	\mathcal{M}^{\mu^+_{\kappa}\mu^-_{\kappa}\rightarrow HZ_{\lambda}}_0=\Lambda^{\kappa}_{L} \,
	\mathcal{M}^{\mu^+_{\kappa}\mu^-_{\kappa}\rightarrow HZ_{L}}_0
	\end{equation}
	since the Born amplitudes for transverse
	polarised $Z$ bosons are suppressed by $M_Z^2/s$
	\cite{Chanowitz:1985hj,Bohm:2001yx}.
	
	For the second approximation, due to the
	ultra-relativistic initial state momenta in the process chirality and helicity of the muons are assumed to coincide. Another consequence is that amplitudes with helicity configurations $(+,+)$ and $(-,-)$ of the muons are assumed to vanish. Due to this the chirality dependent quantities $|\mathcal{M}^{\mu^+_{\kappa}\mu^-_{\kappa}\rightarrow
		HZ_{L}}_0|^2$ can be evaluated easily in the \texttt{WHIZARD} framework using the corresponding polarisation settings of the muon initial-states. 
	For an estimated unpolarised factor we arrive at a factor of the form
	\begin{align}
	\label{unpolsud}
	\Lambda_{\text{est}}^{\text{unpol}} =
	\frac{\sum_{\kappa}\Lambda^{\kappa}_{L}
		\left|\mathcal{M}^{\mu^+_{\kappa}\mu^-_{\kappa}\rightarrow
			HZ_{L}}_0\right|^2}{4\left|\mathcal{M}^{\mu^+\mu^-\rightarrow
			HZ_L}_0\right|^2}
	\end{align}
	which approximates the non-integrated virtual corrections in the high-energy limit for
	unpolarised muon beams when multiplied with the
	unpolarised Born squared amplitudes $\left|\mathcal{M}^{\mu^+\mu^-\rightarrow
		HZ_L}_0\right|^2$. Using the fact that the amplitudes depend on the
	muon chiralities only due to the EW couplings, the angular dependency
	of squared Born amplitudes exclusive in the muon chiralities normalised
	to the unpolarised ones drops out. This implies
	\begin{align}
	\frac{\left|\mathcal{M}^{\mu^+_{\kappa}\mu^-_{\kappa}\rightarrow
			HZ_{L}}_0\right|^2}{\left|\mathcal{M}^{\mu^+\mu^-\rightarrow
			HZ^L}_0\right|^2}=\frac{\left(d\sigma^{\mu^+_{\kappa}\mu^-_{\kappa}\rightarrow
			HZ_{L}}_{B}/d\Omega\right)}{\left(d\sigma^{\mu^+\mu^-\rightarrow
			HZ_{L}}_{B}/d\Omega\right)}=\frac{\sigma^{\mu^+_{\kappa}\mu^-_{\kappa}\rightarrow
			HZ_{L}}_{B}}{\sigma^{\mu^+\mu^-\rightarrow
			HZ_{L}}_{B}}= \frac{\sigma^{\mu^+_{\kappa}\mu^-_{\kappa}\rightarrow
			HZ}_{B}}{\sigma^{\mu^+\mu^-\rightarrow
			HZ}_{B}}
	\quad .
	\end{align}
	Hence, in this approach Eq.~(\ref{unpolsud}) can be rewritten as
	\begin{align}
	\Lambda_{\text{est}}^{\text{unpol}} = \frac{1}{4}
	\frac{\sum_{\kappa} \, \Lambda^{\kappa}_{L} \,
		\sigma^{\mu^+_{\kappa}\mu^-_{\kappa}\rightarrow
			HZ}_{B}}{\sigma^{\mu^+\mu^-\rightarrow
			HZ}_{B}} \qquad.
	\label{finaldeltaunpol}
	\end{align}
	This approximative correction factor is evaluated using integrated polarised and unpolarised Born cross sections $\sigma_B$ computed with \texttt{WHIZARD+RECOLA}.
	The analytical Sudakov factors $\Lambda^{\kappa}_{\lambda}$ of Eq.~(\ref{sudoverall}), the estimated factor $\Lambda^{\text{unpol}}_{\text{est}}$
	of Eq.~(\ref{finaldeltaunpol}) for the scattering angle
	$\theta_{H}=90^{\circ}$ as well as $\Lambda^{\text{unpol}}_{\text{est,c}}$,
	i.~e. Eq.~(\ref{finaldeltaunpol}) with the angular-dependent terms
	$\Lambda^{\kappa}_{\theta,L}$ dropped, are depicted in
	Fig.~\ref{sudpic} as a function of the centre-of-mass energy.
	
	\chapter{WHIZARD steering and output files}
		For the SINDARIN settings
	{\scriptsize
		\begin{verbatim}
		alias pr = b:B:g:A
		alpha_power = 0
		alphas_power = 2
		$nlo_correction_type = "EW"
		process pptt_ew = pr, pr => t,T {nlo_calculation = full}
		\end{verbatim}}
	{\normalsize the following output can be obtained in order to compute NLO EW corrections to the process $pp\to t\bar{t}$ with $p$ defined here for simplicity only by partons $b$, $\bar{b}$, $\gamma$ and $g$.}
	\section{FKS table}
	\label{secFKStable}
	{\normalsize
	FKS regions are set up according to the FKS table written into the \texttt{fks\_regions.out} file. This file is useful for the purpose of understanding the underlying FKS structure of a process, in particular for those involving a multi-dimensional flavour space. The following definitions are used:
	\begin{itemize}
		\setlength{\labelsep}{-1cm}
		\item[\texttt{alr}] \qquad \qquad index assigned to an FKS singular region
		\item[\texttt{flst\_real}] \qquad \qquad real flavour structure
		\item[\texttt{i\_real}] \qquad \qquad index of the real flavour structure
		\item[\texttt{em}] \qquad \qquad position of the emitter in \texttt{flst\_real}
		\item[\texttt{ftuples}] \qquad \qquad FKS pairs corresponding to \texttt{flst\_real}
		\item[\texttt{flst\_born}] \qquad \qquad underlying Born flavour structure
		\item[\texttt{i\_born}] \qquad \qquad index of \texttt{flst\_born}
		\item[\texttt{corr}] \qquad \qquad correction type for the IR splitting associated with the FKS region
	\end{itemize}
	\normalsize
	  Besides the correction types \texttt{qcd} and \texttt{ew}, \texttt{corr} can take the value \texttt{none}. In this case, the subtraction terms are dropped in the calculation as the real flavour structure is not singular for the corresponding FKS parameters. In this context, parameters associated with subtraction terms, e.~g. \texttt{flst\_born}, therefore have no meaning.
	  \quad\\
	  
	 For the exemplary process and settings, stated above, the following output is given by the \texttt{fks\_regions.out} file:}
	{\scriptsize
	\begin{verbatim}
		Total number of regions:    36
		alr ||        flst_real || i_real ||  em ||             ftuples ||     flst_born || i_born || corr
		  1 || [-5, 5, 6,-6,22] ||      1 ||   3 || {(0,5),(3,5),(4,5)} || [-5, 5, 6,-6] ||      1 ||  ew
		  2 || [-5, 5, 6,-6,22] ||      1 ||   4 || {(0,5),(3,5),(4,5)} || [-5, 5, 6,-6] ||      1 ||  ew
		  3 || [-5, 5, 6,-6,22] ||      1 ||   0 || {(0,5),(3,5),(4,5)} || [-5, 5, 6,-6] ||      1 ||  ew
		  4 || [-5, 5, 6,-6,21] ||      2 ||   3 || {(0,5),(3,5),(4,5)} || [-5, 5, 6,-6] ||      1 ||  qcd
		  5 || [-5, 5, 6,-6,21] ||      2 ||   4 || {(0,5),(3,5),(4,5)} || [-5, 5, 6,-6] ||      1 ||  qcd
		  6 || [-5, 5, 6,-6,21] ||      2 ||   0 || {(0,5),(3,5),(4,5)} || [-5, 5, 6,-6] ||      1 ||  qcd
		  7 || [ 5,22, 6,-6,-5] ||      3 ||   2 || {(1,5),(2,5)}       || [-5, 5, 6,-6] ||      1 ||  ew
		  8 || [-5,22, 6,-6,-5] ||      3 ||   1 || {(1,5),(2,5)}       || [21,22, 6,-6] ||      5 ||  qcd
		  9 || [-5,21, 6,-6,-5] ||      4 ||   2 || {(1,5),(2,5)}       || [-5, 5, 6,-6] ||      1 ||  qcd
		 10 || [-5,21, 6,-6,-5] ||      4 ||   1 || {(1,5),(2,5)}       || [21,21, 6,-6] ||      6 ||  none
		 11 || [ 5,-5, 6,-6,22] ||      5 ||   3 || {(0,5),(3,5),(4,5)} || [ 5,-5, 6,-6] ||      2 ||  ew
		 12 || [ 5,-5, 6,-6,22] ||      5 ||   4 || {(0,5),(3,5),(4,5)} || [ 5,-5, 6,-6] ||      2 ||  ew
		 13 || [ 5,-5, 6,-6,22] ||      5 ||   0 || {(0,5),(3,5),(4,5)} || [ 5,-5, 6,-6] ||      2 ||  ew
		 14 || [ 5,-5, 6,-6,21] ||      6 ||   3 || {(0,5),(3,5),(4,5)} || [ 5,-5, 6,-6] ||      2 ||  qcd
		 15 || [ 5,-5, 6,-6,21] ||      6 ||   4 || {(0,5),(3,5),(4,5)} || [ 5,-5, 6,-6] ||      2 ||  qcd
		 16 || [ 5,-5, 6,-6,21] ||      6 ||   0 || {(0,5),(3,5),(4,5)} || [ 5,-5, 6,-6] ||      2 ||  qcd
		 17 || [ 5,22, 6,-6, 5] ||      7 ||   2 || {(1,5),(2,5)}       || [ 5,-5, 6,-6] ||      2 ||  ew
		 18 || [ 5,22, 6,-6, 5] ||      7 ||   1 || {(1,5),(2,5)}       || [21,22, 6,-6] ||      5 ||  qcd
		 19 || [ 5,21, 6,-6, 5] ||      8 ||   2 || {(1,5),(2,5)}       || [ 5,-5, 6,-6] ||      2 ||  qcd
		 20 || [ 5,21, 6,-6, 5] ||      8 ||   1 || {(1,5),(2,5)}       || [21,21, 6,-6] ||      6 ||  none
		 21 || [22,-5, 6,-6,-5] ||      9 ||   1 || {(1,5),(2,5)}       || [ 5,-5, 6,-6] ||      2 ||  ew
		 22 || [22,-5, 6,-6,-5] ||      9 ||   2 || {(1,5),(2,5)}       || [22,21, 6,-6] ||      4 ||  qcd
		 23 || [22, 5, 6,-6, 5] ||     10 ||   1 || {(1,5),(2,5)}       || [-5, 5, 6,-6] ||      1 ||  ew
		 24 || [22, 5, 6,-6, 5] ||     10 ||   2 || {(1,5),(2,5)}       || [22,21, 6,-6] ||      4 ||  qcd
		 25 || [22,21, 6,-6,21] ||     11 ||   3 || {(2,5),(3,5),(4,5)} || [22,21, 6,-6] ||      4 ||  qcd
		 26 || [22,21, 6,-6,21] ||     11 ||   4 || {(2,5),(3,5),(4,5)} || [22,21, 6,-6] ||      4 ||  qcd
		 27 || [22,21, 6,-6,21] ||     11 ||   2 || {(2,5),(3,5),(4,5)} || [22,21, 6,-6] ||      4 ||  qcd
		 28 || [21,-5, 6,-6,-5] ||     12 ||   1 || {(1,5),(2,5)}       || [ 5,-5, 6,-6] ||      2 ||  qcd
		 29 || [21,-5, 6,-6,-5] ||     12 ||   2 || {(1,5),(2,5)}       || [21,21, 6,-6] ||      6 ||  none
		 30 || [21, 5, 6,-6, 5] ||     13 ||   1 || {(1,5),(2,5)}       || [-5, 5, 6,-6] ||      1 ||  qcd
		 31 || [21, 5, 6,-6, 5] ||     13 ||   2 || {(1,5),(2,5)}       || [21,21, 6,-6] ||      6 ||  none
		 32 || [21,22, 6,-6,21] ||     14 ||   3 || {(1,5),(3,5),(4,5)} || [21,22, 6,-6] ||      5 ||  qcd
		 33 || [21,22, 6,-6,21] ||     14 ||   4 || {(1,5),(3,5),(4,5)} || [21,22, 6,-6] ||      5 ||  qcd
		 34 || [21,22, 6,-6,21] ||     14 ||   1 || {(1,5),(3,5),(4,5)} || [21,22, 6,-6] ||      5 ||  qcd
		 35 || [21,21, 6,-6,22] ||     15 ||   3 || {(3,5),(4,5)}       || [21,21, 6,-6] ||      6 ||  ew
		 36 || [21,21, 6,-6,22] ||     15 ||   4 || {(3,5),(4,5)}       || [21,21, 6,-6] ||      6 ||  ew
		------------------------------------------------------------------------
		
	\end{verbatim}}
	\section{BLHA order file}
	\label{secBLHAorder}
	
	{\normalsize \onehalfspacing The output \texttt{pptt\_ew\_SUB.olp} file,  i.~e. the BLHA order file for requesting squared matrix elements which enter the real subtraction terms, contains the following statements:}
	{\scriptsize
	\begin{verbatim}
	# BLHA order written by WHIZARD 3.0.3+
	
	# BLHA interface mode: OpenLoops
	# process: pptt_ew_SUB
	# model: SM
	InterfaceVersion         BLHA2
	CorrectionType           EW
	Extra AnswerFile         pptt_ew_SUB.olc
	IRregularisation         CDR
	CouplingPower QCD        2
	CouplingPower QED        0
	ewscheme                 Gmu
	extra use_cms            1
	extra me_cache           0
	extra IR_on              0
	extra psp_tolerance      10e-7
	
	
	# Process definitions
	
	
	
		AmplitudeType            Tree
		CorrectionType           EW
		CouplingPower QCD        2
		CouplingPower QED        0
		-5   5 ->   6  -6
		
		AmplitudeType            ccTree
		CorrectionType           EW
		CouplingPower QCD        2
		CouplingPower QED        0
		-5   5 ->   6  -6
		
		AmplitudeType            sctree_polvect
		CorrectionType           EW
		CouplingPower QCD        2
		CouplingPower QED        0
		-5   5 ->   6  -6
		
		AmplitudeType            Tree
		CorrectionType           EW
		CouplingPower QCD        2
		CouplingPower QED        0
		5  -5 ->   6  -6
		
		AmplitudeType            ccTree
		CorrectionType           EW
		CouplingPower QCD        2
		CouplingPower QED        0
		5  -5 ->   6  -6
		
		AmplitudeType            sctree_polvect
		CorrectionType           EW
		CouplingPower QCD        2
		CouplingPower QED        0
		5  -5 ->   6  -6
		
		AmplitudeType            Tree
		CorrectionType           QCD
		CouplingPower QCD        1
		CouplingPower QED        1
		-2002  21 ->   6  -6
		
		AmplitudeType            ccTree
		CorrectionType           QCD
		CouplingPower QCD        1
		CouplingPower QED        1
		-2002  21 ->   6  -6
		
		AmplitudeType            sctree_polvect
		CorrectionType           QCD
		CouplingPower QCD        1
		CouplingPower QED        1
		-2002  21 ->   6  -6
		
		AmplitudeType            Tree
		CorrectionType           QCD
		CouplingPower QCD        1
		CouplingPower QED        1
		21 -2002 ->   6  -6
		
		AmplitudeType            ccTree
		CorrectionType           QCD
		CouplingPower QCD        1
		CouplingPower QED        1
		21 -2002 ->   6  -6
		
		AmplitudeType            sctree_polvect
		CorrectionType           QCD
		CouplingPower QCD        1
		CouplingPower QED        1
		21 -2002 ->   6  -6
		
		AmplitudeType            Tree
		CorrectionType           EW
		CouplingPower QCD        2
		CouplingPower QED        0
		21  21 ->   6  -6
		
		AmplitudeType            ccTree
		CorrectionType           EW
		CouplingPower QCD        2
		CouplingPower QED        0
		21  21 ->   6  -6
		
		AmplitudeType            sctree_polvect
		CorrectionType           EW
		CouplingPower QCD        2
		CouplingPower QED        0
		21  21 ->   6  -6
		
	\end{verbatim}}
\newpage
	\section{SINDARIN steering file}
	\label{secSindarin}
	{\normalsize The \texttt{SINDARIN} input commands for the \texttt{.sin} file to steer an NLO EW cross section computation for the process $pp\to e^+e^-j$ read as follows:}
	\begin{verbatim}
		# model
		model = SM (Complex_Mass_Scheme)
		$blha_ew_scheme = "GF"
				
		GF = 1.16639E-5
	
		# protons, jets, definitions ...
		alias jet = u:U:d:D:s:S:c:C:b:B:gl:A:E1:e1
		alias pr = u:U:d:D:s:S:c:C:b:B:A:gl
		alias j = u:U:d:D:s:S:c:C:b:B:A:gl
		alias quarks = u:U:d:D:s:S:c:C:b:B
		
		# masses
		mZ = 91.15348
		mW = 80.35797
		
		ms = 0
		mc = 0
		mb=0
		mtop = 173.34
		
		me = 0
		mmu = 0
		mtau = 0
		
		# widths
		wZ = 2.494266
		wW = 2.084299
		wtop = 1.36918
		wH = 0
		
		# coupling powers
		alpha_power = 2
		alphas_power = 1
		
		# running of alpha_s
		alphas_nf = 5
		alphas_order=2
		?alphas_from_lhapdf = true
		alphas=0.118
		
		# OLP and integrator settings
		$method = "openloops"
		?openloops_use_cms = true
		$integration_method = "vamp2"
		$rng_method = "rng_stream"
		openmp_num_threads = 1
		?omega_openmp = false
		
		# beam settings
		sqrts = 13 TeV
		
		beams = p,p => lhapdf
		$lhapdf_file = "LUXqed_plus_PDF4LHC15_nnlo_100"
		
		# photon recombination
		photon_rec_r0=0.1
		
		# jet clustering algorithm
		jet_algorithm = antikt_algorithm
		jet_r = 0.4
		
		
		cuts =
		
		let subevt @recfermion = photon_recombination [A:e1:E1:quarks] in
		let subevt @dressedleptons = select if abs(real(PDG)) == 11. [@recfermion] in
		let subevt @firstlep = extract index 1 [@dressedleptons] in
		let subevt @secondlep = extract index 2 [@dressedleptons] in
		let subevt @dressedquarks = select if abs(real(PDG)) <= 5. [@recfermion] in
		let subevt @notreco_photon = select if abs(real(PDG)) == 22. [@recfermion] in
		let subevt @clustered_jets = cluster [join[gl, join[@notreco_photon,@dressedquarks]]] in
		let subevt @pt_selected = select if Pt > 30 GeV [@clustered_jets] in
		let subevt @eta_selected = select if abs(Eta) < 4.5 [@pt_selected] in
		count[@eta_selected] >= 1
		and all Pt > 10. [@dressedleptons]
		and all abs(Eta) < 2.5 [@dressedleptons]
		and all Dist > 0.4 [@firstlep, @secondlep]
		and all M >= 30 [collect[@dressedleptons]]
		
		
		
		scale = sum (Pt/2) [jet]
		
		$nlo_correction_type = "EW"
		
		process nlo_ppeej_ew = pr, pr => e1,E1, j {nlo_calculation = full}
		
		
		integrate (nlo_ppeej_ew) { iterations = 1:1000:"gw" mult_call_virt = 0.1}
	\end{verbatim}
	\chapter{Further checks and numerical results}
	\label{secChecksDiffDistr}
	\normalsize
	\section{Checks on differential distributions}
\begin{figure}[h]
		\centering
		\includegraphics[width=0.49\textwidth]{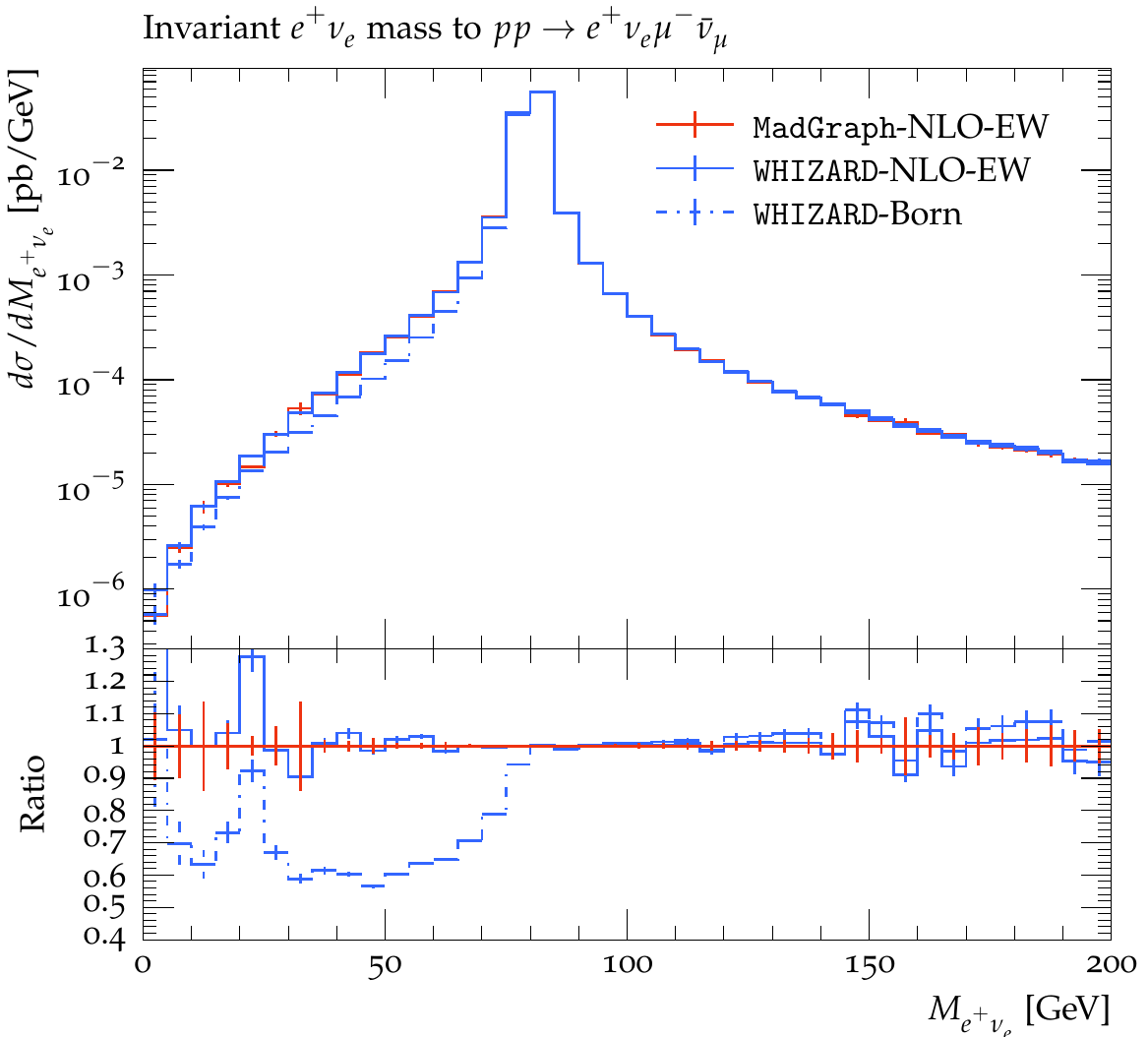}
		\includegraphics[width=0.49\linewidth]{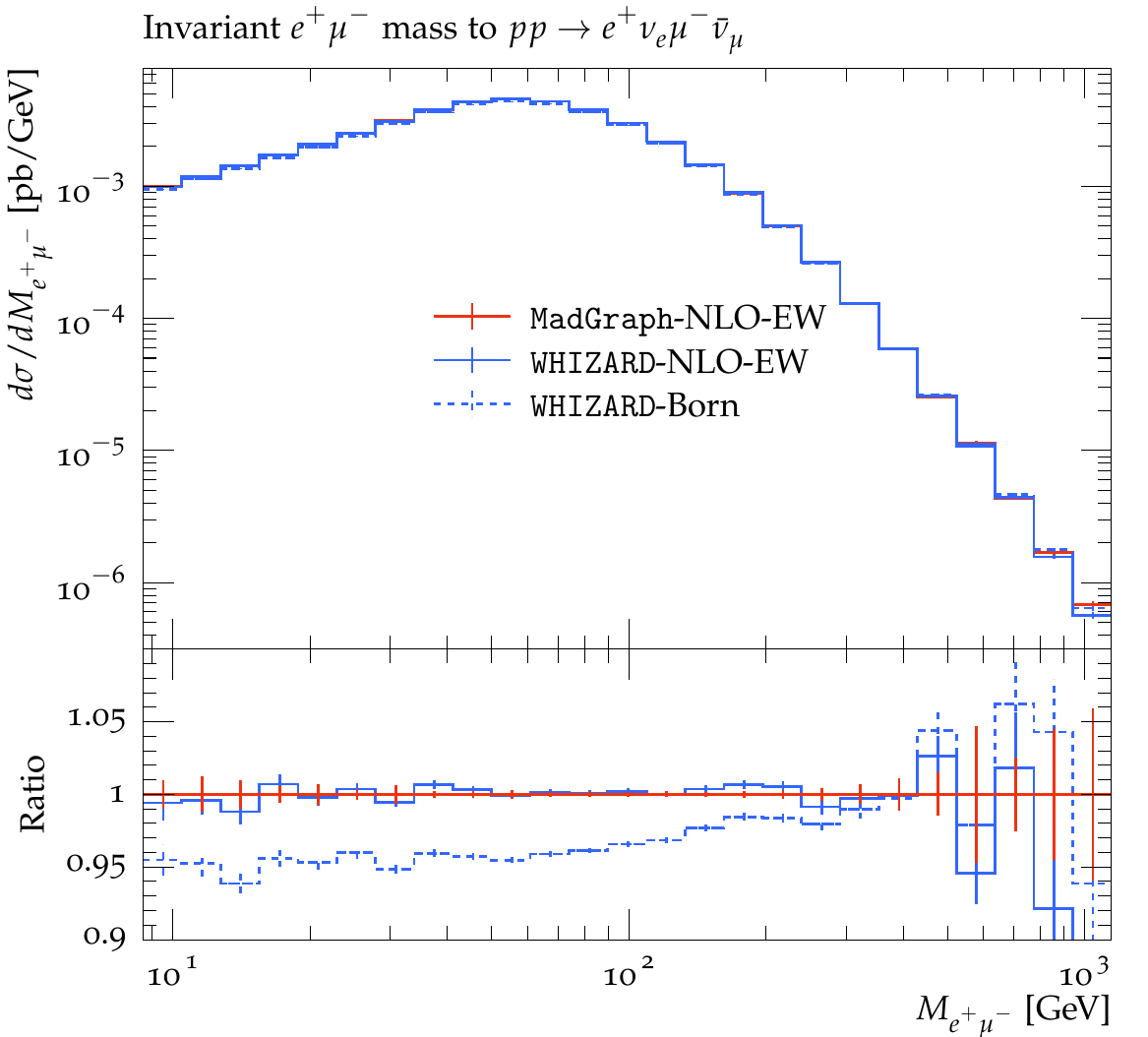}
		\caption{Checks on differential distributions at NLO EW of the invariant mass of the $e^+\nu_e$ system (left) and the $e^+\mu^-$ system (right) obtained with \texttt{WHIZARD+OpenLoops} and \texttt{MG5\_aMC@NLO} and  Born differential distributions obtained with \texttt{WHIZARD+OpenLoops} for the process $pp\to e^+\nu_e\mu^-\nu_{\mu}$ at $\sqrt{s}=13$ TeV}
		\label{diffdistrInvMass}
	\end{figure}

	The differential distributions at NLO EW obtained with fixed-order `events' generated with \texttt{WHIZARD+OpenLoops} are validated using the reference MC generator \texttt{MG5\_aMC@NLO} for cross-checks. In Figs.~\ref{diffdistrInvMass} - \ref{diffdistrMuon} comparisons of NLO EW differential distributions to the process $pp\to e^+\nu_e\mu^-\bar{\nu}_{\mu}$ obtained with both MC generators and the corresponding Born distribution obtained with \texttt{WHIZARD+OpenLoops} are shown. The ratio for each distribution, depicted below the main plots, uses the `\texttt{MadGraph}-NLO-EW' distribution as normalisation.
		\begin{figure}[h]
	\centering
	\includegraphics[width=0.49\linewidth]{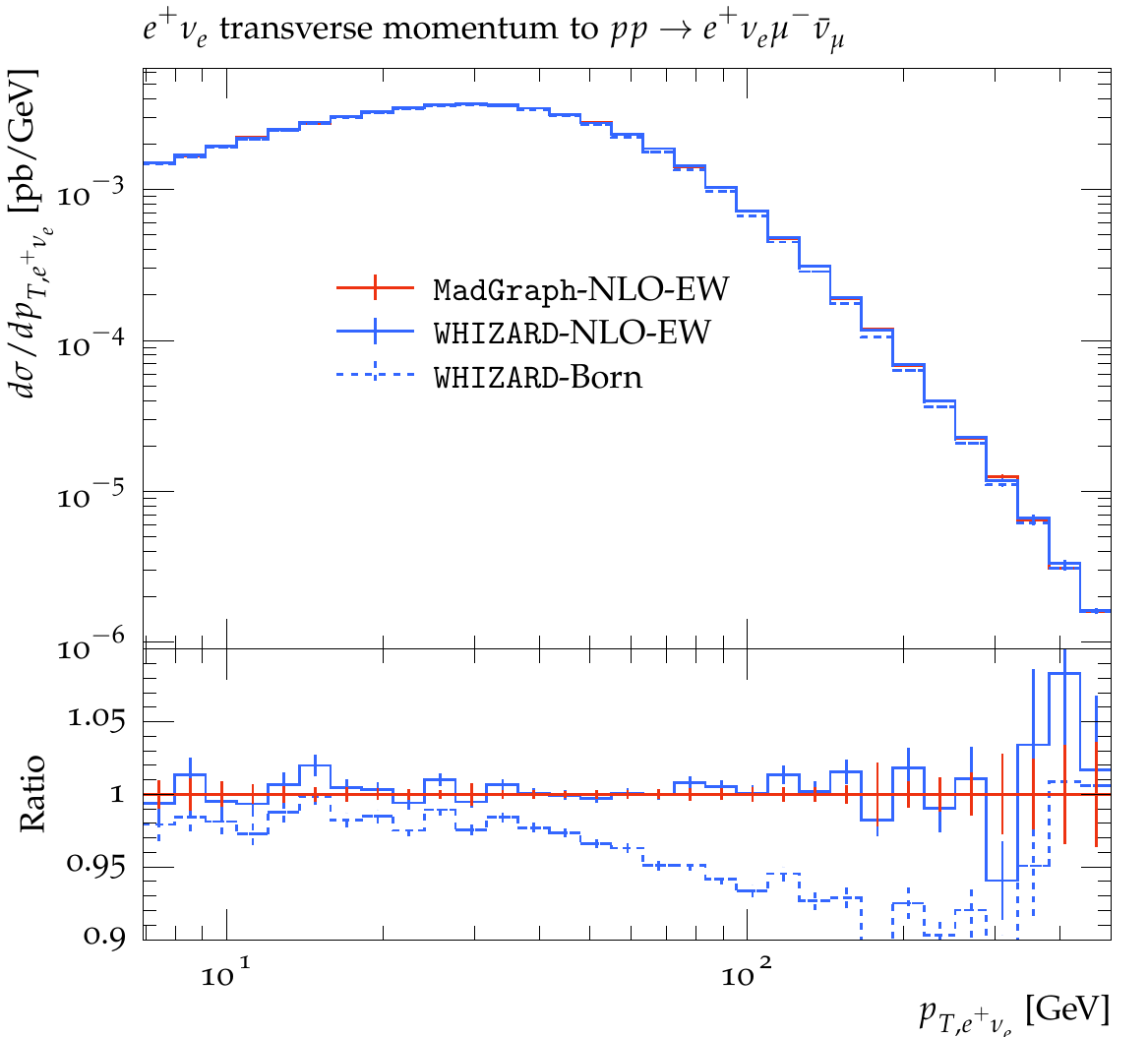}
	\includegraphics[width=0.49\textwidth]{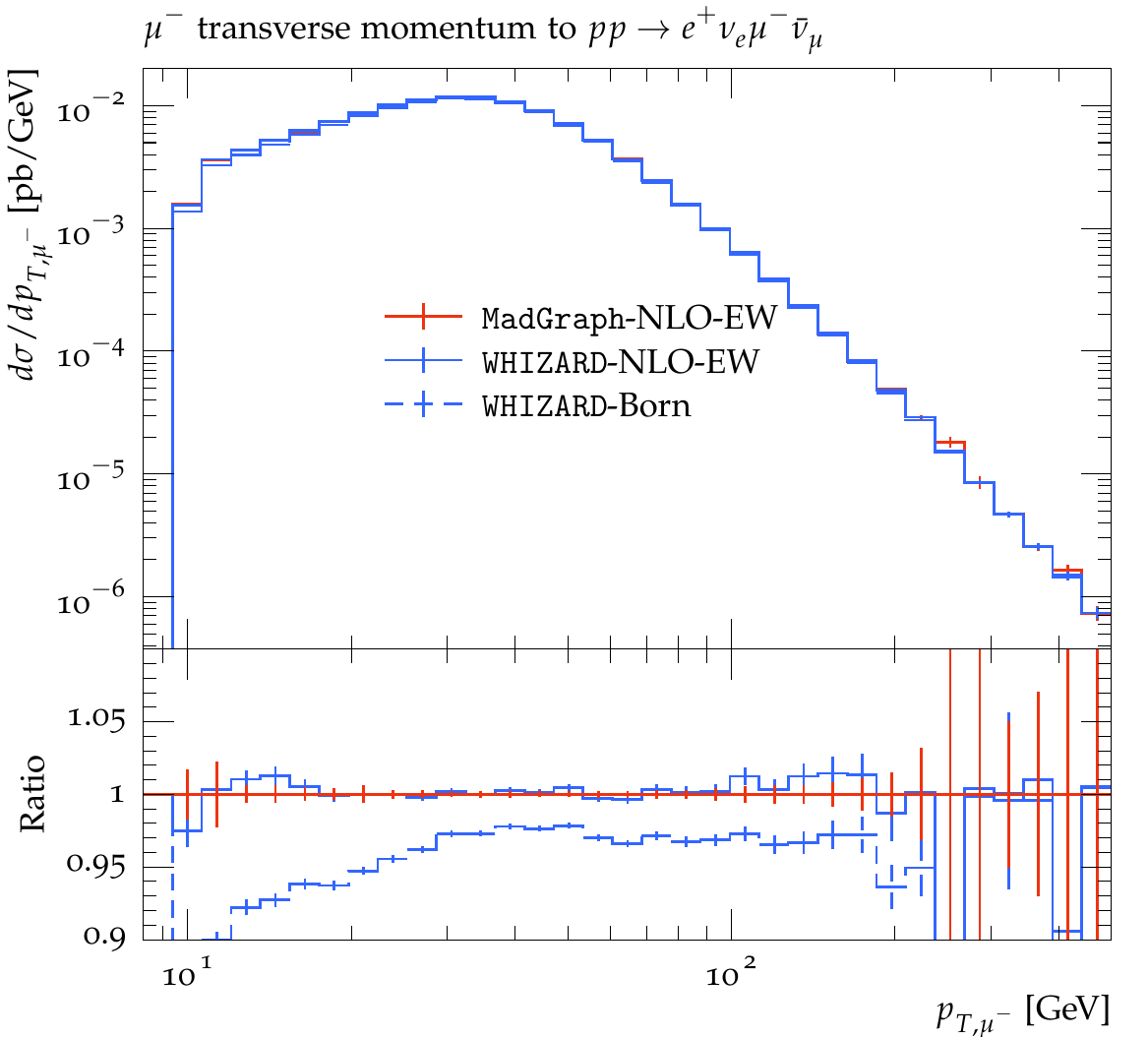}
	\caption{Checks on differential distributions at NLO EW of the transverse momentum of the $e^+\nu_e$ system (left) and the muon (right) obtained with \texttt{WHIZARD+OpenLoops} and \texttt{MG5\_aMC@NLO} and Born differential distributions obtained with \texttt{WHIZARD+OpenLoops} for the process $pp\to e^+\nu_e\mu^-\nu_{\mu}$ at $\sqrt{s}=13$ TeV}
	\end{figure}
	\begin{figure}[h]
	\centering
	\includegraphics[width=0.49\linewidth]{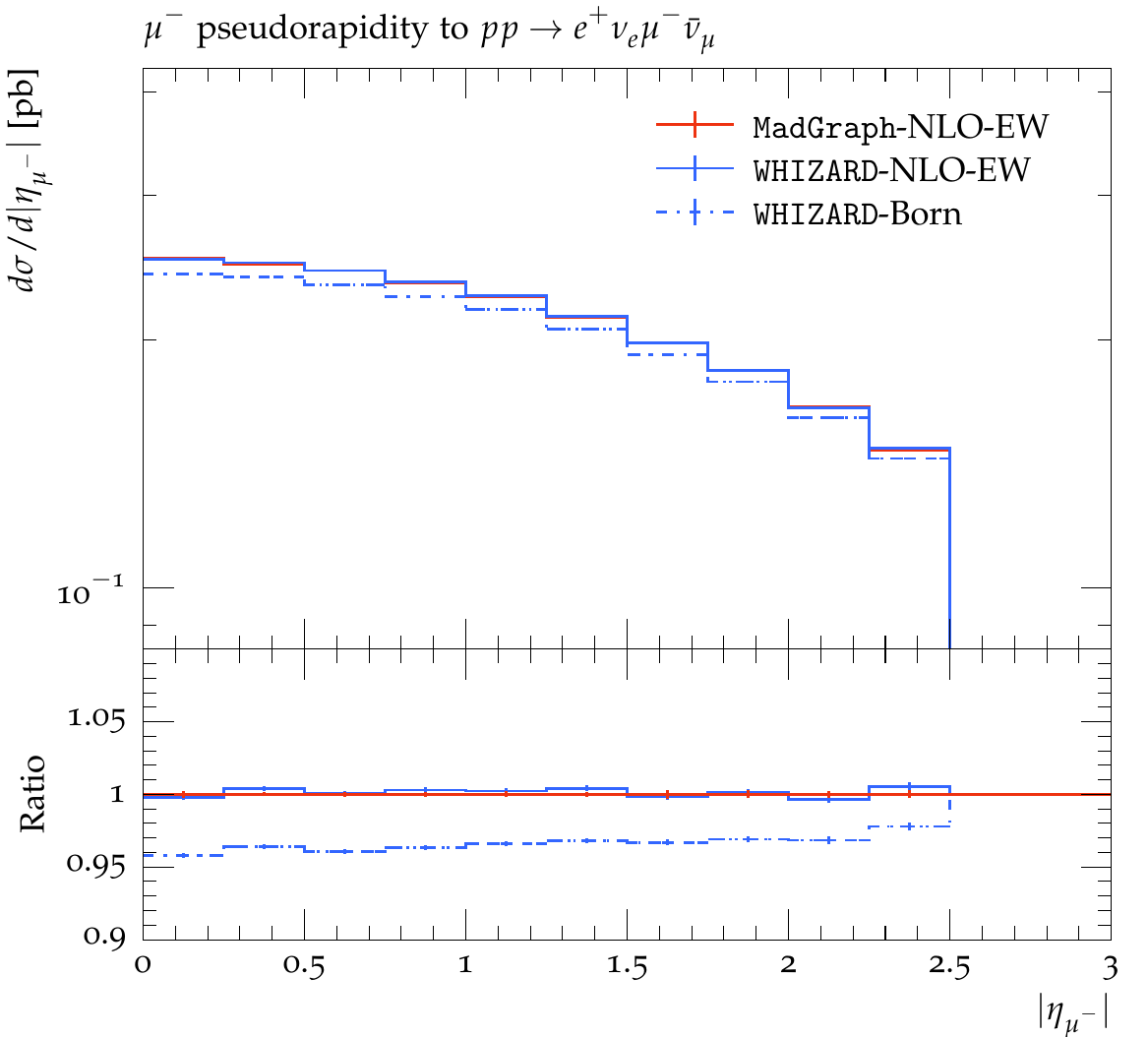}
	\includegraphics[width=0.49\linewidth]{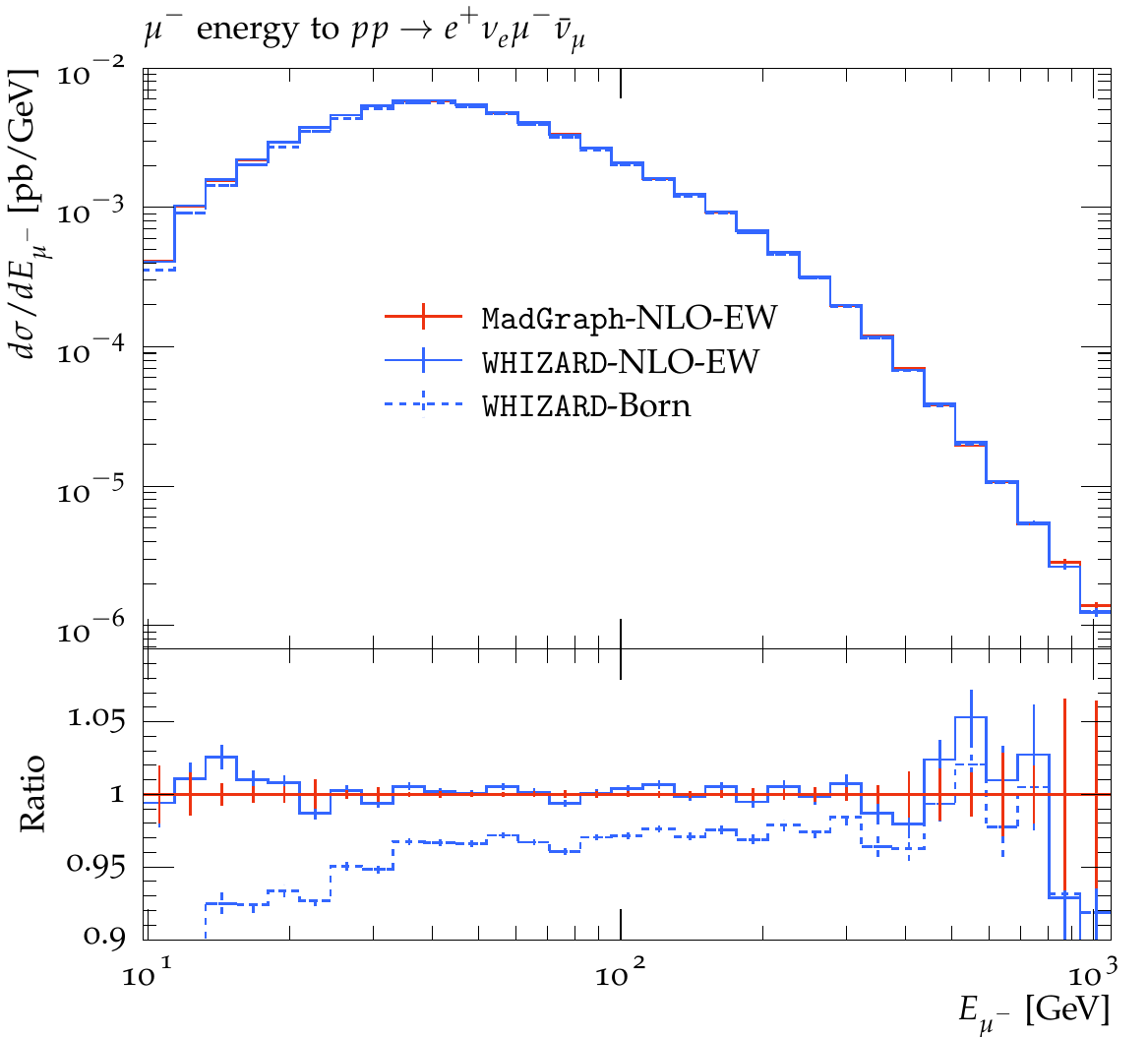}
	\caption{Checks on differential distributions at NLO EW of the muon pseudorapidity (left) and the muon energy (right) obtained with \texttt{WHIZARD+OpenLoops} and \texttt{MG5\_aMC@NLO} and Born differential distributions obtained with \texttt{WHIZARD+OpenLoops} for the process $pp\to e^+\nu_e\mu^-\nu_{\mu}$ at $\sqrt{s}=13$ TeV}
	\label{diffdistrMuon}
\end{figure}
	\section{Technical results on top-pair production}
	The cumulative results of the cross section contributions to $pp\to t\bar{t}$ for each coupling order shown in Table~\ref{ttbarmixed} can be further split into all existing PDF channels of the process. For each channel and coupling power combination the numerical result of the contribution at LO and NLO is checked with reference results of \texttt{MUNICH/MATRIX}. These checks yield agreement for each of these contributions within the MC statistics. For reference in Sec.~\ref{secToppairLHC} with respect to the main discussion of the mixed coupling expansion of this process, the numerical contributions for each PDF channel of $\text{LO}_{20}$, $\text{LO}_{11}$ and $\text{LO}_{02}$ are shown in Table~\ref{pdfchannelsLO} and of $\text{NLO}_{21}$, $\text{NLO}_{12}$ and $\text{NLO}_{03}$ in Table~\ref{pdfchannelsNLO}, respectively. In these tables, `$-$' denotes that the process is forbidden due to the coupling order and `$\times$' due to colour, respectively.
		\begin{table}[h]
		\centering
		\small
		{\onehalfspacing
		\begin{tabularx}{0.78\textwidth}{c|r|r|r}
			PDF channel & $\text{LO}_{20}$& $\text{LO}_{11}$
			 	 &$\text{LO}_{02}$ \\
			\hline\hline
			     $gg$  & $3.7587(1)\cdot 10^5$ & $-$     &    $  - $    \\
			    $q\bar{q}$   & $6.1337(1)\cdot 10^4$ &         $ \times $       &  $ 4.1006(5)\cdot 10^2 $   \\
			     $b\bar{b}$  & $7.5988(2)\cdot 10^2$ & $-1.41859(3)\cdot 10^3$   &    $ 2.3491(5) \cdot 10^3 $     \\
			$\gamma g$&$-$& $3.1913(6)\cdot 10^3$&$-$\\
			$\gamma\gamma$&$-$& $-$&$3.15354(5)\cdot 10^0 $\\
			\hline
			$\sum$ & $4.3797(1)\cdot 10^5$&$1.7727(6)\cdot 10^3$& $2.7623(5)\cdot 10^3$
		\end{tabularx}}
				\caption{Results of LO cross section contributions to $pp\to t\bar{t}$ at $\sqrt{s}=13$ TeV splitted into PDF channels and coupling powers; $q$ is defined without bottom quarks, $q\neq b$}
				\label{pdfchannelsLO}
\end{table}
	\begin{table}[h]
	\centering
	\small
	{\onehalfspacing
		\begin{tabularx}{0.78\textwidth}{c|r|r|r}
			PDF channel & $\delta\text{NLO}_{21}$& $\delta\text{NLO}_{12}$
			&$\delta\text{NLO}_{03}$ \\
			\hline\hline
			$gg$  & $-5.185(1)\cdot 10^3$ & $-$     &    $  - $    \\
			$gq/g\bar{q}$  & $5.095(3)\cdot 10^1$ & $2.1296(2)\cdot 10^3$     &    $  - $    \\
			
			$gb/g\bar{b}$  & $1.8036(4)\cdot 10^2$ & $-2.7506(7)\cdot 10^2$     &    $  - $    \\
			$q\bar{q}$   & $-2.752(1)\cdot 10^2$ &      $1.7312(3)\cdot 10^2$     &  $ 1.5498(6) \cdot 10^1  $   \\
			$b\bar{b}$  & $-2.0766(3)\cdot 10^2$ & $1.684(1)\cdot 10^2$   &    $-3.69(5)\cdot 10^{-1} $     \\
			$\gamma g$&$8.10(2)\cdot 10^2$& $2.0898(2)\cdot 10^2$&$-$\\
			
			$\gamma q/
			\gamma\bar{q}$  & $-6.910(2)\cdot 10^1$ & $\times$     &    $  5.5529(7)\cdot 10^0 $    \\
			
			$\gamma b/\gamma \bar{b}$  & $-3.7932(9)\cdot 10^0$ & $-3.805(8)\cdot 10^{-1}$     &    $  6.17(2)\cdot 10^{-2} $    \\
			$\gamma\gamma$&$-$& $6.421(4)\cdot 10^{-1}$&$1.352(2)\cdot 10^{-1} $\\
			\hline
			$\sum$ & $-4.700(2)\cdot 10^3$&$2.4053(3)\cdot 10^3$& $2.0879(8)\cdot 10^1$
	\end{tabularx}}
	\caption[pure EW onshell]{Results of NLO cross section contributions to $pp\to t\bar{t}$ at $\sqrt{s}=13$ TeV splitted into PDF channels and coupling powers; $q$ is defined without bottom quarks, $q\neq b$}
	\label{pdfchannelsNLO}
\end{table}

\printbibliography[heading=bibintoc, title={References}]
	\chapter*{\Large \underline{Eidesstattliche Versicherung / Declaration on oath}}
	\thispagestyle{empty}
	Hiermit versichere ich an Eides statt, die vorliegende Dissertationsschrift selbst verfasst und keine
	anderen als die angegebenen Hilfsmittel und Quellen benutzt zu haben.

	\vspace{2cm}

	\noindent Hamburg, den 17.10.2022 \hspace{3cm} \makebox[7cm]{\hrulefill}\\
	\mbox{}\hspace{8.5cm}{\small{Pia Mareen Bredt}}
	\chapter*{}
	\thispagestyle{empty}
	Ich versichere, dass dieses gebundene Exemplar der Dissertation und das in elektronischer Form
	eingereichte Dissertationsexemplar (\"uber den Docata-Upload) und das bei der Fakult\"at
	(zust\"andiges Studienbüro bzw. Promotionsb\"uro Physik) zur Archivierung eingereichte gedruckte
	gebundene Exemplar der Dissertationsschrift identisch sind.

	\vspace{2cm}

	\noindent Hamburg, den 17.10.2022 \hspace{3cm} \makebox[7cm]{\hrulefill}\\
	\mbox{}\hspace{8.5cm}{\small{Pia Mareen Bredt}}
\end{document}